\documentclass[10pt, twoside]{article}
\usepackage[paperwidth=170mm, paperheight=244mm]{geometry}
\usepackage{graphicx}
\usepackage{amsmath}
\usepackage{amsfonts}
\usepackage{amssymb}
\usepackage{amsthm}
\usepackage{bbm}
\usepackage{bm}
\usepackage{mathtools}
\usepackage{fancyhdr}
\usepackage{subfig}
\usepackage{lettrine}
\usepackage[explicit]{titlesec}
\usepackage{watermark}
\usepackage{color}
\usepackage[final]{pdfpages}
\usepackage{setspace}
\usepackage{chngcntr}
\usepackage{nameref}
\usepackage{enumitem}
\usepackage[perpage]{footmisc}
\usepackage{lscape}

\usepackage{wrapfig}
\usepackage{float}
\usepackage{cases}
\usepackage{stmaryrd}
\usepackage{tipa}
\usepackage{bbold}
\usepackage{bm}
\usepackage{color}
\usepackage{xcolor}
\usepackage{CJK}
\usepackage{physics}
\usepackage{array}
\usepackage{multirow}
\usepackage{mathtools}
\usepackage{comment}
\definecolor{gg}{RGB}{8, 135, 68}
\DeclareMathOperator{\diff}{d}

\newcommand{\subsubsubsection}[1]{\paragraph{#1}\mbox{}\\}
\usepackage[font=small]{caption} %,labelfont=bf]{caption}
\usepackage[colorlinks, citecolor=blue, linkcolor=black, urlcolor=blue]{hyperref}

\usepackage[T1]{fontenc}
\usepackage{lmodern}

\numberwithin{equation}{section}
\numberwithin{figure}{section}
\numberwithin{table}{section}

\definecolor{grey}{rgb}{0.5,0.5,0.5}
\definecolor{l-grey}{rgb}{0.8,0.8,0.8}
\definecolor{white}{rgb}{1,1,1}
\definecolor{black}{rgb}{0,0,0}
\definecolor{myred}{rgb}{0.831, 0.165, 0.184}
\definecolor{myblue}{rgb}{0.149, 0.471, 0.698}
\definecolor{mypurple}{rgb}{0.443, 0.165, 0.439}

\let\origdoublepage\cleardoublepage
\newcommand{\clearemptydoublepage}{%
  \clearpage
  {\pagestyle{empty}\origdoublepage}%
}

\let\cleardoublepage\clearemptydoublepage

\titleformat{\section}[display]{\vspace*{190pt} \bfseries\sffamily \Huge}
{\begin{picture}(0,0)\put(-60,-30){\textcolor{grey}{\thesection}}\end{picture}}
{0pt}
{#1}
[]
\titlespacing*{\section}{40pt}{10pt}{40pt}[40pt]

\titlespacing*{\subsection}{0pt}{30pt}{20pt}[0pt]
\titleformat{\subsection}[display]{\Large \sffamily}{}{0pt}{\thesubsection \ #1}[]

\begin{document}

\pagestyle{fancy}
\renewcommand{\headrulewidth}{0pt}
\fancyhead{}
\fancyfoot{}

%%%%%%%%%%%%%%%%%%%%%%%%
% THE COVER
%%%%%%%%%%%%%%%%%%%%%%%%

%\includepdf[fitpaper=true, offset= 70 -70]{img/Cover.pdf}
%\includepdf[fitpaper=true, offset= 70 -70]{img/basic_microwaving.pdf}

%%%%%%%%%%%%%%%%%%%%%%%%
% HALF-TITLE PAGE
%%%%%%%%%%%%%%%%%%%%%%%%
%\vspace*{\fill}

% Aqui va una página en blanco donde pone el título, nada más
%\begin{center}
%\linespread{1.7}{\huge Wireless Microwave \\  Quantum Communication \par}
%\end{center}

%\vspace*{\fill}

\setcounter{page}{1}
\clearemptydoublepage

%%%%%%%%%%%%%%%%%%%%%%%%
% TITLE PAGE
%%%%%%%%%%%%%%%%%%%%%%%%

\begin{center}

\hrule

\vspace{16pt}
{\huge Wireless Microwave \\  Quantum Communication \par}
\vspace{16pt}

\hrule

\vspace{35pt}

{\Large {\bf Tasio Gonz\'{a}lez Raya } }

\vspace{35pt}

\emph{Supervised by} \\

\vspace{15pt}

{\large

Dr. Mikel Sanz\\ \vspace{8pt} %and\\ \vspace{8pt} Prof. \'{I}\~{n}igo Egusquiza

}

%\vspace{50pt}
\vspace*{\fill}

\includegraphics[height=2.5cm]{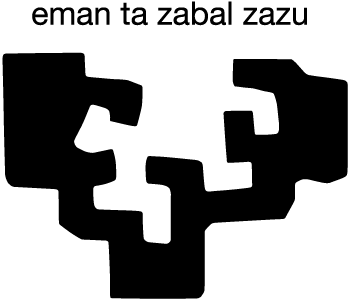}

\vspace{20pt}

Departamento de Qu\'imica F\'isica\\
%y\\
%\vspace{4pt}
%Departamento de F\'isica Te\'orica e Historia de la Ciencia \\
\vspace{4pt}
Facultad de Ciencia y Tecnolog\'ia\\
\vspace{4pt}
Universidad del Pa\'is Vasco

\vspace{15pt}

{\large April 2023}

\end{center}

\pagebreak
%%%%%%%%%%%%%%%%%%%%%%%%
% COPYRIGHTS PAGE
%%%%%%%%%%%%%%%%%%%%%%%%

%This document is a PhD thesis developed during the period from October 2013 to June 2017 at QUTIS (Quantum Technologies for Information Science) group, led by Prof. Enrique Solano. This work was funded by the University of the Basque Country with a PhD fellowship.

This document is a PhD thesis developed during the period from October 2018 to April 2023 at QUTIS (Quantum Technologies for Information Science) group, led by Prof. Enrique Solano, then transformed into NQUIRE center, whose ``Quantum Computing and Architectures'' group is led by Dr. Mikel Sanz. This work was funded by the University of the Basque Country with a PIF contract.

\vspace*{250pt}

\noindent \textcopyright 2023 by Tasio Gonz\'{a}lez Raya. All rights reserved.

\bigskip

\noindent An electronic version of this Thesis can be found at \href{http://www.nquirephysics.com}{www.nquirephysics.com}

\bigskip

\noindent Bilbao, April 2023

\vspace*{\fill}

{\setstretch{1.2}

\noindent This document was generated with the 2022  \LaTeX  \ distribution.

\noindent The \LaTeX \  template is adapted from a \href{https://github.com/iagobaapellaniz/PhD_Thesis}{template} by Iagoba Apellaniz.

\noindent The bibliographic style was created by Sof\'ia Martinez Garaot.

}

\cleardoublepage

%%%%%%%%%%%%%%%%%%%%%%%%
% DEDICATION PAGE
%%%%%%%%%%%%%%%%%%%%%%%%

\vspace*{80pt}
\begin{flushright}
\emph{To my family}
\end{flushright}

\cleardoublepage

%%%%%%%%%%%%%%%%%%%%%%%%
% EPIGRAPH PAGE
%%%%%%%%%%%%%%%%%%%%%%%%

\vspace*{150pt}
\hspace*{\fill}\begin{minipage}{\textwidth-90pt}
%\emph{Ya est\'{a} incorporado a una comunidad de la que, a pesar de todo, forma parte y de la que no podr\'{a} deshacerse con facilidad. Al entrar all\'{i}, la ciudad -- con una de sus conciencias m\'{a}s agudas -- de \'{e}l ha tomado nota: existe.}
%\emph{Cuando las gaviotas siguen el pesquero es porque creen que se echarán las sardinas al mar.}
%\emph{We make our world significant by the courage of our questions and the depth of our answers.}
%\emph{No impondr\'{i}a mis propios asuntos a la atenci\'{o}n de los lectores si no hubiera recibido muchas preguntas y muy concretas por parte de mis conciudadanos en relaci\'{o}n a mi modo de vivir.}
\emph{I should not obtrude my affairs so much on the notice of my readers if very particular inquiries had not been made by my townsmen concerning my mode of life.}
\end{minipage}
\\
%\end{center}
\vspace*{10pt}
\begin{flushright}
%{\setstretch{2} Luis Mart\'{i}n-Santos, Tiempo de silencio}
%{\setstretch{2} Eric Cantona}
%{\setstretch{2} Carl Sagan, Cosmos}
{\setstretch{2} Henry David Thoreau, Walden}
\end{flushright}
%\end{center}

%%%%%%%%%%%%%%%%%%%%%%%%
% TABLE OF CONTENTS
%%%%%%%%%%%%%%%%%%%%%%%%

\titleformat{\section}[display]
{\vspace*{160pt}
\bf\Huge}
{{\textcolor{black}{\thesection}}. #1}
{0pt}
{#1}
[]
\titlespacing*{\section}{75pt}{10pt}{40pt}[40pt]

\textsf{\tableofcontents}

%%%%%%%%%%%%%%%%%%%%%%%%
% ABSTRACT
%%%%%%%%%%%%%%%%%%%%%%%%

\section*{Abstract}
\pagenumbering{roman}
\fancyfoot[LE,RO]{\thepage}
\phantomsection
\addcontentsline{toc}{section}{Abstract}

%In the quest for a global quantum communication network, signals in the optical regime have been established as the main resource, for both fibre and free-space links. Nevertheless, microwaves enjoy a higher atmospheric transparency, and are therefore used for current telecommunications. This is also the natural talking frequency for superconducting circuits, one of the most promising quantum technology platforms.

This Thesis explores the limits in the application of propagating quantum microwaves for quantum communication and quantum sensing, as well as the design of new devices and protocols to fight these limitations. We take advantage of Gaussian quantum states for quantum teleportation and quantum illumination, and studies how these protocols can be improved using entanglement distillation and partial purification, respectively. The Thesis is centered around open-air entanglement distribution, and it follows the steps of state generation inside the cryostat, impedance matching between the cryostat and the open air with a new generation of coplanar antennae, and open air propagation, in the limited framework of current microwave technology. We also address the limitations produced by losses and measurement inefficiencies, and explore the extension to satellite quantum communications. There, we analyze the effects of diffraction and turbulence, studying how the latter affects signals in the optical regime as well. We conclude by studying the teleportation of quantum information in a quantum local area network. To sum up, this Thesis contributes to the development of wireless quantum communications in the microwave regime, studying its technological limitations and how to overcome them. Nevertheless, quantum technologies working in this frequency range are still emergent and plenty of work must be accomplished in order to make them competitive.

\cleardoublepage

%%%%%%%%%%%%%%%%%%%%%%%%
% RESUMEN
%%%%%%%%%%%%%%%%%%%%%%%%

\section*{Resumen}
\fancyfoot[LE,RO]{\thepage}
\phantomsection
\addcontentsline{toc}{section}{Resumen}

Las leyes de la mecánica cuántica, necesarias para una descripción precisa de la naturaleza a nivel microscópico, fueron postuladas al comienzo del siglo pasado. En los inicios, aun así, había controversia; de ahí el famoso trabajo de Einstein, Podolsky y Rosen en 1935, afirmando que ninguna teoría que cumpliera el realismo local podría ser completa, y necesitaría apoyarse en variables (clásicas) adicionales, lo que se conoce como el modelo de variables locales ocultas. Este más bien filosófico escollo, conocido como la paradoja EPR, fue solventado por John S. Bell en 1964, cuando diseñó una serie de experimentos, los cuales indicaban que las predicciones hechas en el marco de la mecánica cuántica eran incompatibles con un modelo de variables ocultas subyacente satisfaciendo requerimientos de localidad. A esto se le llamó el teorema de Bell, e impone restricciones a los resultados de las medidas hechas localmente entre dos partículas que estén correlacionadas, dadas unas variables locales ocultas, mostrando que la mecánica cuántica predice la violación de estas restricciones. Una de las desigualdades de Bell más famosas es la desigualdad CHSH, la cual expuso la posiblidad de una realización experimental de un test de Bell. tras algunos experimentos fallidos en los años 70 llegó la propuesta y los experimentos de Aspect, la primera prueba de la no-separabilidad de la mecánica cuántica. A estos les sucedieron muchos experimentos, la mayoría de los cuales utilizaron medidas con dos resultados, en las cuales los fotones son tratados como sistemas de dos niveles, con dos estados de polarización o de número de fotones. Estos son conocidos como estados de {\it qubit}, la unidad de información codificada en un sistema cuántico, análogo al bit clásico. 

Un qubit está caracterizado por un vector $|\psi\rangle$ en un espacio de Hilbert bidimensional $\mathcal{H}$, cuyos vectores de la base se suelen indicar por $|0\rangle$ y $|1\rangle$, generalmente denominados vectores de la base computacional. Mientras que un estado puro clásico puede estar en ``0'' o ``1'', un estado cuántico puro puede estar en una superposición de ambos,
\begin{equation}
|\psi\rangle = a|0\rangle + b|1\rangle \quad \text{ con } \quad \{a, b\}\in\mathbb{C} \quad \text{ y } \quad | a|^{2}+|b|^{2} = 1.
\end{equation}
Cada elemento de la base todavía corresponde al clásico bit 0 o 1 a través de la medida, pero al medir el estado $|\psi\rangle$, estos valores se distribuirán de acuerdo con los pesos $a$ y $b$, de modo que la probabilidad de medir $|\psi\rangle$ en el estado $|0\rangle$ viene dada por la amplitud $|\langle\psi|0\rangle|^{2}=|a|^{2}$. Esta propiedad de superposición se puede extender a un escenario de $N$-qubits, con un espacio de Hilbert $2^{N}$-dimensional. Otra propiedad es el entrelazamiento, la ``acción espeluznante a distancia'' en la paradoja EPR, y la manifestación más común de las correlaciones cuánticas que no se pueden explicar de forma clásica. Surge en un escenario de varios qubits, donde el estado global del sistema no se puede describir en términos de los estados locales; se puede cuantificar, por ejemplo, a través de la entropía del sistema reducido. Dado un estado máximamente entrelazado, como el estado de Bell $( |0,0\rangle+|1,1\rangle)/\sqrt{2}$, tomar la traza parcial sobre el segundo subsistema conduce al estado $(|0 \rangle\langle 0| + |1\rangle\langle 1|)/2$. Este es un estado máximamente mezclado y tiene máxima entropía, lo que significa que la información contenida en las correlaciones cuánticas se ha perdido y, por lo tanto, no hay ninguna medida que podamos realizar para extraer la información completa contenida en él. Los estados mixtos, en general, se describen mediante una matriz de densidad $\rho$ que es positiva y normalizada, $\tr\rho=1$. Además, las matrices de densidad también pueden describir estados puros, que satisfacen $\tr\rho^{2}=1$, mientras que los estados mixtos siguen $\tr\rho^{2}<1$.

En 1984, hubo una propuesta para utilizar la superposición de estados cuánticos como recurso para transferir información de forma remota. En lo que llegó a conocerse como el protocolo BB84, se afirma que dos partes pueden desarrollar una clave cuántica segura compartiendo estados cuánticos e información clásica. El emisor genera estados ya sea sobre una base computacional o usando una base de superposición, y el receptor mide con la misma elección de base; luego, usan un canal clásico para comunicar su elección de base y mantienen los bits clásicos correspondientes a eventos coincidentes. Además, pueden detectar la presencia de un intruso anunciando públicamente parte de la cadena de bits obtenida; si coincide una cantidad suficiente de elementos, entonces pueden conservarlo para desarrollar una clave segura, y si no, pueden descartarlo y comenzar de nuevo. Aunque no está directamente relacionado con el teorema de Bell, la seguridad del protocolo BB84 contra ataques individuales está relacionada con la desigualdad CHSH. Otra propuesta que se inspiró en el teorema de Bell es el protocolo E91. A su vez, este protocolo tiene a ambas partes compartiendo estados entrelazados y midiendo con un conjunto de bases que no coincide completamente. De esta manera, pueden mantener los bits clásicos que resultan de la medición en la misma base, siempre que los otros resultados de la medición pasen una prueba de realismo local para verificar si hay intrusos. La seguridad de estos protocolos se basa en los postulados de la mecánica cuántica; en virtud del teorema de no clonación y el colapso del estado de un sistema cuántico bajo medición, un espía no puede extraer información sin afectar el proceso y dejar un rastro. Estas dos propuestas allanaron el camino para lo que hoy se conoce como distribución de claves cuánticas. Se han llevado a cabo experimentos, desde la realización de la primera criptografía cuántica, usando fibras ópticas, más adelante por el aire, y finalmente en un satélite a larga distancia.

Poco después del BB84, se propuso un avance crucial en la comunicación cuántica: la teleportación cuántica. Este protocolo tiene como objetivo transferir la información de un estado cuántico desconocido en poder de una parte, a una segunda en una ubicación remota, por medio de un recurso entrelazado previamente compartido, y comunicación clásica. También se ha realizado experimentalmente en numerosas ocasiones: con sistemas fotónicos en el laboratorio, a través de fibras ópticas, por el aire, y en un enlace satelital, así como en variedad de plataformas cuánticas: con resonancia magnética nuclear, con iones atrapados, con circuitos superconductores, e incluso entre objetos macroscópicos. La ventaja que se puede obtener mediante la teletransportación cuántica se basa en la existencia de un entrelazamiento previamente compartido entre ambas partes, al igual que para los protocolos de distribución de claves cuánticas mediados por entrelazamiento. El acto de compartir estados entrelazados entre las partes de la comunicación se conoce como distribución de entrelazamiento, y se ha logrado experimentalmente con fibras ópticas, así como por el aire. Este también es un punto clave para la famosa iniciativa del Internet cuántico.

El entrelazamiento se puede codificar en muchos grados diferentes de libertad de los sistemas cuánticos; los experimentos que hemos mencionado anteriormente utilizan el número de fotones, la polarización y el entrelazamiento de intervalos de tiempo, entre otros. Aparte de los estados cuánticos de variable discreta, el entrelazamiento también se puede definir usando estados bosónicos. Estos estados describen espacios de Hilbert de dimensión infinita y sus operadores de cuadratura tienen un espectro continuo. Los sistemas asociados a espacios de Hilbert de dimensión infinita se conocen como sistemas de "variable continua" y tienen una descripción cuántica particularmente complicada.

Los estados cuánticos Gaussianos son una familia de estados de variable continua que admiten una descripción simple; pueden describirse mediante distribuciones Gaussianas en su representación en el espacio de fases. Por lo general, son fáciles de producir experimentalmente y se pueden usar para describir el estado de los sistemas cuánticos entrelazados. Por lo tanto, sus capacidades de procesamiento de información cuántica han sido ampliamente estudiadas. Además, cualquier evolución cuántica que involucre estados Gaussianos, operaciones Gaussianas y medidas Gaussianas, admite una representación compacta conocida como formalismo simpléctico. Esto permite reemplazar vectores de estado de dimensión infinita y matrices de operadores de un sistema de $N$ modos por un vector $2N$-dimensional y una matriz $2N\times2N$, el vector de desplazamiento y la matriz de covarianza, respectivamente, que pueden caracterizar completamente una evolución cuántica Gaussiana. Con la matriz de covarianza, también podemos calcular las características de estos estados, como la pureza y el entrelazamiento.

A pesar de las múltiples ventajas, el campo de la información cuántica Gaussiana presenta algunas limitaciones; por ejemplo, la imposibilidad de destilar el entrelazamiento o de realizar corrección de errores cuántica con operaciones Gaussianas y medidas Gaussianas. Sin embargo, muchos protocolos de destilación de entrelazamiento con operaciones no Gaussianas se han estudiado en variable continua. De manera similar, la corrección de errores cuántica con variable continua se ve obligada a abandonar el ámbito de los estados Gaussianos. Ejemplos de estados cuánticos Gaussianos incluyen estados coherentes, estados térmicos y estados squeezed, entre otros. El caso paradigmático de estados cuánticos Gaussianos entrelazados bipartitos son los estados squeezed de dos modos, que también se pueden usar para la teleportación cuántica con variable continua.

El formalismo de variable continua se usa frecuentemente para la comunicación cuántica, y especialmente para la teleportación cuántica; de hecho, solo un año después de que apareciera el primer artículo sobre teleportación cuántica, le siguió una versión en variable continua. Luego fue reemplazado por una propuesta más realista, el famoso protocolo de teleportación cuántica de Braunstein-Kimble, seguido de la primera realización experimental. Naturalmente, surgieron otros trabajos a partir de entonces que discutían mejoras en el protocolo y el experimento. También ha habido avances en la distribución de entrelazamiento en variable continua, con experimentos, así como en la distribución de claves cuánticas en este formalismo.

La mayoría de los experimentos de comunicación cuántica utilizan fotones en el rango óptico, principalmente debido a los leves efectos de difracción y al tenue ruido térmico. Sin embargo, en este rango hay muchas fuentes de error e ineficiencia: grandes pérdidas por absorción en el aire libre y el elevado consumo de energía, por nombrar algunos. Al mismo tiempo, las plataformas cuánticas actuales más prometedoras, los circuitos superconductores, los centros de vacantes de nitrógeno o los iones atrapados, funcionan en el régimen de microondas o utilizan señales de microondas. Por lo tanto, para establecer un canal de comunicación cuántica entre unidades de procesamiento basadas en estas tecnologías, se requiere convertir fotones de microondas a óptico o usar señales cuánticas de microondas directamente. El primer enfoque todavía sufre de enormes ineficiencias cuánticas de conversión del orden de $10^{-5}$. En esta Tesis, consideramos el enfoque de comunicación cuántica puramente de microondas, sus ventajas y limitaciones.

Las limitaciones en la capacidad de transporte de información cuántica y la universalidad de los estados cuánticos Gaussianos se derivan de su descripción simple y su fácil generación experimental, y esto limita naturalmente el rendimiento de los protocolos de comunicación cuántica. Otro factor limitante es la imposibilidad de generar estados a temperatura ambiente con microondas cuánticas. Los dispositivos de microondas que funcionan en frecuencias de $1$--$100$ GHz están contaminados con fotones térmicos a temperatura ambiente; este número es de 1250 fotones medios para 5 GHz a temperatura ambiente ($T=300$ K). Esta es una de las principales limitaciones y crea la necesidad de enfriamiento criogénico en los circuitos superconductores, para protegerlos del ruido térmico.

Los dispositivos superconductores de microondas de última generación incluyen el amplificador paramétrico de Josephson, el transistor de electrones de alta movilidad, el conversor paramétrico de Josephson y el circulador, entre otros. Sin embargo, uno de los dispositivos superconductores de microondas más relevantes, y del cual se derivan los amplificadores y los convertidores paramétricos de Josephson, es la unión de Josephson. Este elemento no lineal tiene aplicaciones esenciales en computación cuántica y en el procesamiento de información cuántica, y su desarrollo ha dado lugar a diferentes experimentos de transferencia de estados cuánticos y de preparación remota de entrelazamiento entre varios dispositivos superconductores basados en uniones de Josephson, así como a análisis de ruido. Otra aplicación interesante de este dispositivo es el amplificador paramétrico de Josephson, que puede generar estados squeezed; estos pueden usarse para producir estados entrelazados para la comunicación cuántica de microondas. En criogenia, ha habido varias realizaciones de distribución de entrelazamiento de microondas y de teleportación cuántica. Una propuesta de teleportación cuántica con microondas cuánticas propagantes fue seguida por un experimento reciente, realizado dentro de un criostato. En esta Tesis, intentamos construir un modelo realista para la distribución de entrelazamiento por el aire libre y la teleportación cuántica con microondas para estudiar formalmente los límites de este protocolo. Este modelo debe tener en cuenta los desafíos asociados con las tecnologías cuánticas de microondas, así como los que enfrentan la comunicación cuántica y los estados cuánticos Gaussianos.

Experimentos recientes en el aire han fallado en la preservación eficiente del entrelazamiento al usar antenas comerciales, en parte porque la amplificación ``clásica'' de las señales cuánticas únicamente puede perjudicar a las correlaciones cuánticas. Sin embargo, el aspecto de ``matching'' de impedancias de las antenas clásicas debe imitarse para reducir las reflexiones en las señales que viajan desde el criostato al aire libre. A partir de ahí, el mecanismo de pérdida en el entorno consiste en la absorción de fotones de señal, la termalización de la señal y su difracción, y esto puede superarse mediante técnicas de destilación de entrelazamiento, alejándose del entorno Gaussiano. Protocolos como el de intercambio de entrelazamiento también pueden ser beneficiosos para este tipo de procesos, aunque los avances actuales en este tema carecen de eficiencia y presentan ciertas barreras tecnológicas.

En el camino hacia una red de comunicación cuántica global, la distribución de entrelazamiento y la teleportación cuántica entre satélites representa un alivio de la atenuación atmosférica y el ruido térmico, donde las comunicaciones a través de enlaces tierra-satélite descendentes o ascendentes pueden presentar el mayor desafío. Mientras tanto, la mayoría de los avances en este ámbito se inclinan hacia las aplicaciones en distribución de claves cuánticas. Dejando a un lado la sobrecarga tecnológica, las mejoras pasarán por la comprensión de los diferentes mecanismos de pérdida en el espacio libre, a saber, la difracción, la atenuación atmosférica e incluso los efectos de las turbulencias. Derivadas de pequeñas variaciones de temperatura y presión en el interior de la atmósfera, las turbulencias afectan a las señales en el rango óptico, mientras que las microondas, debido a sus grandes longitudes de onda, son insensibles a ellas. Estos efectos han sido bien estudiados para señales clásicas en el régimen óptico. En el mismo rango de frecuencias, algunos trabajos recientes han estudiado las turbulencias en los canales de transmisión atmosféricos cuánticos, estableciendo es posible preservar la no-clasicalidad de las señales. También se publicaron artículos perspicaces sobre los efectos de la propagación de señales cuánticas en el espacio libre. Mientras tanto, otros se centraron en los límites para la generación de claves y la distribución de entrelazamiento entre estaciones terrestres y entre estaciones terrestres y satélites.

A diferencia de las redes de comunicación global, las redes de área local normalmente requieren una conexión inalámbrica entre diferentes unidades. Los avances en la conexión de estos procesadores con la información cuántica no solo son relevantes para las comunicaciones cuánticas, sino que también pueden encontrar aplicaciones en la computación cuántica. Dadas las limitaciones que presentan los procesadores cuánticos actuales, que caracterizan la era NISQ (siglas en inglés para elementos cuánticos ruidosos de escala intermedia), el enfoque de la computación cuántica distribuida podría reducir el ruido y permitir cálculos más eficientes. Por lo tanto, es interesante explorar protocolos de teleportación cuántica para comunicar múltiples estados de qubit entre diferentes procesadores.

Aunque la tecnología cuántica de microondas está un par de décadas por detrás de la óptica, tiene un futuro brillante por delante. Sería natural predecir un período de coexistencia entre los dos regímenes; si bien la transducción de señales de microondas al rango óptico aún no es eficiente, las comunicaciones ópticas han demostrado ser la opción correcta para largas distancias. Por otro lado, un protocolo de microondas puede funcionar en una red de área local cuántica; una implementación de distribución de claves cuánticas en un entorno de este tipo utilizando enlaces criogénicos representaría un hito importante para la comunicación cuántica de microondas. Si bien los avances recientes en el conteo de fotones probablemente conducirán a nuevos experimentos de iluminación cuántica en criogenia con microondas, las aplicaciones al radar cuántico de microondas están aún fuera de nuestro alcance.

A corto plazo, el enfoque principal debe estar en los experimentos dentro de un criostato, porque las realizaciones al aire libre de la comunicación cuántica y la iluminación cuántica con microondas aún están fuera de nuestro alcance. Una de las principales razones es la falta de colimadores para reducir la difracción, pero esto también se puede mitigar con repetidores cuánticos. El diseño de una antena receptora es otro paso importante a dar, junto con la implementación de antenas tanto emisoras como receptoras. Por último, pero no menos importante, es crucial mejorar la generación de enredos implementando mayores ganancias en los amplificadores paramétricos, mientras se reduce el ruido. Sin embargo, a medida que los circuitos superconductores continúen consolidándose y expandiéndose, la comunicación y detección cuántica de microondas seguirá creciendo; las redes inalámbricas clásicas de microondas estarán ahí esperando.

%\subsection{What you will find in this Thesis}

El camino hacia una red de comunicación cuántica universal pasa por comprender las limitaciones de una extensión de los paradigmas de comunicación clásicos al ámbito cuántico. Mientras que las conexiones por cable se suelen realizar con señales en rango óptico, las microondas se utilizan para enlaces en el aire libre. Por lo tanto, estudiamos la distribución de microondas cuánticas a través del aire libre; estudiamos los estados cuánticos Gaussianos, cómo se generan, cómo se lanzan al aire y cómo se degradan en este ambiente. A diferencia de las señales clásicas, donde el principal recurso es la potencia, nos centramos en cómo se comporta el entrelazamiento, su relación con otras características del estado, como la pureza, y en cómo se puede incrementar para mitigar los efectos del ruido y del entorno. Esta Tesis está estructurada en seis capítulos, además de esta introducción y un capítulo final, y está dedicada al estudio de las diferentes piezas que deben unirse para la comunicación cuántica y la detección cuántica por el aire libre.

Comenzamos explorando las propiedades de los estados Gaussianos, una familia de estados cuánticos en variable continua que se utilizan habitualmente en la comunicación cuántica, ya que son fáciles de producir experimentalmente. Revisamos diferentes características de estos estados que son relevantes desde la perspectiva de la información cuántica, y caracterizamos los elementos de una evolución completamente Gaussiana: los canales cuánticos Gaussianos y las medidas cuánticas Gaussianas. Después, revisamos el protocolo de teleportación cuántica de Braunstein-Kimble, un hito de la comunicación cuántica Gaussiana, y exploramos las técnicas de destilación de entrelazamiento y de intercambio de entrelazamiento para estados Gaussianos bipartitos generales, que pueden mejorar la fidelidad de este protocolo. Continuamos estudiando la purificación de estados Gaussianos usando operaciones Gaussianas y encontramos la imposibilidad de purificación completa de un solo modo de un estado Gaussiano entrelazado de dos modos sin una degradación completa del entrelazamiento. Por lo tanto, nos enfocamos en aumentar la pureza, mientras reducimos el entrelazamiento, y usamos los estados resultantes para un protocolo de iluminación cuántica. Los estados resultantes de las técnicas de purificación parcial de una sola copia y de dos copias muestran una mayor información cuántica de Fisher que los originales, y tienen un número promedio de fotones más bajo, lo que se traduce en una mayor precisión para la iluminación cuántica.

Más adelante, revisamos algunos de los avances recientes en dispositivos cuánticos superconductores, contextualizando sus implicaciones en la comunicación cuántica de microondas. Estudiamos los amplificadores paramétricos de Josephson y el papel que juegan en la generación de estados cuánticos entrelazados, esbozando un modelo de ruido para estados Gaussianos. Luego presentamos el diseño de una antena cuántica para señales entrelazadas de microondas que se propagan desde un criostato, donde se generan estados cuánticos para reducir los efectos térmicos, al aire libre. Este dispositivo tiene como objetivo reducir las reflexiones implementando el ``matching'' de impedancias entre los diferentes medios y maximizando la preservación del entrelazamiento.

Después de analizar la generación de estados y la propagación eficiente fuera del criostato, presentamos un mecanismo de pérdidas en el aire libre compuesto por pérdidas por absorción y termalización, y estudiamos los límites de la distribución del entrelazamiento por el aire libre, calculando el alcance del mismo. Luego presentamos protocolos de intercambio de entrelazamiento y de destilación de entrelazamiento para luchar contra la degradación ambiental, empleando los estados resultantes como recursos para la teleportación cuántica. Como están involucrados en los protocolos discutidos en este capítulo, aquí hablamos también del conteo de fotones y de la detección ``homodyne'' con microondas, y estudiamos el efecto de errores e ineficiencias en estas operaciones.

Luego damos el salto a estudiar los límites de la comunicación cuántica de microondas entre satélites en el espacio, reemplazando la absorción por la difracción como el principal mecanismo de pérdidas. Establecemos las condiciones para la preservación del entrelazamiento relacionando la distancia y el tamaño de las antenas, similar a la categorización del espacio libre. Más adelante, nos centramos en el rango óptico, en el que se han realizado la mayoría de los avances y experimentos. Allí, exploramos los efectos de la difracción, la absorción atmosférica, las ineficiencias del detector y las turbulencias en diferentes escenarios de comunicación cuántica: tierra a tierra, tierra a satélite (enlace ascendente), satélite a tierra (enlace descendente) y satélite a satélite. Se proporciona también una comparación entre ambos regímenes de frecuencias.

Para concluir, investigamos la transmisión de información en forma de estados de qubit entre diferentes procesadores cuánticos, utilizando recursos entrelazados de variable continua, en un entorno de computación cuántica distribuida. Investigamos la ubicación de los estados cuánticos puros en la esfera de Bloch y calculamos las fidelidades, promediando sobre qubits uniformemente distribuidos. Comparamos la teleportación cuántica de Braunstein-Kimble, usando un estado de variable continua con un enfoque híbrido, usando el mismo estado, pero aplicando el protocolo de teleportación característico del formalismo de variable discreta. Estudiamos las pérdidas en el recurso entrelazado para estos dos casos, así como para un estado de Bell con teleportación de variable discreta. Para concluir, investigamos la teleportación de un estado de dos qubits arbitrario, utilizando una pareja de estados entrelazados de dos modos.

\cleardoublepage

%%%%%%%%%%%%%%%%%%%%%%%%
% AKNOWLEDGMENTS
%%%%%%%%%%%%%%%%%%%%%%%%

\section*{Acknowledgements}
\phantomsection
\addcontentsline{toc}{section}{Acknowledgements}

To Laura, to my family and to my friends. To tío Juan and tío Lorenzo. To my flatmates, to my coworkers, and to my collaborators. To all the people who has helped and supported me, and to those who had to bear with me. Thank you. 

\cleardoublepage

%%%%%%%%%%%%%%%%%%%%%%%%
% FIGURES AND PUBLICATIONS
%%%%%%%%%%%%%%%%%%%%%%%%

%---------Publications---------------------------------------
\section*{List of Publications}
\phantomsection
\addcontentsline{toc}{section}{List of Publications}

This Thesis is based on the following publications and preprints:
\\

%{\bf Chapter 2: \nameref{sec2}}

%\begin{enumerate}

%\end{enumerate}

%{\bf Chapter 3: \nameref{sec3}}

%\begin{enumerate}[resume]
\begin{enumerate}

\item {\bf T. Gonzalez-Raya} and M. Sanz, \\
{\it Coplanar Antenna Design for Microwave Entangled Signals Propagating in Open Air},\\
\href{https://doi.org/10.22331/q-2022-08-23-783}{Quantum {\bf 6}, 783 (2022).}

%\end{enumerate}

%{\bf Chapter 4: \nameref{sec4}}

%\begin{enumerate}[resume]

\item {\bf T. Gonzalez-Raya}, M. Casariego, F. Fesquet, M. Renger, V. Salari, M. M\"{o}tt\"{o}nen, Y. Omar, F. Deppe, K. G. Fedorov, and M. Sanz, \\
{\it Open-Air Microwave Entanglement Distribution for Quantum Teleportation},\\
\href{https://doi.org/10.1103/PhysRevApplied.18.044002}{Physical Review Applied {\bf 18}, 044002 (2022).}

%\end{enumerate}

%{\bf Chapter 5: \nameref{sec5}}

%\begin{enumerate}[resume]

%\end{enumerate}

%{\bf Chapter 7: \nameref{sec7}}

%\begin{enumerate}[resume]

\item M. Casariego, E. Z. Cruzeiro, S. Gherardini, {\bf T. Gonzalez-Raya}, R. Andr\'{e}, G. Fraz\~{a}o, G. Catto, M. M\"{o}tt\"{o}nen, D. Datta, K. Viisanen, J. Govenius, M. Prunnila, K. Tuominen, M. Reichert, M. Renger, K. G. Fedorov, F. Deppe, H. van der Vliet, A. J. Matthews, Y. Fern\'{a}ndez, R. Assouly, R. Dassonneville, B. Huard, M. Sanz, Y. Omar, \\
{\it Propagating Quantum Microwaves: Towards Applications in Communication and Sensing},\\
\href{https://doi.org/10.1088/2058-9565/acc4af}{Quantum Sci. Technol. {\bf 8}, 023001 (2023).}

\item {\bf T. Gonzalez-Raya}, S. Pirandola, and M. Sanz, \\
{\it Satellite-based entanglement distribution and quantum teleportation with continuous variables},\\
\href{https://doi.org/10.48550/arXiv.2303.17224}{arXiv:2303.17224 (2023).}

\item {\bf T. Gonzalez-Raya}, G. Giedke, and M. Sanz, \\
{\it Partial purification of microwave Gaussian entangled states for quantum illumination},\\
\href{}{(in preparation)}

\end{enumerate}

\vspace{\fill}

Other publications not included in this Thesis:

\begin{enumerate}[resume]

\item {\bf T. Gonzalez-Raya}, X.-H. Cheng, I. L. Egusquiza, X. Chen, M. Sanz, and E. Solano,\\
{\it Quantized Single-Ion-Channel Hodgkin-Huxley Model for Quantum Neurons},\\
\href{https://doi.org/10.1103/PhysRevApplied.12.014037}{Physical Review Applied {\bf 12}, 014037 (2019).}

\item {\bf T. Gonzalez-Raya}, E. Solano, and M. Sanz,\\
{\it Quantized Three-Ion-Channel Model for Neural Action Potentials},\\
\href{https://doi.org/10.22331/q-2020-01-20-224}{Quantum {\bf 4}, 224 (2020).}

\item {\bf T. Gonzalez-Raya}, J. M. Lukens, L. C. C\'{e}leri, and M. Sanz, \\
{\it Quantum Memristors in Frequency-Entangled Optical Fields},\\
\href{https://doi.org/10.3390/ma13040864}{Materials {\bf 13}, 864 (2020).}

\item {\bf T. Gonzalez-Raya}, R. Asensio-Perea, A. Martin, L. C. C\'{e}leri, M. Sanz, P. Lougovski, and E. F. Dumitrescu, \\
{\it Digital-Analog Quantum Simulations Using The Cross-Resonance Effect},\\
\href{https://doi.org/10.1103/PRXQuantum.2.020328}{Physical Review X Quantum {\bf 2}, 020328 (2021).}

\end{enumerate}

\cleardoublepage

%---------Figures-------------------------------------------

\phantomsection
\listoffigures
\addcontentsline{toc}{section}{List of Figures}
\counterwithin{figure}{section}

\cleardoublepage

%----------Abbreviations-------------------------------------

\section*{Abbreviations and \\ conventions}
\phantomsection
\addcontentsline{toc}{section}{Abbreviations and conventions}

We use the following abbreviations throughout the Thesis
\begin{enumerate}[leftmargin=5cm]
\item [\bf QKD]{Quantum key distribution}
\item [\bf CV]{Continuous variable}
\item [\bf DV]{Discrete variable}
\item [\bf CF]{Characteristic function}
\item [\bf QI]{Quantum illumination}
\item [\bf CFI]{Classical Fisher information}
\item [\bf QFI]{Quantum Fisher information}
\item [\bf EPR]{Einstein-Podolsky-Rosen}
\item [\bf TMSV]{Two-mode squeezed vacuum}
\item [\bf TMST]{Two-mode squeezed thermal}
\item [\bf PS]{Photon-subtracted}
\item [\bf ES]{Entanglement-swapped}
\item [\bf JPA]{Josephson parametric amplifier}
\item [\bf HEMT]{High electron mobility transistor}
\item [\bf SQUID]{Superconducting quantum interference device}
\item [\bf JPC]{Josephson parametric converter}
\item [\bf TL]{Transmission line}
\item [\bf QED]{Quantum electrodynamics}
\item [\bf LEO]{Low Earth orbit}
\item [\bf MEO]{Medium Earth orbit}
\item [\bf HEO]{High Earth orbit}
\item [\bf GEO]{Geostationary orbit}
\item [\bf FSPL]{Free-space path loss}
\end{enumerate}

%We set the reduced Planck constant $\hbar = 1$ throughout the thesis. Additionally, the speed of light $c$ is set to $1$ in chapter~\ref{sec4} and the corresponding appendix~\ref{app:theory4_QFT}.

\cleardoublepage

%%%%%%%%%%%%%%%%%%%%%%%%
% HEADINGS AND PAGE NUM.
%%%%%%%%%%%%%%%%%%%%%%%%

\renewcommand{\headrulewidth}{0.5pt}
\fancyfoot[LE,RO]{\thepage}
\fancyhead[LE]{\rightmark}
\fancyhead[RO]{\leftmark}

%%%%%%%%%%%%%%%%%%%%%%%%
% REDEFINE TITLE FORMAT
%%%%%%%%%%%%%%%%%%%%%%%%

% \titleformat{\section}[display]
% {\vspace*{150pt}
% \bf\Huge}
% {\hspace{-72pt}{\textcolor{grey}{\thesection}} \hspace{38pt} {\textcolor{white}{#1}}}
% {0pt}
% {}
% []
% \titlespacing*{\section}{108pt}{10pt}{60pt}[20pt]
%
% %Intro
% \titleformat{\section}[display]
% {\vspace*{190pt}
% \bfseries\sffamily \huge}
% {\begin{picture}(0,0)\put(-50,-25){\textcolor{grey}{\thesection}}\end{picture}}
% {0pt}
% {\textcolor{white}{#1}}
% []
% \titlespacing*{\section}{80pt}{10pt}{50pt}[20pt]
%
% %ch 2
% \titleformat{\section}[display]
% {\vspace*{190pt}
% \bfseries\sffamily \huge}
% {\begin{picture}(0,0)\put(-50,-28){\textcolor{grey}{\thesection}}\end{picture}}
% {0pt}
% {\textcolor{white}{#1}}
% []
% \titlespacing*{\section}{80pt}{10pt}{50pt}[20pt]
%
% %ch 3, 4 y 5
% \titleformat{\section}[display]
% {\vspace*{190pt}
% \bfseries\sffamily \LARGE}
% {\begin{picture}(0,0)\put(-50,-35){\textcolor{grey}{\thesection}}\end{picture}}
% {0pt}
% {\textcolor{white}{#1}}
% []
% \titlespacing*{\section}{80pt}{10pt}{40pt}[20pt]

%Conclusions
%Julen conclusions
\titleformat{\section}[display]
{\vspace*{190pt}
\bfseries\sffamily \huge}
{\begin{picture}(0,0)\put(-50,-25){\textcolor{grey}{\thesection}}\end{picture}}
{0pt}
{\textcolor{white}{#1}}
[]
\titlespacing*{\section}{80pt}{10pt}{50pt}[20pt]

%%%%%%%%%%%%%%%%%%%%%%%
%INTRODUCTION
%%%%%%%%%%%%%%%%%%%%%%%

\pagenumbering{arabic}
\section{Introduction}

% \thiswatermark{\put(1,-241){\color{l-grey}\rule{84pt}{48pt}}
% \put(84,-241){\color{grey}\rule{410pt}{48pt}}}

%JUlEN INTRO
\thiswatermark{\put(1,-280){\color{l-grey}\rule{70pt}{42pt}}
\put(70,-280){\color{grey}\rule{297pt}{42pt}}}

\lettrine[lines=2, findent=3pt,nindent=0pt]{T}{he} laws of quantum mechanics that are necessary for an accurate microscopic description of nature were postulated at the beginning of the last century. Initially, however, there was still some controversy; hence, the famous statement by Einstein, Podolsky and Rosen~\cite{Einstein1935} in 1935, that no theory satisfying local realism could ever be complete, and would need to be supported by additional (classical) variables, know as the local hidden variable model. This rather philosophical roadblock, known as the EPR paradox, was cleared by John S. Bell~\cite{Bell1964} in 1964, when he designed a series of tests, which indicated that the predictions made in the framework of quantum mechanics are incompatible with an underlying hidden-variable model satisfying a natural requirement of locality. This came to be known as Bell's theorem, and it imposes constraints on the outcomes of measurements performed locally on particles that are correlated, given local hidden variables, showing that quantum mechanics predicts a violation of these constraints. One of the most famous Bell-type inequalities is the CHSH inequality~\cite{Clauser1969}, which exposed the possibility of an experimental realization of a Bell test. After a few failed experiments in the 1970s, came Aspect's proposal~\cite{Aspect1976} and experiments~\cite{Aspect1982,Aspect1982_2}, the first proof of the non-separability of quantum mechanics. Many experiments followed~\cite{Rarity1990,Tittel1998,Weihs1998,Rowe2001}, most of them using two-outcome measurements, in which photons are treated as two level systems, with either two polarization or two photon-number states. 

%~\cite{Clauser1974}

These are known as {\it qubit} states, the unit of information encoded in a quantum system, analogous to the classical bit.

%\lettrine[lines=2, findent=3pt,nindent=0pt]{T}{he} laws of quantum mechanics that are necessary for an accurate microscopic description of nature were postulated at the beginning of the last century; with the development of classical computing machines, that were increasing in power, nonetheless, it was proposed in the 1980s that the best way to simulate the evolution of a quantum system was to use another quantum system. If the simulation of the evolution of a quantum system is so hard to compute with a classical machine, could we use a quantum system to solve other classical computational problems? This idea spurred the development of quantum computation. This new computational paradigm attempts to outperform its classical counterpart by taking advantage of the principles of quantum mechanics. The classical unit of information, the \textit{bit}, is now encoded in the state of a quantum system, and it is renamed as \textit{qubit}.

% Cannot continue without discussing entanglement in the context of the EPR paradox, Bell and the nonlocality test.  

A qubit is characterized by a vector $|\psi\rangle$ in a 2-dimensional Hilbert space $\mathcal{H}$, with basis vectors generally denoted by $|0\rangle$ and $|1\rangle$, usually referred to as the computational basis vectors. While a classical pure state can be in either ``0'' or ``1'', a pure quantum state can be in a superposition of both, 
\begin{equation}
|\psi\rangle = a|0\rangle + b|1\rangle \quad \text{ with } \quad \{a, b\}\in\mathbb{C} \quad \text{ and } \quad |a|^{2}+|b|^{2} = 1.
\end{equation}
Each element of the basis still corresponds to the classical bit 0 or 1 through measurement, but when measuring the state $|\psi\rangle$, these values will be distributed according to the weights $a$ and $b$, so that the probability of measuring $|\psi\rangle$ in state $|0\rangle$ is given by the amplitude $|\langle\psi|0\rangle|^{2}=|a|^{2}$. This superposition property can be extended to a $N$-qubit scenario, with a $2^{N}$-dimensional Hilbert space. Another property is entanglement, the spooky action at distance in the EPR paradox, and the most common manifestation of quantum correlations that cannot be explained classically. It arises in a multi-qubit scenario, where the global state of the system cannot be described in terms of the local states; it can be quantified, for example, through the entropy of the reduced system. Given a maximally entangled state, such as the Bell state $( |0,0\rangle+|1,1\rangle)/\sqrt{2}$, taking the partial trace over the second subsystem leads to the state $(|0\rangle\langle 0| + |1\rangle\langle 1|)/2$. This is a maximally-mixed state, and it has maximum entropy, meaning that information contained in the quantum correlations has been lost, and therefore, there is no measurement we can perform to extract the full information contained in the state. Mixed states, in general, are described by a density matrix $\rho$ which is positive and normalized, $\tr\rho=1$. Moreover, density matrices can also describe pure states, which satisfy $\tr\rho^{2}=1$, while mixed states follow $\tr\rho^{2}<1$. 

%Postulates regarding the information contained in quantum states that can be extracted macroscopically concern the field of quantum information theory. 

In 1984, there was a proposal to use the superposition of quantum states as a resource to transfer information remotely. In what came to be known as the BB84 protocol~\cite{Bennett1984}, it is claimed that two parties can develop a secure quantum key by sharing quantum states and classical information. The sender generates states either on the computational basis or using a superposition basis, and the receiver measures with the same choice of basis; then, they use a classical channel to communicate their choice of basis, and keep the classical bits corresponding to coinciding events. Furthermore, they can detect the presence of an eavesdropper by publicly announcing part of the obtained bit string; if a sufficient amount of elements coincide, then they can keep it to develop a secure key, and if not, they can discard it and start again. Although not directly related to Bell's theorem, the security of the BB84 protocol against individual attacks is related to the CHSH inequality~\cite{Gisin2002}. 

Another proposal that was inspired by Bell's theorem is the E91 protocol~\cite{Ekert1991}. In turn, this protocol has both parties sharing entangled states and measuring with a basis set that does not coincide completely. This way they can keep the classical bits that result from measuring in the same basis, provided that the other measurement results pass a local realism test to check for eavesdroppers.

%Shortly after Feynman's quantum simulation manifesto~\cite{Feynman1982}, there was a proposal to use the superposition of quantum states as a resource to transfer information remotely. In what came to be known as the BB84 protocol~\cite{Bennett1984}, it is claimed that two parties can develop a secure quantum key by sharing quantum states and classical information. The sender generates states either on the computational basis or using a superposition basis, and the receiver measures with the same choice of basis; then, they use a classical channel to communicate their choice of basis, and keep the classical bits corresponding to coinciding events. Furthermore, they can detect the presence of an eavesdropper by publicly announcing part of the obtained bit string; if a sufficient amount of elements coincide, then they can keep it to develop a secure key, and if not, they can discard it and start again. This was followed by the E91 protocol~\cite{Ekert1991}, a similar proposal that, in turn, has both parties sharing entangled states and measuring with a basis set that does not coincide completely. This way they can keep the classical bits that result from measuring in the same basis, provided that the other measurement results pass a local realism (Bell) test to check for eavesdroppers. 

The security of these protocols relies on the postulates of quantum mechanics; by virtue of the no-cloning theorem and the collapse of the state of a quantum system under measurement, an eavesdropper cannot extract information without affecting the process and leaving a trace. These two proposals paved the way for what is know today as quantum key distribution. Experiments have been performed, since the first quantum cryptography realization~\cite{Bennett1992}, using optical fibres~\cite{Hiskett2006,Dixon2008}, then in free space~\cite{SchmittManderbach2007,Ursin2007}, and finally in a large scale satellite link~\cite{Liao2017,Liao2018}. 

%Experiments have been performed in the laboratory~\cite{Minder2019,Wang2021}, in free space~\cite{SchmittManderbach2007,Ursin2007}, and in a satellite link~\cite{Liao2017,Liao2018}. With applications in quantum cryptography, quantum key distribution is considered a staple of quantum communication. 

Not long after the BB84, a crucial breakthrough in quantum communication was proposed: quantum teleportation~\cite{Bennett1993}. This protocol aims at transferring information of an unknown quantum state held by one party, to a second one at a remote location, by means of a previously-shared entangled resource and classical communication. It has also been realized experimentally in numerous occasions: with photonic systems in the laboratory~\cite{Bouwmeester1997,Boschi1998,Kim2001,Marcikic2003,Ursin2004}, through optical fibers~\cite{Takesue2015,Valivarthi2016,Sun2016}, in free space~\cite{Jin2010,Yin2012,Ma2012}, and in a satellite link~\cite{Ren2017}, as well as in a variety of quantum platforms: with nuclear magnetic resonance~\cite{Nielsen1998}, with trapped ions~\cite{Barrett2004,Riebe2004,Olmschenk2009}, with superconducting circuits~\cite{Baur2012,Steffen2013}, and even between macroscopic objects~\cite{Bao2012}. 

The advantage that can be obtained by quantum teleportation relies on the existence of previously-shared entanglement between both parties, same as for entanglement-mediated quantum key distribution protocols. The act of sharing entangled states between communication parties is known as entanglement distribution~\cite{Cirac1997,Chou2007}, and it has been attained experimentally with optical fibres~\cite{Hubel2007,Zhang2008,Wengerowsky2019}, as well as in free space~\cite{Fedrizzi2009,Yin2012,Herbst2015,Yin2017}. This is also a key point for the famous quantum internet initiative~\cite{Kimble2008,Wehner2018,Gyongyosi2019}.

Entanglement can be codified in many different degrees of freedom of quantum systems; the experiments that we have mentioned above use photon number, polarization, and time-bin entanglement, among others. Other than discrete-variable (DV) quantum states, entanglement can also be defined using bosonic states~\cite{Horodecki2000,Bowen2004,Skrovseth2005,Adesso2007}. These states describe infinite-dimensional Hilbert spaces, and their quadrature operators have a continuum spectrum. Systems associated with infinite-dimensional Hilbert spaces are referred to as  ``continuous-variable'' (CV) systems, and have a particularly complicated quantum description~\cite{Lloyd1999,Gottesman2001,Braunstein2005,Menicucci2006,Adesso2014,Serafini2017}. 

%in continuous-variable quantum systems~\cite{Braunstein2005,Serafini2017}. A quantum continuous variable (CV) describes the expectation value of a given quantum observable that is defined over an infinite-dimensional Hilbert space. Whereas a qubit represents the state of a quantum system with two distinct energy levels, a CV quantum state describes the state of a quantum system with infinite energy levels, for example, a quantum harmonic oscillator. Therefore, CV quantum states are useful to describe the electromagnetic field. 

% Gaussian states and typical CV entanglement
Gaussian quantum states~\cite{Ferraro2005,Zhang2006} are a family of CV states which admit a simple description; they can be described by Gaussian distributions in their phase-space representation. These are generally easy to produce experimentally, and can be used to describe the state of entangled quantum systems~\cite{Serafini2004,Adesso2005}. Therefore, their quantum-information-processing capabilities have been widely studied~\cite{Weedbrook2012}. Furthermore, any quantum evolution involving Gaussian states, Gaussian operations and Gaussian measurements, admits a compact representation known as symplectic formalism. This allows one to replace infinite-dimensional state vectors and operator matrices of an $N$-mode system by a $2N$-dimensional vector and a $2N\times2N$ matrix, the displacement vector and the covariance matrix, respectively, which can fully characterize a Gaussian quantum evolution. Using the covariance matrix, we can also compute characteristics of these states, such as the purity and the entanglement~\cite{Kim2002,Paris2003,Giedke2003,Eisert2003,Serafini2004}. 

Despite the multiple advantages, the field of Gaussian quantum information presents some limitations; for example, the impossibility to distill entanglement~\cite{Eisert2002} or to perform quantum error correction~\cite{Niset2009,Namiki2014} with Gaussian operations and Gaussian measurements. Nevertheless, many entanglement distillation protocols with non-Gaussian operations have been studied in CVs~\cite{Browne2003,Eisert2004,Campbell2013,Seshadreesan2019,Mardani2020}. Similarly, CV quantum error correction~\cite{Braunstein1998_2,Noh2020,Wu2021} is forced to abandon the realm of Gaussian states. 

Examples of Gaussian quantum states include coherent states, thermal states, and squeezed states, among others. The paradigmatic case of bipartite entangled Gaussian quantum states are two-mode squeezed states~\cite{Serafini2004}, which can also be used for quantum teleportation with CV~\cite{Pirandola2006,Pirandola2015}.

The CV formalism is frequently used for quantum communication, and especially for quantum teleportation; in fact, only a year after the first quantum teleportation paper appeared, a CV version followed~\cite{Vaidman1994}. It was then replaced by a more realistic proposal~\cite{Braunstein1998}, the famous Braunstein-Kimble quantum teleportation protocol, followed by the first experimental realization~\cite{Furusawa1998}. Naturally, other works came thereafter that discussed improvements in the protocol~\cite{Milburn1999,Opatrny2000,Cochrane2002,Marian2006} and the experiment~\cite{Bowen2003,Yukawa2008,Huo2018}. There have also been advances in CV entanglement distribution~\cite{Johnson2002,Mista2009,Dias2020}, with experiments~\cite{Dynes2009,Inagaki2013}, as well as in CV quantum key distribution~\cite{Korzh2015,Zhang2019,Fesquet2022}. 

Most quantum communication experiments use photons in the optical regime, mainly because of the mild diffraction effects and faint thermal background. Nevertheless, in this range there are many sources of error and inefficiency~\cite{Sanz2018}: large absorption losses in open air and significant power consumption requirements, to name a few. At the same time, the current most promising quantum computing platforms, namely superconducting circuits, nitrogen-vacancy centers, or trapped ions, either work in the microwave regime or use microwave signals. Therefore, in order to establish a quantum communication channel between processing units based on these technologies, one requires either converting microwave photons to the optical domain~\cite{Forsch2020, Rueda2019} or using microwave quantum signals directly. The former approach still suffers from huge conversion quantum inefficiencies of the order of $10^{-5}$. In this Thesis, we consider the purely microwave quantum communication approach, its advantages, and limitations. 

Limitations to the quantum information bearing capabilities and operational universality of Gaussian quantum states stems from their simple description and generation, and this naturally limits the performance of quantum communication protocols. Another limiting factor is the impossibility for room temperature state generation with quantum microwaves. Microwave devices working at $1$--$100$ GHz frequencies are polluted with thermal photons at room temperature; this number is 1250 for 5 GHz at room temperature ($T=300$ K). This is one of the main limitations, and creates the need for cryogenic cooling in superconducting circuits, in order to shield them from thermal noise. 

Current state of the art microwave superconducting devices include the Josephson parametric amplifier (JPA)~\cite{Yurke1988,Yurke1989}, the high electron mobility transistor (HEMT)~\cite{Ardizzi2022}, the Josephson parametric converter (JPC)~\cite{Liu2017,Liu2020,Chien2020}, and the circulator~\cite{Mahoney2017,Chapman2017}, among others. However, one of the most relevant microwave superconducting devices, and from which Josephson parametric amplifiers and converters stem, is the Josephson junction~\cite{Josephson1969}. This nonlinear element has essential applications in quantum computation~\cite{Buttiker1987,Bouchiat1998,Koch2007}, and quantum information processing~\cite{Yamamoto2008,Castellano-Beltran2008}, and its development has led to different experiments in quantum state transfer and remote entanglement preparation between various Josephson junction-based superconducting devices~\cite{Axline2018, Campagne2018, Kurpiers2018, Leung2019, Roch2015, Narla2016,Dickel2018,Magnard2020}, as well as to sensitive noise analysis~\cite{Goetz2017, Goetz2017_2}. Another interesting application of this device is the JPA, which can generate squeezed states~\cite{Fedorov2016}; these can be used to produce entangled states~\cite{Eichler2011,Eichler2012,Fedorov2018,Pogorzalek2019} for microwave quantum communication. 

% Magnard2020 is maybe not entanglement distribution

In cryogenic environments, there have been various realizations of microwave entanglement distribution~\cite{Kurpiers2018,Magnard2020} and quantum teleportation~\cite{Baur2012,Steffen2013}. A proposal for quantum teleportation with propagating quantum microwaves~\cite{DiCandia2015} was followed by a recent experiment~\cite{Fedorov2021}, performed inside a cryostat. In this Thesis, we attempt to build a realistic model for open-air entanglement distribution and quantum teleportation with microwaves to formally study the limits of this protocol. This model should take into account the challenges associated with microwave quantum technologies, as well as those encountered by quantum communication and Gaussian quantum states. 

In open air, recent experiments have failed on efficient entanglement preservation while using commercial antennae~\cite{SandboChang2019,Luong2020,Barzanjeh2020}, partly because ``classical'' amplification of quantum signals is nothing but detrimental to quantum correlations. Nevertheless, the impedance-matching aspect of classical antennae needs to be mimicked in order to reduce reflections on signals traveling from the cryostat into the open air. Thereon, the loss mechanism entails absorption of signal photons, thermalization, and diffraction, and this can be overcome by entanglement distillation techniques, striving away from the Gaussian realm. Protocols like entanglement swapping~\cite{Hoelscher-Obermaier2011} can also be beneficial for this type of process, although the current advances in this topic lack efficiency and present certain technological barriers. 

On the road to a global quantum communication network, entanglement distribution and quantum teleportation between satellites represents a relief from atmospheric attenuation and thermal noise, where downlink or uplink communications may present the biggest challenge. Meanwhile, most advances in this are are leaning towards QKD applications~\cite{Pirandola2020,Sidhu2021}. Setting aside the technological overhead, improvements will go through understanding the different loss mechanisms in free space, namely diffraction, atmospheric attenuation, and even the effects of turbulence. Stemming from small variations of temperature and pressure inside the atmosphere, turbulence affects optical signals, whereas microwaves, due to their large wavelengths, are insensitive to it. These effects have been well studied for classical signals in the optical regime~\cite{Fante1975,Fante1980}. In the same frequency range, some recent works have studied turbulence in quantum atmospheric transmission channels~\cite{Vasylyev2016}, establishing that non-classicality of signals can be preserved~\cite{Vasylyev2012}. Insightful papers into the effects of free space propagation of quantum signals were published in Refs.~\cite{Liorni2019,Vasylyev2019}. Meanwhile, others focused on the limits for key generation and entanglement distribution between ground stations~\cite{Pirandola2021_2} and between ground stations and satellites~\cite{Pirandola2021}. 

As opposed to global communication networks, local area networks normally require wireless connection between different units. Advances in the connection of these processors with quantum information is not only relevant for quantum communications, but it also might find applications in quantum computation. Given the limitations presented by current quantum processors, which characterize the NISQ (Noisy intermediate-scale quantum) era~\cite{Preskill2018}, a distributed quantum computing~\cite{Beals2013,Cuomo2020,DiAdamo2021,Gyongyosi2021} approach could reduce noise and allow for more efficient calculations~\cite{Bharti2022}. Therefore, it is interesting to explore quantum teleportation protocols for communicating multiple qubit states between different processors.

%Roadmap 
Even though microwave quantum technology is a couple of decades behind quantum optics, it has a bright future ahead. It would be natural to predict a period of coexistence between the two regimes~\cite{Casariego2022}; while microwave-to-optical transduction is still not efficient, optical communications have proven to be the correct choice for long distances. On the other hand, an all-microwave protocol can work in a quantum local area network; an implementation of QKD on such a setting using cryolinks would represent an important milestone for microwave quantum communication. While recent advances in photon counting will probably lead to new quantum illumination experiments in cryogenia with microwaves, an application to the microwave quantum radar is still out of reach. 

In the near term, the main focus should be on experiments inside a cryostat, because open-air realizations of microwave quantum communication and quantum illumination are still out of reach. One of the main reasons is the lack of collimators to reduce diffraction, but this can also be mitigated by quantum repeaters. The design of a receiver antenna is another important step to be taken, together with the implementation of both emitting and receiving antennae. Last but not least, improving entanglement generation by implementing larger gains in parametric amplifiers, while reducing the noise. Nevertheless, as superconducting circuits continue to consolidate and expand, microwave quantum communication and sensing will carry on growing; the classical microwave wireless networks will just be there waiting.

\subsection{What you will find in this Thesis}

The path towards a universal quantum communication network goes through understanding the limitations of an extension of the classical communication paradigms to the quantum realm. While cable connections are usually performed with signals in the optical regime, microwaves are used for free-space links. Therefore, we study the distribution of quantum microwaves through free space; we study Gaussian quantum states, how they are generated, how they are launched into open air, and how they degrade under the environment. As opposed to classical signals, where the main resource is power, we focus on how entanglement behaves, its relation to other features of the state, such as the purity, and how it can be increased to mitigate the effects of noise and the environment. This Thesis is structured in six chapters, plus this introduction and a concluding chapter, and it is devoted to the study of the different pieces that need to come together for quantum communication and sensing in open air.

In chapter \ref{sec2}, we explore the properties of Gaussian states, a family of quantum states in continuous variables that are routinely used in quantum communication, as they are easy to produce experimentally. We review different features of such states that are relevant from a quantum information perspective, and characterize the elements of an all-Gaussian evolution: Gaussian quantum channels and gaussian quantum measurements. Then, we review the Braunstein-Kimble quantum teleportation protocol, a milestone of Gaussian quantum communication, and explore entanglement distillation and entanglement swapping techniques for general bipartite Gaussian states, which can improve the fidelity of this protocol. 

In chapter \ref{sec3}, we explore the purification of Gaussian states using Gaussian operations and encounter the impossibility   of complete purification of a single mode of a two-mode entangled Gaussian state without complete entanglement degradation. Therefore, we focus on increasing the purity, while reducing entanglement, and use the resulting states for quantum illumination. States resulting from single-copy and two-copy partial purification techniques show higher quantum Fisher information that the original ones, and have lower average number of photons, which translates in higher precision for quantum illumination. 

In chapter \ref{sec4}, we review some of the recent advances in superconducting quantum devices, contextualizing their implications in microwave quantum communication. We study Josephson parametric amplifiers and the role they play on the generation of entangled quantum states, sketching a noise model for Gaussian states. We then present the design of a quantum antenna for microwave entangled signals propagating from a cryostat, where quantum states are generated to reduce thermal effects, into open air. This device aims at reducing the reflections by implementing impedance matching between the different media, and maximizing entanglement preservation. 

In chapter \ref{sec5}, we study entanglement distribution of microwave entangled signals propagating in open air. After discussing state generation and efficient propagation out of the cryostat, we present a loss mechanism in open air composed of absorption losses and thermalization, and study the limits of entanglement distribution by computing the reach of entanglement. We then introduce entanglement distillation and entanglement swapping protocols to fight environmental degradation, employing the resulting states as resources for quantum teleportation. As they are involved in the protocols discussed in this chapter, here we discuss photocounting and homodyne detection with microwaves, and study the effect of errors and inefficiencies in these operations. 

In chapter \ref{sec6}, we study the limits of microwave quantum communication between satellites in space, replacing absorption with diffraction as the main loss mechanism. We establish the conditions for entanglement preservation by relating the distance and the size of the antennae, similar to the categorization of free space. First, we focus on the optical regime, in which most advances and experiments have been performed. There, we explore the effects of diffraction, atmospheric absorption, detector inefficiencies, and turbulence, on different quantum communication scenarios: ground-to-ground, ground-to-satellite (uplink), satellite-to-ground (downlink), and satellite-to-satellite. A comparison between both frequency regimes is provided. 

In chapter \ref{sec7}, we investigate the transmission of information in the form of qubit states between different quantum processors, using continuous-variable entangled resources, in a distributed quantum computing environment. We investigate the placement of pure quantum states in the Bloch sphere, and compute the average fidelities for a uniformly-distributed qubits. We compare the Braunstein-Kimble quantum teleportation, using a CV state with and without photon subtraction, with a hybrid approach, using the same state, but applying the DV protocol. We study losses in the distribution of the entangled resource for these two cases, as well as for a Bell state with DV quantum teleportation. We conclude by investigating the teleportation of an average two-qubit state using a pair of two-mode entangled states.

%%%%%%%%%%%%%%%%%%%%%%%%
% REDEFINE TITLE FORMAT
%%%%%%%%%%%%%%%%%%%%%%%%

% \titleformat{\section}[display]
% {\vspace*{150pt}
% \bf\Huge}
% {\hspace{-72pt}{\textcolor{grey}{\thesection}} \hspace{38pt} {\textcolor{white}{#1}}}
% {0pt}
% {}
% []
% \titlespacing*{\section}{108pt}{10pt}{60pt}[20pt]
%
% %Intro
% \titleformat{\section}[display]
% {\vspace*{190pt}
% \bfseries\sffamily \huge}
% {\begin{picture}(0,0)\put(-50,-25){\textcolor{grey}{\thesection}}\end{picture}}
% {0pt}
% {\textcolor{white}{#1}}
% []
% \titlespacing*{\section}{80pt}{10pt}{50pt}[20pt]
%
% %ch 2
% \titleformat{\section}[display]
% {\vspace*{190pt}
% \bfseries\sffamily \huge}
% {\begin{picture}(0,0)\put(-50,-28){\textcolor{grey}{\thesection}}\end{picture}}
% {0pt}
% {\textcolor{white}{#1}}
% []
% \titlespacing*{\section}{80pt}{10pt}{50pt}[20pt]
%
% %ch 3, 4 y 5
\titleformat{\section}[display]
{\vspace*{190pt}
\bfseries\sffamily \LARGE}
{\begin{picture}(0,0)\put(-50,-35){\textcolor{grey}{\thesection}}\end{picture}}
{0pt}
{\textcolor{white}{#1}}
[]
\titlespacing*{\section}{80pt}{10pt}{40pt}[20pt]

% %Conclusions
% %Julen conclusions
% \titleformat{\section}[display]
% {\vspace*{190pt}
% \bfseries\sffamily \huge}
% {\begin{picture}(0,0)\put(-50,-25){\textcolor{grey}{\thesection}}\end{picture}}
% {0pt}
% {\textcolor{white}{#1}}
% []
% \titlespacing*{\section}{80pt}{10pt}{50pt}[20pt]

%%%%%%%%%%%%%%%%%%%%%%
% CHAPTER 2
%%%%%%%%%%%%%%%%%%%%%%

\section[Gaussian States in Quantum Communication]{Gaussian States in \\ Quantum Communication}
\label{sec2}

% \thiswatermark{\put(1,-282){\color{l-grey}\rule{84pt}{88pt}}
% \put(84,-282){\color{grey}\rule{410pt}{88pt}}}
%
%
% %Julen ch 2
% \thiswatermark{\put(1,-308){\color{l-grey}\rule{70pt}{64pt}}
% \put(70,-308){\color{grey}\rule{297pt}{64pt}}}
% \fancyhead[RO]{\leftmark}

%Julen ch 3
\thiswatermark{\put(1,-302){\color{l-grey}\rule{70pt}{60pt}}
\put(70,-302){\color{grey}\rule{297pt}{60pt}}}

% %Julen ch 4
% \thiswatermark{\put(1,-302){\color{l-grey}\rule{70pt}{60pt}}
% \put(70,-302){\color{grey}\rule{297pt}{60pt}}}
% \fancyhead[RO]{\small \leftmark}
%
%
% %Julen ch 5
% \thiswatermark{\put(1,-302){\color{l-grey}\rule{70pt}{60pt}}
% \put(70,-302){\color{grey}\rule{297pt}{60pt}}}
% \fancyhead[RO]{\leftmark}

%GAUSSIAN STATES FOR QUANTUM COMMUNICATION
\lettrine[lines=2, findent=3pt,nindent=0pt]{A}{} ``continuous variable'' (CV) refers to a degree of freedom of a quantum system that is described by a continuous-spectrum operator.

%\begin{figure}[t!]
%	\centering
%	\includegraphics[width=\textwidth]{img/chap2/theory2/Figure8.pdf}
%	\caption[Spectral functions in the metallic and insulating phases]{\textbf{Spectral functions in the metallic and insulating phases.} Spectral functions obtained with the $XY$ method with 24 Trotter steps (blue line) and exact solution of the two-site SIAM (red dashed line). The parameters of the two-site SIAM are iterated to self-consistency with $(a)$ $U=5t^*$ and $(b)$ $U=8t^*$.}
%	\label{fig:theory2_spec}
%\end{figure}

Bosonic CV states are those whose quadratures (or, equivalently, their creation and annihilation operators) have a continuous spectrum and, therefore, the complete description of the Hilbert space requires an infinite-dimensional basis (typically, the Fock basis). Gaussian states are CV states associated with Hamiltonians that are, at most, quadratic in the field operators. As such, their full description does not require the infinite-dimensional density matrix, and can be compressed into a vector and a matrix, called the displacement vector and the covariance matrix, respectively. These are related to the first and second moments of a Gaussian distribution; hence their name ``Gaussian states''. For a system with density matrix $\rho$ describing $N$ distinguishable modes, or particles, the displacement vector $\vec{d}$ is a $2N$ vector and the covariance matrix $\Sigma$ is a $2N\times 2N$ square matrix: 
\begin{align}
\vec{d} &=\tr\left[ \rho {\bf r}\right]\\
\Sigma &= \tr\left[ \rho \lbrace ({\bf r} - \vec{d}), ({\bf r} - \vec{d})^\intercal\rbrace\right].
\end{align}
Here ${\bf r}^{\intercal}= ( \hat{x}_1, \hat{p}_1, \hat{x}_2, \hat{p}_2, \ldots, \hat{x}_N, \hat{p}_N)$ defines the so-called ``real basis'', for which canonical commutation relations read $\left[{\bf r}, {\bf r}^\intercal\right] = i\Omega$, where $\Omega = \bigoplus_{j=1}^N \Omega_1$ is the symplectic form (or the symplectic matrix), and 
\begin{equation}
\Omega_1 = \begin{pmatrix} 0 & 1\\ -1 & 0 \end{pmatrix},
\end{equation}
where we have chosen natural units, $\hbar = 1$. Note that the \textit{canonical} position and momentum operators  are defined by the choice $\kappa=2^{-1/2}$ in $\hat{a}_j = \kappa(\hat{x}_j + i \hat{p}_j)$.

The normal mode decomposition theorem~\cite{Arvind1995}, which follows from Williamson's seminal work~\cite{Williamson1936,Williamson1937,Williamson1939}, can be stated as every positive-definite Hermitian matrix $\Sigma$ of dimension $2N\times2N$ can be diagonalized with a symplectic matrix $S_{D}$: $D = S_{D} \Sigma S_{D}^\intercal$, with $D = \text{diag}\left( \nu_1, \nu_1,  \ldots, \nu_N, \nu_N \right)$, where the $\nu_{a}$, for $a \in \{1,...,N\}$, are the symplectic eigenvalues of $\Sigma$, defined as the positive eigenvalues of matrix $i\Omega\Sigma$. A Gaussian state satisfies $\nu_{a} \geq 1$, with equality for all $a$ strictly for the pure state case (which meets $\det\Sigma=1$). This is a consequence of the uncertainty principle, which is enforced by $\Sigma+i\Omega\geq0$. For a two-mode Gaussian states with covariance matrix
\begin{equation}\label{eq:twoMode}
\Sigma = \begin{pmatrix} \Sigma_A & \varepsilon_{AB}\\ \varepsilon_{AB}^\intercal & \Sigma_B \end{pmatrix},
\end{equation}
the Heisenberg principle can be expressed as $\det\Sigma - \Delta +1 \geq 0$, where $\Delta=\det\Sigma_{A}+\det\Sigma_{B}+2\det\varepsilon_{AB}$ is the symplectic invariant. The latter can also be used to compute the symplectic eigenvalues, 
\begin{equation}
\nu_{\pm} = \sqrt{\frac{\Delta \pm \sqrt{\Delta^2- 4 \det\Sigma}}{2}}.
\end{equation}

%%%%% Symplectic diagonalization %%%%%%%%
Williamson's theorem also guarantees the block diagonalization of the covariance matrix, and therefore predicts the existence of a symplectic transformation that takes a two-mode covariance matrix into
\begin{equation}\label{CM_NF}
\Sigma_{\text{NF}} = \begin{pmatrix} a & 0 & c_{1} & 0 \\ 0 & a & 0 & c_{2} \\ c_{1} & 0 & b & 0 \\ 0 & c_{2} & 0 & b \end{pmatrix},
\end{equation}
which is often referred to as Simon normal form~\cite{Serafini2017}. Here, we describe a possible symplectic transformation that can be used to obtain this normal form. Assume we start with a covariance matrix 
\begin{equation}
\Sigma = \begin{pmatrix} a_{11} & a_{12} & c_{11} & c_{12} \\ a_{12} & a_{22} & c_{21} & c_{22} \\ c_{11} & c_{21} & b_{11} & b_{12} \\ c_{12} & c_{22} & b_{12} & b_{22} \end{pmatrix},
\end{equation}
where we have already enforced that diagonal blocks are symmetric. We propose the following symplectic transformation
\begin{equation}
S = \begin{pmatrix} S_{C}^{(1)} & 0 \\ 0 & S_{C}^{(2)} \end{pmatrix} \begin{pmatrix} S_{A} & 0 \\ 0 & S_{B} \end{pmatrix},
\end{equation}
such that $S\Sigma S^{\intercal}=\Sigma_{\text{NF}}$. Here, we have
\begin{eqnarray}
\nonumber S_{A} &=& \sqrt{\frac{a}{a_{11}+2a_{12}+a_{22}}} \begin{pmatrix} \frac{a_{12}+a_{22}}{a} & -\frac{a_{12}+a_{11}}{a} \\ 1 & 1 \end{pmatrix}, \\
S_{B} &=& \sqrt{\frac{b}{b_{11}+2b_{12}+b_{22}}} \begin{pmatrix} \frac{b_{12}+b_{22}}{b} & -\frac{b_{12}+b_{11}}{b} \\ 1 & 1 \end{pmatrix},
\end{eqnarray}
identifying $a = \sqrt{a_{11}a_{22}-a_{12}^{2}}$ and $b = \sqrt{b_{11}b_{22}-b_{12}^{2}}$. Furthermore, we have
\begin{eqnarray}
\nonumber S_{C}^{(1)} &=& \begin{pmatrix} \sqrt{1-v^{2}} & -v \\ v & \sqrt{1-v^{2}} \end{pmatrix}, \\
S_{C}^{(2)} &=& \begin{pmatrix} \sqrt{1-w^{2}} & -w \\ w & \sqrt{1-w^{2}} \end{pmatrix},
\end{eqnarray}
where we have identified
\begin{eqnarray}
\nonumber z^{2} &=& (l_{11}^{2}+l_{12}^{2}+l_{21}^{2}+l_{22}^{2})^{2}-4(l_{11}l_{22}-l_{12}l_{21})^{2}, \\
v^{2} &=& \frac{1}{2}\left( 1 + \frac{l_{11}^{2}+l_{12}^{2}-l_{21}^{2}-l_{22}^{2}}{z} \right), \\
\nonumber w^{2} &=& \frac{1}{2}\left( 1 + \frac{l_{11}^{2}-l_{12}^{2}+l_{21}^{2}-l_{22}^{2}}{z} \right).
\end{eqnarray}
This last transformation acts after the off-diagonal blocks have been transformed by $S_{A}$ and $S_{B}$, so we require the following redefinitions
\begin{eqnarray}
\nonumber l_{11} &=& \frac{[c_{11}(a_{12}+a_{22})-c_{21}(a_{11}+a_{12})](b_{12}+b_{22})}{\sqrt{a b (a_{11}+2a_{12}+a_{22})(b_{11}+2b_{12}+b_{22})}} \\
\nonumber &+& \frac{[c_{22}(a_{11}+a_{12})-c_{12}(a_{12}+a_{22})](b_{11}+b_{12})}{\sqrt{a b (a_{11}+2a_{12}+a_{22})(b_{11}+2b_{12}+b_{22})}}, \\
l_{12} &=& \sqrt{\frac{b}{a}} \frac{(a_{12}+a_{22})(c_{11}+c_{12})-(a_{11}+a_{12})(c_{21}+c_{22})}{\sqrt{(a_{11}+2a_{12}+a_{22})(b_{11}+2b_{12}+b_{22})}}, \\
\nonumber l_{21} &=& \sqrt{\frac{a}{b}} \frac{(b_{12}+b_{22})(c_{11}+c_{21})-(b_{11}+b_{12})(c_{12}+c_{22})}{\sqrt{(a_{11}+2a_{12}+a_{22})(b_{11}+2b_{12}+b_{22})}}, \\
\nonumber l_{22} &=& \sqrt{ab} \frac{c_{11}+c_{12}+c_{21}+c_{22}}{\sqrt{(a_{11}+2a_{12}+a_{22})(b_{11}+2b_{12}+b_{22})}}.
\end{eqnarray}
Eventually, we will obtain
\begin{eqnarray}
\nonumber c_{1} &=& \frac{1}{2}\left( c_{11} + c_{22} + \sqrt{(c_{11}-c_{22})^{2} + 4c_{12}c_{21}}\right), \\
c_{2} &=& \frac{1}{2}\left( c_{11} + c_{22} - \sqrt{(c_{11}-c_{22})^{2} + 4c_{12}c_{21}}\right).
\end{eqnarray}

Furthermore, the symplectic-diagonal form of $\Sigma_{\text{NF}}$ is
\begin{equation}
D = \begin{pmatrix} \nu_{+} & 0 & 0 & 0 \\ 0 & \nu_{+} & 0 & 0 \\ 0 & 0 & \nu_{-} & 0 \\ 0 & 0 & 0 & \nu_{-} \end{pmatrix},
\end{equation}
where the symplectic eigenvalues are
\begin{equation}
\nu_{\pm} = \sqrt{\frac{a^{2}+b^{2}+2c_{1}c_{2}\pm\sqrt{(a^{2}-b^{2})^{2}+4(a c_{1}+b c_{2})(a c_{2}+b c_{1})}}{2}}.
\end{equation}
We present here one example of a matrix that achieves symplectic diagonalization, starting from $\Sigma_{\text{NF}}$. We can write
\begin{equation}
S_{D} = \begin{pmatrix} u_{1} & 0 & v_{1} & 0 \\ 0 & u_{2} & 0 & v_{2} \\ w_{1} & 0 & z_{1} & 0 \\ 0 & w_{2} & 0 & z_{2} \end{pmatrix},
\end{equation}
where we have defined 
\begin{eqnarray}
\nonumber u_{1} &=& \sqrt{\frac{\nu_{+}}{\nu_{-}}} \sqrt{\frac{b \nu_{-}}{a b-c_{1}^{2}}-w_{1}^{2}}, \\
\nonumber u_{2} &=& \sqrt{\frac{\nu_{+}}{\nu_{-}}} \sqrt{\frac{b \nu_{-}}{a b-c_{2}^{2}}-w_{2}^{2}}, \\
v_{1} &=& -\frac{1}{b}\sqrt{\frac{\nu_{+}}{\nu_{-}}} \left( w_{1}\sqrt{a b-c_{1}^{2}} + c_{1}\sqrt{\frac{b \nu_{-}}{a b-c_{1}^{2}}-w_{1}^{2}} \right), \\
\nonumber v_{2} &=& -\frac{1}{b}\sqrt{\frac{\nu_{+}}{\nu_{-}}} \left( w_{2}\sqrt{a b-c_{2}^{2}} + c_{2}\sqrt{\frac{b \nu_{-}}{a b-c_{2}^{2}}-w_{2}^{2}} \right), \\
\nonumber z_{1} &=& \frac{1}{b} \left( -c_{1}w_{1} + \sqrt{a b-c_{1}^{2}}\sqrt{\frac{b \nu_{-}}{a b-c_{1}^{2}}-w_{1}^{2}} \right), \\
\nonumber z_{2} &=& \frac{1}{b} \left( -c_{2}w_{2} + \sqrt{a b-c_{2}^{2}}\sqrt{\frac{b \nu_{-}}{a b-c_{2}^{2}}-w_{2}^{2}} \right),
\end{eqnarray}
%\begin{equation}
%S_{D} = \begin{pmatrix} \sqrt{\frac{\nu_{+}}{\nu_{-}}} \sqrt{\frac{b \nu_{-}}{a b-c_{1}^{2}}-w_{1}^{2}} & 0 & -\frac{1}{b}\sqrt{\frac{\nu_{+}}{\nu_{-}}} \left( w_{1}\sqrt{a b-c_{1}^{2}} + c_{1}\sqrt{\frac{b \nu_{-}}{a b-c_{1}^{2}}-w_{1}^{2}} \right) & 0 \\ 0 & \sqrt{\frac{\nu_{+}}{\nu_{-}}} \sqrt{\frac{b \nu_{-}}{a b-c_{2}^{2}}-w_{2}^{2}} & 0 & -\frac{1}{b}\sqrt{\frac{\nu_{+}}{\nu_{-}}} \left( w_{2}\sqrt{a b-c_{2}^{2}} + c_{2}\sqrt{\frac{b \nu_{-}}{a b-c_{2}^{2}}-w_{2}^{2}} \right) \\ w_{1} & 0 & \frac{1}{b} \left( -c_{1}w_{1} + \sqrt{a b-c_{1}^{2}}\sqrt{\frac{b \nu_{-}}{a b-c_{1}^{2}}-w_{1}^{2}} \right) & 0 \\ 0 & w_{2} & 0 & \frac{1}{b} \left( -c_{2}w_{2} + \sqrt{a b-c_{2}^{2}}\sqrt{\frac{b \nu_{-}}{a b-c_{2}^{2}}-w_{2}^{2}} \right) \end{pmatrix},
%\end{equation}
where $w_{1}$ and $w_{2}$ are free parameters. For example, if we set $w_{1}=w_{2}=0$, we can get a simpler expression,
\begin{equation}
S_{D} = \begin{pmatrix} \sqrt{\frac{b \nu_{+}}{a b-c_{1}^{2}}} & 0 & -\sqrt{\frac{\nu_{+}}{b}} \frac{c_{1}}{\sqrt{a b-c_{1}^{2}}} & 0 \\ 0 & \sqrt{\frac{b \nu_{+}}{a b-c_{2}^{2}}} & 0 & -\sqrt{\frac{\nu_{+}}{b}} \frac{c_{2}}{\sqrt{a b-c_{2}^{2}}} \\ 0 & 0 & \sqrt{\frac{\nu_{-}}{b}} & 0 \\ 0 & 0 & 0 & \sqrt{\frac{\nu_{-}}{b}}\end{pmatrix}.
\end{equation}
For a general, systematic way to obtain the symplectic diagonalization of a covariance matrix, see for example Ref.~\cite{Pereira2021}.

%%%%%%%%%%%%%%%%%%%%%%%%%%%%

As a measure of bipartite, mixed state entanglement, the negativity is the most commonly used entanglement monotone, and is defined as $2\mathcal{N}(\rho)=\norm{\tilde{\rho}}_1-1$, where $\norm{\tilde{\rho}}_{1}=\tr\sqrt{\tilde{\rho}^\dagger \tilde{\rho}}$ is the trace norm of the partially transposed density operator. In general, $\mathcal{N}(\rho) =\abs{ \sum_j \lambda_j}$ with the $\lambda_j$ the negative eigenvalues of $\tilde{\rho}$. For a bipartite Gaussian state with covariance matrix $\Sigma$, one defines the two partially transposed symplectic eigenvalues~\cite{Serafini2004} as
\begin{equation}\label{eq:ptseigen}
\tilde{\nu}_{\pm}=\sqrt{\frac{\tilde{\Delta} \pm \sqrt{\tilde{\Delta}^2- 4 \det\Sigma}}{2}},
\end{equation}
where the partially transposed symplectic invariant is $\tilde{\Delta} = \det \Sigma_A + \det \Sigma_B - 2\det \varepsilon_{AB}$, and $\det\tilde{\Sigma}=\det\Sigma$. The negativity can then be obtained as
\begin{equation}\label{eq:negativity}
\mathcal{N}(\rho) =\max \left\{0, \frac{1-\tilde{\nu}_{-}}{2\tilde{\nu}_{-}}\right\}.
\end{equation} 
Hence, a bipartite Gaussian state is separable~\cite{Simon2000,Duan2000} when the smaller partially transposed symplectic eigenvalue meets the condition $\tilde{\nu}_{-} \geq 1$, which can also be expressed as
\begin{equation}
\det\Sigma -\tilde{\Delta} +1 \geq 0.
\end{equation}
Notice that this inequality cannot be violated if $\det\varepsilon_{AB}\geq 0$~\cite{Serafini2017}. This is because $\tilde{\Delta}\leq\Delta$, which implies that $\det\Sigma -\tilde{\Delta} +1 \geq \det\Sigma -\Delta +1$, and the latter must be positive due to the uncertainty principle. On the other hand, $\det\varepsilon_{AB}<0$ is not sufficient for entanglement; only the violation of the above inequality can indicate it. Alternatively, the state is entangled when $\tilde{\nu}_{-} < 1$ is met. The former is only valid for bipartite Gaussian states; a more general separability condition for covariance matrices is~\cite{Werner2001,Gittsovich2008}
\begin{equation}
\Sigma \geq \sigma_{A}\oplus\sigma_{B}.
\end{equation}
This means that, if there exists two covariance matrices $\sigma_{A}$, $\sigma_{B}$, such that the above condition is satisfied, then $\Sigma$ is a covariance matrix describing a separable state. 

The most famous case of Gaussian entangled states are two-mode squeezed vacuum (TMSV) states, described by the covariance matrix 
\begin{equation}
\Sigma_{\text{TMSV}} = \begin{pmatrix} \cosh 2r \mathbb{1}_{2}& \sin2r\sigma_{z} \\ \sinh2r\sigma_{z} & \cosh2r\mathbb{1}_{2} \end{pmatrix},
\end{equation}
with $\mathbb{1}_{2} = \text{diag}(1,1)$, $\sigma_{z} = \text{diag}(1,-1)$, and where $r$ is the squeezing parameter. In this Thesis, we will consider a more realistic kind of entangled Gaussian states, which consider a thermal contribution in each mode of the state. These are two-mode squeezed thermal (TMST) states, and their covariance matrix is 
\begin{equation}
\Sigma_{\text{TMST}} = (1+2n)\begin{pmatrix} \cosh 2r \mathbb{1}_{2}& \sin2r\sigma_{z} \\ \sinh2r\sigma_{z} & \cosh2r\mathbb{1}_{2} \end{pmatrix},
\end{equation}
where $n$ is not the total number of photons, but the number of thermal photons in each mode. 

A paradigmatic case of Gaussian states are coherent states, yet they cannot be entangled by Gaussian operations~\cite{Kim2002}. Nevertheless, their description presents a simple introduction to the phase-space formalism. Coherent states~\cite{Glauber1963} $\lbrace\ket{\alpha}\rbrace_{\alpha\in \mathbb{C}}$ are defined as the eigenstates of the annihilation operator $\hat{a}$ with eigenvalue $\sqrt{2}\alpha =x + ip$, where $x, p\in \mathbb{R}$ are the eigenvalues of the canonical position and momentum  operators, respectively. They play an important role in quantum CVs, as they allow for a straightforward phase space description of Gaussian states. The displacement operator $\hat{D}(\vec{\alpha}) = \exp[i\vec{\alpha}^{\intercal}\Omega_1^{\intercal}{\bf r}] \equiv \hat{D}(\alpha) = \exp \left[ \alpha \hat{a}^\dagger - \bar{\alpha} \hat{a}\right]$ acts on the vacuum as $\hat{D}(\alpha)|0\rangle = |\alpha\rangle$, and satisfies $\hat{D}(\alpha)^\dagger = \hat{D}(-\alpha)$. In this context, it is common to use, for a coherent state $\vec{\alpha}^{\intercal} = (x, p) = \sqrt{2}(\mathbb{Re}\alpha, \mathbb{Im}\alpha)$. Coherent states are not orthogonal, and their overlap can be computed as $\bra{\beta}\ket{\alpha} = \exp\left[ -\frac{1}{2}(\alpha\bar{\beta} -\bar{\alpha}\beta -\abs{\alpha-\beta}^2)\right]$. This does not prevent the set of all coherent states from forming a basis, which, though overcomplete, allows one to find the coherent states resolution of the identity $\mathbb{1} = \pi^{-1}\int \diff^2 \alpha \ket{\alpha}\bra{\alpha}$, where $2 \diff^2 \alpha = 2 \diff \mathbb{Re}\alpha \diff \mathbb{Im}\alpha = \diff x \diff p$, enabling the computation of traces of operators in an integral fashion: $\tr \hat{O} =\pi^{-1} \int \diff^2 \alpha \bra{\alpha}\hat{O}\ket{\alpha}$. More generally, an $N$-mode displacement operator may be defined via $\hat{D}(\vec{\alpha}) =\bigotimes_{j=1}^N  \hat{D}(\alpha_j) =\hat{D}\left({\bigoplus_{j=0}^N}\vec{\alpha}_j\right) $, where $\vec{\alpha}_j^{\intercal}=(x_j, p_j)$, and $\Omega =\bigoplus_{j=1}^N \Omega_1$.

A complete representation of states that is closely related to coherent states is given by the (Wigner) characteristic function, normally referred to simply as the characteristic function (CF), and for an $N$-mode state $\rho$ (not necessarily Gaussian), is given by 
\begin{equation}
\chi(\vec{\alpha}) = \tr\left[ \rho\hat{D}(\vec{\alpha})\right],
\end{equation}
with normalization condition given by $\chi(\vec{0}) =1$. Alternatively, we will write
\begin{equation}
\chi(\alpha_{1}, \ldots, \alpha_{N}) = \tr\left[ \rho\hat{D}(\alpha_{1})\otimes\ldots\otimes\hat{D}(\alpha_{N})\right],
\end{equation}
which therefore sets
\begin{equation}
\rho = \frac{1}{(2\pi)^{N}}\int \diff^{2N}\alpha \chi(\alpha_{1}, \ldots, \alpha_{N})\hat{D}(-\alpha_{1})\otimes\ldots\otimes\hat{D}(-\alpha_{N}).
\end{equation}
A Gaussian state of first and second moments $(\vec{d}, \Sigma)$ has a CF given by
\begin{equation}
\chi(\alpha_{1}, \ldots, \alpha_{N}) = e^{-\frac{1}{4}\vec{\alpha}^\intercal \Omega^\intercal \Sigma \Omega \vec{\alpha}}e^{-i\vec{\alpha}^\intercal \Omega \vec{d}},
%\chi_{G}(\vec{r}) = e^{-\frac{1}{4}\vec{r}^\intercal \Omega^\intercal \Sigma \Omega \vec{r}e^{-i\vec{r}^\intercal \Omega \vec{d}},
\end{equation}
where $\vec{\alpha} \equiv (\vec{\alpha}_{1}, \ldots, \vec{\alpha}_{N})= (x_1, p_1, \ldots, x_N, p_N) \in \mathbb{R}^{2N}$.

Another feature that can characterize quantum states is the purity. The purity of a Gaussian state with covariance matrix $\Sigma$ is given by $\mu = 1/\sqrt{\det\Sigma}$~\cite{Paris2003}. Furthermore, the number of elements in the kernel of $\Sigma+i\Omega$ indicates the number of ``noise-free'' modes of the state characterized by $\Sigma$, or alternatively, the difference between the dimension of $\Sigma+i\Omega$ and its rank. A state can be considered pure if all of the modes it describes are noise-free, and in that case $\mu = 1$. The uncertainty principle is imposed by the condition $\Sigma+i\Omega\geq 0$, and we know that pure states saturate this inequality. This means that $\Sigma+i\Omega$ has one null eigenvalue for each pure mode, which implies that its kernel has one element for each pure (noise-free) mode.

%%%%%%%%%%%%%%%%%%%%%%%%%%%%%%%%%%%%%%%%%%%%%%%%

\subsection{Gaussian Quantum Channels}
A quantum channel  $\phi(\cdot)$ is a completely positive trace-preserving map acting on quantum states. These are described by density matrices $\rho$ acting on a Hilbert space $\mathcal{H}$, such that $\phi(\rho)$ is also a quantum state. This map can be described as the local manifestation of a unitary evolution occurring on an enlarged Hilbert space, in which our system interacts with an environment, and is represented as
\begin{equation}
\phi(\rho) = \tr_{E}\left[ \hat{U}(\rho\otimes\rho_{E})\hat{U}^{\dagger}\right].
\end{equation} 
A quantum channel is a Gaussian channel~\cite{Holevo1999,Eisert2007,Caruso2008} when, in this representation, $\rho_{E}$ is a Gaussian state, and $\hat{U}$ is a Gaussian unitary operator. The latter is generated by a quadratic Hamiltonian $H$, such that $\hat{U}=e^{\frac{i}{2}\sum_{k,l}{\bf r}_{k}H_{k,l}{\bf r}_{l}}$, where ${\bf r}$ represents the canonical coordinates in a system with $N$ degrees of freedom. Unitary matrices of this kind constitute a representation of the real symplectic group $Sp(2N,\mathbb{R})$, and satisfy $S\Omega S^{\intercal}=\Omega$. This means that they generate a transformation that preserves canonical commutation relations. The relation between this generator and a unitary matrix in a Hilbert space is $S = e^{H\Omega}$. 

% This introduces again the characteristic function
%Any quantum state can be described by the characteristic function, $\chi_{\rho}(\xi) = \tr[\rho D(\xi)]$, which is the Fourier transform of the Wigner function. Here, $D(\xi) = D^{\dagger}(-\xi) = e^{i\vec{\xi}^{\intercal}\Omega \hat{\vec{r}}}$ is the displacement operator, with $\xi\in\mathbb{R}^{2N}$. Alternatively, we can write
%\begin{equation}
%\rho = \frac{1}{(2\pi)^{N}}\int d^{2N}\xi \chi_{\rho}(\xi)D(-\xi).
%\end{equation}
%Consequently, a Gaussian state has a Gaussian characteristic function, such that
%\begin{equation}
%\chi_{\rho}(\xi) = e^{-\frac{1}{4}\vec{\xi}^{\intercal}\Omega^{\intercal}\Sigma\Omega\vec{\xi}+i\vec{\xi}^{\intercal}\Omega^{\intercal}\vec{d}}.
%\end{equation}
%This implies that a Gaussian state is completely characterized by the displacement vector $\vec{d}$, containing the first moments, and the covariance matrix $\Sigma$, which comprises the second moments. The elements of these objects can be seen as $d_{k} = \tr[\hat{r}_{k}\rho]$ and $\Sigma_{k,l} = 2\mathbb{Re}\langle(\hat{r}_{k}-d_{k})(\hat{r}_{l}-d_{l})\rangle$, for $k,l=1,...,2N$. Since it describes a real state, the covariance matrix has to satisfy a condition of positivity, $\Sigma\geq 0$, as well as the Heisenberg uncertainty principle $\Sigma+i\Omega \geq 0$.

The action of a Gaussian channel in the Schr{\"o}dinger picture can be observed directly on the covariance matrix
\begin{equation}
\Sigma' = \phi(\Sigma) =  \tr_{E}\left[ S(\Sigma\oplus\Sigma_{\text{E}})S^{\intercal} \right],
\end{equation}
where $\Sigma_{\text{E}}$ is the covariance matrix which characterizes the state of the environment, and $S$ is a symplectic operator. In general, we can write this as
\begin{equation}
S = \begin{pmatrix} X_{A} & V_{A} & 0 & 0 \\ W_{A} & Z_{A} & 0 & 0 \\ 0 & 0 & X_{B} & V_{B} \\ 0 & 0 & W_{B} & Z_{B} \end{pmatrix},
\end{equation}
and given this form of the symplectic operator, $\Sigma\oplus\Sigma_{\text{E}}$ needs to be expressed as
\begin{equation}
\Sigma\oplus\Sigma_{\text{E}} = \begin{pmatrix} \Sigma_{A} & 0 & \varepsilon_{AB} & 0 \\ 0 & E_{A} & 0 & E_{AB} \\ \varepsilon_{AB}^{\intercal} & 0 & \Sigma_{B} & 0 \\ 0 & E_{AB} & 0 & E_{B} \end{pmatrix}.
\end{equation}
Here, $E_{AB}$ does not need to be zero, although the state of the environments coupled to the subsystems $A$ and $B$, represented by the covariance matrix $\Sigma_{\text{E}}$, needs to be separable, which can be enforced by $\det E_{AB}\geq 0$. This expression can also be written as
\begin{equation}
\Sigma' = \phi(\Sigma) =  X\Sigma X^{\intercal} + Y,
\end{equation}
and this notation implies that $X = X_{A}\oplus X_{B}$ and $Y=V_{A}E_{A}V_{A}^{\intercal}\oplus V_{B}E_{B}V_{B}^{\intercal}$. Here, $X$ describes amplification, attenuation, or rotation in phase space, while $Y$ represents a noise contribution, and they need to satisfy the positivity condition $Y+i\Omega-iX\Omega X^{\intercal} \geq 0$. Notice that we can convert this into
\begin{equation}
\Sigma'+i\Omega = X(\Sigma+i\Omega)X^{\intercal} + Y + i\Omega - iX\Omega X^{\intercal},
\end{equation}
which can also be expressed as
\begin{equation}
\Sigma'+i\Omega = X(\Sigma+i\Omega)X^{\intercal} + V(\Sigma_{\text{E}}+i\Omega)V^{\intercal}.
\end{equation}
A typical example of a Gaussian quantum channel is the attenuation channel. Assume we have a Gaussian state characterized by a single-mode covariance matrix $\Sigma$. Now, consider this state gets mixed with Gaussian environmental noise, described by covariance matrix $\Sigma_{\text{E}}$, in a beam splitter with transmittivity $\cos^{2}\theta = \tau$. The unitary operator describing the action of the beam splitter can be written as
\begin{equation}
\hat{U} = e^{\theta(\hat{a}\hat{a}^{\dagger}_{E}-\hat{a}^{\dagger}\hat{a}_{E})} = e^{i\theta(\hat{p}\hat{x}_{E}-\hat{x}\hat{p}_{E})}.
\end{equation}
If we compare this expression with the general expression for a Gaussian unitary, $\hat{U} = e^{\frac{i}{2}{\bf r}^{\intercal}H {\bf r}}$, we can identify
\begin{equation}
H = \theta\begin{pmatrix} 0 & -\Omega \\ \Omega & 0 \end{pmatrix},
\end{equation}
with $\Omega$ the symplectic matrix. With this, we compute the symplectic operator using $S = e^{H\Omega}$,
\begin{equation}
S = \begin{pmatrix} \cos\theta\mathbb{1}_{2} & \sin\theta\mathbb{1}_{2} \\ -\sin\theta\mathbb{1}_{2} & \cos\theta\mathbb{1}_{2} \end{pmatrix} = \begin{pmatrix} \sqrt{\tau}\mathbb{1}_{2} & \sqrt{1-\tau}\mathbb{1}_{2} \\ -\sqrt{1-\tau}\mathbb{1}_{2} & \sqrt{\tau}\mathbb{1}_{2} \end{pmatrix}.
\end{equation}
With it, we compute the covariance matrix of the outcoming state,
\begin{equation}
\Sigma'= \tr_{E}\left[ S(\Sigma\oplus\Sigma_{\text{E}})S^{\intercal} \right] = \tau\Sigma + (1-\tau)\Sigma_{\text{E}},
\end{equation}
from which we can identify the matrices $X = \sqrt{\tau}\mathbb{1}_{2}$ and $Y = (1-\tau)\Sigma_{\text{E}}$.

Generally, the interaction with an environment implies a loss of information. A way to repurpose that information, while maintaining the Gaussian nature of the state, is by projecting a mode into a Gaussian state, i.e. through Gaussian measurement. 

%%%%%%%%%%%%%%%%%%%%%%%%%%%%%%%%%%%%%%%%%%%%%%%%

\subsection{Gaussian Quantum Measurements}
Imagine that we apply a quantum measurement on mode $B$ of a bipartite quantum state with density matrix $\rho_{AB}$. We describe the measurement as the projection of our quantum state into a positive operator valued measure (POVM) $\Pi$ with a certain outcome $i$. A POVM is a set of positive operators $\Pi_{i}\geq0$, each associated with a measurement outcome $i\in\{1,\ldots,n\}$, for which probability distribution is set by the Born rule, $\tr[\rho\Pi_{i}]=P(i|\rho)$, such that $\sum_{i=1}^{n}\Pi_{i}=\mathbb{1}$. We can express the reduced density matrix after the measurement as $\rho_{A}=P_{i}^{-1}\tr_{B}\left[\rho_{AB}\Pi_{B}^{i}\right]$, with $P_{i}=\tr\left[\rho_{AB}\Pi_{B}^{i}\right]$. 

We assume that $\rho_{AB}$ describes a Gaussian state, with displacement vector $\vec{d}^{\intercal} = \begin{pmatrix} \vec{d}_{A}^{\intercal} & \vec{d}_{B}^{\intercal} \end{pmatrix}$ and covariance matrix
\begin{equation}
\Sigma = \begin{pmatrix} \Sigma_{A} & \varepsilon_{AB} \\ \varepsilon_{AB}^{\intercal} & \Sigma_{B} \end{pmatrix}.
\end{equation}
We now take a Gaussian POVM, such that $\Pi_{i}$ is associated with a Gaussian projection, characterized by a displacement vector $\vec{\upsilon}_{B}^{i}$ and a covariance matrix $\Upsilon_{B}^{i}$, such that $\left(\Upsilon_{B}^{i}\right)_{k,l} = \tr\left[ \Pi_{i}\left\{{\bf r}_{k},{\bf r}_{l}\right\} \right]$. Then, we can write the characteristic functions of the state and the POVM element as
\begin{eqnarray}
\nonumber \chi_{AB}(\alpha,\beta) &=& \exp\bigg[ -\frac{1}{4}\Big( \vec{\alpha}^{\intercal}\Omega^{\intercal}\Sigma_{A}\Omega\vec{\alpha} + \vec{\alpha}^{\intercal}\Omega^{\intercal}\varepsilon_{AB}\Omega\vec{\beta} + \vec{\beta}^{\intercal}\Omega^{\intercal}\varepsilon_{AB}^{\intercal}\Omega\vec{\alpha} \\
\nonumber &+& \vec{\beta}^{\intercal}\Omega^{\intercal}\Sigma_{B}\Omega\vec{\beta} \Big) + i\vec{\alpha}^{\intercal}\Omega^{\intercal}\vec{d}_{A} +i\vec{\beta}^{\intercal}\Omega^{\intercal}\vec{d}_{B} \bigg], \\
\chi_{B}^{i}(\gamma) &=& \exp\left[ -\frac{1}{4}\vec{\gamma}^{\intercal}\Omega^{\intercal}\Upsilon_{B}^{i}\Omega\vec{\gamma} + i\vec{\gamma}^{\intercal}\Omega^{\intercal}\vec{\upsilon}_{B}^{i} \right],
\end{eqnarray}
respectively. Then, we compute
\begin{eqnarray}
\nonumber \tr_{B}\left[\rho_{AB}\Pi_{B}^{i}\right] &=& \frac{1}{\pi^{3}}\int \diff^{2}\alpha \, \diff^{2}\beta \, \diff^{2}\gamma \, \chi_{AB}(\alpha,\beta)\chi_{B}^{i}(\gamma) \times \\
&& \hat{D}_{A}(-\alpha) \tr_{B}\left[\hat{D}_{B}(-\beta)\hat{D}_{B}(-\gamma)\right],
\end{eqnarray}
knowing that $\tr\left[\hat{D}(-\beta)\hat{D}(-\gamma)\right]=\pi\delta^{(2)}(\beta+\gamma)$. Then, we are left with 
\begin{equation}
\tr_{B}\left[\rho_{AB}\Pi_{B}^{i}\right] = \frac{1}{\pi^{2}}\int \text{d}^{2}\alpha \hat{D}(-\alpha) \int\text{d}^{2}\beta \chi_{AB}(\alpha,\beta)\chi_{B}^{i}(-\beta).
\end{equation}
The characteristic functions can combine to
\begin{eqnarray}
\nonumber \chi_{AB}(\alpha,\beta) \chi_{B}^{i}(-\beta) &=& \exp\bigg[ -\frac{1}{4} \vec{\alpha}^{\intercal}\Omega^{\intercal}\Sigma_{A}\Omega\vec{\alpha} -\frac{1}{4} \vec{\beta}^{\intercal}\Omega^{\intercal}\left(\Sigma_{B}+\Upsilon_{B}^{i}\right)\Omega\vec{\beta} \\
&+& i\vec{\alpha}^{\intercal}\Omega^{\intercal}\vec{d}_{A} + i\vec{\beta}^{\intercal}\Omega^{\intercal}\left( \frac{i}{2}\varepsilon_{AB}^{\intercal}\Omega\vec{\alpha} + \vec{d}_{B}-\vec{\upsilon}_{B}^{i}\right) \bigg].
\end{eqnarray}
After computing the integral, we can identify the displacement vector and the covariance matrix,
\begin{eqnarray}
\nonumber \tilde{\vec{d}}_{A} &=& \vec{d}_{A} - \varepsilon_{AB}\left( \Sigma_{B} + \Upsilon_{B}^{i} \right)^{-1}\left(\vec{d}_{B}-\vec{\upsilon}_{B}^{i}\right), \\
\tilde{\Sigma}_{A} &=& \Sigma_{A} - \varepsilon_{AB}\left( \Sigma_{B} + \Upsilon_{B}^{i} \right)^{-1}\varepsilon_{AB}^{\intercal}
\end{eqnarray}
of the resulting state. Notice also that
\begin{equation}
P_{i} = \frac{2}{\sqrt{\det\left(\Sigma_{B} + \Upsilon_{B}^{i}\right)}}\exp\left[ -\left(\vec{d}_{B}-\vec{\upsilon}_{B}^{i}\right)^{\intercal}\left( \Sigma_{B} + \Upsilon_{B}^{i} \right)^{-1}\left(\vec{d}_{B}-\vec{\upsilon}_{B}^{i}\right) \right]. 
\end{equation}
A general (ideal) Gaussian measurement is described as a projection onto the most general Gaussian pure state, a displaced squeezed single-mode vacuum state, with covariance matrix
\begin{equation}
\Upsilon = \begin{pmatrix} \cosh2\xi - \sinh2\xi\cos\varphi & -\sinh2\xi\sin\varphi \\ -\sinh2\xi\sin\varphi & \cosh2\xi + \sinh2\xi\cos\varphi \end{pmatrix}.
\end{equation}
With this, we can obtain the covariance matrix associated to homodyne measurements of the x-quadrature
\begin{equation}
\Upsilon = \lim_{\xi\rightarrow\infty}\begin{pmatrix} e^{-2\xi} & 0 \\ 0 & e^{2\xi} \end{pmatrix},
\end{equation}
by setting $\varphi=0$, or alternatively, of the p-quadrature with $\varphi=\pi$,
\begin{equation}
\Upsilon = \lim_{\xi\rightarrow\infty}\begin{pmatrix} e^{2\xi} & 0 \\ 0 & e^{-2\xi} \end{pmatrix}.
\end{equation}
This represents an abuse of notation, since the limit of infinite squeezing must be taken on the resulting after the measurement process. These measurements correspond to projecting onto an eigenstate of either quadrature, described by an infinitely-squeezed state in the corresponding direction of phase space. If we are measuring two modes simultaneously, projecting onto orthogonal quadratures in different modes (let's say, x in one mode and p in the other) corresponds to double-homodyne measurements. In this fashion, heterodyne measurements are constructed by measuring one mode with an auxiliary system in a coherent state. The covariance matrix associated to heterodyne measurements can be obtained for $\xi=0$, $\Upsilon = \mathbb{1}_{2}$. Equivalently, it corresponds to projecting onto a given coherent state. 

%%%%%%%%%%%%%%%%%%%%%%%%%%%%%%%%%%%%%%%%%%%%%%%%
\begin{figure}[H]
\centering
\includegraphics[width=\textwidth]{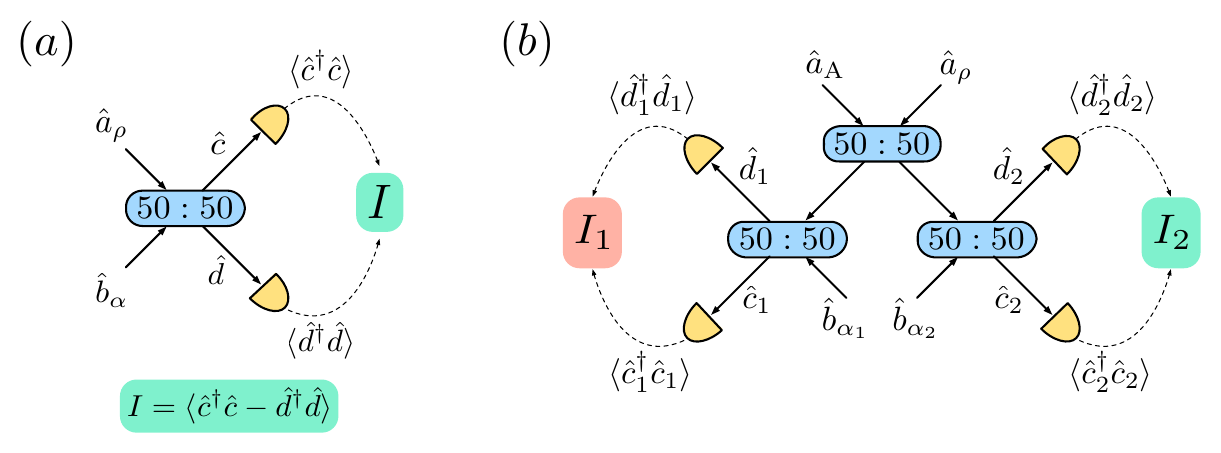}
\caption[Graphical representation of the experimental setup used for measuring quadratures of quantum states]{\textbf{Graphical representation of the experimental setup used for measuring quadratures of quantum states}. (a) Homodyne detection: a mode with annihilation operator $\hat{a}_{\rho}$ is combined with a laser mode $\hat{b}_{\alpha}$ in a 50:50 beam splitter. The difference between the photocurrents of the output modes, $I$, is proportional to a quadrature of the target state $\rho$. (b) Heterodyne detection: a mode with annihilation operator $\hat{a}_{\rho}$ is combined with an ancillary mode $\hat{a}_{A}$ in a 50:50 beam splitter. Homodyne detection is applied on the two output modes, each with a different laser mode. $I_{1}$ and $I_{2}$ each contain the information about one of the quadratures of $\rho$, so knowing the quadratures of the ancillary state we can measure both quadratures simultaneously.}
\label{fig2_1}
\end{figure}
%%%%%%%%%%%%%%%%%%%%%%%%%%%%%%%%%%%%%%%%%%%%%%%%

In practice, homodyne detection is implemented using a 50:50 beam splitter, where the state $\rho$ whose quadrature we want to measure is combined with a laser in a coherent state $|\alpha\rangle$, with $\alpha=|\alpha|e^{i\varphi}$. This setup can be seen in Fig.~\ref{fig2_1}~(a). At each output of the beam splitter, the photocurrents are measured, and by subtracting one from the other, we obtain~\cite{Serafini2017}
\begin{equation}
I = \sqrt{2}|\alpha|\langle\hat{x}_{\rho}\cos\varphi + \hat{p}_{\rho}\sin\varphi\rangle.
\end{equation}
Then, by setting the phase of the laser, we can measure either of the quadratures. For heterodyne detection, we require an ancillary state, which is combined with the target state in a 50:50 beam splitter, as shown in Fig.~\ref{fig2_1}~(b). Then, homodyne detection is applied on each of the output modes, using two coherent states, $|\alpha_{1}\rangle$ and $|\alpha_{2}\rangle$; the resulting differences in photocurrents are
\begin{eqnarray}
I_{1} &=& |\alpha_{1}|\langle\left(\hat{x}_{\rho}+\hat{x}_{A}\right)\cos\varphi_{1} + \left(\hat{p}_{\rho}+\hat{p}_{A}\right)\sin\varphi_{1}\rangle, \\
I_{2} &=& |\alpha_{2}|\langle\left(\hat{x}_{A}-\hat{x}_{\rho}\right)\cos\varphi_{2} + \left(\hat{p}_{A}-\hat{p}_{\rho}\right)\sin\varphi_{2}\rangle.
\end{eqnarray}
By setting $\varphi_{1}=\pi/2$ and $\varphi_{2}=\pi$, we obtain
\begin{eqnarray}
I_{1} &=& |\alpha_{1}|\langle\hat{p}_{\rho}+\hat{p}_{A}\rangle \equiv |\alpha_{1}|\langle\hat{p}_{+}\rangle, \\
I_{2} &=& |\alpha_{2}|\langle\hat{x}_{\rho}-\hat{x}_{A}\rangle \equiv |\alpha_{2}|\langle\hat{x}_{-}\rangle,
\end{eqnarray}
such that $[\hat{x}_{-},\hat{p}_{+}]=0$. This means that these two quadratures can be measured simultaneously, and previous knowledge of the quadratures of the ancillary state, we can infer $\langle\hat{x}_{\rho}\rangle$ and $\langle\hat{p}_{\rho}\rangle$.

%Quantum sensing and quantum metrology are fields where the use of entangled signals can bring a quantum enhancement of the measurement capabilities, by increasing precision or by reducing the number of resources employed. Naturally, measurements here play a crucial role, but we see many examples where the optimization of these measurements leads us outside of the Gaussian realm. This is the case, for example, of quantum illumination.

As we have seen, a proper measurement not only can preserve the Gaussian nature of the evolution, but also preserve the information contained in quantum entanglement in the remaining state. This trait has been exploited by quantum teleportation. 

%%%%%%%%%%%%%%%%%%%%%%%%%%%%%%%%%%%%%%%%%%%%%%%%

\subsection{Quantum Teleportation with Gaussian States}
Quantum teleportation is a quantum communication protocol that, in principle, allows us to achieve perfect transfer of quantum information between two parties by means of previously shared entanglement, combined with local operations and classical communication. The protocol was first proposed in 1993 by Bennett and collaborators~\cite{Bennett1993}, as a way to take advantage of an entangled resource for the task of sending an unknown quantum state from one place to another, using discrete-variable quantum states. The original idea was simple, yet powerful: assuming that a maximally entangled, bipartite Bell state was shared between two parties (Alice and Bob) prior to the start of the protocol, Alice, in possession of some \textit{unknown} state $\ket{\psi}=\alpha\ket{0} + e^{i\beta}\sqrt{1-|\alpha|^2}\ket{1}$ couples her part of the Bell state to $\ket{\psi}$ by means of a Bell measurement, whose 2-bit output she communicates classically to Bob. Upon receiving the message, Bob performs a conditional unitary on his part of the shared Bell state, recovering $\ket{\psi}$ modulo a global phase in his location.

A year later, Vaidman extended the idea to the transmission of a CV state by means of a perfectly correlated (singular) position-momentum EPR state shared by Alice and Bob~\cite{Vaidman1994}. In 1998, Braunstein and Kimble~\cite{Braunstein1998} made this idea more realistic by relaxing the correlation condition to more experimentally accessible states, such as finitely squeezed states. Their protocol, known as the  Braunstein-Kimble protocol, was first realised in 1998 by A. Furusawa \textit{et al.} in the optical domain~\cite{Furusawa1998}. Let us review the protocol here for convenience.

Kimble and Braunstein derived an expression for fidelity between an unknown state of a single-mode bosonic field and a teleported copy, when imperfect quantum entanglement is shared between the two parties. A generalization to a broadband version, where the modes have finite bandwidths, followed quite directly~\cite{Braunstein2005}. In the Braunstein-Kimble protocol, Alice and Bob share a TMSV state, which enables them to teleport the complete state of a single mode of the electromagnetic field, where two orthogonal field quadratures play the role of position and momentum. Shortly after, quantum teleportation of an unknown coherent state was demonstrated, showing an average fidelity (see Eq. \eqref{eq:Pure-PS-fidelity} below)  $\overline{F}=0.58 \pm 0.02$~\cite{Furusawa1998}, which beat the maximum classical fidelity of $\overline{F} = 0.5$ for Gaussian states~\cite{Braunstein2005, Serafini2017, Weedbrook2012}. Other works followed, where the Bell measurement of two orthogonal quadratures was replaced by the photon-number difference and phase sum, and the question of an optimal quantum teleportation protocol depending on the entangled resource was raised~\cite{Milburn1999}. Subtraction of single photons from two-mode squeezed states has been shown to enhance the fidelity of teleportation~\cite{Opatrny2000, Cochrane2002}. We review here the Braunstein-Kimble protocol, replacing the Wigner function approach with its Fourier transform, the characteristic function. This approach has also been followed, for example, in Refs.~\cite{Marian2006,Kitagawa2006}.

%%%%%%%%%%%%%%%%%%%%%%%%%%%%%%%%%%%%%%%%%%%%%%%%
\begin{figure}[H]
\centering
\includegraphics[width=0.5 \textwidth]{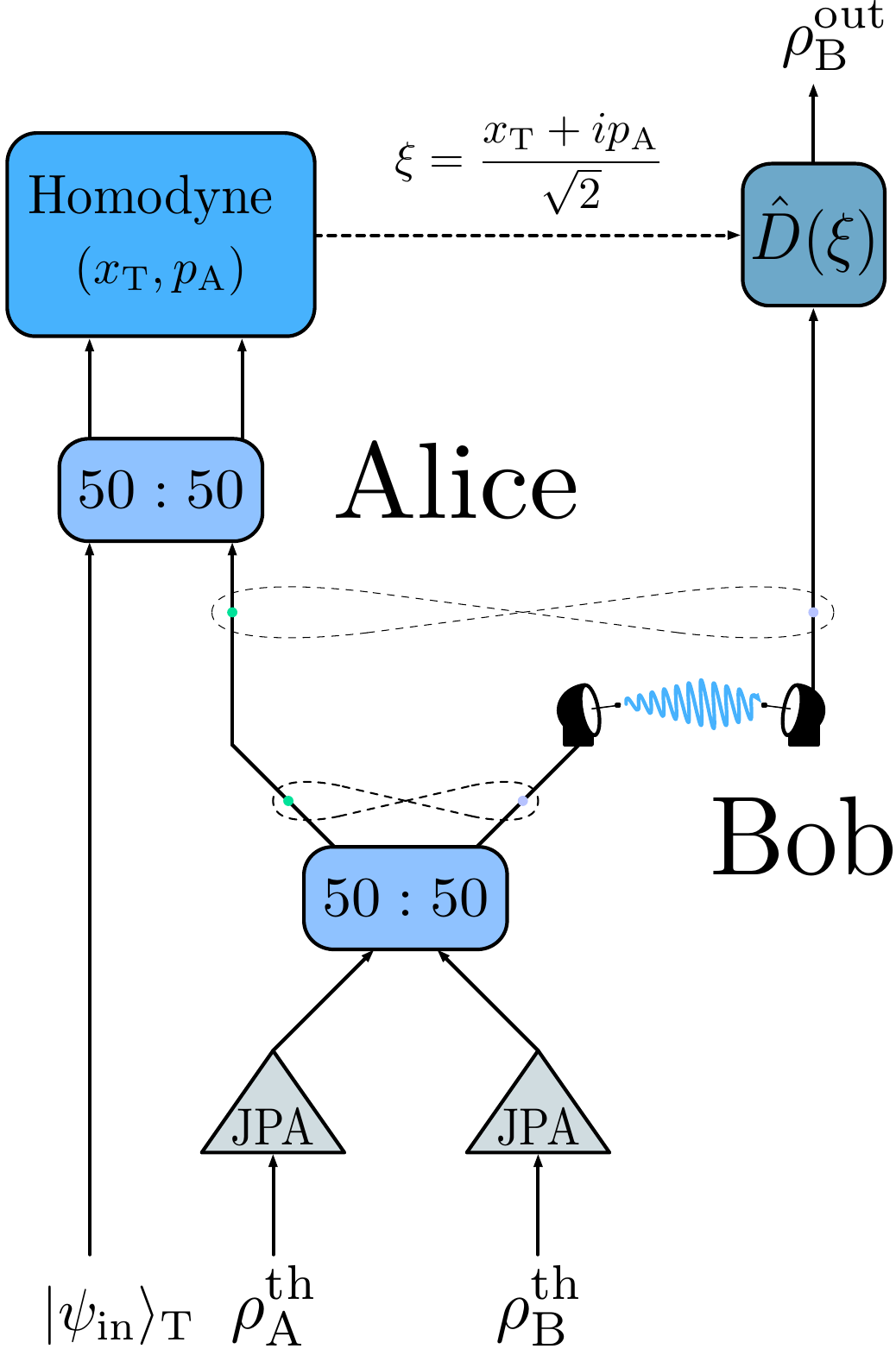}
\caption[Circuit representation of a CV microwave quantum teleportation protocol with Gaussian states]{\textbf{Circuit representation of a CV microwave quantum teleportation protocol with Gaussian states}. The entangled resource is harvested from two single-mode squeezed thermal states, generated from identical JPAs, which are then combined on a balanced beam splitter. Assuming this state is generated by Alice, one of the modes has to be sent to Bob, represented here by the presence of antennae, in order for the two parties to share the entangled resource. Following this, Alice combines the target state to be teleported $|\psi_{\text{in}}\rangle_{\text{T}}$ with the mode of the entangled state she holds in a balanced beam splitter, which is then subject to two homodyne detections, $x_\text{T}$ and $p_\text{A}$. The measurement results $\xi$ are communicated to Bob, who applies a displacement $\hat{D}(\xi)$ on his part of the entangled resource, resulting in the state $\rho^{\text{out}}_{\text{B}}$.}
\label{fig2_2}
\end{figure}
%%%%%%%%%%%%%%%%%%%%%%%%%%%%%%%%%%%%%%%%%%%%%%%%

The protocol works as follows: 
\begin{enumerate}
\item Alice uses a 50:50 beam splitter to couple her part of the resource state $\rho_{AB}$ with an incoming unknown state $\rho^{\text{in}}_{T}$. The output Hilbert spaces of this beam splitter are labeled $A$ and $T$.
\item Alice performs two homodyne detections, where each of the local oscillator phases are set in order to measure photocurrents, whose differences are integrated over some time, and proportional to quadratures $\hat{x}_T = (\hat{x}_1  + \hat{x}_\text{in})/\sqrt{2}$ and $\hat{p}_A = (\hat{p}_1  - \hat{p}_\text{in})/\sqrt{2}$. She sends the outcomes $(x_T, p_A)$ to Bob via a classical communication channel.
\item Bob, upon reception of the signal $(x_T, p_A)$, performs a displacement $\hat{D}(\xi)$ to his part of $\rho_{AB}$, with $\xi = (x_T + ip_A)/\sqrt{2}$. The state at Bob's location is now, in average, closer to $\rho_{\text{in}}$ than what it would be if no entanglement was present in $\rho_{AB}$.
\end{enumerate}
This protocol is depicted in Fig.~\ref{fig2_2}, where we also sketch the sequence that leads to an entangled resource shared through open air by Alice and Bob, which is then consumed in the teleportation process. For simplicity, we define $(x_T, p_A) = (x,p)$. The conditional state that Bob has after knowing the outcomes of Alice's homodyne measurements is
\begin{equation}
\rho_{B}(x,p) = \frac{1}{P_{B}(x,p)} \langle \Pi(x,p)| \rho_{T}^{\text{in}}\otimes \rho_{AB}|\Pi(x,p)\rangle_{TA},
\end{equation}
with $P_{B}(x,p) = \Tr_{B}\langle \Pi(x,p)| \rho_{\text{in}}\otimes \rho_{AB}|\Pi(x,p)\rangle_{TA}$ and 
\begin{equation}
\ket{\Pi(x,p)}_{TA} = \frac{1}{\sqrt{2\pi}}\int_\mathbb{R}\diff y e^{ipy}\ket{x+y}_T\ket{y}_A,
\end{equation}
which is an element of the maximally entangled basis corresponding to Alice's Bell-like measurement~\cite{Kitagawa2006}. Now, this expectation value over the teleported ($T$) and the senders ($A$) modes is computed as 
\begin{eqnarray}
\nonumber \langle \Pi(x,p)| \rho_{\text{in}}\otimes \rho_{AB}|\Pi(x,p)\rangle_{TA} &=& \frac{1}{2\pi}\int_{-\infty}^{\infty}\int_{-\infty}^{\infty} \diff y\diff y' e^{ip(y-y')}\times \\
&& \langle x+y'|\rho_{\text{in}}|x+y\rangle_{T} \langle y'|\rho_{AB}| y\rangle_{A}.
\end{eqnarray}
Once we have computed $\rho_{B}(x,p)$, we need to compute the outcoming state after the receiver applies the displacements, and average over all possible measurement outcomes
\begin{equation}
\rho_{B}^{\text{out}} = \int_{-\infty}^{\infty} \diff x \int_{-\infty}^{\infty} \diff p P_{B}(x,p)\hat{D}_{B}\left(\xi\right)\rho_{B}(x,p)\hat{D}^{\dagger}_{B}\left(\xi\right).
\end{equation}
As a measure of the quality of the protocol, one typically uses an overlap fidelity $F(\rho_\text{in}, \rho_\text{out}) = \tr[\rho_{\text{in}}\rho_{\text{out}}]$, which represents a simplified version of the Uhlmann fidelity $\left(\tr[\sqrt{\sqrt{\rho_{\text{in}}}\rho_{\text{out}}\sqrt{\rho_{\text{in}}}}]\right)^{2}$ in the case when $\rho_{\text{in}}$ is pure. 

The figure of merit in quantum teleportation is the \textit{average} fidelity, which refers to the fact that we have averaged over all possible measurement outcomes,
\begin{equation}\label{eq:Pure-PS-fidelity}
\overline{F} = \tr[\rho_{T}^{\text{in}}\rho_{B}^{\text{out}}].
\end{equation}
Sometimes it can be useful to have it written in terms of the CF~\cite{Marian2006}:
\begin{equation}
\overline{F} =  \frac{1}{\pi}\int \diff ^{2}\beta \chi_{T}^{\text{in}}(-\beta)\chi_{B}^{\text{out}}(\beta),
\end{equation}
with the average over $(x,p)$ having already been performed in $\rho_{B}^{\text{out}}$. Find the step-by-step derivation in Appendix~\ref{app_B}.

If the resource state $\rho_{AB}$ is a Gaussian state with the covariance matrix given in Eq.~\eqref{eq:twoMode}, and the teleported state is a coherent state $|\alpha_{0}\rangle\langle\alpha_{0}|$, the average fidelity can be written as
\begin{equation}\label{eq:Gaussian-fidelity}
\overline{F} = \frac{1}{\sqrt{\det[\mathbb{1}_{2} + \frac{1}{2}\Gamma]}},
\end{equation}
with $\Gamma \equiv \sigma_{z} \Sigma_A \sigma_{z} + \Sigma_B - \sigma_{z} \varepsilon_{AB}-\varepsilon_{AB}^\intercal \sigma_{z}$, and $\sigma_{z}=\text{diag}(1,-1)$. Coherent states are typically those chosen to be teleported due to the ease of their experimental generation. In theory, the result of the average fidelity does not depend on the displacement $\alpha_{0}$; therefore, it will suffice to use an unknown coherent state for a demonstration of quantum teleportation. In experiments, however, the teleportation fidelity may depend on $\alpha_{0}$. 

It is also interesting to see the average fidelity of a process in which $k$ teleportation protocols are concatenated, i.e.,
\begin{equation}
\overline{F}^{(k)} = \frac{1}{\sqrt{\det[\mathbb{1}_{2} + \left(k-\frac{1}{2}\right)\Gamma)]}},
\end{equation}
assuming that, in each step, an entangled Gaussian resource with the covariance matrix that characterizes $\Gamma$ is used. 

Consider a symmetric covariance matrix with $\Sigma_{A}=\Sigma_{B}=\alpha\mathbb{1}_{2}$ and $\varepsilon_{AB}=\gamma \sigma_{z}$. Then, we have $\Gamma = 2(\alpha-\gamma)\mathbb{1}_{2}$ and $\tilde{\nu}_{-}=\alpha-\gamma$, which leads to
\begin{equation}\label{eq:Gaussian-fidelity}
\overline{F} = \frac{1}{1+\tilde{\nu}_{-}}.
\end{equation}
It is easy to check the two following limits for the average teleportation fidelity of an arbitrary coherent state: $\lim_{\tilde{\nu}_{-}\rightarrow 1} \overline{F} = 1/2$ and $\lim_{\tilde{\nu}_{-}\rightarrow 0} \overline{F} = 1$. The first limit corresponds to using no entanglement ($\tilde{\nu}_{-}\geq1$), and is interpreted as the `classical teleportation' threshold, meaning that any approach giving an average fidelity of 0.5 or less does not demonstrate quantum teleportation. The second limit corresponds to an idealized case of an infinite two-mode squeezing level ($\tilde{\nu}_{-}=0$), i.e., an EPR state, which realizes perfect quantum teleportation.

As we can see, the quantum teleportation fidelity is closely related with entanglement; therefore, the fidelity should increase with it, what is known as entanglement distillation. Let us discuss different protocols for entanglement distillation with Gaussian resources. 

\subsection{Entanglement Distillation of Gaussian States}

Entanglement distillation is a technique that aims at increasing entanglement in quantum states by means of local operations and classical communication. Normally, the goal is to convert many copies of a noisy entangled state into as many copies as possible of a pure state with higher entanglement. This particular approach is often referred to as entanglement purification, and in discrete-variable systems, the goal is to obtain a certain form of a maximally-entangled state, or Bell state. In CV systems, this is generally out of reach, since infinite entanglement requires infinite energy for state generation.

Distillation of Gaussian entanglement is not possible with Gaussian operations~\cite{Eisert2002,Fiurasek2002}. This implies that any Gaussian state, after entanglement distillation, will become non Gaussian. This no-go theorem is partly a consequence of the fact that, among all quantum states that share the same covariance matrix, the negativity is minimized by Gaussian states~\cite{Wolf2006}. On the other hand, it also may be a consequence of the CV extension of the Gottesman-Knill theorem, which claims that the application of Gaussian quantum operations on Gaussian quantum states, with quadrature measurements is a process that can be simulated efficiently on a classical computer~\cite{Bartlett2002}. The latter has many implications; for example, it leads to the unattainability of Gaussian quantum error correction~\cite{Niset2009}.

Let us briefly review different techniques to distill entanglement. One of them is noiseless linear amplification, a nondeterministic operation~\cite{Ralph2009,Xiang2010} that requires nonincreasing distinguishability of amplified states, as well as efficient photon counting, which is where the nondeterministic part comes into play. The latter has recently been achieved in the microwave regime~\cite{Dassonneville2020}. At the core of noiseless linear amplification lies a process based on the quantum scissors~\cite{Pegg1998}; the gain of this procedure is inversely proportional to the success probability, which also decreases as the number of resources increases, making it very costly. 

Another widely known protocol is Gaussian distillation, which is also nondeterministic, but it requires only two initial copies of a state, as well as efficient photodetection. If the incoming entangled state is Gaussian then it is initially de-Gaussified by combining two copies of said state with balanced beam splitters and keeping the transmitted mode when any number of photons has been detected at the reflected modes~\cite{Browne2003}. Another possible de-Gaussification protocol applies an operation $\hat{V}=(1-\omega)a^{\dagger}a+\omega a a^{\dagger}$~\cite{Fiurasek2010} on a quantum state without requiring a copy. Gaussian distillation begins when two copies of the resulting state are mixed by $50:50$ beam splitters and, if no photons are reflected, the operation is applied again~\cite{Browne2003,Eisert2004}. Provided that the initial states were entangled, this process leads to a non-Gaussian state with higher entanglement. However, it is also costly in terms of the number of resources, and it only produces a state that is Gaussian (and with higher entanglement) in an infinite-application limit of the Gaussification channel.  

We focus on another nondeterministic protocol, which does not require the storage or production of simultaneous copies of a quantum state, and whose gain is also inversely proportional to the success probability. This protocol is called photon subtraction~\cite{Opatrny2000, Wenger2004}, and it utilizes non-Gaussian operations in order to distill entanglement, as we have seen in the previous protocols. However, in this situation, we do not look to re-Gaussify the state afterwards. We distinguish between heuristic and probabilistic photon subtraction. The former, a more theoretical approach, considers the application of annihilation operators to each mode of the state, whereas the latter uses high-transmissivity beam splitter to non-deterministically subtract photons from each mode of the state, by mixing each mode with a vacuum state, as performed in Ref.~\cite{Takahashi2010}. In the following, we will study both for completeness. 

%%%%%%%%%%%%%%%%%%%%%%%%%%%
\begin{figure}[t]
\centering
\includegraphics[width=0.7\textwidth]{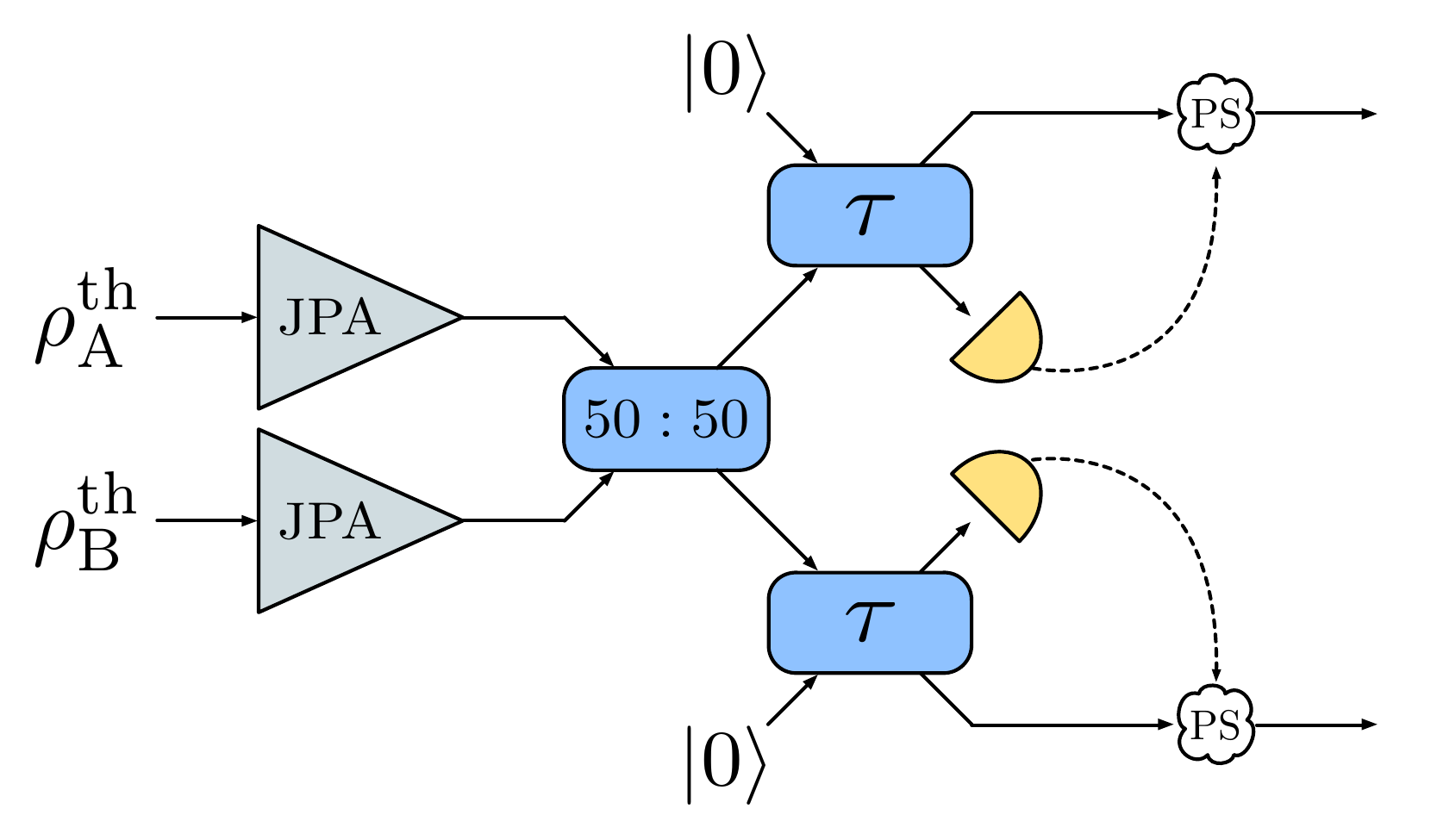}
\caption[Photon subtraction scheme with beam splitters and photocounters, applied to TMST states]{\textbf{Photon subtraction scheme with beam splitters and photocounters, applied to TMST states}, which are generated from single-mode thermal states that are squeezed by a Josephson parametric amplifier (JPA), subsequently combined in a balanced beam splitter. Each mode of the resulting state is then combined with an ancillary vacuum state in high-transmissivity (we consider $\tau = 0.95$) beam splitters, with photocounters placed at each reflected path. The resulting state shows higher entanglement for low values of the squeezing parameter, where the limit for enhancement will vary depending on the number of photons detected.}
\label{fig2_3}
\end{figure}
%%%%%%%%%%%%%%%%%%%%%%%%%%%

\subsubsection{Photon subtraction for two-mode squeezed vacuum states}
We first study probabilistic photon subtraction applied on a two-mode squeezed vacuum state, as done in Ref.~\cite{Kitagawa2006}, which is an easy to generate an entangled Gaussian state in CV. It can be produced by two single-mode squeezed states with squeezing parameter $r$, which are combined by a $50:50$ beam splitter, as shown in Fig.~\ref{fig2_3}, resulting in a TMSV state,
\begin{equation}
\sqrt{1-\lambda^{2}}\sum_{n=0}^{\infty} \lambda^{n}|n,n\rangle_{AB}
\end{equation}
with $\lambda = \tanh(r)$. The next step of the protocol is to mix each mode with an ancillary vacuum state at two highly transmitting, identical beam splitters. The output photon-subtracted (PS) state is postselected depending on the outcome of the photocounts performed at each beam splitter. Here, we focus on PS TMSV states where the same number of photons is subtracted from each mode. The resulting $2k$ PS TMSV state is then
\begin{equation}\label{eq:kPSstate}
|\psi^{(2k)}\rangle_{AB} = P_{2k}^{-1/2}\sum_{n=0}^{\infty} a^{(k)}_n\ket{n,n}_{AB},
\end{equation}
with $a^{(k)}_n \equiv  \sqrt{1-\lambda^{2}}\lambda^{n+k} \begin{pmatrix}n+k\\k\end{pmatrix} (1-\tau)^{k} \tau^{n}$, and $P_{2k} \equiv \sum_{n=0}^\infty \abs{a^{(k)}_n}^2$, which can be interpreted as the probability of successfully subtracting $k$ photons from each mode of a TMSV state. The sum converges to $P_{2k} = \left(1-\lambda ^2\right) (\lambda -\lambda_\tau)^{2 k} \, _2F_1\left(k+1,k+1;1;\lambda_\tau ^2\right)$ where $\, _2F_1\left(a,b;c;z\right)$ is the Gaussian hypergeometric function and $\lambda_\tau\equiv \tau \lambda$.

Let us now focus on the cases $k=1,2$, which correspond to two-photon subtraction (2PS) and four-photon subtraction (4PS), respectively, and whose corresponding success probabilities are
\begin{eqnarray}\label{eq:success-prob-TMSV}
\nonumber P_{2} &=& (1-\lambda^{2})\lambda^{2}(1-\tau)^{2}\frac{(1+\lambda_{\tau}^{2})}{(1-\lambda_{\tau}^{2})^{3}}, \\
P_{4} &=& 4\left(1-\lambda ^2\right) \lambda ^4  (1-\tau)^4\frac{\left(1+\lambda_{\tau}^4+4 \lambda_{\tau} ^2\right)}{\left(1-\lambda_{\tau}^2\right)^5}.
\end{eqnarray}

%%%%%%%%%%%%%%%%%%%%%%%%%%%
\begin{figure}[t]
\centering
\includegraphics[width=0.75\textwidth]{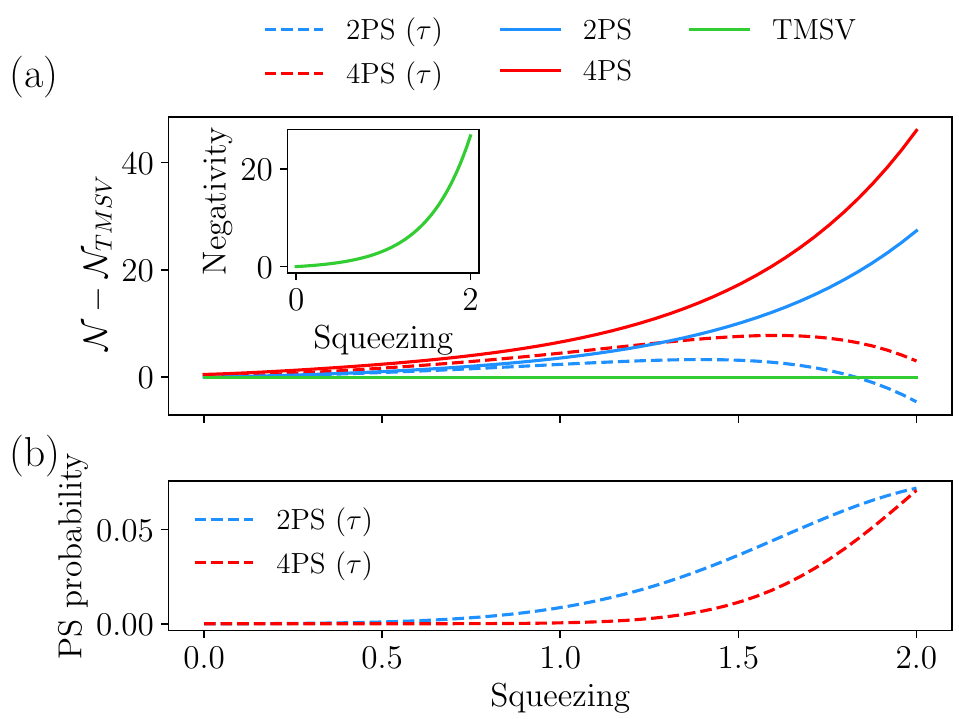}
\caption[Negativity and success probability of photon subtraction on TMSV states]{\textbf{Negativity and success probability of photon subtraction on TMSV states}: (a) Negativity difference between PS and bare TMSV states, represented against the initial squeezing parameter. The blue and red curves represent two- and four- photon subtraction, respectively, whereas the green curve represents the no-gain line, above which any point represents an improvement in negativity. Curves associated with probabilistic photon subtraction appear dashed, whereas the solid ones are associated with heuristic photon subtraction. We have considered the transmissivity of the beam splitters involved in probabilistic photon subtraction to be $\tau=0.95$. In the inset, we represent the negativity curve for the TMSV state versus the initial squeezing parameter. (b) Success probability of symmetric photon-subtraction schemes: two-photon subtraction (2PS) is represented with a blue dashed line, and four-photon subtraction (4PS) is represented with a red dashed line.}
\label{fig2_4}
\end{figure}
%%%%%%%%%%%%%%%%%%%%%%%%%%%

If photon subtraction is successful for any (nonzero) number of photons, the resulting state shows increased entanglement with respect to the TMSV state in a certain interval. This can be seen by computing the negativity $\mathcal{N}(\rho^{(2k)})$ of the family of states~\eqref{eq:kPSstate}. We find that, for $\rho^{(2k)} \equiv \ket{\psi^{(2k)}}\bra{\psi^{(2k)}}$, the negativity is
\begin{equation}
\mathcal{N}\left(\rho^{(2k)}\right) =  \frac{A_k-1}{2},
\end{equation}
where \begin{equation}
A_k \equiv \frac{\left(\sum_{n=0}^\infty a^{(k)}_n\right)^2}{P_{2k}}.
\end{equation}
Performing the sum, we obtain
\begin{equation}
\mathcal{N}\left(\rho^{(2k)}\right)=\frac{1}{2} \left(\frac{(1- \lambda_\tau)^{-2 (k+1)}}{\, _2F_1\left(k+1,k+1;1; \lambda_\tau^2\right)}-1\right),
\end{equation}
which describes the negativity of the heuristic photon-subtraction protocol (see below) in the limit $\tau\rightarrow 1$, while reproducing the negativity of the TMSV state, $\mathcal{N}_{\text{TMSV}}=\lambda/(1-\lambda)$, in the case $\tau\rightarrow1$ and $k=0$.

In Fig.~\ref{fig2_4}~(a), we represent negativity differences as a function of the initial squeezing $r$. We subtract the negativity of the TMSV state from those of the two-photon (blue) and four-photon (red), heuristic (solid) and probabilistic (dashed), subtraction with a beam splitter transmissivity $\tau = 0.95$. Note that probabilistic photon subtraction works for lower squeezing, while heuristic photon subtraction is always advantageous. In Fig.~\ref{fig2_4}~(b), we display the success probability of two-photon (blue, dashed line) and four-photon (red, dashed line) subtraction. Observe that 2PS shows higher probability than 4PS, whereas the latter shows higher improvement than the former. As the squeezing parameter increases, both probabilities grow closer, as probabilistic PS loses its advantage.

The rate of two-mode squeezed state generation is defined by the effective bandwidth of JPAs. In the case of conventional resonator-based JPAs, these bandwidths are typically of the order of about $10$\, MHz~\cite{Pogorzalek2017}. By exploiting more advanced designs based on traveling-wave Josephson parametric amplifiers, one can hope to increase these bandwidth to about $1$\,GHz. However, the price for this increase is typically lower squeezing levels and higher noise photon numbers.

%When dealing with our lossy TMST states, we consider a two-photon-subtraction protocol that is applied right before the teleportation experiment, in order to prepare the entangled resource for enhanced performance. This means that we apply photon subtraction on states that have already traveled through open air. In the case of TMSV states, photon subtraction is beneficial for low squeezing, which translates into low entanglement. In the case of lossy TMST states, entanglement is affected by the initial squeezing, but also by the distance, since reducing it means reducing photon loses and the presence of thermal noise in the state. Consequently, better performance of the teleportation protocol using photon-subtracted entangled states as the resource occurs for small distances. 

\subsubsection{Photon subtraction for bipartite Gaussian states}
We explore an heuristic photon-subtraction protocol performed on a general bipartite Gaussian state. The application of the single-photon annihilation operators on both modes of a bipartite quantum state $\rho$ modifies its characteristic function~\cite{Serafini2017} as
\begin{equation}
\Theta_{A}\Theta_{B}\chi(\alpha,\beta) \equiv \chi(\alpha,\beta)^{(-1,-1)},
\end{equation}
with
\begin{equation}
\Theta_{i} = \partial_{x_{i}}^{2} + \partial_{p_{i}}^{2} + \frac{x_{i}^{2}}{4} + \frac{p_{i}^{2}}{4} + x_{i}\partial_{x_{i}} + p_{i}\partial_{p_{i}} + 1,
\end{equation}
for $i=\{A,B\}$. Given that $\rho = \frac{1}{\pi^{2}}\int \diff^{2}\alpha\int \diff^{2}\beta \chi(\alpha,\beta)\hat{D}_{A}(-\alpha)\hat{D}_{B}(-\beta)$, and assuming that $\rho$ is a Gaussian state with covariance matrix 
\begin{equation}
\Sigma = \begin{pmatrix} \Sigma_{A} & \varepsilon_{AB} \\ \varepsilon^{\intercal}_{AB} & \Sigma_{B} \end{pmatrix},
\end{equation}
then we can write
\begin{eqnarray}\label{2PS_heur_CF}
&& \chi(\alpha,\beta)^{(-1,-1)} = \frac{e^{-\frac{1}{4}\left[\vec{\alpha}^{\intercal}\Omega^{\intercal} \Sigma_{A}\Omega\vec{\alpha} + \vec{\beta}^{\intercal}\Omega^{\intercal} \Sigma_{B}\Omega\vec{\beta} + 2\vec{\alpha}^{\intercal}\Omega^{\intercal} \varepsilon_{AB}\Omega\vec{\beta} \right]}}{m_{A}m_{B}+m_{C}} \times \\
\nonumber && \Big[ \left( m_{B} + \vec{\beta}^{\intercal} M_{B} \vec{\beta} + \vec{\alpha}^{\intercal}M_{BC}\vec{\beta} + \vec{\alpha}^{\intercal} M_{C} \vec{\alpha} \right) \left( m_{A} + \vec{\alpha}^{\intercal} M_{A} \vec{\alpha} + \vec{\alpha}^{\intercal} M_{AC}\vec{\beta} + \vec{\beta}^{\intercal} M_{C} \vec{\beta} \right) \\
\nonumber && + m_{C} - \vec{\alpha}^{\intercal} M_{AC}\Omega^{\intercal} \varepsilon_{AB}^{\intercal}\Omega\vec{\alpha} + 2\vec{\beta}^{\intercal}M_{C}\left(\mathbb{1}_{2} - \Omega^{\intercal} \Sigma_{B}\Omega\right)\vec{\beta} \\
\nonumber && + \vec{\alpha}^{\intercal} \left[M_{AC}\left(\mathbb{1}_{2}-\Omega^{\intercal} \Sigma_{B}\Omega\right) - 2\Omega^{\intercal}\varepsilon_{AB}\Omega M_{C}\right]\vec{\beta} \Big],
\end{eqnarray}
where we have defined 
\begin{eqnarray}
\nonumber m_{A} &=& 1 - \frac{1}{2}\tr \Sigma_{A}, \\
\nonumber m_{B} &=& 1 - \frac{1}{2}\tr \Sigma_{B}, \\
\nonumber m_{C} &=& \frac{1}{2}\tr \left(\varepsilon_{AB}^{\intercal}\varepsilon_{AB}\right), \\
M_{A} &=& \frac{1}{4}\Omega^{\intercal}\left( \mathbb{1}_{2} -2\Sigma_{A} + \Sigma_{A}^{2}\right)\Omega, \\
\nonumber M_{B} &=& \frac{1}{4}\Omega^{\intercal}\left( \mathbb{1}_{2} -2\Sigma_{B} +  \Sigma_{B}^{2}\right)\Omega, \\
\nonumber M_{C} &=& \frac{1}{4}\Omega^{\intercal} \varepsilon_{AB}^{\intercal}\varepsilon_{AB}\Omega, \\
\nonumber M_{AC} &=& \frac{1}{2}\Omega^{\intercal}\left( \Sigma_{A} - \mathbb{1}_{2} \right)\varepsilon_{AB}\Omega, \\
\nonumber M_{BC} &=& \frac{1}{2}\Omega^{\intercal}\varepsilon_{AB}\left( \Sigma_{B}- \mathbb{1}_{2} \right)\Omega.
\end{eqnarray}
Keep in mind that we have assumed that the submatrices $\Sigma_{A}$, $\Sigma_{B}$, and $\varepsilon_{AB}$ of the covariance matrix are symmetric.

This is a theoretic approach to photon subtraction; we also present a more applied approach, in which photon subtraction is performed non-deterministically by combining each of our modes with an ancillary mode in a vacuum state, using a low-reflectivity beam splitter. We refer to this as ``probabilistic'' photon subtraction. We consider again single-photon subtraction in each mode of a bipartite Gaussian state, assuming that $\Sigma_{A}$, $\Sigma_{B}$, and $\varepsilon_{AB}$ are symmetric. The characteristic function of the resulting state becomes
\begin{eqnarray}
\nonumber && \chi^{(-1,-1)}(\alpha,\beta) = \frac{e^{-\frac{1}{4}\left[\vec{\alpha}^{\intercal}\Omega^{\intercal} \tilde{\Sigma}_{A}\Omega\vec{\alpha} + \vec{\beta}^{\intercal}\Omega^{\intercal} \tilde{\Sigma}_{B}\Omega\vec{\beta} + 2\vec{\alpha}^{\intercal}\Omega^{\intercal} \tilde{\varepsilon}_{AB}\Omega\vec{\beta} \right]}}{m_{1}m_{2}+m_{3}} \times \\
\nonumber && \bigg[ \left( m_{1} + \vec{\alpha}^{\intercal}P_{1}\vec{\alpha} + \vec{\beta}^{\intercal}P_{2}\vec{\beta} + \vec{\alpha}^{\intercal}P_{12}\vec{\beta} \right) \left(m_{2} + \vec{\alpha}^{\intercal}Q_{1}\vec{\alpha} + \vec{\beta}^{\intercal}Q_{2}\vec{\beta} + \vec{\alpha}^{\intercal}Q_{12}\vec{\beta} \right) \\
&& + m_{3} + \vec{\alpha}^{\intercal}R_{1}\vec{\alpha} + \vec{\beta}^{\intercal}R_{2}\vec{\beta} + \vec{\alpha}^{\intercal}R_{12}\vec{\beta}\bigg].
\end{eqnarray} 
The submatrices of the covariance matrix of the resulting state transform into
\begin{eqnarray}
\nonumber \tilde{\Sigma}_{A} &=& \tau \Sigma_{A} + (1-\tau)\mathbb{1}_{2} - 2\left( J_{1}X_{A}^{-1}J_{1}^{\intercal} + K_{1}Y^{-1}K_{1}^{\intercal}\right), \\
\tilde{\Sigma}_{B} &=& \tau \Sigma_{B} + (1-\tau)\mathbb{1}_{2} - 2\left( J_{2}X_{A}^{-1}J_{2}^{\intercal} + K_{2}Y^{-1}K_{2}^{\intercal}\right), \\
\nonumber \tilde{\varepsilon}_{AB} &=& \tau \varepsilon_{AB} - 2\left( J_{1}X_{A}^{-1}J_{2}^{\intercal} + K_{1}Y^{-1}K_{2}^{\intercal} \right),
\end{eqnarray}
where $\tau$ is the transmissivity of the beam splitters involved, and the success probability of the protocol is given by
\begin{equation}
P = \frac{m_{1}m_{2}+m_{3}}{\sqrt{\det X_{A}\det Y}}. 
\end{equation}
Here, we have defined
\begin{eqnarray}
\nonumber X_{A} &=& \frac{1}{2}\Omega^{\intercal} \left[ (1-\tau)\Sigma_{A} + (1+\tau)\mathbb{1}_{2} \right]\Omega, \\
\nonumber X_{B} &=& \frac{1}{2}\Omega^{\intercal} \left[ (1-\tau)\Sigma_{B} + (1+\tau)\mathbb{1}_{2} \right]\Omega, \\
H &=& -\frac{1}{2}(1-\tau)\Omega^{\intercal} \varepsilon_{AB}\Omega, \\
\nonumber Y &=& X_{B} - HX_{A}^{-1}H,
\end{eqnarray}
together with
\begin{eqnarray}
\nonumber m_{1} &=& 1-\frac{1}{2}\tr Y^{-1}, \\
m_{2} &=& 1-\frac{1}{2}\tr X_{A}^{-1} -\frac{1}{2}\tr\left(Y^{-1}HW_{X_{A},\mathbb{1}_{2}}H\right), \\
\nonumber m_{3} &=& \frac{1}{2}\tr\left(W_{Y,\mathbb{1}_{2}}HW_{X_{A},\mathbb{1}_{2}}H\right),
\end{eqnarray}
as well as
\begin{eqnarray}
\nonumber K_{1} &=& \frac{1}{2}\sqrt{\tau(1-\tau)}\left[ \varepsilon_{AB}\Omega^{\intercal} + (\Sigma_{A}-\mathbb{1}_{2})\Omega^{\intercal}X_{A}^{-1}H \right], \\
\nonumber K_{2} &=& \frac{1}{2}\sqrt{\tau(1-\tau)}\left[ (\Sigma_{B}-\mathbb{1}_{2})\Omega^{\intercal} + \varepsilon_{AB}\Omega^{\intercal}X_{A}^{-1}H \right], \\
J_{1} &=& \frac{1}{2}\sqrt{\tau(1-\tau)}(\Sigma_{A}-\mathbb{1}_{2})\Omega^{\intercal}, \\
\nonumber J_{2} &=& \frac{1}{2}\sqrt{\tau(1-\tau)}\varepsilon_{AB}\Omega^{\intercal}.
\end{eqnarray}
We used these, together with 
\begin{equation}
W_{X,M} = X^{-1}\tr(X^{-1}M) - \frac{\Omega^{\intercal} M \Omega}{\det X},
\end{equation}
to define
\begin{eqnarray}
\nonumber P_{1} &=& -\frac{1}{2}\Omega K_{1}W_{Y,\mathbb{1}_{2}}K_{1}^{\intercal}\Omega^{\intercal}, \\
P_{2} &=& -\frac{1}{2}\Omega K_{2}W_{Y,\mathbb{1}_{2}}K_{2}^{\intercal}\Omega^{\intercal}, \\
\nonumber P_{12} &=& -\Omega K_{1}W_{Y,\mathbb{1}_{2}}K_{2}^{\intercal}\Omega^{\intercal},
\end{eqnarray}

\begin{eqnarray}
\nonumber Q_{1} &=& -\frac{1}{2}\Omega\left( J_{1}W_{X_{A},\mathbb{1}_{2}}J_{1}^{\intercal} +2J_{1}W_{X_{A},\mathbb{1}_{2}}HY^{-1}K_{1}^{\intercal} + K_{1}W_{Y,HW_{X_{A},\mathbb{1}_{2}}H}K_{1}^{\intercal}\right)\Omega^{\intercal}, \\
\nonumber Q_{2} &=& -\frac{1}{2}\Omega \left(J_{2}W_{X_{A},\mathbb{1}_{2}}J_{2}^{\intercal} +2J_{2}W_{X_{A},\mathbb{1}_{2}}HY^{-1}K_{2}^{\intercal} + K_{2}W_{Y,HW_{X_{A},\mathbb{1}_{2}}H}K_{2}^{\intercal}\right)\Omega^{\intercal}, \\
Q_{12} &=& -\Omega \Big(J_{1}W_{X_{A},\mathbb{1}_{2}}J_{2}^{\intercal} + J_{1}W_{X_{A},\mathbb{1}_{2}}HY^{-1}K_{2}^{\intercal} + K_{1}Y^{-1}HW_{X_{A},\mathbb{1}_{2}}J_{2}^{\intercal}  \\
\nonumber &+& K_{1}W_{Y,HW_{X_{A},\mathbb{1}_{2}}H}K_{2}^{\intercal}\Big)\Omega^{\intercal},
\end{eqnarray}

\begin{eqnarray}
\nonumber R_{1} &=& \frac{1}{2}\Omega\bigg[ J_{1} W_{X_{A},\mathbb{1}_{2}}HW_{Y,\mathbb{1}_{2}}K_{1}^{\intercal} + K_{1}\Big(W_{Y,\mathbb{1}_{2}}\tr\left(Y^{-1}HW_{X_{A},\mathbb{1}_{2}}H\right) \\
\nonumber &+& Y^{-1}\tr\left(W_{Y,\mathbb{1}_{2}}HW_{X_{A},\mathbb{1}_{2}}H\right) - \frac{\Omega H W_{X_{A},\mathbb{1}_{2}}H\Omega^{\intercal}}{\det Y}\tr Y^{-1} \Big) K_{1}^{\intercal} \bigg]\Omega^{\intercal}, \\
R_{2} &=& \frac{1}{2}\Omega\bigg[ J_{2} W_{X_{A},\mathbb{1}_{2}}HW_{Y,\mathbb{1}_{2}}K_{2}^{\intercal} + K_{2}\Big(W_{Y,\mathbb{1}_{2}}\tr\left(Y^{-1}HW_{X_{A},\mathbb{1}_{2}}H\right)  \\
\nonumber &+& Y^{-1}\tr\left(W_{Y,\mathbb{1}_{2}}HW_{X_{A},\mathbb{1}_{2}}H\right) - \frac{\Omega HW_{X_{A},\mathbb{1}_{2}}H\Omega^{\intercal}}{\det Y}\tr Y^{-1} \Big) K_{2}^{\intercal} \bigg]\Omega^{\intercal}, \\
\nonumber R_{12} &=& \frac{1}{2}\Omega\bigg[ J_{1} W_{X_{A},\mathbb{1}_{2}}HW_{Y,\mathbb{1}_{2}}K_{2}^{\intercal} + K_{1} W_{Y,\mathbb{1}_{2}}HW_{X_{A},\mathbb{1}_{2}} J_{2}^{\intercal} \\
\nonumber &+& 2 K_{1}\Big(W_{Y,\mathbb{1}_{2}}\tr\left(Y^{-1}HW_{X_{A},\mathbb{1}_{2}}H\right) + Y^{-1}\tr\left(W_{Y,\mathbb{1}_{2}}HW_{X_{A},\mathbb{1}_{2}}H\right) \\
\nonumber &-& \frac{\Omega HW_{X_{A},\mathbb{1}_{2}}H\Omega^{\intercal}}{\det Y}\tr Y^{-1} \Big) K_{2}^{\intercal} \bigg]\Omega^{\intercal}.
\end{eqnarray}

\subsubsection{Photon subtraction for quantum teleportation}
One interesting application of entanglement distillation techniques is the improvement of quantum teleportation. By using distilled resources, we are able to increase the fidelity of teleporting an unknown coherent state. 

For heuristic photon subtraction, the average quantum teleportation fidelity is 
\begin{equation}
\overline{F} = \frac{1+ h}{\sqrt{\det\left[\mathbb{1}_{2}+\frac{1}{2}\Gamma\right]}},
\end{equation}
with $\Gamma \equiv \sigma_{z} \Sigma_{A}\sigma_{z} + \Sigma_{B} -\sigma_{z} \varepsilon_{AB}-\varepsilon_{AB}^{\intercal} \sigma_{z}$. Here, we have defined $h$ as
\begin{eqnarray}
\nonumber h &=& \frac{1}{E_{0}}\bigg\{\tr\left[\Omega^{\intercal}\left(\mathbb{1}_{2}+\frac{1}{2}\Gamma\right)^{-1}\Omega E_{1}\right] -\frac{2}{\det\left(\mathbb{1}_{2}+\frac{1}{2}\Gamma\right)}\tr\left(\Omega^{\intercal} E_{2}^{A}\Omega E_{2}^{B}\right) \\
&+& 3\tr\left[ \Omega^{\intercal}\left(\mathbb{1}_{2}+\frac{1}{2}\Gamma\right)^{-1}\Omega E_{2}^{A}\right]\tr\left[ \Omega^{\intercal}\left(\mathbb{1}_{2}+\frac{1}{2}\Gamma\right)^{-1}\Omega E_{2}^{B}\right]\bigg\},
\end{eqnarray}
together with
\begin{eqnarray}\label{PS_QT_fid_def}
\nonumber E_{0} &=& m_{A}m_{B} + m_{C}, \\
\nonumber E_{1} &=& m_{A}\left( M_{B} + \sigma_{z} M_{C}\sigma_{z}+ \sigma_{z} M_{BC}\right) + m_{B}\left( \sigma_{z} M_{A}\sigma_{z}+ M_{C} + \sigma_{z} M_{AC}\right) \\
&+& \left(2M_{C}+\sigma_{z} M_{AC}\right)\Omega\left( \mathbb{1}_{2} + \sigma_{z} \varepsilon_{AB} - \Sigma_{B} \right)\Omega^{\intercal} \\
\nonumber E_{2}^{A} &=& M_{C} + \sigma_{z} M_{AC} + \sigma_{z} M_{A} \sigma_{z}, \\
\nonumber E_{2}^{B} &=& M_{B} + \sigma_{z} M_{BC} + \sigma_{z} M_{C} \sigma_{z}.
\end{eqnarray}
We can identify $h$ as the non-Gaussian corrections to the fidelity. In turn, probabilistic photon subtraction leads to the teleportation fidelity
\begin{equation}
\overline{F} = \frac{1+ g}{\sqrt{\det\left[\mathbb{1}_{2}+\frac{1}{2}\tilde{\Gamma}\right]}},
\end{equation}
with $\tilde{\Gamma} \equiv \sigma_{z} \tilde{\Sigma}_{A}\sigma_{z} + \tilde{\Sigma}_{B} -\sigma_{z} \tilde{\varepsilon}_{AB}-\tilde{\varepsilon}_{AB}^{\intercal} \sigma_{z}$ and
\begin{eqnarray}
\nonumber g &=& \frac{1}{m_{1}m_{2}+m_{3}}\bigg[ m_{1}\tr\left[\Omega^{\intercal}\left(\mathbb{1}_{2}+\frac{1}{2}\tilde{\Gamma}\right)^{-1}\Omega\left( \sigma_{z} Q_{1} \sigma_{z} + Q_{2} +  \sigma_{z}Q_{12}\right)\right] \\
\nonumber &+& m_{2}\tr\left[\Omega^{\intercal}\left(\mathbb{1}_{2}+\frac{1}{2}\tilde{\Gamma}\right)^{-1}\Omega\left( \sigma_{z} P_{1} \sigma_{z} + P_{2} +  \sigma_{z}P_{12}\right)\right] \\
\nonumber && +\tr\left[\Omega^{\intercal}\left(\mathbb{1}_{2}+\frac{1}{2}\tilde{\Gamma}\right)^{-1}\Omega\left( \sigma_{z} P_{1} \sigma_{z} + P_{2} +  \sigma_{z}P_{12}\right)\right]\times \\
\nonumber && \tr\left[\Omega^{\intercal}\left(\mathbb{1}_{2}+\frac{1}{2}\tilde{\Gamma}\right)^{-1}\Omega\left( \sigma_{z} Q_{1}\sigma_{z} + Q_{2} + \sigma_{z}Q_{12}\right)\right] \\
\nonumber && + \tr\left[\Omega^{\intercal}\left(\mathbb{1}_{2}+\frac{1}{2}\tilde{\Gamma}\right)^{-1}\Omega\left(\sigma_{z} R_{1}\sigma_{z} + R_{2} + \sigma_{z}R_{12}\right)\right] \\
&+& 2\tr\left[ W_{\Omega^{\intercal}\left(\mathbb{1}_{2}+\frac{1}{2}\tilde{\Gamma}\right)\Omega,\sigma_{z} P_{1}\sigma_{z} + P_{2} + \sigma_{z}P_{12}}\left(\sigma_{z} Q_{1}\sigma_{z} + Q_{2} + \sigma_{z}Q_{12}\right)\right] \bigg],
\end{eqnarray}
where the non-Gaussian corrections are collected in $g$.

A step-by-step derivation of the quantum teleportation fidelities for general Gaussian states with both heuristic and probabilistic photon subtraction can be found in Appendix~\ref{app_B}. For completeness, let us present the results for (heuristic) photon addition applied to general Gaussian states.  

\subsubsection{Photon addition for bipartite Gaussian states}
Similar to photon subtraction, photon addition represents another non-deterministic entanglement distillation protocol that only requires one copy of the state. In analogy to photon subtraction, we briefly discuss heuristic photon-addition performed on a general bipartite Gaussian. The characteristic function of a bipartite quantum state after the application of single-photon creation operators on each mode is~\cite{Ryl2017}
\begin{equation}
\Theta'_{A}\Theta'_{B}\chi(\alpha,\beta) \equiv \chi(\alpha,\beta)^{(1,1)},
\end{equation}
with
\begin{equation}
\Theta'_{i} = \partial_{x_{i}}^{2} + \partial_{p_{i}}^{2} + \frac{x_{i}^{2}}{4} + \frac{p_{i}^{2}}{4} - x_{i}\partial_{x_{i}} - p_{i}\partial_{p_{i}} - 1,
\end{equation}
for $i=\{A,B\}$. The two-photon-added characteristic function for a general bipartite Gaussian state can be written as
\begin{eqnarray}
\nonumber && \chi(\alpha,\beta)^{(1,1)} = \frac{e^{-\frac{1}{4}\left[\vec{\alpha}^{\intercal}\Omega^{\intercal} \Sigma_{A}\Omega\vec{\alpha} + \vec{\beta}^{\intercal}\Omega^{\intercal} \Sigma_{B}\Omega\vec{\beta} + 2\vec{\alpha}^{\intercal}\Omega^{\intercal} \varepsilon_{AB}\Omega\vec{\beta} \right]}}{m_{A}m_{B}+m_{C}} \times \\
\nonumber && \Big[ \left( m_{B} + \vec{\beta}^{\intercal} M_{B} \vec{\beta} + \vec{\alpha}^{\intercal}M_{BC}\vec{\beta} + \vec{\alpha}^{\intercal} M_{C} \vec{\alpha} \right) \left( m_{A} + \vec{\alpha}^{\intercal} M_{A} \vec{\alpha} + \vec{\alpha}^{\intercal} M_{AC}\vec{\beta} + \vec{\beta}^{\intercal} M_{C} \vec{\beta} \right) \\
\nonumber && + m_{C} - \vec{\alpha}^{\intercal} M_{AC}\Omega^{\intercal} \varepsilon_{AB}^{\intercal}\Omega\vec{\alpha} - 2\vec{\beta}^{\intercal}M_{C}\left(\mathbb{1}_{2} + \Omega^{\intercal} \Sigma_{B}\Omega\right)\vec{\beta} \\
&& - \vec{\alpha}^{\intercal} \left[M_{AC}\left(\mathbb{1}_{2}+\Omega^{\intercal} \Sigma_{B}\Omega\right) - 2\Omega^{\intercal}\varepsilon_{AB}\Omega M_{C}\right]\vec{\beta} \Big],
\end{eqnarray}
where we have defined 
\begin{eqnarray}
\nonumber m_{A} &=& - 1 - \frac{1}{2}\tr \Sigma_{A}, \\
\nonumber m_{B} &=& - 1 - \frac{1}{2}\tr \Sigma_{B}, \\
\nonumber m_{C} &=& \frac{1}{2}\tr \left(\varepsilon_{AB}^{\intercal}\varepsilon_{AB}\right), \\
M_{A} &=& \frac{1}{4}\Omega^{\intercal}\left( \mathbb{1}_{2} +2\Sigma_{A} + \Sigma_{A}^{2}\right)\Omega, \\
\nonumber M_{B} &=& \frac{1}{4}\Omega^{\intercal}\left( \mathbb{1}_{2} +2\Sigma_{B} +  \Sigma_{B}^{2}\right)\Omega, \\
\nonumber M_{C} &=& \frac{1}{4}\Omega^{\intercal} \varepsilon_{AB}^{\intercal}\varepsilon_{AB}\Omega, \\
\nonumber M_{AC} &=& \frac{1}{2}\Omega^{\intercal}\left( \Sigma_{A} + \mathbb{1}_{2} \right)\varepsilon_{AB}\Omega, \\
\nonumber M_{BC} &=& \frac{1}{2}\Omega^{\intercal}\varepsilon_{AB}\left( \Sigma_{B} + \mathbb{1}_{2} \right)\Omega.
\end{eqnarray}

Photon subtraction and photon addition, as well as other entanglement distillation techniques, can be the focal point of a quantum repeater strategy. This can also be the case of entanglement swapping. 

\subsection{Entanglement Swapping}
In this section, we contemplate the CV version of entanglement swapping~\cite{Hoelscher-Obermaier2011}, a procedure characteristic of quantum repeaters that attempt to reduce the distance that states have to travel through the environment, and hence attenuate the effects of entanglement degradation. We consider the case in which there are two entangled states, shared by three parties pairwisely. That is, between Alice and Charlie, and between Charlie and Bob. Entanglement swapping is a technique that allows for the conversion of two bipartite entangled states into a single one shared by initially unconnected parties. By performing measurements in a maximally entangled basis, Charlie is able to transform the entangled resources he shares with Alice and with Bob into a single entangled state shared only by Alice and Bob. In the CV formalism, these measurements are described by homodyne detection, and their effect on the state is computed as we have seen in the CV teleportation protocol. Consider that these states are Gaussian, with covariance matrices
\begin{eqnarray}
\nonumber \Sigma_{1} &=& \begin{pmatrix} \Sigma_{A} & \varepsilon_{AB} \\ \varepsilon_{AB}^{\intercal} & \Sigma_{B} \end{pmatrix}, \\
\Sigma_{2} &=& \begin{pmatrix} \Sigma_{C} & \varepsilon_{CD} \\ \varepsilon_{CD}^{\intercal} & \Sigma_{D} \end{pmatrix},
\end{eqnarray}
and null displacement vectors. Then, the covariance matrix of the entanglement-swapped (ES) state is
\begin{equation}
\Sigma^{\text{ES}}_{AD} = \begin{pmatrix} \Sigma^{\text{cond}}_{A} & \varepsilon^{\text{cond}}_{AD} \\ \varepsilon_{AD}^{\text{cond}\intercal} & \Sigma^{\text{cond}}_{D} \end{pmatrix},
\end{equation}
conditioned by the measurement results is characterized by
\begin{eqnarray}
\nonumber \Sigma_{A}^{\text{cond}} &=& \Sigma_{A} - \varepsilon_{AB}^{\intercal}\left(\Sigma_{B}+\sigma_{z}\Sigma_{C}\sigma_{z}\right)^{-1}\varepsilon_{AB}, \\
\Sigma_{D}^{\text{cond}} &=& \Sigma_{D} - \varepsilon_{CD}^{\intercal}\left(\Sigma_{C}+\sigma_{z}\Sigma_{B}\sigma_{z}\right)^{-1}\varepsilon_{CD}, \\
\nonumber \varepsilon_{AD}^{\text{cond}} &=& \varepsilon_{AB}^{\intercal}\left(\Sigma_{B}\sigma_{z}+\sigma_{z}\Sigma_{C}\right)^{-1}\varepsilon_{CD}.
\end{eqnarray}
In Appendix~\ref{app_C}, we present a step-by-step derivation of these identities. Please, see that these formulas appear incorrectly in Eqs. 38 a-c in Ref.~\cite{GonzalezRaya2022}. There, we used the identity for $2\times2$ symmetric matrices $A^{-1}=\Omega^{\intercal}A\Omega/\det A$, but the symplectic matrices are missing. This does not modify any of the results represented.

\begin{figure}[t]
\centering
\includegraphics[width=0.75\textwidth]{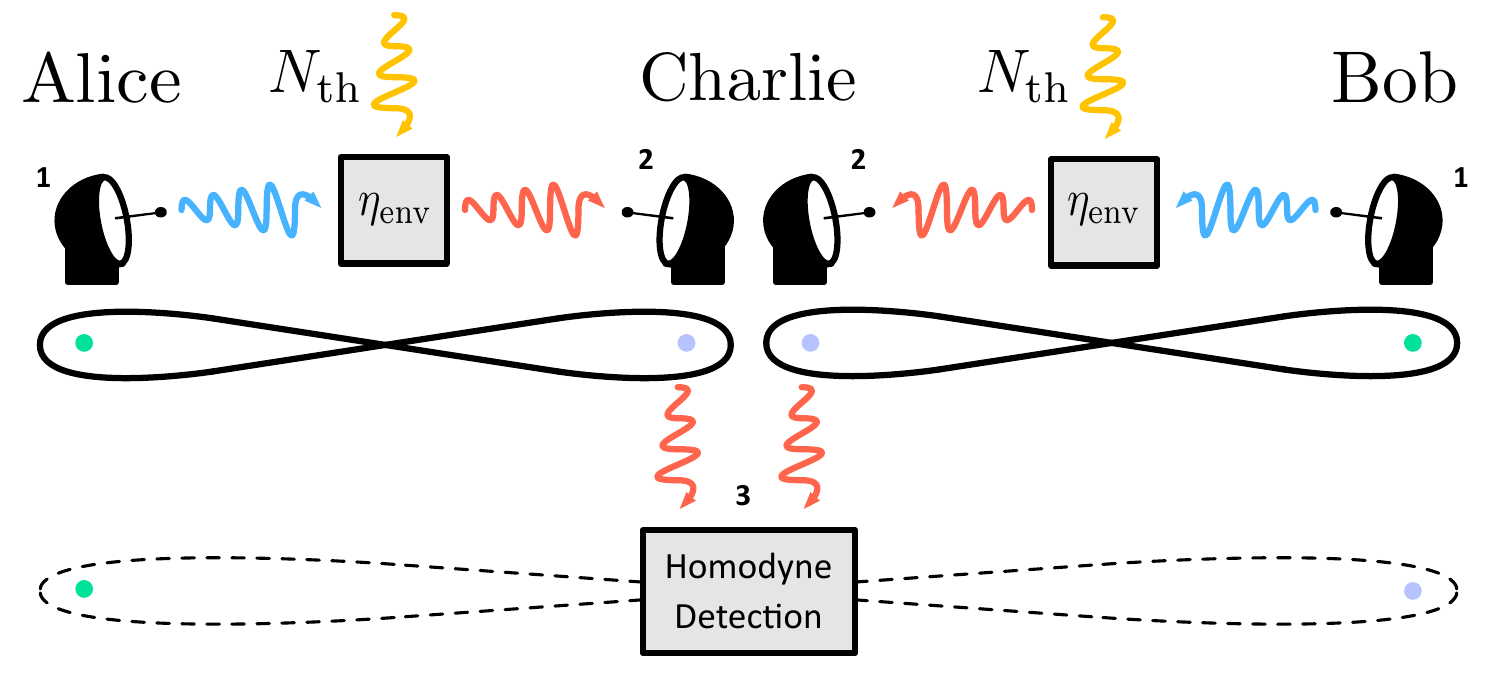}
\caption[Sketch of the optimal Gaussian entanglement swapping scheme involving three parties]{\textbf{Sketch of the optimal Gaussian entanglement swapping scheme involving three parties}, and three key steps. First, Alice and Bob generate two-mode squeezed thermal states and, while keeping one of the modes each, send the others through open air, where they are subject to photon loss and get mixed with thermal noise. Second, Charlie receives and processes both modes, and third, he uses them to perform homodyne detection. In the end, Charlie is able to transform the pairwisely entangled states he shares with Alice and with Bob independently into an entangled state held solely between Alice and Bob.}
\label{fig2_5}
\end{figure}

We observe that, in the setup we are considering, the only protocol that presents an improvement in negativity with respect to the bare states is that in which Alice and Bob generate the two-mode entangled states, and each send one of the modes to Charlie. This setup is represented in Fig.~\ref{fig2_5}. Then, the two modes used for entanglement swapping are those that have become mixed with environmental noise. Nevertheless, this enhancement occurs for large distances, which implies low negativities, and works significantly better in low-temperature environments, where $N_{\text{th}}$ is reduced. Considering $\Sigma_{A} = \Sigma_{D} = \alpha\mathbb{1}_{2}$, $\Sigma_{B} = \Sigma_{C} = \beta\mathbb{1}_{2}$, and $\varepsilon_{AB} = \varepsilon_{CD} = \gamma \sigma_{z}$, then we can characterize the covariance matrix of the ES state by
\begin{eqnarray}\label{ES_submatrices}
\nonumber \Sigma_{A}^{\text{cond}} &=& \left( \alpha - \frac{\gamma^{2}}{2\beta}\right)\mathbb{1}_{2}, \\
\Sigma_{D}^{\text{cond}} &=& \left( \alpha - \frac{\gamma^{2}}{2\beta}\right)\mathbb{1}_{2}, \\
\nonumber \varepsilon_{AD}^{\text{cond}} &=& \frac{\gamma^{2}}{2\beta} \sigma_{z}.
\end{eqnarray}
The condition for this characterization to be appropriate is given by
\begin{equation}
\left|\sqrt{\det\Sigma_{1}}-\frac{\beta}{\alpha}\right| \geq 0.
\end{equation}
As shown in Appendix~\ref{app_C}, if we average over all possible measurement results, we obtain
\begin{eqnarray}
\nonumber \tilde{\Sigma}_{A} &=& \Sigma_{A} + \Sigma_{B} + \sigma_{z}\Sigma_{C}\sigma_{z} + 2\varepsilon_{AB}, \\
\tilde{\Sigma}_{D} &=& \Sigma_{D} + \Sigma_{B} + \sigma_{z}\Sigma_{C}\sigma_{z} - 2 \sigma_{z} \varepsilon_{CD}, \\
\nonumber \tilde{\varepsilon}_{AD} &=& \Sigma_{B}+\sigma_{z}\Sigma_{C}\sigma_{z} + \varepsilon_{AB} - \sigma_{z}\varepsilon_{CD}.
\end{eqnarray}
If the initial states are Gaussian, and we replace again $\Sigma_{A} = \Sigma_{D} = \alpha\mathbb{1}_{2}$, $\Sigma_{B} = \Sigma_{C} = \beta\mathbb{1}_{2}$, and $\varepsilon_{AB} = \varepsilon_{CD} = \gamma \sigma_{z}$, we find that the resulting state is separable. This is shown by $\det\tilde{\varepsilon}_{AD}=4\beta(\beta-\gamma)\geq0$. Therefore, in order to preserve entanglement in this protocol, we must retain the information about the measurement results. 

%%%%%%%%%%%%%%%%%%%%%%%%%%%
%%%%%%%%%%%%%%%%%%%%%%%%%%%

\subsubsection{Entanglement swapping for quantum teleportation}

Taking $\Sigma_{A} = \Sigma_{D}$, $\Sigma_{B} = \Sigma_{C}$, and $\varepsilon_{AB} = \varepsilon_{CD}$, and knowing that entanglement swapping is Gaussian-preserving, the quantum teleportation fidelity using an ES resource is given by
\begin{equation}
\overline{F} = \frac{1}{\sqrt{\det\left[\mathbb{1}_{2}+\frac{1}{2}\Gamma_{\text{ES}}\right]}}.
\end{equation} 
Here, we have defined
\begin{eqnarray}
\nonumber \Gamma_{\text{ES}} &\equiv& \Sigma_{A} - \varepsilon_{AB}^{\intercal}(\Sigma_{B} + \sigma_{z}\Sigma_{B}\sigma_{z})^{-1}\varepsilon_{AB} + \sigma_{z}\left[ \Sigma_{A} - \varepsilon_{AB}^{\intercal}(\Sigma_{B} + \sigma_{z}\Sigma_{B}\sigma_{z})^{-1}\varepsilon_{AB}\right]\sigma_{z} \\
&-& \sigma_{z}\varepsilon_{AB}^{\intercal}(\Sigma_{B}\sigma_{z} + \sigma_{z}\Sigma_{B})^{-1}\varepsilon_{AB} - \varepsilon_{AB}^{\intercal}(\Sigma_{B}\sigma_{z} + \sigma_{z}\Sigma_{B})^{-1}\varepsilon_{AB}\sigma_{z}.
\end{eqnarray}

% CONCLUSIONS

Gaussian states are the paradigmatic example of the state of a bosonic CV system, partly due to their compact description in phase space using the symplectic formalism. They also embrace some of the most well-known CV quantum states, and are easy to prepare experimentally. We have then introduced an all-Gaussian toolbox of operations and protocols: Gaussian quantum channels, Gaussian measurements, and Gaussian quantum teleportation. We have also seen that all these perks do not come without disadvantages; for example, the fact that entanglement cannot be distilled with local Gaussian operations. Therefore, we introduced photon subtraction, and followed the same formalism, now dealing with non-Gaussian resources. Finally, we discussed another protocol that can also be crucial for quantum repeaters; entanglement swapping.

%%%%%%%%%%%%%%%%%%%%%%%%
% REDEFINE TITLE FORMAT
%%%%%%%%%%%%%%%%%%%%%%%%

%%%%%%%%%%%%%%%%%%%%%%
% CHAPTER 3
%%%%%%%%%%%%%%%%%%%%%%

\section[Partial Purification for Gaussian Quantum Illumination]{Partial Purification for \\ Gaussian Quantum Illumination}
\label{sec3}

% \thiswatermark{\put(1,-282){\color{l-grey}\rule{84pt}{88pt}}
% \put(84,-282){\color{grey}\rule{410pt}{88pt}}}
%
%
% %Julen ch 2
% \thiswatermark{\put(1,-308){\color{l-grey}\rule{70pt}{64pt}}
% \put(70,-308){\color{grey}\rule{297pt}{64pt}}}
% \fancyhead[RO]{\leftmark}

%Julen ch 3
\thiswatermark{\put(1,-302){\color{l-grey}\rule{70pt}{60pt}}
\put(70,-302){\color{grey}\rule{297pt}{60pt}}}

% %Julen ch 4
% \thiswatermark{\put(1,-302){\color{l-grey}\rule{70pt}{60pt}}
% \put(70,-302){\color{grey}\rule{297pt}{60pt}}}
% \fancyhead[RO]{\small \leftmark}
%
%
% %Julen ch 5
% \thiswatermark{\put(1,-302){\color{l-grey}\rule{70pt}{60pt}}
% \put(70,-302){\color{grey}\rule{297pt}{60pt}}}
% \fancyhead[RO]{\leftmark}

%GAUSSIAN PARTIAL PURIFICATION

\lettrine[lines=2, findent=3pt,nindent=0pt]{G}{}aussian states are very versatile resources for quantum communication tasks, that also present a very convenient description using the symplectic formalism. Despite presenting a classical Gaussian profile in their phase space distributions, they can exhibit quantum entanglement in a bipartite setting. For that matter, there exist Gaussian-preserving (symplectic) transformations that can generate entanglement, the most famous one being the beam splitter transformation. Nevertheless, as we have discussed in the previous chapter, distillation of Gaussian entanglement using local Gaussian operations is not possible~\cite{Eisert2002}, a statement that can also be said about error correction~\cite{Niset2009,Namiki2014}. Despite this, many entanglement distillation protocols have been studied in CVs~\cite{Browne2003,Eisert2004,Campbell2013,Seshadreesan2019,Mardani2020}, that use non-Gaussian operations. One of them is photon subtraction, which we discussed in chapter~\ref{sec2}. The same thing happens to CV quantum error correction~\cite{Braunstein1998,Noh2020,Wu2021}, also forced to abandon the realm of Gaussian states. 

All these results are focusing on entanglement as a resource, but we want to investigate the purity as well, its relation with entanglement, and to develop a toolbox of operations that allow us to increase the purity. Naturally, increasing the purity of an entangled state leads to entanglement degradation; gaining local information about two systems implies losing non-local information. Given this statement, we want to investigate what are the limitations for the purification of Gaussian states under Gaussian local operations. 

%\subsection{Gaussian Purification No-Go}
First, we formulate a no-go theorem for the purification of two-mode Gaussian states. We study the generic problem of purifying, using local operations, at least one mode of a noisy Gaussian state that is entangled. Imagine that we apply a Gaussian channel on our initial state, characterized by a covariance matrix $\Sigma$, such that the outcome is 
\begin{equation}
\Sigma'=X\Sigma X^{\intercal} + Y.
\end{equation}
The purification has been successful if the rank of $\Sigma'+i\Omega$ has been reduced and, assuming that $Y$ described uncorrelated noise, this can only happen if:
\begin{enumerate}
\item $X$ is not full rank.
\item Both $X(\Sigma+i\Omega)X^{\intercal}$ and $Y+i\Omega -i X\Omega X^{\intercal}$ have a common kernel.
\end{enumerate}
In the case in which $X$ is not full rank, we see that the resulting state cannot be entangled. Since $X$ describes the action of a local channel, $X=X_{A}\oplus X_{B}$, and then, for it to be rank-reduced, either $X_{A}$ or $X_{B}$ must have a kernel. This implies either $\det X_{A}=0$ or $\det X_{B}=0$. Since we can write
\begin{equation}
\Sigma' = \begin{pmatrix} X_{A}\Sigma_{A}X_{A}^{\intercal}+Y_{A} & X_{A}\varepsilon_{AB}X_{B}^{\intercal} \\ X_{B}\varepsilon_{AB}^{\intercal}X_{A}^{\intercal} & X_{B}\Sigma_{B}X_{B}^{\intercal}+Y_{B} \end{pmatrix},
\end{equation}
the off-diagonal block matrices will satisfy $\det(X_{A}\varepsilon_{AB}X_{B}^{\intercal})=0$, which implies the state described by $\Sigma'$ is separable. Therefore, we conclude that purification of a single mode of a two-mode Gaussian entangled state cannot happen by means of local operations without the loss of entanglement, i.e. the projection onto a separable state. 

%%%%%%%%%%%%%%%%%%%%%%%%%%%%%%%%%%%%%%%%%%%%%%%%

\subsection{Partial Purification}
Now that we know that, by means of local operations, we cannot purify a mode on a mixed state without completely degrading entanglement, we want to address the question of whether we can reduce the thermal noise of a given Gaussian state. That is, we want to reduce the smallest eigenvalue of $\Sigma+i\Omega$, or alternatively, increase the purity $\mu = 1/\sqrt{\det\Sigma}$. Assuming we start from a TMST state with
\begin{equation}
\Sigma = \begin{pmatrix} c\mathbb{1}_{2} & s \sigma_{z} \\ s \sigma_{z} & c\mathbb{1}_{2} \end{pmatrix},
\end{equation}
where we have defined $c=(1+2n)\cosh2r$ and $s=(1+2n)\sinh2r$, we can obtain a final Gaussian state
\begin{equation}
\Sigma' = \begin{pmatrix} c'\mathbb{1}_{2} & s' \sigma_{z} \\ s' \sigma_{z} & c'\mathbb{1}_{2} \end{pmatrix},
\end{equation}
with $n'<n$ and $r'<r$. This is done by applying a local Gaussian channel characterized by $\Sigma'=X\Sigma X^{\intercal}$, with $X=\sqrt{\tau}\mathbb{1}_{4}$ and $Y=0$. This represents an attenuation channel, since $\tau=s'/s=c'/c$, and then $0<\tau<1$. However, notice that with this transformation, the symplectic eigenvalue is reduced, $\tilde{\nu}_{-}\rightarrow\tau^{2}\tilde{\nu}_{-}$, which results in higher entanglement, while the purity is increased, $\mu\rightarrow\mu/\tau^{2}$.

Of course, this type of operation is not a Gaussian channel, since $X$ alone does not represent a symplectic transformation. Having set $Y=0$, we need $V=0$, since $Y=V\Sigma_{\text{E}}V^{\intercal}$ and $\Sigma_{\text{E}}$ cannot be zero. Then, the symplectic condition $X\Omega X^{\intercal}+V\Omega V^{\intercal} = \Omega$ is only met in the case of $\tau=1$, in which the application of $X$ represents the evolution of an isolated system.

If we consider a general Gaussian quantum channel, with $Y\neq 0$, taking into account that $\det X + \det V = 1$, and imposing that both $X$ and $V$ must be invertible, we end up with an attenuation channel. This is imposed by $X=\sqrt{\tau}\mathbb{1}_{4}$ and $Y=(1-\tau)m\mathbb{1}_{4}$, where we have set $V = \sqrt{1-\tau}\mathbb{1}_{4}$ and $\Sigma_{\text{E}} = m\mathbb{1}_{4}$. Here, $m=1+2n_{\text{E}}$, and $n_{\text{E}}$ is the number of thermal photons in the environment. Then, the resulting state is characterized by 
\begin{equation}
\Sigma' = \begin{pmatrix} [\tau c + (1-\tau)m] \mathbb{1}_{2} & \tau s \sigma_{z} \\ \tau s \sigma_{z} & [\tau c + (1-\tau)m] \mathbb{1}_{2} \end{pmatrix}.
\end{equation}
If we impose a form for this state, as we had done in the previous case, such that
\begin{equation}
\Sigma' = \begin{pmatrix} c'\mathbb{1}_{2} & s' \sigma_{z} \\ s' \sigma_{z} & c'\mathbb{1}_{2} \end{pmatrix},
\end{equation}
the first condition we find is that $\tau=s'/s$. This leads to a condition on the state of the environment, 
\begin{equation}
m = \frac{sc'-cs'}{s-s'},
\end{equation}
and since we know that $n_{\text{E}}\geq0$, and thus $m\geq 1$, we have that $sc'-cs' \geq s-s'$. This can be reduced to $(c'-1)/(c-1)\geq s'/s$, which is
\begin{equation}
\frac{(1+2n')\cosh2r'-1}{(1+2n)\cosh2r-1} \geq \frac{(1+2n')\sinh2r'}{(1+2n)\sinh2r}.
\end{equation}
The symplectic eigenvalue of the partially-transposed covariance matrix is
\begin{equation}
\tilde{\nu}_{-} = \tau(c-s) + (1-\tau)m = \tau\tilde{\nu}_{-}^{(0)}+(1-\tau)\tilde{\nu}_{-}^{\text{E}},
\end{equation}
with the partially-transposed symplectic eigenvalues of a TMST state and a thermal state are given by $\tilde{\nu}_{-}^{(0)}=(1+2n)e^{-2r}$ and $\tilde{\nu}_{-}^{\text{E}} = m$, respectively. The resulting state is entangled if $\tilde{\nu}_{-}<1$, which imposes the condition
\begin{equation}
s' > \frac{(m-1)s}{m-(c-s)}.
\end{equation}
If we connect this with the previous one, we can get
\begin{equation}
c' > 1 + \frac{(m-1)(c-1)}{m-(c-s)}.
\end{equation}
Furthermore, the purity of this state is given by
\begin{equation}
\mu = \left\{ \left[ \tau c + (1-\tau)m\right]^{2} - \tau^{2}s^{2} \right\}^{-1},
\end{equation}
which is increased if $c^{2}-s^{2} > \left[ \tau c + (1-\tau)m\right]^{2} - \tau^{2}s^{2}$. If we rearrange this, we can obtain
\begin{equation}
c^{2}-s^{2}-m^{2} > \tau(s-c+m)(s+c-m),
\end{equation}
and using $\tau(s-c+m)>m-1$ from the partially-transposed symplectic eigenvalue, we arrive at
\begin{equation}
m < (c+s)\left(\frac{c-s+1}{c+s+1}\right) = (1+2n)e^{-2r}\left( \frac{1+2n + e^{2r}}{1+2n + e^{-2r}}\right).
\end{equation}
Writing $m=1+2n_{\text{E}}$, this can be rearranged into
\begin{equation}
n_{\text{E}} < \frac{2n(1+n)}{1+(1+2n)e^{2r}},
\end{equation}
which imposes a more restrictive condition than just $n_{\text{E}}<n$. This implies that purification can only happen if the purity of the environment is higher than that of the initial TMST state. 

\subsubsection{Partial purification with ideal measurements}
We now consider that we have access to making measurements on the environment, and attempt to obtain information from them. Our Gaussian measurement is described by a positive operator $\Pi_{i}$ taken from a POVM, with covariance matrix $\Upsilon_{i}$. If we project this onto a partition of modes labelled by 2 of a bipartite Gaussian state with covariance matrix 
\begin{equation}
\Sigma = \begin{pmatrix} \Sigma_{1} & \varepsilon_{12} \\ \varepsilon_{12}^{\intercal} & \Sigma_{2} \end{pmatrix},
\end{equation}
the covariance matrix of the resulting state of the partition of modes labelled by 1 is
\begin{equation}
\tilde{\Sigma}_{1} = \Sigma_{1} - \varepsilon_{12}\left( \Sigma_{2} + \Upsilon_{i} \right)^{-1}\varepsilon_{12}^{\intercal}.
\end{equation}
Recall that homodyne measurements of the x-quadrature are characterized by
\begin{equation}
\Upsilon_{i} = \lim_{\xi\rightarrow\infty}\begin{pmatrix} e^{-2\xi} & 0 \\ 0 & e^{2\xi} \end{pmatrix},
\end{equation}
by setting $\varphi=0$, or alternatively, of the p-quadrature with $\varphi=\pi$,
\begin{equation}
\Upsilon_{i} = \lim_{\xi\rightarrow\infty}\begin{pmatrix} e^{2\xi} & 0 \\ 0 & e^{-2\xi} \end{pmatrix}.
\end{equation}
Another way to describe the action of homodyne measurements on the covariance matrix of the remaining state is to write
\begin{equation}
\tilde{\Sigma}_{1} = \Sigma_{1} - \varepsilon_{12}\left[ \begin{pmatrix} 1 & 0 \\ 0 & 0 \end{pmatrix} \Sigma_{2} \begin{pmatrix} 1 & 0 \\ 0 & 0 \end{pmatrix} \right]^{\text{MP}}\varepsilon_{12}^{\intercal}
\end{equation}
for the x-quadrature, where $\text{MP}$ indicates the pseudo inverse operation~\cite{Weedbrook2012}, since the resulting matrix is singular. For the p-quadrature, we just compute
\begin{equation}
\tilde{\Sigma}_{1} = \Sigma_{1} - \varepsilon_{12}\left[ \begin{pmatrix} 0 & 0 \\ 0 & 1 \end{pmatrix} \Sigma_{2} \begin{pmatrix} 0 & 0 \\ 0 & 1 \end{pmatrix} \right]^{\text{MP}}\varepsilon_{12}^{\intercal}.
\end{equation}
For x-quadrature homodyne measurements on the environment, we obtain the purity of the remaining state
\begin{equation}
\mu = \frac{1}{m}\sqrt{\frac{(1-\tau)(c+s)+m\tau}{(c+s)(\tau(c+s)+m(1-\tau))}}\times\sqrt{\frac{(1-\tau)(c-s)+m\tau}{(c-s)(\tau(c-s)+m(1-\tau))}}.
\end{equation}
For the partially-transposed symplectic eigenvalue, we get
\begin{equation}
\tilde{\nu}_{-} = \sqrt{m(c-s)\frac{\tau(c-s) + m(1-\tau)}{(1-\tau)(c-s)+m\tau}}.
\end{equation}
Alternatively, if we measure orthogonal quadratures in different modes of the environment through homodyne detection (let's say, x in the first and p in the second), we obtain
\begin{equation}
\mu = \frac{(1-\tau)c+m\tau}{m[m(1-\tau)c+\tau(c^{2}-s^{2})]},
\end{equation}
and for the partially-transposed symplectic eigenvalue
\begin{equation}
\tilde{\nu}_{-} = \frac{\sqrt{m c [((m-c)^{2}-s^{2})(1-\tau)\tau + m c]}- m\tau s}{(1-\tau)c+m\tau}.
\end{equation}
This kind of measurements are often referred to as ``double-homodyne''. Heterodyne detection is constructed through double-homodyne measurements with an auxiliary system in a coherent state. The covariance matrix associated to heterodyne measurements can be obtained for $\xi=0$, $\Upsilon_{i} = \mathbb{1}_{2}$. Equivalently, it corresponds to projecting onto a given coherent state. We compute the purity of the remaining state after this type of measurements,
\begin{equation}
\mu = \left[\frac{(1-\tau)(c+s)+1+m\tau}{(m+\tau)(c+s)+m(1-\tau)}\right] \times \left[\frac{(1-\tau)(c-s)+1+m\tau}{(m+\tau)(c-s)+m(1-\tau)}\right],
\end{equation}
as well as the symplectic eigenvalue of the partial transposition,
\begin{equation}
\tilde{\nu}_{-} = \frac{(m+\tau)(c-s)+m(1-\tau)}{(1-\tau)(c-s)+1+m\tau}.
\end{equation}
We observe in all cases that the purity of the resulting state is larger than that of the initial state, provided that $m^{2}<c^{2}-s^{2}$, meaning that the purity of the environment has to be larger than that of the initial state, as we saw earlier. We also see that the remaining state is entangled if
\begin{equation}
\tau > \frac{1}{2}\left( 1 - \frac{1+m(c-s)}{m-c+s}\right).
\end{equation}
Naturally, however, the negativity of the state after measuring the environment can never be larger than that of the initial state. 

The improvement shown by the application of ideal measurements is caused by the fact that these represent a projection of the state of the environment onto a pure Gaussian state, which naturally purifies the corresponding system coupled to it.

%%%%%%%%%%%%%%%%%%%%%%%%%%%%%%%%%%%%%%%%%%%%%%%%
\subsubsection{Quantum Fisher information}
In order to test the efficiency of these purification schemes, we need to use a metric; we need to find a function that characterizes how useful the resulting states are for a given protocol. In the following, we will use the quantum Fisher information to test how the partially-purified states perform in a typical quantum illumination problem.

Quantum illumination (QI)~\cite{Lloyd2008} is a quantum sensing protocol in which one attempts to detect a low-reflectivity object using two entangled quantum modes. One mode is sent at the object, and it is either reflected or lost. The other mode is kept in the laboratory, and measured jointly with the part of the signal that arrives after being reflected. Naturally, this can be attained using Gaussian quantum states~\cite{Tan2008}. The performance of this protocol is generally evaluated using the Fisher information. 

% These are about microwave quantum illumination
%\cite{Barzanjeh2015,Barzanjeh2020}

The classical Fisher information (CFI) is a distinguishability measure between two probability distributions, related to the maximum knowledge that can be obtained about a given parameter that characterizes a random probability distribution. It's quantum correspondent, the quantum Fisher information (QFI)~\cite{Braunstein1994,Paris2009}, is formulated similarly, in the context of density matrices and observables, and is related to the quantum Cram\'{e}r-Rao bound~\cite{Cramer1946,Rao1992}, $\text{var}(\hat{O}_{\epsilon})\geq 1/MH(\epsilon)$. This bounds the error in estimating a parameter $\epsilon$, given the QFI $H(\epsilon)$, an optimal observable $\hat{O}$, and a number of repetitions $M$. 

We will use it to estimate the performance of our states to be used to detect a low-reflectivity object. Starting with a bipartite entangled state, one of the modes is sent into open-air to detect the presence of a highly-transparent object, with which it interacts. The reflected photons are detected, and a combined measurement with the mode kept in the laboratory is performed. Our guess is that, in certain regimes, the partially-purified states will present higher QFI than the bare states. This is true in the case of TMST states, for which the QFI (in this type of problems) increases monotonically with the purity. 

For Gaussian quantum states, the QFI can be computed using only the displacement vector and the covariance matrix. The QFI for a bipartite Gaussian quantum state with covariance matrix $\Sigma$, related to a parameter $\epsilon$, is given by~\cite{Safranek2015}
\begin{eqnarray}
\nonumber H(\epsilon) &=& \frac{1}{2(\det E-1)}\bigg\{ \left(\det E\right) \tr\left[\left(E^{-1}\dot{E}\right)^{2}\right] + 4(\nu_{+}^{2}-\nu_{-}^{2})\left( -\frac{\dot{\nu}_{+}^{2}}{\nu_{+}^{4}-1} + \frac{\dot{\nu}_{-}^{2}}{\nu_{-}^{4}-1}\right) \\
&+& \sqrt{\det\left(\mathbb{1}+E^{2}\right)} \tr\left[\left(\left(\mathbb{1}+E^{2}\right)^{-1}\dot{E}\right)^{2}\right] \bigg\} +2\dot{\vec{d}}^{\dagger} \Sigma^{-1}\dot{\vec{d}},
\end{eqnarray}
where $\dot{\vec{d}}$, $\dot{E}$ indicate the element-wise derivatives of $\vec{d}$, $E$ with respect to $\epsilon$, respectively.  Here, $\nu_{\pm}$ are the symplectic eigenvalues of $\Sigma$, which can also be obtained as $2\nu_{\pm} =\sqrt{\tr E^{2} \pm \sqrt{(\tr E^{2})^{2}-16 \det E}}$. We have also defined $E = iT\Omega\Sigma T^{\intercal}$, with
\begin{equation}
T = \begin{pmatrix} 1 & 0 & 0 & 0 \\ 0 & 0 & 1 & 0 \\ 0 & 1 & 0 & 0 \\ 0 & 0 & 0 & 1 \end{pmatrix}, \quad
\Omega = \begin{pmatrix} 0 & 1 & 0 & 0 \\ -1 & 0 & 0 & 0 \\ 0 & 0 & 0 & 1 \\ 0 & 0 & -1 & 0 \end{pmatrix}.
\end{equation}
Here, $\vec{d}$ indicates the displacement vector. Provided that we are in the context of QI, $\epsilon$ plays the role of the reflectivity of the object.

In order to provide a fair comparison between the initial and the partially-purified states, we need to equate the amount of resources used. In a quantum illumination context, this means setting the same number of photons. For the task of detecting low-reflectivity objects, we can claim that one scheme is better than another if we make sure they are using the same number of photons. More specifically, if we are shining the object with a single mode, we need to fix the quantity corresponding to the number of photons in that mode times the number of repetitions, for a fair comparison. 

To claim a quantum advantage in QI, we compute the ratio of the QFI and the CFI. The typical case studied in CVs is a TMSV state versus a single-mode coherent state, which plays the role of a classical signal. The number of signal photons, corresponding to the mode of the TMSV state that is sent at the object, is given by $N_{s}=\sinh^{2} r$, and we define $N_{\text{th}}$ as the average number of thermal photons in the environment through which the signal mode propagates. Then, we obtain
\begin{equation}
\lim_{\epsilon\rightarrow0}\frac{H_{Q}(\epsilon)}{H_{C}(\epsilon)} = 1+\frac{N_{\text{th}}}{1+N_{s}+N_{\text{th}}+2N_{s}N_{\text{th}}},
\end{equation}
in the limit of low reflectivity. This quantity is always greater than 1, and it indicates a higher advantage of the quantum strategy in the range of low signal-to-noise ratio. 

We can compute the same metric for a general two-mode Gaussian quantum state, characterized by a null displacement vector and a covariance matrix given in normal form, as in Eq.~\eqref{CM_NF}. Then, the ratio between the quantum and the classical Fisher information, in the limit of low reflectivity, considering that we are probing the object with mode A, is given by
\begin{equation}
\lim_{\epsilon\rightarrow0}\frac{H_{Q}(\epsilon)}{H_{C}(\epsilon)} = \frac{(1+2N_{\text{th}})[2c_{1}c_{2}+b(1+2N_{\text{th}})(c_{1}^{2}+c_{2}^{2})]}{4\alpha^{2}[b^{2}(1+2N_{\text{th}})^{2}-1]}.
\end{equation}
Here, we have considered a coherent state with $\alpha\in\mathbb{R}$, therefore, $\alpha^{2}$ indicates the number of photons in the coherent state. If we wanted to use the same number of photons, then we could replace $\alpha^{2}$ by $(a-1)/2 = N_{s}$, which corresponds to the number of signal photons from the Gaussian quantum state that arrive at the object.

%Entanglement plays a crucial role in the virtue of QI; however, the purity of the states is also relevant. As we have seen, the advantage of using TMSV states grows with the signal to noise ratio; when dealing with mixed states, less signal photons can mean higher purity, while also implying lower entanglement. 

Nevertheless, in our case, we want to compare the performance of the states before and after partial purification. We will take into account the number of repetitions, and equate $M_{2}\langle n_{2}\rangle = M_{1}\langle n_{1}\rangle$. Then, we want to look at the ratio
\begin{equation}\label{eq:efficiency}
\lim_{\epsilon\rightarrow0}\frac{M_{2}H_{2}(\epsilon)}{M_{1}H_{1}(\epsilon)}=\lim_{\epsilon\rightarrow0}\frac{\langle n_{1}\rangle H_{2}(\epsilon)}{\langle n_{2}\rangle H_{1}(\epsilon)},
\end{equation}
which we want to identify as the efficiency. This is chosen because of the Cram\'{e}r-Rao bound, and if the ratio given before is larger than 1, the we would have reduced the bound with the partially-purified states, meaning that a lower estimation error could be achieved. 

\begin{figure}[t]
\centering
\includegraphics[width=\textwidth]{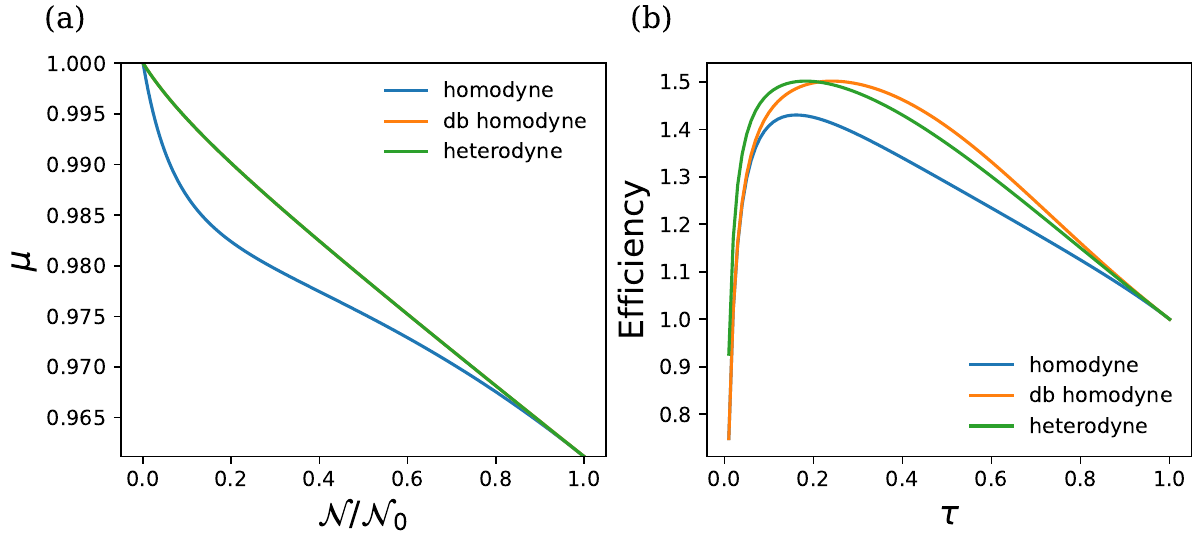}
\caption[Partial purification of a TMST state in a pure-loss channel with meaasurement]{\textbf{Partial purification of a TMST state in a pure-loss channel with meaasurement}. We take a TMST state with $n=10^{-2}$ and $r=1$, assuming the environment is in a vacuum state ($m=1$). The blue, orange and green curves correspond to the result of homodyne, double homodyne, and heterodyne ideal measurements, respectively, performed on the environment. (a) Purity versus negativity of the output state, represented for attenuation coefficients $\tau$ between 0 and 1. We notice that, while the negativity of the output state cannot be increased with respect to that of the initial state, the purity can be improved by homodyne and heterodyne measurements. (b) Efficiency of the purification scheme versus $\tau$. We can observe a region of small $\tau$ were there is no advantage, but in the remaining region partial purification with measurements can reduce the estimation error, compared with the initial TMST states.}
\label{fig3_1}
\end{figure}

In Fig.~\ref{fig3_1}, we represent the purity versus the negativity for the output state of an attenuation channel, for $\tau\in[0,1]$. We have considered an initial TMST state with $n=10^{-2}$ thermal photons and squeezing parameter $r=1$, and an environment in the vacuum state ($m=1$). The blue, orange, and green curves correspond to performing homodyne, double homodyne, and heterodyne measurements, respectively. In Fig.~\ref{fig3_1}~(a), we represent the purity versus the negativity of the output state, normalized by dividing it by the negativity of the initial state. We can observe how there an almost straight line that delimits the region of allowed partial purification with the available entanglement. Fig.~\ref{fig3_1}~(b) shows the efficiency of the purification scheme, given by the ratio in Eq.~\eqref{eq:efficiency}. All three instances here have equal amount of resources. We can observe that double homodyne and heterodyne measurements perform better than homodyne ones, reaching a maximum efficiency of 1.5. This means that, with the resulting partially-purified states, the error for estimating a given observable using quantum illumination is reduced by 2/3. 

%%%%%%%%%%%%%%%%%%%%%%%%%%%%%%%%%
%%%%%%%%%%%%%%%%%%%%%%%%%%%%%%%%%

\subsection{Partial Purification with Two Copies}
Let us now investigate the case in which two copies of the same state are used to increase the purity of the output state, by means of local Gaussian operations. We will consider states that are initially entangled, and will require that the output state is not completely separable. 

\subsubsection{Swapping-like protocol}
\begin{figure}[t]
{\includegraphics[width=\textwidth]{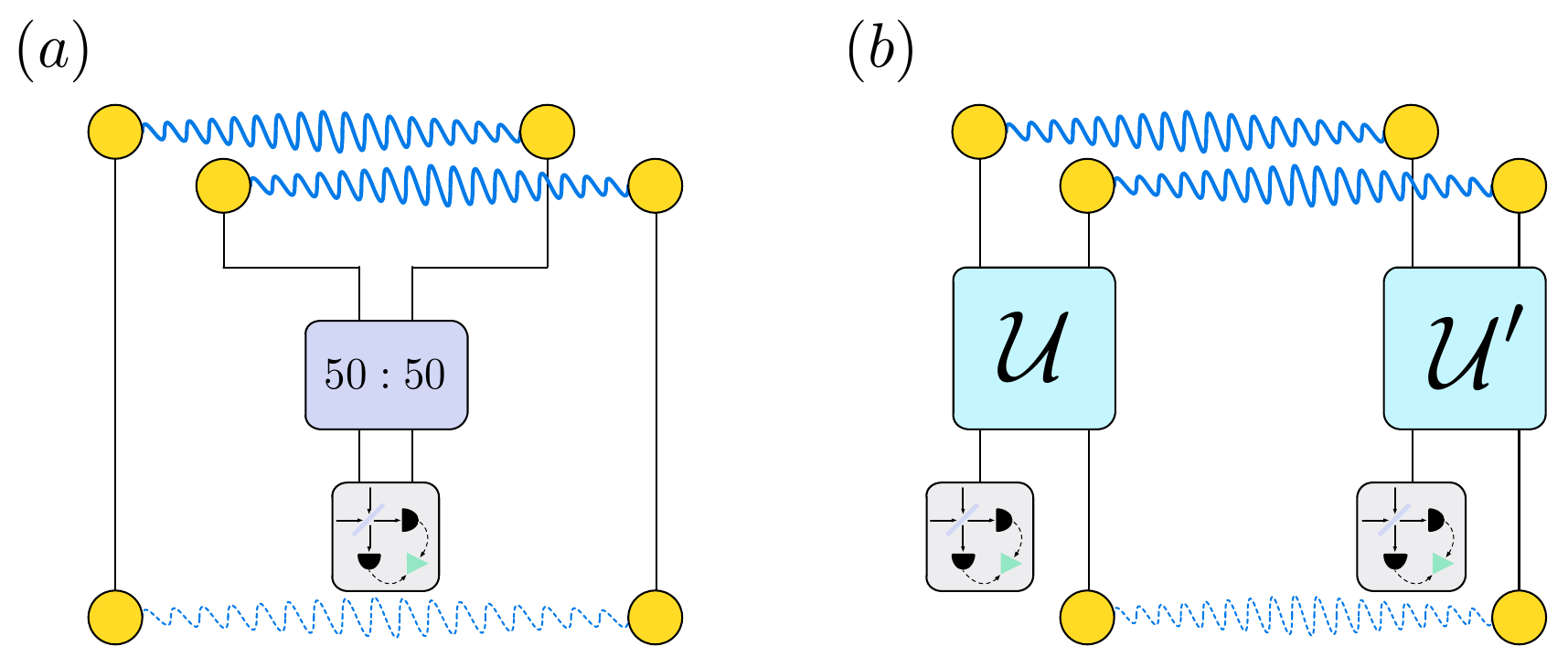}}
\caption[Partial-purification schemes that use two copies of a TMST state]{\textbf{Partial-purification schemes that use two copies of a TMST state}. (a) We represent a swapping-like protocol, in which two uncorrelated modes, one held by Alice and the other by Bob, are combined in a 50:50 beam splitter. Then, both modes are subject to Gaussian measurements. (b) We illustrate a distillation-inspired protocol, where Alice and Bob apply locally two-mode Gaussian operations ($\mathcal{U}$, $\mathcal{U}'$) that resemble the CV-equivalent of CNOT gates. Then, each party measures one of their modes. }
\label{fig3_2}
\end{figure}

We consider a protocol in which one mode of each copy of the state is combined in a beam splitter with transmissivity $\tau$, by a party that is located between Alice and Bob. These two modes are taken as the environment, and a Gaussian measurement is applied to each one. The corresponding scheme is depicted in Fig.~\ref{fig3_2}~(a). We will choose measurements that have an orthogonal representation in phase space by setting $\varphi_{2}=\varphi_{1}+\pi=\varphi+\pi$, while we will set $\xi_{1}=\xi_{2}=\xi$. If measurements are not taken as orthogonal in phase space, the resulting state will be separable. 

%%%%%%%%%%%%%%%%%%%%%%%%%%%%%%%%%%%%
%%%%%%%%%%%%%%%%%%%%%%%%%%%%%%%%%%%%

By considering $\tau=1/2$, we obtain
\begin{equation}
\mu = \frac{1}{c^{2}-s^{2}-\frac{s^{2}(c^{2}-s^{2}-1)}{1+c^{2}+2c\cosh2\xi}}
\end{equation}
and
\begin{equation}
\tilde{\nu}_{-} = \frac{c+e^{2\xi}(c^{2}-s^{2})}{1+ce^{2\xi}}.
\end{equation}
In the case $\xi\rightarrow\infty$, these represent the result of double-homodyne measurements, and we reproduce the results of an entanglement-swapping operation. In this scenario, the purity remains the same, but entanglement is reduced. Notice that we need a finite $\xi$ to reduce the purity of the output state. This protocol effectively teleports entanglement to the remaining, also previously-uncorrelated modes, held by Alice and Bob. 

\begin{figure}[t]
{\includegraphics[width=0.75\textwidth]{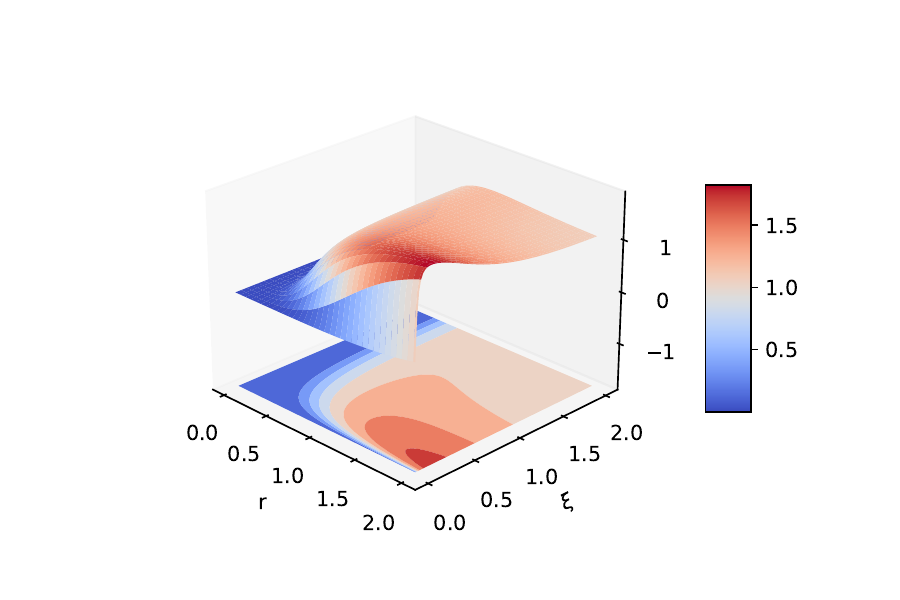}}
\caption[Efficiency of a two-copy purification scheme, based on entanglement swapping, of an initial TMST state]{\textbf{Efficiency of a two-copy purification scheme, based on entanglement swapping, of an initial TMST state} with $n=10^{-2}$, represented against the squeezing parameter of the initial state and the squeezing associated to the Gaussian projective measurements. The latter are taken orthogonal in their phase-space representation, as double homodyne with finite squeezing. We have also considered $\tau=1/2$, and $\varphi = 0$.}
\label{fig3_3}
\end{figure}

In Fig.~\ref{fig3_3}, we represent the efficiency for different values of the squeezing of the initial TMST state, and for different squeezing parameters of the Gaussian projective measurement. We observe that the maximum efficiency happens for small $\xi$, and it increases for larger squeezing, which is natural, since we are fixing the initial number of thermal photons. For $r=1$, the maximum efficiency is 1.40, which is smaller than the maximum values obtained in the single-copy case, represented in Fig.~\ref{fig3_1}. Nevertheless, in that case we were assuming that the environment was in a pure state, while now we are considering two mixed states. 

%%%%%%%%%%%%%%%%%%%%%%%%%%%%%%%%%%%%
%%%%%%%%%%%%%%%%%%%%%%%%%%%%%%%%%%%%

%Let us look at another two-copy protocol that can give an advantage.

\subsubsection{Distillation-inspired protocol}
We envision the case in which Alice and Bob share two copies of an entangled state, and each decide to combine their modes of such states by applying some unitary operation. In the discrete-variable formalism, by choosing this operation to be the CNOT gate, bit-flip errors could be corrected~\cite{Bennett1996,Dur2007}. Here, we will apply the continuous-variable equivalent~\cite{Wang2001} of such gate, which effectively copies the information of one quadrature into another,
\begin{eqnarray}
\nonumber \mathcal{U}|x_{1},x_{2}\rangle &=& |x_{1}\mp x_{2},x_{2}\rangle, \\
\mathcal{U}|p_{1},p_{2}\rangle &=& |p_{1},p_{2}\pm p_{1}\rangle.
\end{eqnarray}
Switching to the symplectic convention, we will identify
\begin{eqnarray}
\nonumber \mathcal{U} \begin{pmatrix} x_{1} \\ p_{1} \\ x_{2} \\ p_{2} \end{pmatrix} &=& \begin{pmatrix} 1 & 0 & -\omega & 0 \\ 0 & 1 & 0 & 0 \\ 0 & 0 & 1 & 0 \\ 0 & \omega & 0 & 1 \end{pmatrix} \begin{pmatrix} x_{1} \\ p_{1} \\ x_{2} \\ p_{2} \end{pmatrix} =  \begin{pmatrix} x_{1} - \omega x_{2} \\ p_{1} \\ x_{2} \\ \omega p_{1} + p_{2} \end{pmatrix}, \\
\mathcal{U}' \begin{pmatrix} x_{1} \\ p_{1} \\ x_{2} \\ p_{2} \end{pmatrix} &=& \begin{pmatrix} 1 & 0 & 0 & 0 \\ 0 & 1 & 0 & -\omega \\ \omega & 0 & 1 & 0 \\ 0 & 0 & 0 & 1 \end{pmatrix} \begin{pmatrix} x_{1} \\ p_{1} \\ x_{2} \\ p_{2} \end{pmatrix} =  \begin{pmatrix} x_{1} \\ p_{1} - \omega p_{2} \\ \omega x_{1} + x_{2} \\ p_{2} \end{pmatrix}.
\end{eqnarray}
Below, we give the explicit combination of operations that give rise to the CNOT-like gate we use in this section. This gate is actually known as the quantum nondemolition gate~\cite{Filip2005,Yoshikawa2008}, and we can identify several variations of it through the operators
\begin{eqnarray}
\nonumber && \mathcal{U} = \begin{pmatrix} 1 & 0 & -\omega & 0 \\ 0 & 1 & 0 & 0 \\ 0 & 0 & 1 & 0 \\ 0 & \omega & 0 & 1 \end{pmatrix}, \quad \mathcal{U}^{\intercal} = \begin{pmatrix} 1 & 0 & 0 & 0 \\ 0 & 1 & 0 & \omega \\ -\omega & 0 & 1 & 0 \\ 0 & 0 & 0 & 1 \end{pmatrix}, \\
&& \mathcal{U}' = \begin{pmatrix} 1 & 0 & 0 & 0 \\ 0 & 1 & 0 & -\omega \\ \omega & 0 & 1 & 0 \\ 0 & 0 & 0 & 1 \end{pmatrix}, \quad \mathcal{U}'^{\intercal} = \begin{pmatrix} 1 & 0 & \omega & 0 \\ 0 & 1 & 0 & 0\\ 0 & 0 & 1 & 0 \\ 0 & -\omega & 0 & 1 \end{pmatrix}.
\end{eqnarray}
The essence of this transformation lies in two beam splitter transformations, interrupted by a squeezing operation. Consider the beam splitter operator 
\begin{equation}
\mathcal{B}(\tau,\phi) = \begin{pmatrix} \sqrt{\tau}\mathbb{1}_{2}& \sqrt{1-\tau}R(\phi) \\ -\sqrt{1-\tau}R^{\intercal}(\phi) & \sqrt{\tau}\mathbb{1}_{2}  \end{pmatrix},
\end{equation}
where we have defined the phase-shifter operator
\begin{equation}
R(\phi) = \begin{pmatrix} \cos\phi & -\sin\phi \\ \sin\phi & \cos\phi \end{pmatrix},
\end{equation}
and the single-mode squeezing operator
\begin{equation}
\mathcal{S}(\xi,\varphi) = \begin{pmatrix} \cosh\xi - \sinh\xi\cos\varphi & -\sinh\xi\sin\varphi \\ -\sinh\xi\sin\varphi & \cosh\xi + \sinh\xi\cos\varphi \end{pmatrix}.
\end{equation}
By identifying $\omega = 2\sinh\xi$, we can express
\begin{equation}
\mathcal{U}_{i} = \mathcal{B}(\tau_{i},\phi_{i})\begin{pmatrix} \mathcal{S}(\xi,0) & 0 \\ 0 & \mathcal{S}(\xi,\pi) \end{pmatrix} \mathcal{B}^{\intercal}(1-\tau_{i},\phi_{i}),
\end{equation}
and characterize each gate by the parameters
\begin{eqnarray}
\nonumber \mathcal{U} \, &\longrightarrow& \,  \tau = \frac{1}{1+e^{2\xi}}\, , \, \phi = \pi, \\
\nonumber \mathcal{U}^{\intercal} \, &\longrightarrow& \,  \tau = \frac{1}{1+e^{-2\xi}}\, , \, \phi = \pi, \\
\nonumber \mathcal{U}' \, &\longrightarrow& \,  \tau = \frac{1}{1+e^{-2\xi}}\, , \, \phi = 0, \\
\mathcal{U}'^{\intercal} \, &\longrightarrow& \,  \tau = \frac{1}{1+e^{2\xi}} \, , \, \phi = 0.
\end{eqnarray}
With this toolbox at hand, we can devise a distillation-inspired purification scheme for thermal states. These operations generate entanglement, and by making use of that resource, together with heterodyne measurements, we will be able to distill purity in the outcome states. Starting from thermal states with covariance matrix $\Sigma_{\text{th}} = m\mathbb{1}_{2}$, we can apply these operations onto two copies, such that
\begin{equation}
\mathcal{U}\begin{pmatrix} \Sigma_{\text{th}} & 0 \\ 0 & \Sigma_{\text{th}} \end{pmatrix} \mathcal{U}^{\intercal} =  m\mathcal{U}\mathcal{U}^{\intercal} =  m\begin{pmatrix} 1+\omega^{2} & 0 & -\omega & 0 \\ 0 & 1 & 0 & \omega \\ -\omega & 0 & 1 & 0 \\ 0 & \omega & 0 & 1+\omega^{2} \end{pmatrix}.
\end{equation}
By considering heterodyne measurements on one of the modes of the resulting state, we get the covariance matrix for the remaining state
\begin{equation}
m \begin{pmatrix} 1 + \frac{\omega^{2}}{1+m} & 0 \\ 0 & \frac{1+m}{1+m(1+\omega^{2})} \end{pmatrix}.
\end{equation}
The purity is given by
\begin{equation}
\mu = \frac{1}{m}\sqrt{\frac{1+m(1+\omega^{2})}{1+m+\omega^{2}}},
\end{equation}
and this is larger that the initial purity of $1/m$ for $\omega\in\mathbb{R}$. If we take the limit of diverging $\omega$, we find
\begin{equation}
\lim_{\omega\rightarrow\infty} \mu = \frac{1}{\sqrt{m}},
\end{equation}
a significative improvement in the purity of the thermal state. If we perform $k$ rounds of purification and use $2^{k}$ copies of the same thermal state, in the limit $\omega\rightarrow\infty$, we end up with the purity
\begin{equation}
\mu = \sqrt{\frac{k(1+m)-m}{k(1+m)-1}} = \sqrt{1 - \frac{m-1}{k(1+m)-1}},
\end{equation}
which tends to 1 as $k$ goes to infinity. 

Let us move on to the case of partial purification of entangled states. We assume that Alice applies $\mathcal{U}$ on modes 1 and 3, and Bob applies $\mathcal{U}'$ on modes 2 and 4. Then, we consider that Alice projects onto a Gaussian state with squeezing $\xi$ in the x direction, whereas Bob does the same in the p direction. The purity of the resulting state is
\begin{equation}
\mu = \frac{1}{c^{2}-s^{2}-\frac{c^{2}\omega^{2}(c^{2}-s^{2}-1)}{1+c^{2}-s^{2}+2c\cosh2\xi+c\omega^{2}(c+e^{-2\xi})}}.
\end{equation}
Given that the second term in the denominator is positive, the purity can increase. The symplectic eigenvalue of the partial transposition is given by
%\begin{eqnarray}\label{sym_eig_2copy_xp}
%\nonumber \tilde{\nu}_{-} &=& \frac{c e^{2\xi}\sqrt{\left[(1+c^{2}-s^{2})(1+\omega^{2})+c(e^{2\xi}+e^{-2\xi}(1+\omega^{2})^{2}) \right](1+c^{2}-s^{2}+2c\cosh2\xi)}}{c e^{4\xi}+c(1+\omega^{2})+e^{2\xi}(1+c^{2}(1+\omega^{2})-s^{2})} \\
%&-& \frac{s\left[c e^{4\xi}+c(1+\omega^{2})+e^{2\xi}(1+c^{2}-s^{2})\right]}{c e^{4\xi}+c(1+\omega^{2})+e^{2\xi}(1+c^{2}(1+\omega^{2})-s^{2})}
%\end{eqnarray}
\begin{equation}\label{sym_eig_2copy_xp}
\tilde{\nu}_{-} = \frac{c \sqrt{u\left[u +\omega^{2}(u + c \omega^{2}e^{-2\xi} - 2c\sinh2\xi ) \right]}-s\left( u + c \omega^{2}e^{-2\xi}\right)}{u + c \omega^{2}\left( c+e^{-2\xi}\right)}
\end{equation}
with $u=1+c^{2}-s^{2} + 2c \cosh2\xi$. Through this operation and measurement, the first moments $(x_{1},p_{1})$ and $(x_{2},p_{2})$ of the initial state, corresponding to Alice and to Bob, respectively, transform as follows:
\begin{equation}
\begin{pmatrix} x_{1} \\ p_{1} \\ x_{2} \\ p_{2} \end{pmatrix} \longrightarrow \begin{pmatrix} x_{1} \left( 1 - \frac{\omega}{1+c e^{2\xi}} \right) \\ p_{1}\left( 1 - \frac{c\omega(1+\omega)}{c(1+\omega^{2}) + e^{2\xi}}\right) \\ x_{2}\left( 1 - \frac{c\omega(1+\omega)}{c(1+\omega^{2}) + e^{2\xi}}\right) \\ p_{2} \left( 1 - \frac{\omega}{1+c e^{2\xi}} \right)\end{pmatrix},
\end{equation}
showing that the information of the quadratures of one party is not used by the other.

Furthermore, notice that
\begin{equation}
\lim_{\omega\rightarrow\infty} \mu = \frac{1}{1+\frac{c^{2}-s^{2}-1}{1+c e^{2\xi}}},
\end{equation}
which means that we can completely purify the state with perfect homodyne measurements ($\xi\rightarrow\infty$). Naturally, this will come at the cost of negativity. See that, in this limit, the symplectic eigenvalue of the partial transposition goes to
\begin{equation}
\lim_{\omega\rightarrow\infty} \tilde{\nu}_{-} = \frac{e^{\xi}\sqrt{c(1+c^{2}-s^{2}+2c\cosh2\xi)}-s}{1+c e^{2\xi}},
\end{equation}
which goes to 1 with $\xi\rightarrow\infty$, yielding a separable state. 

\begin{figure}[t]
{\includegraphics[width=\textwidth]{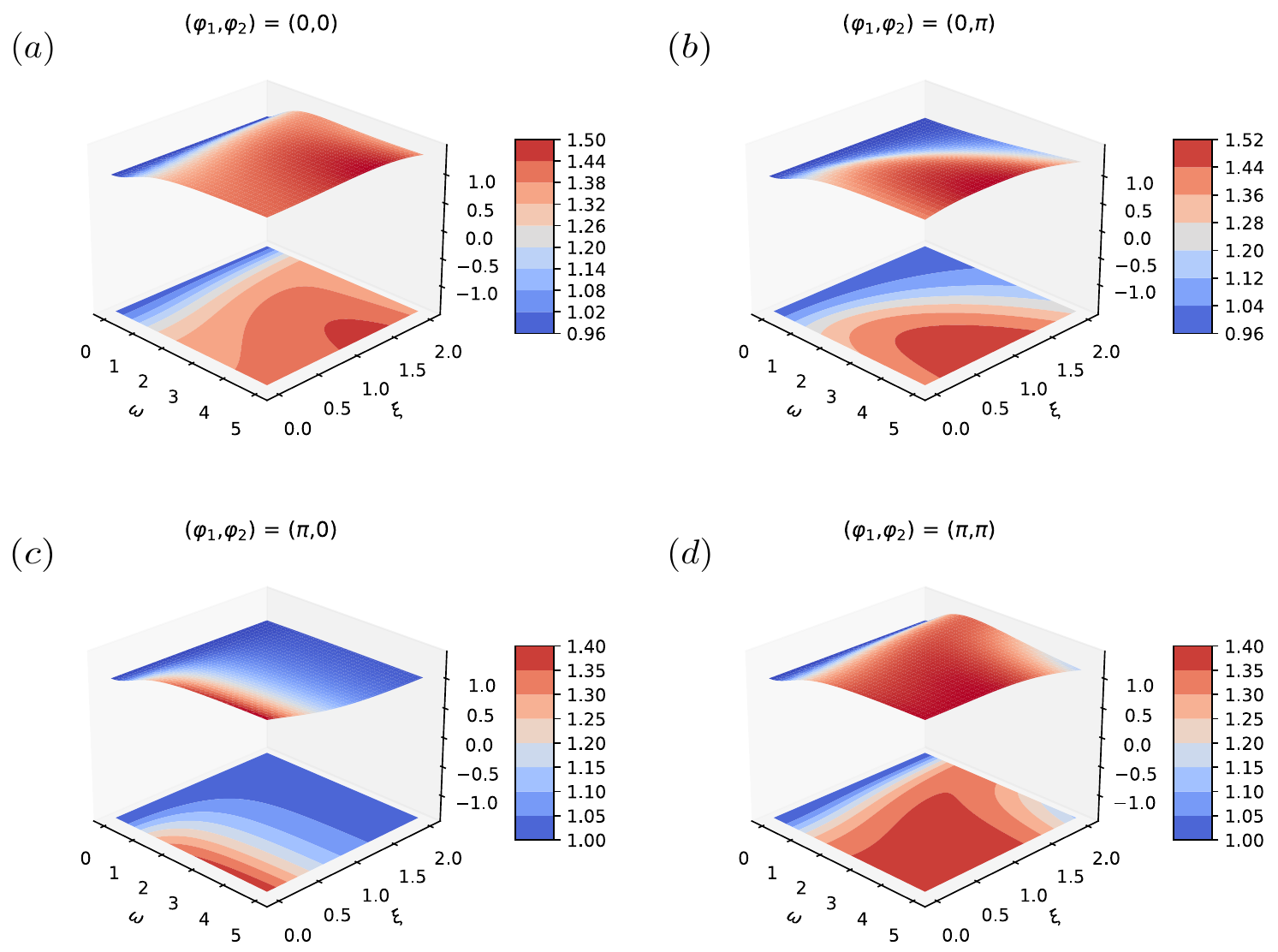}}
\caption[Efficiency of a two-copy distillation-inspired purification scheme of an initial TMST state]{\textbf{Efficiency of a two-copy distillation-inspired purification scheme of an initial TMST state} with $n=10^{-2}$ and $r=1$, represented against $\omega$ and against $\xi$. We consider measurements with specific directions in phase space: (a) Both in the x-direction; (b) First in the x-direction, second in the p-direction; (c) First in the p-direction, second in the x-direction; (d) Both in the p-direction.}
\label{fig3_4}
\end{figure}

The efficiency of this scheme is represented in Fig.~\ref{fig3_4}~(b), for different values of $\omega$ and $\xi$. This is actually the instance that yields a maximum efficiency of 1.5, in the range of parameters represented here. In Fig.~\ref{fig3_4}, we show that the same maximum efficiency can be obtained with other measurement configurations: (a) Both Alice and Bob project onto a Gaussian state with squeezing $\xi$ in the x direction; (b) Alice projects onto a Gaussian state with squeezing $\xi$ in the x direction, whereas Bob does the same in the p direction; (c) Alice projects onto a Gaussian state with squeezing $\xi$ in the p direction, whereas Bob does the same in the x direction; (d) Both Alice and Bob project onto a Gaussian state with squeezing $\xi$ in the p direction. We see that the common measurement scheme that can lead to a maximum efficiency is heterodyne detection ($\xi=0$), which is independent of the phase space direction of the projection. 

We have investigated using two copies of the already purified states (with lower squeezing), which resource-wise would amount to using four copies of the initial state. We found that the maximum efficiency that can be obtained increases barely in cases any of the four cases. Then, we believe that it is not interesting to investigate this protocol assuming that we have infinite copies available. Nonetheless, if our goal is simply to purify states, consuming entanglement as the resource, a multi-copy scenario can lead to interesting results, as we have seen with thermal states. There, we have shown that we can completely purify thermal states in an infinite-copy case, with $\omega\rightarrow\infty$. In this same regime, we observe that just two copies lead to a purity that increases from $1/m$ to $1/\sqrt{m}$, with $m\geq 1$.

%%%%%%%%%%%%%%%%%%%%%%%%%%%%%%%%%%%%

\section[Superconducting Devices for Microwave Quantum Communication]{Superconducting Devices for \\ Microwave Quantum Communication}
\label{sec4}

% \thiswatermark{\put(1,-282){\color{l-grey}\rule{84pt}{88pt}}
% \put(84,-282){\color{grey}\rule{410pt}{88pt}}}
%
%
% %Julen ch 2
% \thiswatermark{\put(1,-308){\color{l-grey}\rule{70pt}{64pt}}
% \put(70,-308){\color{grey}\rule{297pt}{64pt}}}
% \fancyhead[RO]{\leftmark}

%Julen ch 3
\thiswatermark{\put(1,-302){\color{l-grey}\rule{70pt}{60pt}}
\put(70,-302){\color{grey}\rule{297pt}{60pt}}}

% %Julen ch 4
% \thiswatermark{\put(1,-302){\color{l-grey}\rule{70pt}{60pt}}
% \put(70,-302){\color{grey}\rule{297pt}{60pt}}}
% \fancyhead[RO]{\small \leftmark}
%
%
% %Julen ch 5
% \thiswatermark{\put(1,-302){\color{l-grey}\rule{70pt}{60pt}}
% \put(70,-302){\color{grey}\rule{297pt}{60pt}}}
% \fancyhead[RO]{\leftmark}

% NEW DEVICES FOR MICROWAVE QUANTUM COMMUNICATION

\lettrine[lines=2, findent=3pt,nindent=0pt]{S}{uperconducting} quantum devices working in the microwave regime have become extremely relevant due to the advances in the field of circuit quantum electrodynamics. These are very valuable for quantum computing, as superconducting circuits have been established as the running quantum platform. But these advances not only pertain to quantum computation, but also to entanglement generation, state reconstruction, squeezing, and quantum teleportation, to name a few. 

The main drawback for the use of propagating microwave photons is the difficulty to detect them, due to their low energy. Therefore, amplification of signals has been one of the most studied subjects in this topic. In classical microwave communication, especially in free space, amplification plays a crucial role, as increasing the number of photons improves the chances the receiver has of detecting that signal. On the other hand, in the quantum regime, the photons introduced by the amplifier are a threat to the quantumness of the signal~\cite{Caves1982}. For example, let us examine the case of cryogenic high electron mobility transistor (HEMT) amplifiers. These devices are able to greatly enhance signals in a large frequency spectrum, while introducing a significant amount of thermal photons. This noise is reflected in the input-output relation 
\begin{equation}
\hat{a}_{\text{out}}=\sqrt{g_H}\hat{a}_{\text{in}}+\sqrt{g_H-1}\hat{h}_H,
\end{equation}
with $a_{\text{in}}$, $a_{\text{out}}$, and $h_H$ the annihilation operators of the input field, output field, and noise added by the amplifier, respectively. From this formula, we can see that amplification is a procedure that acts individually on the modes of a quantum state, which means that we can increase the number of photons of that mode, but they will not be entangled with the others. Therefore, amplification cannot increase quantum correlations. In fact, the introduction of thermal noise can almost certainly lead to entanglement degradation. 

HEMTs normally work at 4 K temperatures, which implies that the number of thermal photons they introduce is around $n_{H} \sim 10 - 20$, for 5 GHz frequencies. It is the number of thermal photons that determines the thermally-radiated power~\cite{Clerk2010}, $P = \hbar \omega N B$, meaning that the excess output noise produces a flux of $N$ photons per second in a bandwidth of $B$ Hz. The gain is given by the ratio between output and input powers, and for a constant bandwidth, it is just $g_{H}=n_{H}/n$, the ratio between the number of thermal photons introduced by the HEMT and the number of photons in the input state. A HEMT of these characteristics produces a gain of $g_{H}=2\times 10^{3}$ when acting on a TMST state with $n\sim 10^{-2}$, which completely destroys entanglement. In order for entanglement to survive this amplification process, the HEMT must be placed at temperatures below 100 mK. However, it has been shown that placing a HEMT at cryogenic temperatures does not suffice to reduce thermal noise originated from self heating, and more sophisticated techniques are required~\cite{Ardizzi2022}.

Nevertheless, it has been proven that HEMTs actually can still be used for quantum state reconstruction~\cite{Menzel2010}, for the detection of two-mode squeezing~\cite{Eichler2011}, and even for path entanglement~\cite{Menzel2012}. This is understandable, given that these amplifiers preserve the phase and do not introduce correlated noise that would otherwise most likely destroy photon-number entanglement. Furthermore, for certain tasks, it has been proven an advantage by using a HEMT combined with a Josephson parametric amplifier, or JPA~\cite{Teufel2009}. The latter is a phase-sensitive amplifier, so in this case, one of the quadratures of a quantum state can be squeezed below the quantum limit set by vacuum fluctuations. Then, by placing a HEMT after the JPA in the detection process, the effects of the noise introduce by the former are mitigated. This can also be achieved with two JPAs placed sequentially~\cite{Mallet2011}.

Architectures based on Josephson junctions~\cite{Josephson1969} have been proven to be very versatile, as they can be used to build various devices with many different applications. Arguably the most popular of them is the JPA~\cite{Yurke1988,Yurke1989}, which is nothing but a Josephson-junction-based architecture working as a phase-sensitive parametric amplifier. Some implementations of the JPA rely on superconducting quantum interference devices (SQUIDs), which are two Josephson junctions with identical critical currents placed in a loop. Normally, the magnetic flux threading the SQUID is controlled to change the system's resonance frequency; this system can be a coplanar-waveguide resonator either with SQUIDs along the inner conductor~\cite{CastellanosBeltran2007,CastellanosBeltran2008}, or short-circuited to ground by the SQUID~\cite{Yamamoto2008}. The latter is known as a flux-driven JPA.

Other implementations of a parametric amplifier are based on the Josephson ring modulator~\cite{Bergeal2010,Bergeal2010_2,Roch2012}, a ring of four identical Josephson junctions threaded by a magnetic flux which induces coupling between three modes. These devices are known as Josephson parametric converters (JPCs)~\cite{Liu2017,Liu2020,Chien2020}, and they can work as phase-preserving amplifiers and frequency upconverters. They can also generate two mode squeezing, as proven in Refs.~\cite{Bergeal2012,Flurin2012}, and by combining two JPCs with different pump tones, one can implement a three-port non-reciprocal device called the circulator~\cite{Abdo2013,Sliwa2015}. It is used to shield other devices in the cryostat from thermal radiation~\cite{Pozar2012}, and is a key component in the readout of qubits in superconducting platforms; therefore, many experimental implementations have followed~\cite{Mahoney2017,Chapman2017}.

In most cases, the amplifiers described here were developed for measuring in superconducting quantum platforms; the design of efficient photocounters has probably been the most relevant milestone for propagating quantum microwaves. Although many theoretical proposals for photodetectors~\cite{Helmer2009,Sathyamoorthy2014,Fan2014,Sathyamoorthy2016,Leppakangas2018,Royer2018,Grimsmo2021} and photocounters~\cite{Romero2009,Romero2009_2,Sokolov2020} have been made based on circuit QED, and many photodetection experiments have been realized~\cite{Chen2011,Inomata2016,Besse2018,Kono2018,Lescanne2020}, it was not until recently that the first experimental realization of a (number-resolved) photocounter with propagating quantum microwaves was achieved~\cite{Dassonneville2020,Essig2021}, employing Josephson ring modulators. 

These are a key component in quantum communications, and their development for quantum microwave signals paves the way for more advances in quantum devices working in this frequency regime. Given that one of the main sources of imperfection for microwaves, due to the large wavelengths, is diffraction, the development of proper collimators is also due.

Here, we have illustrated how important Josephson junctions are to quantum information processing devices working in the microwave regime. Arguably the most famous application of Josephson junctions is the Josephson parametric amplifier; particularly, we are interested in its squeezing capability for state generation.

\subsection{Josephson Parametric Amplifiers}

One of the most important applications of JPAs is the generation of squeezed states. Squeezing is an operation in which one of the variances of the electromagnetic field quadratures of a quantum state is reduced below the level of vacuum fluctuations, while the conjugate quadrature is amplified, satisfying the uncertainty principle. This can be achieved by sending the vacuum state to a JPA, a coplanar waveguide resonator line terminated by a direct-current superconducting interference device, known as a dc SQUID. The dc SQUID provides magnetic flux tunability to the resonator and enables parametric phase-sensitive amplification, which is the key for generating squeezed microwave states~\cite{Zhong2013,Pogorzalek2017}. By combining two single-mode squeezed states, with squeezing in orthogonal directions in phase space, in a balance beam splitter, we can generate two-mode squeezed states. These are Gaussian bipartite entangled states, and the most common resource used for CV quantum communication. 

The relation between the frequency of the external magnetic flux, $\Omega$, and the fundamental frequency of the JPA, $\omega_{c}$, determines whether the JPA operates in the phase-insensitive or phase-sensitive regime. The latter is achieved in the so-called degenerate regime, $\Omega = 2\omega_{c}$. A corresponding three-wave mixing process, when one pump photon splits into two signal photons, is described by the Hamiltonian
\begin{equation}
\hat{H} = g\left( \bar{\beta}\hat{a}^{2} - \beta \hat{a}^{\dagger 2} \right).
\end{equation}
It can be shown that the aforementioned Hamiltonian corresponds to  a single-mode squeezing operator
\begin{equation}
\hat{S}(\xi) = \exp\left[ \frac{1}{2}(\bar{\xi}\hat{a}^{2}-\xi \hat{a}^{\dagger 2})\right],
\end{equation}
with the squeezing parameter given by $|\xi| \propto 2g|\beta|t$. 

A pair of states with the same squeezing levels, but squeezed in perpendicular directions in phase space, can be combined in a 50:50 beam splitter to produce two-mode squeezed states~\cite{Kim2002}. In the microwave regime, the action of the symmetric 50:50 beam splitter is carried out by a hybrid ring~\cite{Mariantoni2010}, and squeezed states are produced by JPAs. Due to the bright thermal background at microwave frequencies, which is not negligible even at cryogenic temperatures, these devices are subject to various sources of imperfections and noise. Therefore, the output states can be effectively modelled as TMST states~\cite{Serafini2004}, whose second moments differ from those of ideal two-mode squeezed vacuum states by a factor of $1+2n$, where $n$ is the average number of thermal photons. This is equivalent to considering ideal JPAs with identical $n$-photon thermal state inputs, instead of vacuum states.

Nevertheless, thermal photons in squeezed thermal states may have various physical origins. One of the most trivial reasons for noise in the two-mode squeezed states is finite temperatures of the input JPA modes, which lead to the fact that one applies a squeezing operator to a thermal state rather than to a vacuum. Another important source of noise in squeezed states produced by flux-driven JPAs arises from Poisson photon-number fluctuations in the pump mode, which lead to extra quasithermal photons in the output squeezed states~\cite{Renger2020}. Last but not least, higher-order nonlinear effects also contribute to additional effective noise under the Gaussian approximation~\cite{Boutin2017}.  More experimental details on the microwave squeezing and related imperfections can be found elsewhere~\cite{Fedorov2018}.

The action of a JPA on the quadratures of an input state, as presented in the suplementary material in Ref.~\cite{Fedorov2018}, can be modelled by
\begin{eqnarray}
\nonumber \hat{x} &\longrightarrow& \hat{x} \left( \frac{2\chi + \kappa - \gamma}{2\chi - \kappa - \gamma}\right) + \hat{x}_{\text{JPA}} \left( \frac{2\sqrt{\kappa\gamma}}{2\chi - \kappa - \gamma}\right), \\
\hat{p} &\longrightarrow& \hat{p} \left( \frac{2\chi - \kappa + \gamma}{2\chi + \kappa + \gamma}\right) - \hat{p}_{\text{JPA}} \left( \frac{2\sqrt{\kappa\gamma}}{2\chi + \kappa + \gamma}\right).
\end{eqnarray}
Here, $\kappa$ is the coupling rate of the input field to the resonator, $\gamma$ is the internal rate of loss into the thermal bath, and $\chi$ is the three-wave-mixing strength. We can interpret this transformation as $x$-squeezing for $\chi<0$, and $p$-squeezing for $\chi>0$. We have obtained estimations of the order of these parameters in JPAs from Ref~\cite{Parker2022}, such that $\gamma/\kappa < 10^{-4}$ and $|\chi|/\kappa \in [0,1/2]$. We will define $\bar{\chi} \equiv - 2\chi/\kappa$ and $\bar{\gamma}\equiv \gamma/\kappa$, so that $\bar{\chi}\in[0,1]$ and $\bar{\gamma}<10^{-4}$. Now, we have $x$ squeezing for positive $\bar{\chi}$. With this, we can rewrite
\begin{eqnarray}
\nonumber \hat{x} &\longrightarrow& \hat{x} \left( -\frac{ 1 - \bar{\chi} - \bar{\gamma} }{ 1 + \bar{\chi} + \bar{\gamma}}\right) + \hat{x}_{\text{JPA}} \left( -\frac{2\sqrt{\bar{\gamma}}}{1 + \bar{\chi} + \bar{\gamma}}\right), \\
\hat{p} &\longrightarrow& \hat{p} \left( -\frac{1 + \bar{\chi} - \bar{\gamma}}{1 - \bar{\chi} + \bar{\gamma}}\right) - \hat{p}_{\text{JPA}} \left( \frac{2\sqrt{\bar{\gamma}}}{1 - \bar{\chi} + \bar{\gamma}}\right).
\end{eqnarray}

Given that this device, acting on a single-mode Gaussian state, also couples to a thermal bath, we use the quantum channel formalism to describe the operation of squeezing a general Gaussian quantum state using a JPA. The symplectic transformation that achieves this can be written as 
\begin{equation}
\mathcal{S}_{\text{JPA}} = \begin{pmatrix} \alpha_{1} & 0 & \alpha_{2} & 0 \\ 0 & \beta_{1} & 0 & -\beta_{2} \\ \alpha_{2}\frac{\beta_{2}}{\beta_{4}} & 0 & \alpha_{2}\frac{\beta_{1}}{\beta_{4}} & 0 \\ 0 & -\beta_{4} & 0 & \beta_{4}\frac{\alpha_{1}}{\alpha_{2}} \end{pmatrix},
\end{equation}
where we have defined
\begin{eqnarray}
\nonumber \alpha_{1} &=& -\frac{ 1 - \bar{\chi} - \bar{\gamma} }{ 1 + \bar{\chi} + \bar{\gamma}}, \\
\nonumber \alpha_{2} &=&  -\frac{2\sqrt{\bar{\gamma}}}{1 + \bar{\chi} + \bar{\gamma}}, \\
\beta_{1} &=& -\frac{1 + \bar{\chi} - \bar{\gamma}}{1 - \bar{\chi} + \bar{\gamma}}, \\
\nonumber \beta_{2} &=& \frac{2\sqrt{\bar{\gamma}}}{1 - \bar{\chi} + \bar{\gamma}},
\end{eqnarray}
whereas $\beta_{4}$ is related to the coupling of the bath to the system, which is not for us to determine. Here, $\bar{\chi}$ and $\bar{\gamma}$ are related to the three-wave-mixing strength and to the loss rate, as we have seen.

In the symplectic formalism, the single-mode squeezing operator is expressed as
\begin{equation}
\mathcal{S}(\xi,\varphi) = \begin{pmatrix} \cosh\xi - \cos\varphi\sinh\xi & -\sin\varphi\sinh\xi \\ -\sin\varphi\sinh\xi & \cosh\xi + \cos\varphi\sinh\xi \end{pmatrix}.
\end{equation}
Applying single-mode squeezing on a thermal state, produces 
\begin{equation}
\mathcal{S}(\xi,\varphi) m\mathbb{1}_{2}\mathcal{S}^{\intercal}(\xi,\varphi) =  m \begin{pmatrix} \cosh2\xi - \cos\varphi\sinh2\xi & -\sin\varphi\sinh2\xi \\ -\sin\varphi\sinh2\xi & \cosh2\xi + \cos\varphi\sinh2\xi \end{pmatrix}.
\end{equation} 
Connecting to this ``noiseless'' ideal JPA, we can identify the squeezing gain by
\begin{equation}
e^{-2\xi} = \left( \frac{1-\bar{\chi}}{1+\bar{\chi}} \right)^{2} =\left.\alpha_{1}^{2}\right|_{\bar{\gamma}=0} = \left.\frac{1}{\beta_{1}^{2}}\right|_{\bar{\gamma}=0}
\end{equation}
in the case in which $\bar{\gamma}=0$, which implies $\alpha_{2}=\beta_{2}=0$. In reality, however, we will have that
\begin{equation}
\alpha_{1} = \frac{1+\alpha_{2}\beta_{2}}{\beta_{1}},
\end{equation}
in the case $\bar{\gamma}\neq0$, in order for the transformation to be symplectic. Notice that, when we change $\bar{\chi}\rightarrow-\bar{\chi}$, we obtain $p$ squeezing instead of $x$ squeezing, and this amounts to changing $\alpha_{1}\leftrightarrow\beta_{1}$ and $\alpha_{2}\leftrightarrow-\beta_{2}$.

We assume that our state and the state of the bath have covariance matrices $\Sigma = m\mathbb{1}_{2}$ and $\Sigma_{\text{JPA}} = m_{\text{JPA}}\mathbb{1}_{2}$ respectively. The action of this transformation on both states is
\begin{equation}
\mathcal{S}_{\text{JPA}} \begin{pmatrix} \Sigma & 0 \\ 0 & \Sigma_{\text{JPA}} \end{pmatrix}\mathcal{S}_{\text{JPA}} ^{\intercal} ,
\end{equation}
and by tracing out the contribution from the bath, the covariance matrix of the state is
\begin{equation}
\Sigma' = \begin{pmatrix} \alpha_{1}^{2} m + \alpha_{2}^{2} m_{\text{JPA}} & 0 \\ 0 & \beta_{1}^{2} m + \beta_{2}^{2} m_{\text{JPA}} \end{pmatrix}.
\end{equation}
If we write this state as the outcome of a quantum channel, $\Sigma' = X\Sigma X^{\intercal} + Y$, we can identify
\begin{eqnarray}
\nonumber X &=& \begin{pmatrix} \alpha_{1} & 0 \\ 0 & \beta_{1} \end{pmatrix}, \\
Y &=& m_{\text{JPA}} \begin{pmatrix} \alpha_{2}^{2} & 0 \\ 0 & \beta_{2}^{2} \end{pmatrix}.
\end{eqnarray}
Keep in mind that this characterizes the action of the JPA channel onto a thermal state, but it can also hold for a more general Gaussian state, whose covariance matrix, by virtue of Williamson's theorem, can be brought into diagonal form, $\Sigma = m\mathcal{1}_{2}$. In any case, however, we are considering the bath to be in a thermal state. 

We impose that the quantum state resulting from this channel is a squeezed thermal state, and from the result of it acting on a Gaussian state with covariance matrix $\Sigma = m\mathcal{1}_{2}$, we read its number of thermal photons and squeezing
\begin{eqnarray}
\nonumber m' &=& \frac{\sqrt{\left[ m(1-\bar{\chi} - \bar{\gamma})^{2} + 4m_{\text{JPA}}\bar{\gamma}\right]\left[ m(1+\bar{\chi} - \bar{\gamma})^{2} + 4m_{\text{JPA}}\bar{\gamma}\right]}}{(1-\bar{\chi} + \bar{\gamma})(1+\bar{\chi} + \bar{\gamma})}, \\
e^{-2\xi'} &=& \left(\frac{1-\bar{\chi} + \bar{\gamma}}{1+\bar{\chi} + \bar{\gamma}}\right)\sqrt{\frac{m(1-\bar{\chi} - \bar{\gamma})^{2} + 4m_{\text{JPA}}\bar{\gamma}}{m(1+ \bar{\chi} - \bar{\gamma})^{2} + 4m_{\text{JPA}}\bar{\gamma}}}.
\end{eqnarray}

The reason for cryogenia is to shield the superconducting devices and quantum signals from the bright microwave thermal background at room temperature. This is a relevant source of decoherence for quantum microwaves, as it contributes to the classicalization of signals. The impedance associated to propagation of signals inside a cryostat is 50 $\Omega$, whereas in open air it is 377 $\Omega$; therefore, a signal travelling between both media is susceptible to enduring reflections due to an impedance mismatch. An antenna is then needed for impedance matching, while avoiding any kind of amplification of the signal. Given that classical antennae perform signal amplification, adding thermal photons that can compromise quantum correlations, as we discussed earlier in this chapter, we have to design a quantum antenna that performs impedance matching, but that can also preserve entanglement. 

%%%%%%%%%%%%%%%%%%%%%%%%%%%%%%%%%%%%%%%%%%%%%%%

\subsection{Coplanar Antenna Design for Microwave Signals}
We address this problem by considering the quantum antenna as a coplanar waveguide with a position-dependent impedance. We observe that the shape of the antenna defines its reflectivity, and this affects entanglement. Therefore, our goal is to optimize the impedance function in order to minimize the reflectivity of the antenna. As a paradigmatic case, we study the transmission of two-mode squeezed states into open air, since they are easy to generate and robust to photon losses. We employ a numerical optimization method through interpolation, which repurposes each solution, as well as an ansatz for the impedance, qualitatively-based on the solution from the numerical case. We find that the reflectivity can be reduced below $10^{-9}$, while entanglement preservation with real-life experimental parameters would require values below $10^{-4}$. To conclude, we investigate how errors in the optimal impedance affect the output entanglement to illustrate the impact that small fabrication imperfections could have on the performance of the antenna.

%%%%%%%%%%%%%%%%%%%%%%%%%%%%%%%%%%%%%%%%%%%%%%%
\begin{figure}[t]
\centering
{\includegraphics[width=0.75 \textwidth]{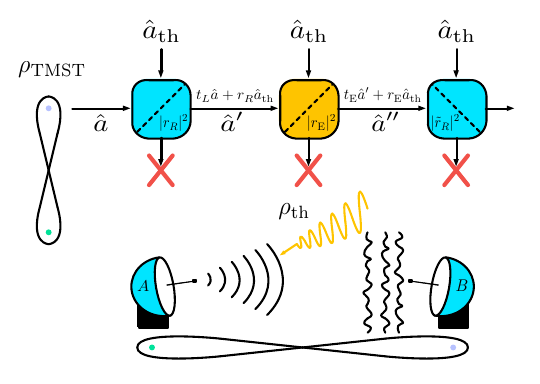}}
\caption[Sketch describing an open-air microwave quantum communication protocol]{\textbf{Sketch describing an open-air microwave quantum communication protocol}, in which a party A generates a two-mode entangled quantum state and sends one of the modes to a second party, B, through an environment dominated by thermal noise, keeping the other mode. The effect of the antennae, as well as the transmission in open air, are modeled by beam splitters, which allow for the description of the deterioration of the state due to thermal noise. These have reflectivities $|r_{R}|^{2}$ and $|\tilde{r}_{R}|^{2}$ for the antennae, and $|r_{E}|^{2}$ for the imperfect open-air transmission.}
\label{fig4_1}
\end{figure}
%%%%%%%%%%%%%%%%%%%%%%%%%%%%%%%%%%%%%%%%%%%%%%%

\subsubsection{Antenna model}
\begin{figure}[t]
\centering
{\includegraphics[width=0.95 \textwidth]{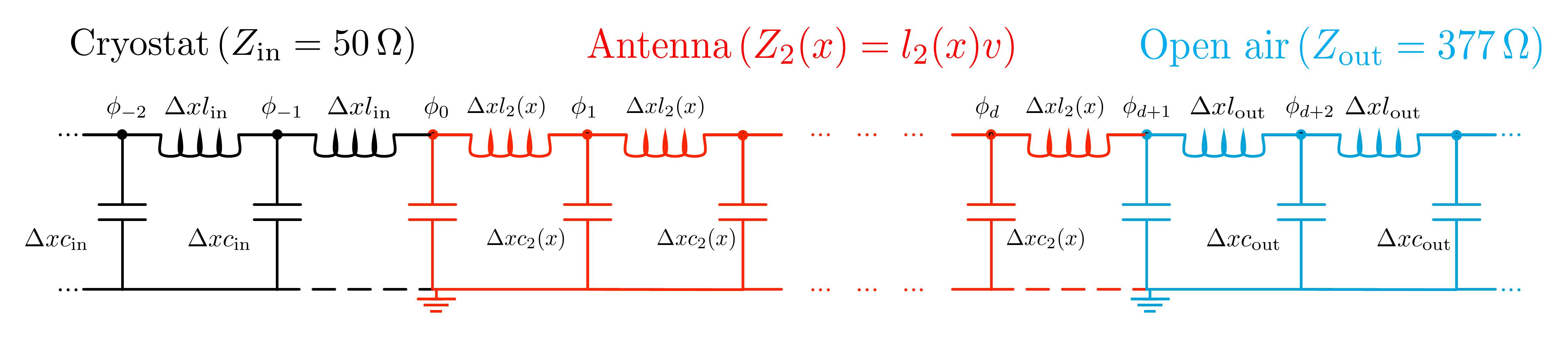}}
\caption[Circuit graph representation of a quantum antenna as an inhomogeneous cavity]{\textbf{Circuit graph representation of a quantum antenna as an inhomogeneous cavity}, in red, connecting a cryostat (black) with the open air (blue), both represented by transmission lines.}
\label{fig4_2}
\end{figure}

We attempt to design an antenna for an open-air microwave quantum communication protocol, in which an entangled state is produced by a source A, keeping one mode and sending another through a waveguide into open air, to be received at a remote location B, while preserving the entanglement between both modes. This scenario is represented in Fig.~\ref{fig4_1}. For this, we propose to use a transmission line (TL) as a waveguide that sends out the state, then a finite inhomogeneous TL as the antenna, and then another TL to represent propagation in open air~\cite{Sanz2018}. This circuit is sketched in Fig.~\ref{fig4_2}. The TL on the left has an impedance of 50 $\Omega$, whereas that on the right has an impedance of 377 $\Omega$. Then, the antenna serves as an inhomogeneous medium that achieves a smooth transition from two very different impedances. The Lagrangian describing this circuit is
\begin{eqnarray}
\nonumber \mathcal{L} &=& \sum_{i=-N}^{-1} \left[ \frac{\Delta x \, c_{\text{in}}}{2}\dot{\phi}_{i}^{2} - \frac{(\phi_{i+1}-\phi_{i})^{2}}{2\Delta x \, l_{\text{in}}} \right] +  \sum_{j=0}^{d} \left[ \frac{\Delta x \, c_{2}(x)}{2}\dot{\phi}_{j}^{2} - \frac{(\phi_{j+1}-\phi_{j})^{2}}{2\Delta x \, l_{2}(x)} \right] \\
 &+& \sum_{k=d+1}^{N} \left[ \frac{\Delta x \, c_{\text{out}}}{2}\dot{\phi}_{k}^{2} - \frac{(\phi_{k+1}-\phi_{k})^{2}}{2\Delta x \, l_{\text{out}}} \right],
\end{eqnarray}
where we have defined $l_{\text{in}}, c_{\text{in}}$ as the inductance and capacitance densities of the transmission line inside the cryostat, $l_{2}(x), c_{2}(x)$ as the inductance and capacitance densities of the antenna, and $l_{\text{out}}, c_{\text{out}}$ as the inductance and capacitance densities of the second transmission line. See that, inside the antenna, the inductances and capacitances depend on the position. This is necessary for a smooth change of impedance. 
%%%%%%%%%%%%%%%%%%%%%%%%%%%%%%%%%%%%%%%%%%%%%%%
%%%%%%%%%%%%%%%%%%%%%%%%%%%%%%%%%%%%%%%%%%%%%%%
The reflectivity of the antenna entirely depends on the impedance between the different media. As the impedance is defined by the densities of inductance and capacitance, $Z = \sqrt{l/c}$, which can be independently manipulated in nano fabrication, we can choose without a loss of generality the propagation velocity through the antenna, $v=1/\sqrt{lc}$, to be constant. Consequently, the dependence on the position falls entirely onto the inductance, and we can express the impedance in the antenna as
\begin{equation}
Z_{2}(x) = \sqrt{\frac{l_{2}(x)}{c_{2}(x)}} = l_{2}(x) v.
\end{equation}

Taking $N\rightarrow\infty$ in order to consider semi-infinite transmission lines, amounts to taking the continuum limit $\Delta x\rightarrow 0$. Then, we rewrite the Lagrangian,
\begin{equation}
\mathcal{L} = \int_{-\infty}^{\infty} dx \left[ \frac{c(x)}{2} (\partial_{t}\phi(x,t))^{2} - \frac{1}{2l(x)} (\partial_{x}\phi(x,t))^{2} \right],
\end{equation}
defining the capacitances and inductances as
\begin{equation}
l(x) = \begin{cases} 
l_{\text{in}} & \text{if } x < 0 \\ 
l_{2}(x) & \text{if } 0 \leq x \leq d \\ 
l_{\text{out}} & \text{if } x > d 
\end{cases}
\end{equation}
and 
\begin{equation}
c(x) = \begin{cases} 
c_{\text{in}} & \text{if } x < 0 \\ 
c_{2}(x) & \text{if } 0 \leq x \leq d \\ 
c_{\text{out}} & \text{if } x > d 
\end{cases}
\end{equation}
From the minimal action principle, we obtain the Euler-Lagrange equations for this Lagrangian, 
\begin{equation}\label{E-L}
c(x) \partial_{t}^{2}\phi(x,t) = \partial_{x}\left(\frac{\partial_{x}\phi(x,t)}{l(x)}\right).
\end{equation}
For the left and right transmission lines, $l(x)$ and $c(x)$ are constant, and Eq.~\eqref{E-L} is just the wave equation. This means that for the left and right TLs, the solutions to the equations of motion are plane waves. However, the solution for the antenna is not as straightforwardly obtained. To do so, we employ the variable separation method; we then propose the solution $\phi(x,t)=\sum_{n}\varphi_{n}(t)u_{n}(x)$ and expand 
\begin{equation}
c(x) \ddot{\varphi}_{n}(t)u_{n}(x) = \varphi_{n}(t)\left( \frac{l(x)u''_{n}(x)-l'(x)u'_{n}(x)}{l^{2}(x)} \right),
\end{equation}
which leads to the expression
\begin{equation}
c(x) l(x) \frac{\ddot{\varphi}_{n}(t)}{\varphi_{n}(t)} = \frac{1}{u_{n}(x)}\left( u''_{n}(x) - \frac{l'(x)}{l(x)} u'_{n}(x) \right).
\end{equation}
On both sides of the equation, the solutions are constants, 
\begin{eqnarray}
&& \ddot{\varphi}_{n}(t) = -\zeta v^{2} \varphi_{n}(t), \\
&& u''_{n}(x) - \frac{Z'(x)}{Z(x)} u'_{n}(x) = -\zeta u_{n}(x),
\end{eqnarray}
where we have used $c(x)l(x)=1/v^{2}$ and $Z(x)=l(x)v$. From this, we see that $\zeta=k_{n}^{2}=(\omega_{n}/v)^{2}$ is the wavenumber. Then, the equation that we need to solve is that for $u_{n}(x)$, which can be written as the Sturm-Liouville problem. In order to solve this equation for the antenna, we need to fix $Z(x)$. 

\subsubsubsection{Linear antenna}
For a simple case of study, we consider that $Z(x)$ for $0\leq x\leq d$ is a linear function of the position,
\begin{equation}\label{Z_lin}
Z(x) = \left(1-\frac{x}{d}\right)Z_{\text{in}} + \frac{x}{d} Z_{\text{out}},
\end{equation}
which implies that the inductance in the antenna is also linear. From now on, we will focus on a single mode of the wavefunction, and we will drop the subscript notation by changing $u_{n}(x)\rightarrow u(x)$. Now, the equation we want to solve is
\begin{equation}
u''(x)-\frac{Z'(x)}{Z(x)}u'(x) + k^{2}u(x) = 0.
\end{equation}
First, we multiply by $(Z(x)/Z'(x))^{2}$, which results in 
\begin{equation}
\left(\frac{Z(x)}{Z'(x)}\right)^{2}u''(x)-\frac{Z(x)}{Z'(x)}u'(x) + k^{2}\left(\frac{Z(x)}{Z'(x)}\right)^{2}u(x) = 0.
\end{equation}
For the linear impedance in Eq.~\eqref{Z_lin}, we have 
\begin{equation}
\frac{Z(x)}{Z'(x)} = x + d\frac{Z_{\text{in}}}{Z_{\text{out}}-Z_{\text{in}}}.
\end{equation}
Let us define $y = k Z(x)/Z'(x)$, such that $u'(x) = k u'(y)$, $u''(x) = k^{2} u''(y)$. With this, we can rewrite it as
\begin{equation}
y^{2}u''(y)-y u'(y) + y^{2}u(y) = 0.
\end{equation}
If we introduce $y\varphi(y)=u(y)$, with $u'(y)=\varphi(y)+y\varphi'(y)$ and $u''(y)=2\varphi'(y)+y\varphi''(y)$, we obtain 
\begin{equation}
y^{2}\varphi''(y)+y \varphi'(y) + (y^{2}-1)\varphi(y) = 0,
\end{equation}
which is the first order Bessel differential equation. The solution to this equation is
\begin{equation}
\varphi(y) = b_{1} J_{1}(y) + b_{2} Y_{1}(y),
\end{equation}
where $J_{1}(\cdot)$, $Y_{1}(\cdot)$ are the Bessel functions of the first and second kind, respectively, and $b_{1}$, $b_{2}$ are arbitrary constants. If we undo all the variable changes, we find
\begin{eqnarray}
\nonumber u(x) &=& k\left( x + d\frac{Z_{\text{in}}}{Z_{\text{out}}-Z_{\text{in}}}\right) \bigg[ b_{1} J_{1}\left( k x + k d\frac{Z_{\text{in}}}{Z_{\text{out}}-Z_{\text{in}}}\right) \\
&+& b_{2} Y_{1}\left( k x + k d\frac{Z_{\text{in}}}{Z_{\text{out}}-Z_{\text{in}}}\right) \bigg].
\end{eqnarray}
Finally, we redefine $b_{i} k/(Z_{\text{out}}-Z_{\text{in}}) = c_{i}$, with $i=\{1,2\}$, to write the solution as
\begin{eqnarray}\label{SL_solution}
\nonumber u(x) &=& \left( Z_{\text{in}}(d-x) + Z_{\text{out}}x\right) \bigg[ c_{1} J_{1}\left(kx + kd\frac{Z_\text{in}}{Z_\text{out}-Z_\text{in}}\right) \\
&+& c_{2} Y_{1}\left(kx + kd\frac{Z_\text{in}}{Z_\text{out}-Z_\text{in}}\right) \bigg].
\end{eqnarray}
Then, our problem can be translated into a scattering problem, 
%\begin{equation}
%\begin{matrix} u_{1}(x) = A e^{ikx} + B e^{-ikx} \\ u_{2}(x) = u(x) \\ u_{3}(x) = F e^{iqx} + G e^{-iqx} \end{matrix} \qquad \text{for } \begin{matrix} x < 0 \\ 0 \leq x \leq d \\ x > d \end{matrix},
%\end{equation}
\begin{eqnarray}
\nonumber && u_{1}(x) = A e^{ikx} + B e^{-ikx}  \qquad \text{for } x<0, \\
\nonumber && u_{2}(x) = u(x) \qquad  \qquad \qquad \text{for } 0 \leq x \leq d, \\
&& u_{3}(x) = F e^{iqx} + G e^{-iqx}  \qquad \text{for } x > d,
\end{eqnarray}
where $k=\omega/(c/3)$ is the wavenumber inside the cryostat and the antenna, considering that the propagation velocity is $v=c/3$ inside these two circuits and that $q=\omega/c$ is the wavenumber in open air, with $v = c$. Imposing the continuity of voltage
\begin{equation}
\left.\dot{\phi}(x,t)\right|_{x^{-}} = \left.\dot{\phi}(x,t)\right|_{x^{+}}
\end{equation}
and current
\begin{equation}
\left. \partial_{x}\left(\frac{\phi(x,t)}{l(x)}\right)\right|_{x^{-}} =\left. \partial_{x}\left(\frac{\phi(x,t)}{l(x)}\right)\right|_{x^{+}},
\end{equation}
is equivalent to imposing the continuity of these functions and their derivatives. Boundary conditions are imposed at frontier points $x=0$ and $x=d$, such that 
\begin{eqnarray*}
\nonumber \lim_{x\rightarrow0^{-}}u(x) &=& u_{1}(0), \qquad \lim_{x\rightarrow d^{-}}u(x) = u_{2}(d), \\
\lim_{x\rightarrow0^{+}}u(x) &=& u_{2}(0), \qquad \lim_{x\rightarrow d^{+}}u(x) = u_{3}(d),
\end{eqnarray*}
considering that the impedance is constant across the boundaries, $\lim_{x\rightarrow0^{-}}Z(x) = \lim_{x\rightarrow0^{+}}Z(x) = Z_{\text{in}}$ and $\lim_{x\rightarrow d^{-}}Z(x) = \lim_{x\rightarrow d^{+}}Z(x) = Z_{\text{out}}$. Notice that, since we have imposed that the velocity is constant throughout the antenna, it will jump from $v_{\text{in}} = c/3$ to $v_{\text{out}} = c$ when moving from the antenna to open air. Therefore, the continuity of the current at $x=d$ will be expressed as
\begin{equation}
v_{\text{in}}\left. \partial_{x}\left(\frac{u_{2}(x)}{Z(x)}\right)\right|_{x=d^{-}} =\frac{v_{\text{out}}}{Z_{\text{out}}}\left. \partial_{x}u_{3}(x)\right|_{x=d^{+}}.
\end{equation}
If we do this, we find 
\begin{eqnarray}
 A+B &=& d Z_{\text{in}} \left[ c_{1} J_{1}\left(kd\frac{Z_{\text{in}}}{Z_{\text{out}}-Z_{\text{in}}}\right) + c_{2} Y_{1}\left(kd\frac{Z_{\text{in}}}{Z_{\text{out}}-Z_{\text{in}}}\right) \right], \\
\nonumber A - B &=& -i d Z_{\text{in}} \left[ c_{1} J'_{1}\left(kd\frac{Z_{\text{in}}}{Z_{\text{out}}-Z_{\text{in}}}\right) + c_{2} Y'_{1}\left(kd\frac{Z_{\text{in}}}{Z_{\text{out}}-Z_{\text{in}}}\right) \right],
\end{eqnarray}
and also
\begin{eqnarray}
 F e^{iqd}+Ge^{-iqd} &=& d Z_{\text{out}} \left[ c_{1} J_{1}\left(kd\frac{Z_{\text{out}}}{Z_{\text{out}}-Z_{\text{in}}}\right) +c_{2} Y_{1}\left(kd\frac{Z_{\text{out}}}{Z_{\text{out}}-Z_{\text{in}}}\right) \right], \\
\nonumber F e^{iqd}-Ge^{-iqd} &=& -i d Z_{\text{out}} \left[ c_{1} J'_{1}\left(kd\frac{Z_{\text{out}}}{Z_{\text{out}}-Z_{\text{in}}}\right) + c_{2} Y'_{1}\left(kd\frac{Z_{\text{out}}}{Z_{\text{out}}-Z_{\text{in}}}\right) \right],
\end{eqnarray}
where it will be useful to know that
\begin{equation}
J_{1}(x)Y'_{1}(x) - Y_{1}(x)J'_{1}(x) = \frac{2}{\pi x}.
\end{equation}
The transfer matrix $T$ is defined as
\begin{equation}
\begin{pmatrix} F \\ G \end{pmatrix} = T  \begin{pmatrix} A \\ B \end{pmatrix},
\end{equation}
and it can be used to construct the scattering matrix $S$, defined as
\begin{equation}
\begin{pmatrix} F \\ B \end{pmatrix}_{\text{out}} = S  \begin{pmatrix} A \\ G \end{pmatrix}_{\text{in}}  = \begin{pmatrix} S_{11} & S_{12} \\ S_{21} & S_{22} \end{pmatrix} \begin{pmatrix} A \\ G \end{pmatrix}_{\text{in}},
\end{equation}
which will not be normalized ($S S^{\dagger} \neq \mathbb{1}$). For that, we can redefine $S$ as
\begin{equation}
S' = \begin{pmatrix} \theta_{1} & 0 \\ 0 & \theta_{2} \end{pmatrix} S \begin{pmatrix} \theta_{1} & 0 \\ 0 & \theta_{2} \end{pmatrix} = \begin{pmatrix} \theta_{1}^{2} S_{11} & \theta_{1}\theta_{2} S_{12} \\ \theta_{1}\theta_{2} S_{21} & \theta_{2}^{2} S_{22} \end{pmatrix},
\end{equation}
and find the parameters $\theta_{1}$, $\theta_{2}$, with which the matrix $S'$ satisfies unitarity conditions. First of all, the determinant must be equal to one (in modulus). This implies that
\begin{equation}
\text{det}\,S' =  \theta_{1}^{2}\theta_{2}^{2} \text{det}\,S = e^{i\varphi}.
\end{equation}
Also, the rows of the matrix must represent orthonormal vectors,
\begin{equation}
\begin{pmatrix} \theta_{1}^{2} S_{11} & \theta_{1}\theta_{2} S_{12} \end{pmatrix} \begin{pmatrix} \theta_{1}\theta_{2} \bar{S}_{21} \\ \theta_{2}^{2} \bar{S}_{22} \end{pmatrix} = \theta_{1}\theta_{2}\left[ \theta_{1}^{2}S_{11}\bar{S}_{21} + \theta_{2}^{2}S_{12}\bar{S}_{22} \right],
\end{equation} 
and, from these two conditions, we can obtain the parameters
\begin{eqnarray}
&& \theta_{1}^{4} = -\frac{e^{i\varphi}}{\text{det}\,S} \, \frac{S_{12}\bar{S}_{22}}{S_{11}\bar{S}_{21}}, \\
&& \theta_{2}^{4} = -\frac{e^{i\varphi}}{\text{det}\,S} \, \frac{S_{11}\bar{S}_{21}}{S_{12}\bar{S}_{22}}.
\end{eqnarray}
For this scattering problem, we find that the unitary scattering matrix is given by
\begin{equation}
S' = \frac{e^{i\varphi/2}}{\sqrt{\text{det}\,S}}\begin{pmatrix} -\sqrt{\frac{ q Z_{\text{in}}}{ k Z_{\text{out}}}} S_{11} & S_{12} \\ S_{21} & -\sqrt{\frac{ k Z_{\text{out}}}{ q Z_{\text{in}}}} S_{22} \end{pmatrix},
\end{equation}
where $\varphi$ is a free parameter of the system, and thus can be set to zero. Also we have that, for this problem, $|\text{det}\,S|=1$. In the entries of this matrix, we can identify the transmission and reflection coefficients,
\begin{equation}\label{scattering_matrix}
S' = \begin{pmatrix} t_{L} & r_{R} \\ r_{L} & t_{R} \end{pmatrix}.
\end{equation}
In the case in which there is no antenna, we recover the usual formulas in optics when there is an abrupt change in the impedance,
\begin{eqnarray}
&& |t_{L}|^{2} = |t_{R}|^{2} \rightarrow \left(\frac{2\sqrt{Z_{\text{in}}Z_{\text{out}}}}{Z_{\text{in}}+Z_{\text{out}}}\right)^{2}, \\
&& |r_{R}|^{2} = |r_{L}|^{2} \rightarrow \left(\frac{Z_{\text{in}}-Z_{\text{out}}}{Z_{\text{in}}+Z_{\text{out}}}\right)^{2}.
\end{eqnarray}
Notice that, when both impedances are equal, there are no reflections. In the opposite limit, having an infinitely-long antenna, we find that
\begin{eqnarray}
&& |t_{L}|^{2} = |t_{R}|^{2} \rightarrow 1, \\
&& |r_{R}|^{2} = |r_{L}|^{2} \rightarrow 0.
\end{eqnarray}
This limit corresponds to an infinitesimally-slow (adiabatic) change of impedance, generating no reflections in a wave propagating through it into another medium. Now, we want to apply the scattering matrix to a given state propagating through the antenna, and study the entanglement of the output state. 

%obtain the transmission coefficient for a given state, depending on the size of the antenna, $d$. 

\subsubsection{Entanglement through the antenna}
We study the performance of the antenna for two-mode squeezed states, which are the best candidate for Gaussian entangled quantum states with CVs due to the stability and simplicity with which they are generated. 

Consider a TMST state and a thermal state with $N_{\text{th}}$ photons coming from the environment. The covariance matrix describing these three modes is
\begin{equation}
\Sigma_{\text{th-in}} = (1+2n) \begin{pmatrix} \varrho\mathbb{1}_{2} & 0 & 0 \\ 0 & \cosh 2r \mathbb{1}_{2} & \sinh 2r \sigma_{z} \\ 0 & \sinh 2r \sigma_{z} & \cosh 2r \mathbb{1}_{2} \end{pmatrix},
\end{equation}
where we have defined $\varrho = (1+2N_{\text{th}})/(1+2n)$. This is the state prepared in the cryostat. Now, we want to send one of the modes of the TMST state through the antenna to another party, and optimize the entanglement between the remaining mode and the transmitted one. The latter goes through the antenna while mixing with the thermal noise coming from the environment, a process characterized by the scattering matrix, while the former remains untouched. This process is described by the action of the operator
\begin{equation}
T = \begin{pmatrix} S' & 0 \\ 0 & \mathbb{1}_{2} \end{pmatrix}
\end{equation}
on the matrix above. We then trace out the reflected part coming from the scattering matrix, and obtain the covariance matrix of the output state,
\begin{equation}
\Sigma_{\text{out}} = tr_{B} \left[ T\Sigma_{\text{th-in}} T^{\dagger}\right].
\end{equation}
Given the order in which we have written the states in the covariance matrix, the scattering matrix $S'$ is just a reshuffling of the one in Eq.~\eqref{scattering_matrix},
\begin{equation}
S' =  \begin{pmatrix} r_{R}\mathbb{1}_{2} & t_{L}\mathbb{1}_{2} \\ t_{R}\mathbb{1}_{2} & r_{L}\mathbb{1}_{2} \end{pmatrix}.
\end{equation}
This way, we find the covariance matrix of the output state
\begin{equation}\label{CMout}
\Sigma_{\text{out}} = (1+2n) \begin{pmatrix} \left(\varrho |r_{R}|^{2} + |t_{L}|^{2} \cosh 2r\right)\mathbb{1}_{2} & t_{L} \sinh 2r \sigma_{z} \\ \bar{t}_{L} \sinh 2r \sigma_{z} & \cosh 2r \mathbb{1}_{2} \end{pmatrix}.
\end{equation}
As a measure of entanglement, we use the negativity, computed through the smallest symplectic eigenvalue of the partially-transposed covariance matrix. For the initial TMST state, this is
\begin{equation}
\tilde{\nu}_{-}^{\text{in}} = (1+2n)e^{-2r},
\end{equation}
and the condition for entanglement, $\tilde{\nu}_{-}^{\text{in}}<1$, is expressed as $r > \frac{1}{2}\log(1+2n)$. Notice that this condition is $r>0$ for two-mode squeezed vacuum states ($n=0$). 

The outgoing state will also be a TMST state, up to unitary transformations. Thus, its symplectic eigenvalue will have the same form as that of the initial state. If we set the initial squeezing to zero, no squeezing can be generated through the beam splitter, and we have $\tilde{\nu}_{-}^{\text{in}}  = 1+2n$ for the initial state, and $\tilde{\nu}_{-}^{\text{out}}  = 1+2n'$ for the final state. Comparing this formula with the symplectic eigenvalue obtained from Eq.~\eqref{CMout}, we find that $n'=n$. Then, $\tilde{\nu}_{-}^{\text{out}}  = (1+2n)e^{-2r'}$, and
\begin{equation}
r' = -\frac{1}{2}\log\left( \frac{\tilde{\nu}_{-}^{\text{out}} }{1+2n}\right).
\end{equation}
If we write out explicitly the symplectic eigenvalue of the output state, we obtain
\begin{equation}
\tilde{\nu}_{-}^{\text{out}} = \frac{1}{\sqrt{2}} \sqrt{\tilde{\Delta}(\Sigma_{\text{out}})- \sqrt{\tilde{\Delta}^{2}(\Sigma_{\text{out}})-4\det\Sigma_{\text{out}}}}
\end{equation}
where we have identified
\begin{equation}
\tilde{\Delta}(\Sigma_{\text{out}}) = (1+2n)^{2} \left[ (\varrho|r_{R}|^{2}+|t_{L}|^{2}\cosh2r)^{2} + \cosh^{2}2r + 2|t_{L}|^{2}\sinh^{2}2r \right],
\end{equation}
together with
\begin{eqnarray}
\nonumber \sqrt{\tilde{\Delta}^{2}(\Sigma_{\text{out}})-4\det\Sigma_{\text{out}}} &=&  (1+2n)^{2} \Big[ \varrho |r_{R}|^{2}+(1+|t_{L}|^{2})\cosh 2r \Big] \times \\
&& \sqrt{(\varrho-\cosh2r)^{2}|r_{R}|^{4} + 4|t_{L}|^{2}\sinh^{2}2r}.
\end{eqnarray}
The number of thermal photons is estimated from the Bose-Einstein distribution
\begin{equation}
n(f) \propto \frac{1}{e^{\frac{\hbar f}{k_{B}T}}-1}.
\end{equation}
The number of thermal photons of frequency $f = \omega/2\pi = 5$ GHz is $8\times10^{-3}$ at temperature $T\sim50$ mK, whereas at room temperature ($T\sim300$ K), the number of thermal photons is approximately $1250$, which implies $\varrho\sim 2500$. Now, we can approximate $\tilde{\nu}_{-}^{\text{out}}$ depending on the relation between $\varrho|r_{R}|^{2}$ and $|t_{L}|^{2}$, and obtain a simplified form in the different regimes. The first case we study is $\varrho|r_{R}|^{2} \gg 1$ with $|r_{R}|\ne0$. Here, we find that
\begin{equation}
\tilde{\nu}_{-}^{\text{out}} = \tilde{\nu}_{-}^{\text{in}}+ (1+2n)\sinh2r \left[ 1 - \frac{|t_{L}|^{2}(1+|t_{L}|^{2})\sinh^{2}2r}{2\varrho|r_{R}|^{2}\cosh2r} \right].
\end{equation}
Total reflection by the antenna is achieved by taking $|r_{R}|\rightarrow 1$, then $\tilde{\nu}_{-}^{\text{out}} = (1+2n)\cosh2r$, which is always greater or equal to 1. This means that there cannot be entanglement, because we are neglecting the reflected mode, and the transmitted one only has thermal noise from the environment. Then, we just have two thermal states. Total transmission, $|r_{R}|\rightarrow 0$, breaks the approximation we have made here.  Furthermore, see that we recover the result $1+2n$ as $r\rightarrow 0$. In this case, $\tilde{\nu}_{-}^{\text{out}}$ is smaller for larger $|t_{L}|$, only showing entanglement for $|r_{R}|<0.1$. We will focus on this regime in the following case.

The second case describes the scenario in which $\varrho|r_{R}|^{2} \ll 1$ with $|t_{L}|\approx 1$. This regime is more restrictive and it is close to total transmission, because we need $|r_{R}| < 10^{-2}$ in order to have $\varrho|r_{R}|^{2} \ll 1$. Here, we find
\begin{equation}
\tilde{\nu}_{-}^{\text{out}} = \tilde{\nu}_{-}^{\text{in}} \left( 1 + \frac{\varrho|r_{R}|^{2}}{2}e^{2r} \right) = \tilde{\nu}_{-}^{\text{in}}+ \left( \frac{1}{2}+N_{\text{th}}\right)|r_{R}|^{2}.
\end{equation}
When $r\rightarrow 0$, the approximation breaks down and we would have to substitute before the approximation. For total transmission, $|r_{R}|=0$, we recover the initial state, since no thermal noise from the environment is mixed with the mode transferred through the antenna. The condition for entanglement on the initial state is
\begin{equation}
r > \frac{1}{2}\log(1+2n)
\end{equation}
and, in this case, for the output state we find
\begin{equation}
r > \frac{1}{2}\log(1+2n) -\frac{1}{2}\log\left[ 1-  \left( \frac{1}{2}+N_{\text{th}}\right)|r_{R}|^{2}\right],
\end{equation}
which is, of course, more restrictive. The first inequality imposes $\left( \frac{1}{2}+N_{\text{th}}\right)|r_{R}|^{2} < 1$. If we approximate $\log(1\pm x) \simeq \pm x$ for $x \ll 1$, then we can write the condition for entanglement on the input state's squeezing parameter as $r > n$ for the initial state, and $r > n + \frac{1}{2}N_{\text{th}}|r_{R}|^{2}$ for the output state.

We have found that it is not possible to achieve values of the reflection coefficient lower than $|r_{R}| \sim 0.08$ with a linear antenna. For the squeezing of the initial state around $r=1$, we need $|r_{R}| < 0.026$ in order for the output state to be entangled. An antenna in which the impedance grows linearly with the position is not sufficient for entanglement preservation, and for this we explore the stepwise antenna.

%%%%%%%%%%%%%%%%%%%%%%%%%%%%%%%%%%%%%%%%%%%%%%%
\begin{figure}[h]
\centering
{\includegraphics[width=0.95 \textwidth]{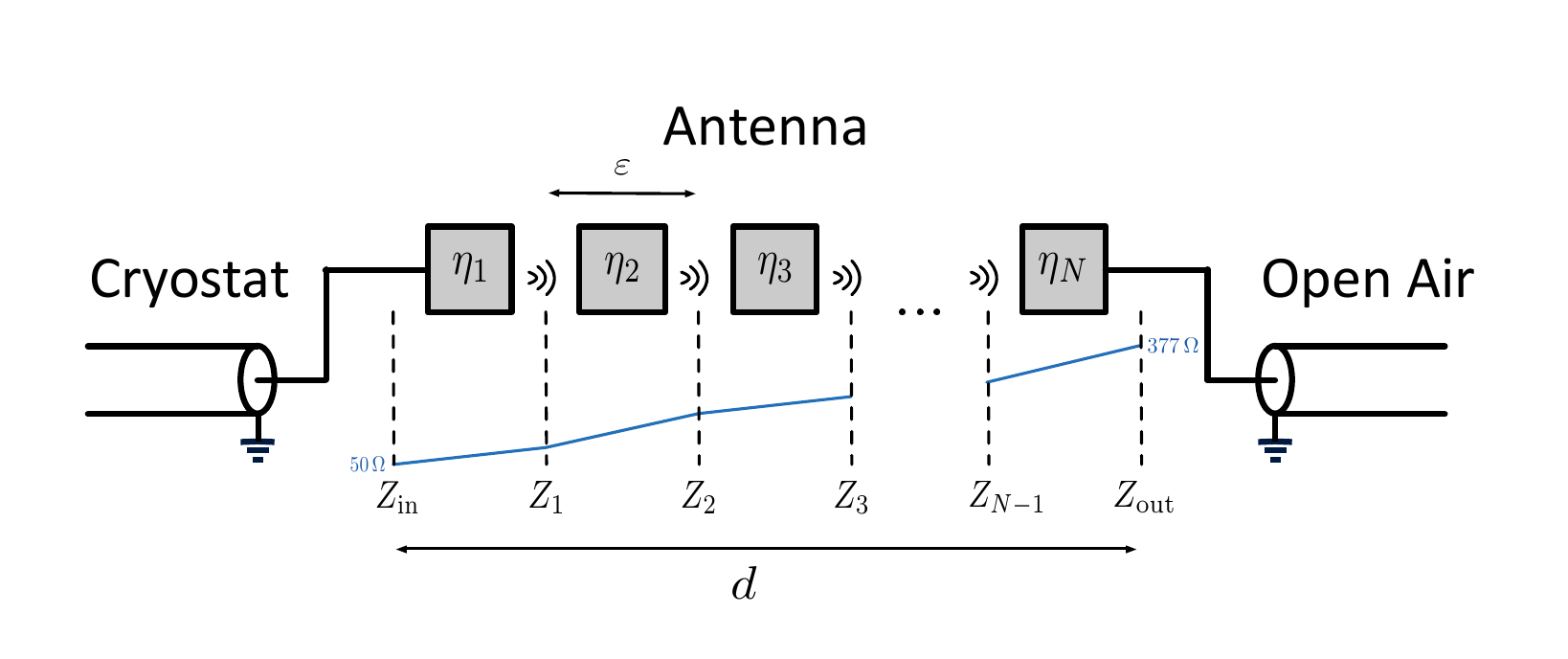}}
\caption[Circuit connecting the cryostat and the open air through a stepwise antenna]{\textbf{Circuit connecting the cryostat and the open air through a stepwise antenna}. It is divided in $N$ slices of length $\varepsilon$, inside which the impedance changes linearly, corresponding to a beam splitter with reflectivity $\eta_{i}=|r_{R}^{(i)}|^{2}$, setting $N+1$ scattering problems. Globally, it allows one to implement a general function of the impedance.}
\label{fig4_3}
\end{figure}
%%%%%%%%%%%%%%%%%%%%%%%%%%%%%%%%%%%%%%%%%%%%%%%
 
\subsubsection{Stepwise antenna}\label{sec_4}
We propose a different approach to study the circuit: consider the division of the antenna in $N$ infinitesimally small slices, in which the impedance changes linearly with the position. All these slices together yield an impedance that changes with the position. This setup can be seen in Fig.~\ref{fig4_3}. The difference with the previous approach is that now we have $N-1$ new parameters, i.e. the impedances of the intermediate slices, which we can use to optimize step by step the transfer of the quantum state in the antenna, together with the size of the TL. This is a similar derivation to the one leading to Eq.~\ref{SL_solution}. In this case, we have an impedance
\begin{equation}
Z(x) = \left( m+1-\frac{x}{\varepsilon}\right)Z_{m} + \left( \frac{x}{\varepsilon} -m \right)Z_{m+1}
\end{equation}
at a slice $m$ in the TL. Then, parameter $y$ will be defined as
\begin{equation}
y = k\frac{Z(x)}{Z'(x)} = k(x-m\varepsilon) + k\varepsilon\frac{Z_{m}}{Z_{m+1}-Z_{m}}.
\end{equation}
Hence, the arbitrary parameters of the solution need to be redefined as $b_{i}^{(m)} k/(Z_{m+1}-Z_{m}) = c_{i}^{(m)}$, for $i=\{1,2\}$. Then, the spatial component of the wavefunction for slice $m$ in the antenna is given by
\begin{eqnarray}\label{SL_solution_m}
\nonumber u_{m}(x) &=& [\varepsilon Z_{m}+(x-m\varepsilon)(Z_{m+1}-Z_{m})] \times \\
&&\bigg[ c_{1}^{(m)}J_{1}\left(k(x-m\varepsilon)+k\varepsilon\frac{Z_{m}}{Z_{m+1}-Z_{m}}\right) \\
 \nonumber &+& c_{2}^{(m)}Y_{1}\left(k(x-m\varepsilon)+k\varepsilon\frac{Z_{m}}{Z_{m+1}-Z_{m}}\right)\bigg],
\end{eqnarray}
where $\varepsilon=d/N$ indicates the size of each slice, $x\in[\varepsilon m,\varepsilon (m+1)]$ and $m\in\{0, ..., N-1\}$. See that, for $N=1$, we recover the result of the linear antenna studied above. This system allows us to construct a transfer matrix for each of the $N$ scattering problems, such that the global transfer matrix will be the result of an ordered product of these $N$ matrices. In this problem, 
\begin{equation}
\begin{pmatrix} F \\ G \end{pmatrix} = T_{N}  \begin{pmatrix} c_{1}^{(N)} \\ c_{2}^{(N)} \end{pmatrix} = T_{N}  T_{N-1}\begin{pmatrix} c_{1}^{(N-1)} \\ c_{2}^{(N-1)} \end{pmatrix} = \ldots = T_{N}\ldots T_{0}\begin{pmatrix} A \\ B \end{pmatrix},
\end{equation}
and the global transfer matrix is $T = T_{N}T_{N-1}\ldots T_{0}$. From this global transfer matrix, we can obtain the global scattering matrix, and make it unitary in the same way as we did for the linear antenna. This technique allows us to implement different continuous piecewise functions for the impedance, and provides more freedom in the optimization process. Eventually, the design of this circuit is oriented towards optimizing the resource that is shared between two parties. Thus, the optimization process will involve the minimization of the reflection coefficient $|r_{R}|$, in order to maximize the entanglement in the output state. 
%%%%%%%%%%%%%%%%%%%%%%%%%%%%%%%%%%%%%%%%%%%%%%%
\begin{figure}[t]
\centering
{\includegraphics[width=0.75 \textwidth]{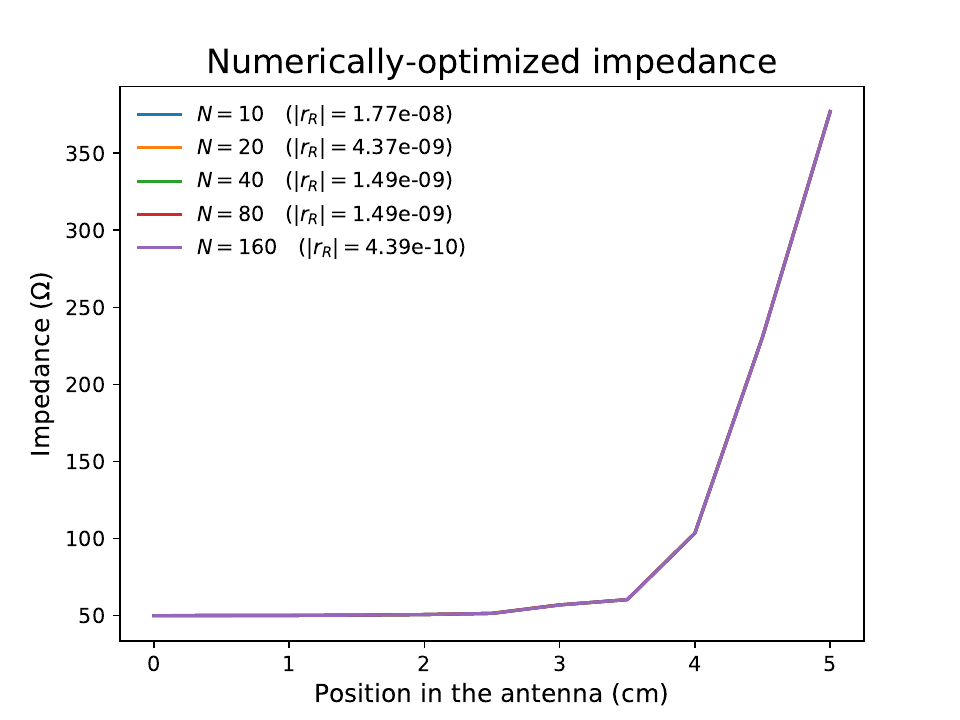}}
\caption[Numerically-optimized impedance curves against the position inside an antenna of length $d = 5$ cm]{\textbf{Numerically-optimized impedance curves against the position inside an antenna of length $d = 5$ cm}, for different values of the number of subdivisions in the antenna, $N$. Starting with the optimal solution for $N=10$, we compute the successive solutions through interpolation. All of the impedances appear overlapped, and for each one, the optimal value of the reflectivity $|r_{R}|$ is shown.}
\label{fig4_4}
\end{figure}
%%%%%%%%%%%%%%%%%%%%%%%%%%%%%%%%%%%%%%%%%%%%%%%
We are facing a global optimization problem that we will perform locally, step by step. Starting from random impedance arrays as initial guesses, we optimize the reflectivity with respect to the first impedance before $Z_{\text{out}} = 377 \, \Omega$, while keeping the rest of the impedances fixed. Once the optimal impedance value for the first point has been found, we update its value and optimize with respect to the previous point. We repeat the process until the point before $Z_{\text{in}} = 50 \, \Omega$ is reached. Of course, $Z_{\text{in}}$ and $Z_{\text{out}}$ must remain fixed. As a criterion for the stability of the solutions, we consider that the optimization process is successful when the difference between the reflectivities computed with the impedance solutions after two consecutive optimization sweeps is smaller than $10^{-10}$. 

Even with just one subdivision ($N=2$), we are able to find small enough values of the reflection coefficient to have an entangled output state. In the solutions presented in Fig.~\ref{fig4_4}, we start from a small number of subdivisions ($N=10$) and optimize the reflectivity. We then interpolate the optimal impedance by doubling the number of slices, adding the average impedance value of every pair of points in the original array in between said points. This means splitting each slice in half, while keeping the same linear impedance function. Because of this, the new impedance array, for $N=20$, gives the same reflectivity as the optimal impedance array we found for $N=10$. Now, taking the interpolated array as the initial guess, we optimize the reflectivity for $N=20$, and continue in the same fashion until we reach $N=160$. In Fig.~\ref{fig4_4}, we can observe how the optimal impedance curves are shaped for different values of $N$, starting at $N=10$, and doubling it through interpolation, until $N=160$. For each value, we also give the value of the reflection coefficient that such an antenna could achieve. Notice that these values decrease as $N$ becomes larger, while the interpolation method leads to very small changes in the impedances, such that the curves overlap and cannot be distinguished.

%%%%%%%%%%%%%%%%%%%%%%%%%%%%%%%%%%%%%%%%%%%%%%%
\begin{figure}[t]
\centering
{\includegraphics[width=0.75 \textwidth]{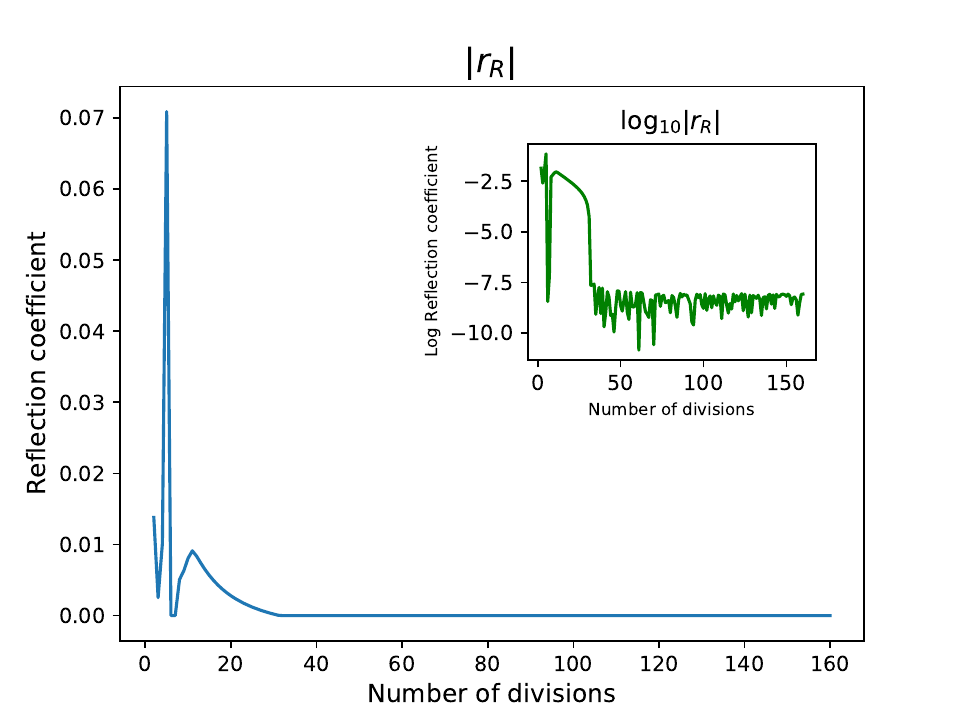}}
\caption[Reflection coefficient computed with the impedance function ansatz]{\textbf{Reflection coefficient computed with the impedance function ansatz} proposed in Eq.~\eqref{ansatz}, represented for different values of the number of subdivisions inside the antenna, for an antenna of size $d=5$ cm. As an inset, we show $\log_{10}|r_{R}|$ to illustrate how this impedance function reduces the reflection coefficient down to $10^{-8}$.}
\label{fig4_5}
\end{figure}
%%%%%%%%%%%%%%%%%%%%%%%%%%%%%%%%%%%%%%%%%%%%%%%

To speed up the optimization process and try to better recognize the behavior of the optimal impedance, we propose an ansatz to describe it,
\begin{equation}\label{ansatz}
Z(x) = Z_{\text{in}} + \alpha \left[ e^{\left(\frac{x}{d}\right)^{\beta}\log\left( 1+\frac{Z_{\text{out}}-Z_{\text{in}}}{\alpha}\right)} - 1\right]
\end{equation}
where $d$ is the size of the antenna, $x$ indicates the position inside it, and $\alpha$, $\beta$ are free parameters that we can optimize. This ansatz is inspired by the qualitative behavior of the curves in Fig.~\ref{fig4_4}, and does not correspond to an actual fit of the numerical data. Our goal is to rewrite $N-1$ local numerical optimization problems as a global optimization problem with just two parameters, $\alpha$ and $\beta$, in order to improve convergence and stability of the solutions. Notice that the results that we will find using this function will differ from those obtained with numerical optimization. In fact, since this is only an approximation of the optimal solution, the reflectivities we compute with this exponential impedance will be larger than those we can obtain with numerical optimization. We have found the optimal values to be $\alpha \sim 10.31$ and $\beta \sim 0.69$, for $d = 5$ cm. See that, for $\alpha \rightarrow \infty$, we recover the linear antenna. 

This function approximates the behavior of the optimal impedance inside the antenna, but the values of the reflection coefficient obtained are not sufficiently small. However, they improve as we increase $N$, as can be seen in Fig.~\ref{fig4_5}, oscillating around $|r_{R}| \sim 10^{-8}$ for $N$ approaching 160. We observe that minimal values of $|r_{R}|$ are achieved for $N>30$, which must represent a regime where $\varepsilon = d/N \ll \lambda$, approaching the continuum limit. This promising result suggests that we could employ the same treatment of the antenna as we did for a linear impedance, but solving the Sturm-Liouville problem with the impedance given by Eq.~\eqref{ansatz}, in the limit $N=1$.

\begin{figure}[t]
\centering
{\includegraphics[width=0.75 \textwidth]{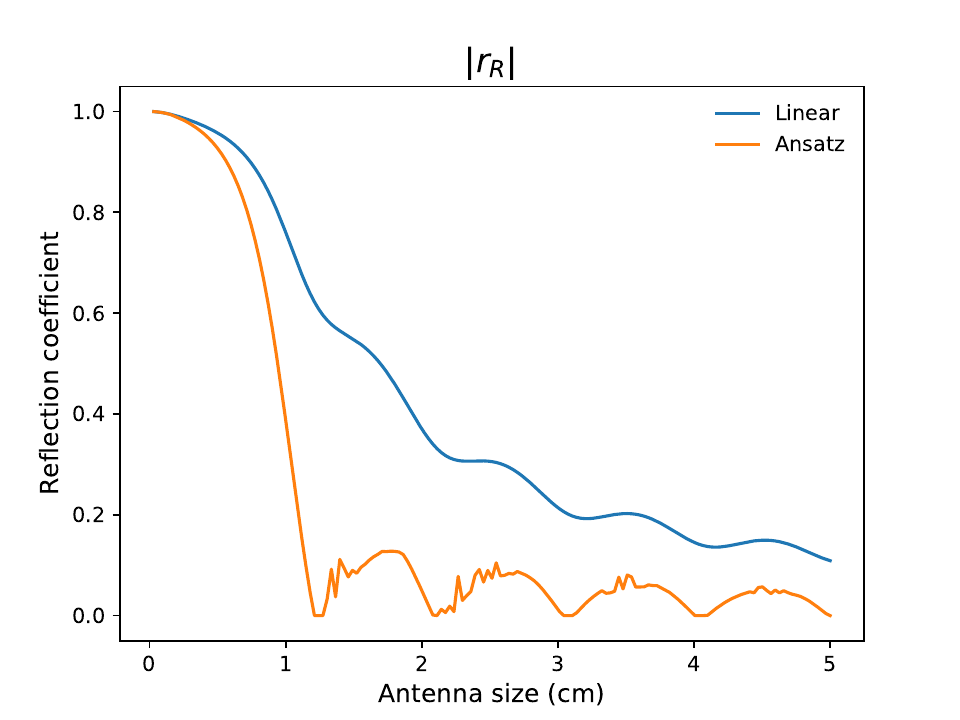}}
\caption[Reflection coefficient computed with a linear impedance versus the impedance in the ansatz, against the size of the antenna]{\textbf{Reflection coefficient computed with a linear impedance versus the impedance in the ansatz, against the size of the antenna}. In blue, the reflection coefficient is obtained with a linear function of the impedance, whereas in orange, we represent the reflection coefficient computed with the impedance ansatz proposed in Eq.~\eqref{ansatz}, for $N=160$. Notice that the reflectivity decreases further as we increase the size of the antenna, continuously for a linear impedance, and jumping between minimum values for the ansatz.}
\label{fig4_6}
\end{figure}

Taking $N=160$, we represent the reflection coefficient versus the antenna size in Fig.~\ref{fig4_6}. In blue, we plot the reflectivity of the antenna with a linear impedance function and, in orange, the result of the reflectivity corresponding to the impedance function proposed in Eq.~\eqref{ansatz}, with optimized parameters. We observe that minimal values of $|r_{R}|$ are achieved for particular values of the antenna size, which approximately coincide with multiples of half the wavelength inside the antenna. Also, we observe that, in order to find optimal values of the reflectivity, we require $d>\lambda/2$.

%%%%%%%%%%%%%%%%%%%%%%%%%%%%%%%%%%%%%%%%%%%%%%%
%%%%%%%%%%%%%%%%%%%%%%%%%%%%%%%%%%%%%%%%%%%%%%%

\begin{figure}[t]
\centering
{\includegraphics[width=0.75 \textwidth]{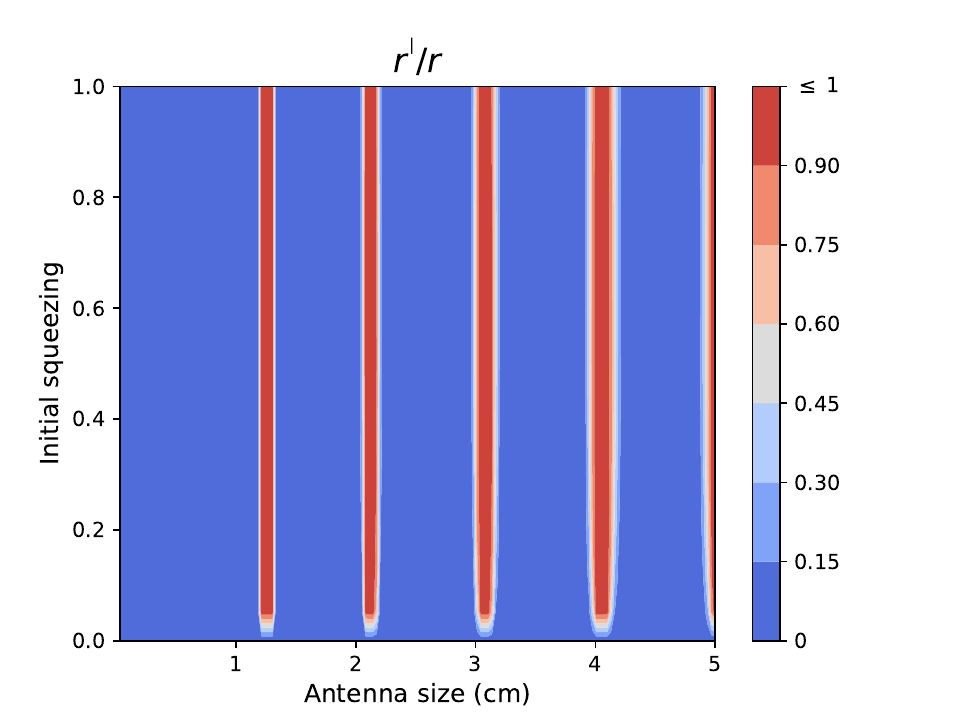}}
\caption[Ratio between the squeezing of the output state and the squeezing of the input state]{\textbf{Ratio between the squeezing of the output state and the squeezing of the input state}, represented in terms of $r$ and the size of the antenna for $N=160$ subdivisions. This shows that at least $90$~\% of the initial squeezing can be recovered with an antenna of size equal to a multiple of half a wavelength, with initial squeezing $r>0$.}
\label{fig4_7}
\end{figure}

Finally, we investigate the squeezing of the output state, in terms of the initial squeezing and the size of the antenna. In Fig.~\ref{fig4_7}, we represent the quotient between the squeezing parameters of output and input states, showing that it is possible to preserve squeezing in the multiples of the half-wavelength of the signal, the same spots for the size of the antenna observed in Fig.~\ref{fig4_6}, for which the reflectivity is minimal.

Furthermore, in order to illustrate the sensitivity of the reflection coefficient to the shape of the antenna, we introduce errors to the numerically-optimized impedance. We do this by drawing random values from a normal distribution, where the variance is a percentage of the value of the function at each point. Using the modified impedance, we compute the reflection coefficient, and then calculate the ratio between the negativity of the output state and the negativity of the input state, $\mathcal{N}_{\text{out}}/\mathcal{N}_{\text{in}}$. This study indicates a limit on manufacturing errors oriented towards the fabrication of such a device. 

%%%%%%%%%%%%%%%%%%%%%%%%%%%%%%%%%%%%%%%%%%%%%%%
%%%%%%%%%%%%%%%%%%%%%%%%%%%%%%%%%%%%%%%%%%%%%%%

We also compute the $n$-average values of the mean negativity ratio, a function towards which the mean should tend to in an infinite-trial scenario. This function is computed as follows: take a discrete function $f_{0}$, evaluated over a grid of points labelled by $x_{k}$, for $k\in[0,L]$. This function results from an average over many trials, given that it has a stochastic component based on a normal distribution. The function still presents traces of stochastic behavior, since the number of trials we can perform is finite. Our goal is to find the value towards which the infinite average of the function tends. For that, we propose the computation of the average of the function on a given point, such that 
\begin{equation}
f_{1}(x_{k}) = \frac{f_{0}(x_{k+1})+2f_{0}(x_{k})+f_{0}(x_{k-1})}{4},
\end{equation}
where $f_{1}$ is the 1-averaged function. Then, the $n$-averaged function is
\begin{equation}
f_{n}(x_{k}) = 2^{-2n} \sum_{m=0}^{2n} \begin{pmatrix} 2n \\ m \end{pmatrix} f_{0}(x_{k+n-m})\theta(k+n-m),
\end{equation}
with $\theta(0)=1$ and $\begin{pmatrix} 2n \\ m \end{pmatrix}=\frac{2n!}{m!(2n-m!)}$. Here, $n$ indicates the number of times the average has been performed, $k$ represents a point where the function is evaluated, and $m$ is a dummy index of the sum that goes through all the values that contributed to the $n$-average of the function at a point $x_{k}$. If $n > k$, then $m\in[0,n+k]$, and if $n < k$, $m\in[0,2n]$. From our definition of average we have taken $f_{n}(x_{0})=...=f_{1}(x_{0})=f_{0}(x_{0})$ and $f_{n}(x_{L})=...=f_{1}(x_{L})=f_{0}(x_{L})$. The largest binomial coefficient, $\begin{pmatrix} n \\ m \end{pmatrix}$, occurs at $m=n/2$, and then the largest contribution to the weighted sum that represents the $n$-average is
\begin{equation}
f_{n}(x_{k}) \approx 2^{-2n} \begin{pmatrix} 2n \\ n \end{pmatrix} f_{0}(x_{k}).
\end{equation}
This process exemplifies a discrete, binomial convolution, which in the continuum limit becomes a Gaussian convolution. 

%%%%%%%%%%%%%%%%%%%%%%%%%%%%%%%%%%%%%%%%%%%%%%%
\begin{figure}[t]
\centering
{\includegraphics[width=0.75 \textwidth]{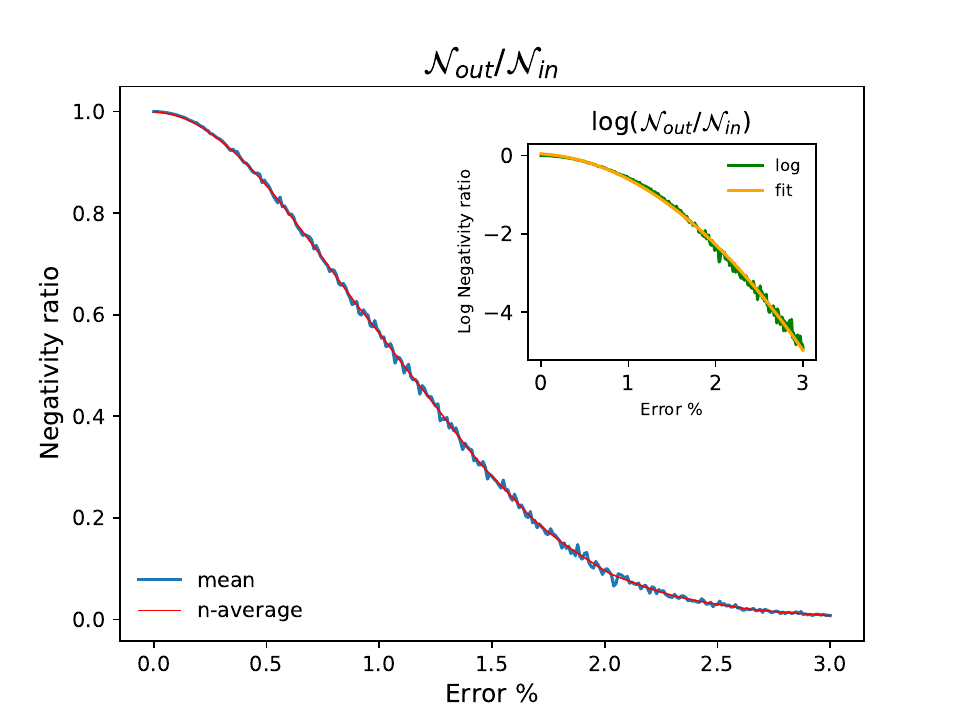}}
\caption[Assessment of the sensibility of the antenna to impedance fabrication errors]{\textbf{Assessment of the sensibility of the antenna to impedance fabrication errors} through the ratio between negativity of the output state $\mathcal{N}_{\text{out}}$ and negativity of the input state $\mathcal{N}_{\text{in}}$, averaged over many iterations in which the impedance function is modified with a random error proportional to a percentage of the value of the impedance at each point. In blue, we represent the mean value of the negativity ratio over different error percentages, and in red we display the smoothing of the mean by applying a n-average technique, for $n=50$. In green, the inset shows the logarithm of the ratio between negativities and, in orange, we show a quadratic fit of the logarithm of the negativity ratio. The latter corresponds to a function $ax^{2}+bx+c$ with $a\sim -0.51$, $b\sim -0.14$, and $c\sim 0.04$, and with variance $\sim 0.01$.}
\label{fig4_8}
\end{figure}
%%%%%%%%%%%%%%%%%%%%%%%%%%%%%%%%%%%%%%%%%%%%%%%

In Fig.~\ref{fig4_8}, we represent the average ratio of negativities for different values of the error percentage (blue), and we observe that it decreases as the error increases, for an initial squeezing $r=1$, $N=160$ subdivisions and antenna size $d = 5$ cm. As an inset, we represent the logarithm of the ratio of negativities (green), which we fit by a quadratic function (orange), as the function seems to follow a Gaussian. In red, we represent the $n$-average of the mean negativity ratio. Here, each error percentage step is averaged $10^{3}$ times, and we have taken $n=50$ for the $n$-average. The results show that the negativity ratio goes to zero for errors over 3\% of the impedance values, and from the quadratic fit of the logarithm, we can extract a function $ax^{2}+bx+c$ with $a\sim -0.51$, $b\sim -0.14$, and $c\sim 0.04$, and with variance $\sim 0.01$.

\subsubsection{Antenna design}
In this chapter, we have proposed an antenna based on a coplanar waveguide, and the characteristics of this waveguide will depend on the impedance we want to implement. Consider a coplanar waveguide, whose central conducting plate has a width of $2a$ and a height much smaller than the total depth of the film, and in which the distance between the middle of the conducting plate and the start of the grounded plates is $b$. By defining $\Theta=a/b$, we can write the density of inductance and the density of capacitance for such a waveguide as~\cite{Goppl2008,Clem2013}
\begin{eqnarray}
\label{CPW_L} l &=& \frac{\mu_{0}}{4}\frac{K\left(\sqrt{1-\Theta^{2}}\right)}{K(\Theta)}, \\ 
\label{CPW_C} c &=& 4\varepsilon_{0}\varepsilon_{\text{eff}}\frac{K(\Theta)}{K\left(\sqrt{1-\Theta^{2}}\right)},
\end{eqnarray}
where $\mu_{0}$ and $\varepsilon_{0}$ are the magnetic permeability and the electric permittivity of the vacuum, respectively, and $\varepsilon_{\text{eff}}$ is the effective dielectric constant; it is a function of the geometry of the waveguide, but also of the permittivities of the substrate and the oxide layers. Here, we have defined $K(y)$ as the complete elliptic integral of the first kind with modulus $y$, such that
\begin{equation}
K(y) = \int_{0}^{\pi/2} \text{d}\theta \frac{1}{\sqrt{1-y^{2}\sin^{2}\theta}}.
\end{equation}
From Eqs.~\eqref{CPW_L},~\eqref{CPW_C}, the characteristic impedance of the waveguide is straightforwardly obtained,
\begin{equation}
Z = \bar{z}\frac{K\left(\sqrt{1-\Theta^{2}}\right)}{K(\Theta)},
\end{equation}
with $\bar{z} = \frac{1}{4}\sqrt{\frac{\mu_{0}}{\varepsilon_{0}\varepsilon_{\text{eff}}}} = 10\pi$. For the cryostat impedance $Z_{\text{in}} = 50\,\Omega$, this requires that $\Theta_{\text{in}}\approx 0.32$, and for the impedance of open air, $Z_{\text{out}} = 377\,\Omega$, $\Theta_{\text{out}}\approx 2.60 \times 10^{-8}$.

In order to implement the kind of antenna proposed here, a coplanar waveguide has to be designed with a varying ratio $\Theta$. One way to do this is to solve the equation above for each value of $Z$. The dependence of $\Theta$ on the position inside the antenna could be inferred by substituting the values of $Z$ by those given in the ansatz proposed in Eq.~\eqref{ansatz}. Alternatively, we could directly propose an ansatz for $\Theta$, targeting a function of the position in the antenna that leads to a technologically-feasible design. Similarly, some parameters in this ansatz can remain free, such that an optimization over them allows us to obtain the ideal impedance. A simple example would be to consider
\begin{equation}
\Theta(x) = \Theta_{\text{in}} + \alpha \left[ e^{\left(\frac{x}{d}\right)^{\beta}\log\left( 1+\frac{\Theta_{\text{out}}-\Theta_{\text{in}}}{\alpha}\right)} - 1\right].
\end{equation}
Usual values of $a$ and $b$ are 5 $\mu$m and 7 $\mu$m, respectively. Fixing the value of $a_{\text{in}}$ to 5 $\mu$m, we would need $b_{\text{in}}=15.63$ $\mu$m in order to obtain $\Theta_{\text{in}}$. To get $\Theta_{\text{out}}$ at the termination of the antenna, we could for example set $a_{\text{out}} = 10$ nm and $b_{\text{out}} = 38.46$ cm. In principle, this may be achieved, given that the electron-beam lithography can achieve a precision below 10 nm for the fabrication of coplanar waveguides. However, the London depth of the material will impose a lower bound on the value of $a$ we would ideally want to set. Different realizations of such a device could be based on carbon-nanotube ink deposits on the gap of the coplanar waveguide, as described in Ref.~\cite{DePaolis2011}, or on coplanar waveguides with width-varying superconducting plate, studied in Ref.~\cite{Shamaileh2019}.

%On a waveguide, the density of inductance is related with the distance between the superconducting islands and the insulating cable, and the density of capacitance is related with the distance between superconducting islands. Then, a coplanar antenna such as the one we have designed has to be constructed on a superconducting waveguide which is not homogeneous. The distances between superconducting islands and the insulating cable need to vary such that the inductance and capacitance densities change according to Eqs.~\ref{ansatz_ind} \& \ref{ansatz_cap} respectively. Finally, these functions need to be optimized, meaning that $\alpha$ and $\beta$ need to be fixed, for a particular set of quantum states of a given frequency. 

Throughout this chapter, we have assumed that the antenna is implemented in a superconducting TL, meaning that the temperature inside it is in the range of mK (or at least below $4$ K). However, this would be very difficult to implement, since the end of this line is connected to the open air, whose temperature is 300 K. In order to maintain a low temperature in the antenna, with a constant propagation velocity of $v_{\text{in}}=c/3$, and still be able to connect it to the open air, we could study the addition of a subsequent waveguide. It would have the impedance of open air, 377 $\Omega$, while presenting a temperature gradient, as well as a velocity gradient, from $c/3$ to $c$. 

We consider modelling absorption losses due to loss of superconductivity in a transmission line of length $L$ that connects the antenna, at cryogenic temperatures, with the open air at $300$~K, by an infinite array of beam splitters. Each beam splitter has a reflectivity $\eta_{i}$ that represents absorption probability, and incorporates thermal noise at a given temperature $T_{i}$ inside the TL, characterized by a number of thermal photons $n(T_{i})$. The output mode of a $N$-beam splitter array of this kind is given by
\begin{equation}
\hat{a}_{N} = \hat{a}_{1}\prod_{i=1}^{N-1}\sqrt{1-\eta_{i}} + \sum_{k=1}^{N-1}\hat{h}_{k}^{\text{in}} \sqrt{\eta_{k}}\prod_{i=k+1}^{N-1}\sqrt{1-\eta_{i}}
\end{equation}
for an input signal mode $\hat{a}_{1}$, where the number of thermal photons incorporated by beam splitter $k$ is given by $n_{k}=\langle \hat{h}_{k}^{\text{in}\dagger}\hat{h}_{k}^{\text{in}}\rangle$. We aim at representing the action of this infinite array of beam splitters as a single beam splitter with effective reflectivity and effective number of thermal photons. In this expression, we can identify effective reflection and transmission coefficients,
\begin{equation}
\hat{a}_{N} = \hat{a}_{1}\sqrt{1-\eta_{\text{eff}}} + \hat{h}_{\text{eff}}^{\text{in}} \sqrt{\eta_{\text{eff}}}.
\end{equation}
Consider the reflectivity of a beam splitter as $\eta_{i}=\mu L/N$, where $\mu$ is the reflectivity per unit length. For very large $N$, assume $L/N = \Delta x$. Then, we could write $\eta_{i}=\mu_{i} \Delta x$, and then the effective reflectivity is simplified by
\begin{equation}
\log(1-\eta_{\text{eff}}) = \sum_{i=1}^{N-1}\log(1-\eta_{i}) = \sum_{i=1}^{N-1}\log(1-\mu_{i}\Delta x).
\end{equation}
For $\Delta x \ll 1$, we can expand this as $\log(1-\mu_{i}\Delta x) \approx -\mu_{i}\Delta x$, and taking the continuum limit, 
\begin{equation}
-\sum_{i=1}^{N-1}\mu_{i}\Delta x \longrightarrow -\int_{0}^{L} dx \mu(x).
\end{equation}
Then, we write $\eta_{\text{eff}} = 1-e^{-\int_{0}^{L} dx \mu(x)}$. Let us now compute the effective number of thermal photons, 
\begin{equation}
\eta_{\text{eff}} n_{\text{eff}} = \langle (\sqrt{\eta_{\text{eff}}}\hat{h}_{\text{eff}})^{\dagger}(\sqrt{\eta_{\text{eff}}}\hat{h}_{\text{eff}})\rangle = \sum_{k=1}^{N-1}\eta_{k} n_{k} \left[\prod_{i=k+1}^{N-1}(1-\eta_{i})\right],
\end{equation}
which can be expressed as
\begin{equation}
\eta_{\text{eff}}n_{\text{eff}} = \sum_{k=1}^{N-1}\mu_{k}\Delta x \, n_{k} e^{-\int_{x}^{L}dx' \mu(x')} = \int_{0}^{L} dx \mu(x)n(x) e^{-\int_{x}^{L}dx' \mu(x')}.
\end{equation}
Then, the effective number of thermal photons that this beam splitter incorporates to the system is
\begin{equation}
n_{\text{eff}} = \frac{\int_{0}^{L} dx \mu(x)n(x) e^{-\int_{x}^{L}dx' \mu(x')}}{1-e^{-\int_{0}^{L} dx \mu(x)}}.
\end{equation}
This expression is general and can be applied to any case in which we know the profile of temperatures. Let us now choose a simple but useful profile which allows us to find a closed expression. Indeed, if we consider that the TL can be kept at temperatures below the critical one for a length $L_{0} < L$, then we can choose
\begin{eqnarray}
n(x) &=& n(T_{\text{in}}) + [n(T_{\text{out}})-n(T_{\text{in}})]\theta(x-L_{0}), \\
\mu(x) &=& \mu_{\text{in}} + (\mu_{\text{out}}-\mu_{\text{in}})\theta(x-L_{0}),
\end{eqnarray}
where $\mu_{\text{in}}$ describes absorption losses at cryogenic temperatures and $\mu_{\text{out}}$ describes absorption losses of the material at room temperature. Then, the effective number of thermal photons becomes
\begin{eqnarray}
\nonumber n_{\text{eff}} &=& n(T_{\text{in}}) \frac{e^{-\mu_{\text{out}}(L-L_{0})}(1-e^{-\mu_{\text{in}}L_{0}})}{1-e^{-\mu_{\text{in}}L_{0}}e^{-\mu_{\text{out}}(L-L_{0})}} \\
&+& n(T_{\text{out}})\frac{1-e^{-\mu_{\text{out}}(L-L_{0})}}{1-e^{-\mu_{\text{in}}L_{0}}e^{-\mu_{\text{out}}(L-L_{0})}}.
\end{eqnarray}
Notice that, when $L_{0}=0$, then $n_{\text{eff}} = n(T_{\text{out}})$ and, when $L_{0}=L$, then $n_{\text{eff}} = n(T_{\text{in}})$. Consequently, for $T_{\text{out}}=300$~K and $\omega/2\pi = 5$~GHz, and by using the Bose-Einstein distribution, we obtain that $n(T_{\text{out}}) \approx N_{\text{th}}\sim1250$, which is considered as the input thermal noise into the antenna. The number of thermal photons at cryogenic temperatures is $n\sim8\times10^{-3}$, corresponding to $T_{\text{in}}=50$~mK and the same frequency. Given that $n/N_{\text{th}}\sim10^{-6}$, we have
\begin{equation}
\frac{n_{\text{eff}}}{N_{\text{th}}} \approx \frac{1-e^{-\mu_{\text{out}}(L-L_{0})}}{1-e^{-\mu_{\text{in}}L_{0}}e^{-\mu_{\text{out}}(L-L_{0})}} \leq 1,
\end{equation}
since $e^{-\mu_{\text{in}}L_{0}}\leq 1$. This implies that, considering this approach, the effect of thermal noise in the antenna is reduced when compared with respect to the study we present here. The reason is that we were considering before the thermal state as the incoming state of the antenna from the right, while it is now substantially reduced since part of the thermal photons are also absorbed in the cryostat before arriving at the antenna. Therefore, the introduction of these losses is a tradeoff between the effect of the effective beam splitter on the entanglement, and the improvement on the performance of the antenna due to the lower number of photons corresponding to the effective thermal state. Of course, these effects will substantially depend on the exact profile of temperatures along the TL. When this is obtained, one should repeat the optimization procedure for the impedance and then add the effective beam splitter after the antenna to take into account the entanglement degradation. 

%As discussed in the introduction of this chapter, although it is a crucial part of classical microwave communication, amplification of signals is not relevant in this setup. Consider a cryogenic HEMT amplifier, currently used in quantum microwave experiments, which produces large gains in a relatively large frequency spectrum, but also introduces a significant amount of noise. Thermal noise added by commercial HEMTs is counted in the range $n\sim10-100$ photons in the considered frequency regime~\cite{DiCandia2015}. Since the amplification is applied to the modes individually, it results ideal to enhance classical signals, but it cannot increase quantum correlations. To sum up, the goal of this work is to preserve quantum correlations in open air and traditional amplification does not provide an advantage in this objective, on the contrary, it could lead to entanglement degradation due to the introduction of thermal noise. 

%%%%% Conclusions %%%%%

In this chapter, we have studied recent advances in superconducting quantum technologies, in the form of devices that are used for quantum communication and for quantum computation. We have discussed different kinds of amplifiers, but focused on Josephson parametric amplifiers due to their state generation capabilities, although the most important milestone for microwaves is the photocounter, achieved through Josephson ring modulators. As the elementary step following state generation in a cryostat, we have presented a quantum antenna that attempts to reduce reflections of a signal travelling from here into open air, and therefore to maximize entanglement preservation. We have treated this device as a finite cavity that connects a waveguide which transports the state out of the cryostat, and a waveguide representing the transmission of that state in open air. Therefore, the antenna realizes a smooth impedance matching between the two environments, maximizing the transmission of energy.

\titleformat{\section}[display]
{\vspace*{190pt}
\bfseries\sffamily \LARGE}
{\begin{picture}(0,0)\put(-50,-35){\textcolor{grey}{\thesection}}\end{picture}}
{0pt}
{\textcolor{white}{#1}}
[]
\titlespacing*{\section}{80pt}{10pt}{40pt}[20pt]

% %Conclusions
% %Julen conclusions
% \titleformat{\section}[display]
% {\vspace*{190pt}
% \bfseries\sffamily \huge}
% {\begin{picture}(0,0)\put(-50,-25){\textcolor{grey}{\thesection}}\end{picture}}
% {0pt}
% {\textcolor{white}{#1}}
% []
% \titlespacing*{\section}{80pt}{10pt}{50pt}[20pt]

%%%%%%%%%%%%%%%%%%%%%%
%CHAPTER 5
%%%%%%%%%%%%%%%%%%%%%%
\section[Wireless Microwave Quantum Teleportation]{Wireless Microwave \\ Quantum Teleportation}
\label{sec5}

% \thiswatermark{\put(1,-282){\color{l-grey}\rule{84pt}{88pt}}
% \put(84,-282){\color{grey}\rule{410pt}{88pt}}}
%
%
% %Julen ch 2
% \thiswatermark{\put(1,-308){\color{l-grey}\rule{70pt}{64pt}}
% \put(70,-308){\color{grey}\rule{297pt}{64pt}}}
% \fancyhead[RO]{\leftmark}

%Julen ch 3
\thiswatermark{\put(1,-302){\color{l-grey}\rule{70pt}{60pt}}
\put(70,-302){\color{grey}\rule{297pt}{60pt}}}

% %Julen ch 4
% \thiswatermark{\put(1,-302){\color{l-grey}\rule{70pt}{60pt}}
% \put(70,-302){\color{grey}\rule{297pt}{60pt}}}
% \fancyhead[RO]{\small \leftmark}
%
%
% %Julen ch 5
% \thiswatermark{\put(1,-302){\color{l-grey}\rule{70pt}{60pt}}
% \put(70,-302){\color{grey}\rule{297pt}{60pt}}}
% \fancyhead[RO]{\leftmark}

% LIMITS OF MICROWAVE ENTANGLEMENT DISTRIBUTION IN OPEN AIR
\lettrine[lines=2, findent=3pt,nindent=0pt]{E}{ntanglement} distribution lies at the core of many quantum communication protocols; it is essential to quantum teleportation, and is required by many QKD protocols. The entangled resources that can be consumed to obtain an advantage in these quantum communication protocols have to be shared between the parties involved, and that step alone represents a challenge. 

In the optical regime, entanglement distribution has been achieved with optical fibres~\cite{Hubel2007,Zhang2008,Dynes2009,Inagaki2013,Wengerowsky2019} and in free space~\cite{Fedrizzi2009,Yin2012,Herbst2015,Yin2017}. 

In the microwave regime, entanglement distribution~\cite{Kurpiers2018,Magnard2020} and quantum teleportation~\cite{Baur2012,Steffen2013} have been achieved inside a cryostat. There has also been a theoretical proposal for CV quantum teleportation~\cite{DiCandia2015}, followed by the realization of entanglement distribution with coaxial cables, subsequently used for quantum teleportation~\cite{Fedorov2021}. Currently, there are no microwave entanglement distribution or quantum teleportation experiments in open air. The main reason behind this is that microwave quantum signals suffer from a bright thermal background at room temperature, which induces photon-absorption losses and thermalization of the signal. Diffraction losses, which we will not study in this chapter, are also expected to play an important role, given the lack of proper collimators for microwave quantum signals. Efficient quantum repeaters, on the path towards quantum communication networks, can help overcome this problem; in the current landscape, however, quantum repeaters use, in most cases, multiple copies of a quantum state, and therefore require quantum memories for storage purposes. Other repeaters for quantum communication are non-deterministic, and the magnitude for improvement decreases with the success probability. 

% Cite some papers on quantum repeaters

In this chapter, we address two pragmatic questions: which is the maximum distance for open-air microwave Gaussian entanglement distribution in a realistic scenario, and which technological and engineering challenges remain to be faced? In particular, we adapt the Braunstein-Kimble quantum teleportation protocol employing entangled resources previously distributed through open air, adapted to microwave technology. Performing this protocol is possible due to the recent breakthrough in the development of microwave homodyning~\cite{Fedorov2021} and photocounting~\cite{Dassonneville2020} schemes. When formulated in continuous variables, teleportation assumes a previously shared entangled state, ideally a TMSV state with infinite squeezing. In real life, however, only a finite squeezing level can be produced, making the state sensitive to entanglement degradation whenever either one or both modes are exposed to decoherence processes like thermal noise and/or photon losses. We study the generation of two-mode squeezed states and the challenges of their subsequent distribution through open air, and compute maximum distances of entanglement preservation for various physical situations. We consider recent advances in microwave photodetection and homodyning, and address their current limitations, and investigate open-air microwave quantum teleportation fidelities using the various quantum states derived in this chapter.

%\begin{figure}[th!]
%	\centering
%	\includegraphics[width=0.8\textwidth]{img/chap5/fig1}
%	\caption[Circuit QED schematic representation and operation sequence for simulating two-qubit gates coupled to the continuum of bosonic modes in a single Trotter step]{\textbf{Circuit QED schematic representation and operation sequence for simulating two-qubit gates coupled to the continuum of bosonic modes in a single Trotter step.} $(a)$ Schematic representation of our proposal for simulating fermion-fermion scattering in QFTs. An open transmission line supporting the continuum of bosonic modes interacts with three superconducting qubits. The second one-dimensional waveguide, forming a resonator due to the capacitors at each edge, supports a single mode of the microwave field and interacts with two superconducting qubits. Each qubit can be individually addressed through on-chip flux lines producing fluxes $\Phi^j_{\rm ext}$ and $\bar{\Phi}^j_{\rm ext}$ to tune the coupling strength and its corresponding energies. $(b)$ Sequence of multiple and single qubit gates, inside a Trotter step, acting on superconducting qubits to generate two-qubit interactions coupled to the continuum.}
%	\label{fig:theory5_Fig1}
%\end{figure}

\subsection{Wireless entanglement distribution}
\begin{figure}[t]
%\vspace{0.5cm}
\centering
\includegraphics[width=0.75\textwidth]{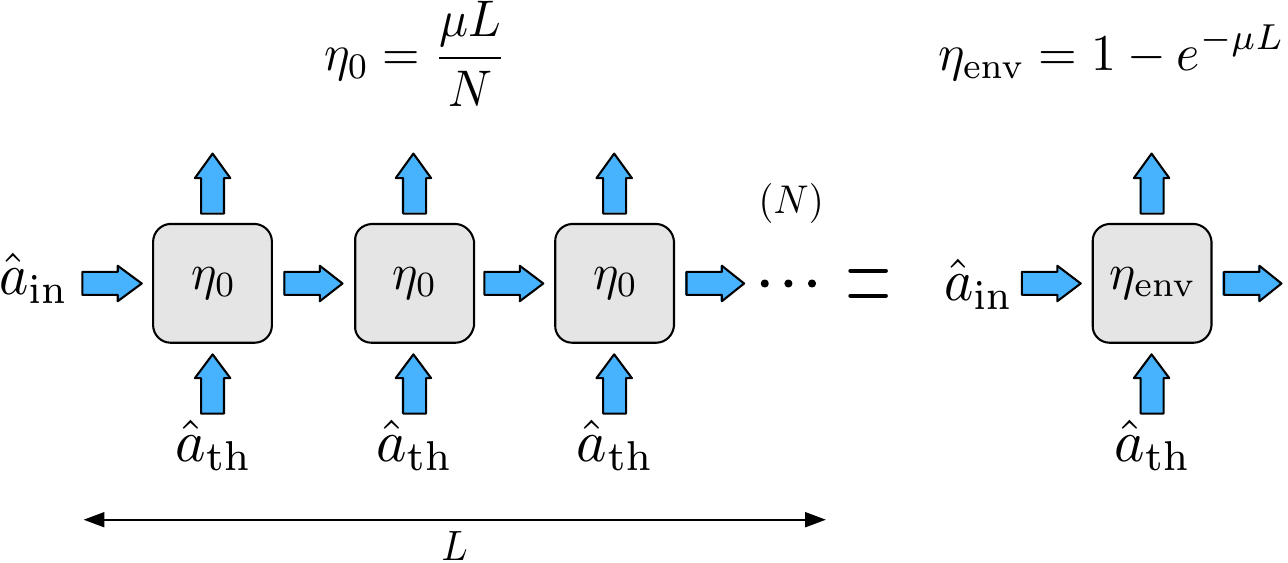}
\caption[Sketch of a beam splitter loss model of an open air quantum channel]{\textbf{Sketch of a beam splitter loss model of an open air quantum channel}. Entanglement degradation of a state propagating in open air at constant temperature is modeled by an array of $N$ beam splitters, each one mixing one signal mode and one thermal mode, represented by $\hat{a}_{\text{in}}$ and $\hat{a}_{\text{th}}$ respectively. The latter introduces thermal noise characterized by $\langle\hat{a}^{\dagger}_{\text{th}}\hat{a}_{\text{th}}\rangle=N_{\text{th}}$ thermal photons, assuming constant temperature throughout the path. An infinite array of beam splitters ($N\rightarrow\infty$) can be approximated by a single beam splitter with reflectivity $\eta_{\text{env}} = 1 - e^{-\mu L}$, where $L$ is the total channel length and $\mu$ is the reflectivity per unit length.}
\label{fig5_1}
\end{figure}

Once we have generated our entangled resource, and once that state has been successfully sent out of the cryostat, we have to address the effects of entanglement degradation in open air. Considering directed transmission in open air, we envision an infinite array of beam splitters to describe losses in open air, as represented in Fig.~\ref{fig5_1}. Each one of these beam splitters represents the probability of an absorption event with probability $\eta_{0}$, such that a propagating signal mode is transformed as
\begin{equation}
\hat{a}_{\text{in}} \longrightarrow \sqrt{1-\eta_{0}}\hat{a}_{\text{in}} + \sqrt{\eta_{0}}\hat{a}_{\text{th}},
\end{equation}
mixing with thermal noise from the environment characterized by $\langle\hat{a}^{\dagger}_{\text{th}}\hat{a}_{\text{th}}\rangle=N_{\text{th}}$ thermal photons. Assuming constant temperature throughout the sequence of possible absorption events, meaning that the thermal noise in each of the beam splitters is characterized by $N_{\text{th}}$, we can obtain the reflectivity of an effective beam splitter based on an attenuation channel~\cite{Serafini2017}, which represents the decay of quantum correlations and amplitudes,
\begin{equation}
\eta_{\text{env}} = 1-e^{-\mu L}.
\end{equation}
Here, $\mu$ represents a density of reflectivity, which in turn models photon losses per unit length, and $L$ is the traveled distance. This density of reflectivity can be interpreted as an attenuation coefficient that quantifies the specific attenuation of signals in a given environment. In this work we consider $\mu = 1.44\times 10^{-6}\text{ m}^{-1}$ for the specific attenuation of 5 GHz signals caused by the presence of oxygen molecules in the environment (see Refs.~\cite{Ho2004,ITU-R}).

%%%%%%%%%%%%%%%%%%%%%%%%%%%
%%%%%%%%%%%%%%%%%%%%%%%%%%%

We could go further and assume that, attached to the antenna (at constant temperature), there is another transmission line where the temperature is not constant throughout the trajectory, which leads to an inhomogeneous absorption probability. This is represented by the density of reflectivity $\mu(x)$, and the number of thermal photons $n(x)$. The latter still follows the Bose-Einstein distribution. An infinite array of beam splitters that reproduce these features (see Ref.~\cite{GonzalezRaya2020}) can be replaced by a single beam splitter with an effective reflectivity and number of thermal photons given by
\begin{eqnarray}
\eta_{\text{env}} &=& 1-e^{-\int_{0}^{L} \diff x \mu(x)}, \\
n_{\text{th}} &=& \frac{\int_{0}^{L} \diff x \mu(x)n(x) e^{-\int_{x}^{L}\diff x' \mu(x')}}{1-e^{-\int_{0}^{L} \diff x \mu(x)}},
\end{eqnarray}
where $L$ represents the total length of the array. Given that we are extending the length in which the transmission line remains at cryogenic temperatures, we see that $n_{\text{th}} \leq N_{\text{th}}$.

%%%%%%%%%%%%%%%%%%%%%%%%%%%
\begin{figure}[t]
\centering
\includegraphics[width=0.75\textwidth]{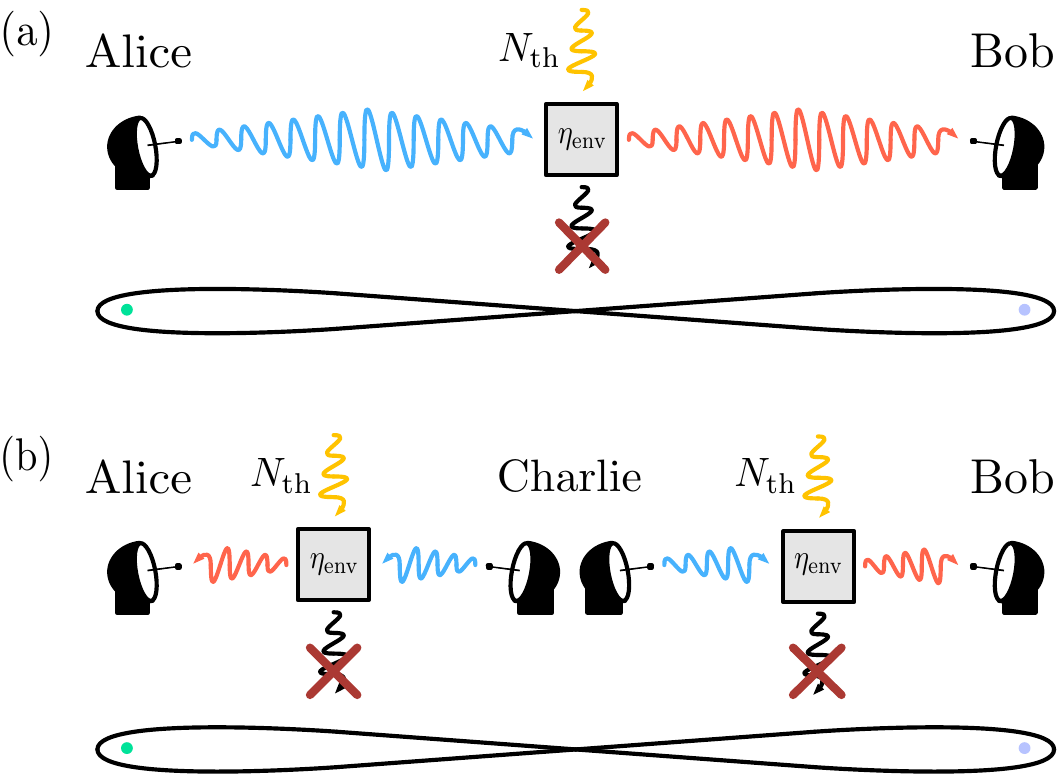}
\caption[Representation of an entanglement distribution protocol that uses antennae to efficiently transmit the quantum states into open air, where photon losses and thermal noise effects are described with a beam splitter with reflectivity $\eta_{\text{env}}$]{\textbf{Representation of an entanglement distribution protocol that uses antennae to efficiently transmit the quantum states into open air, where photon losses and thermal noise effects are described with a beam splitter with reflectivity $\eta_{\text{env}}$}. We analyze two different scenarios: (a) Alice generates the entangled state, and attempts to share one of its modes with Bob by sending it through a noisy and lossy open-air channel that degrades the entanglement strength; (b) Charlie generates a two-mode entangled state, and sends one entangled mode to Alice and another to Bob. In this case, although both modes go through the same noisy and lossy channel, they travel half the distance compared to the previous case.}
\label{fig5_2}
\end{figure}
%%%%%%%%%%%%%%%%%%%%%%%%%%%

Now that we have discussed how the signal is processed into the environment, let us characterize the resulting states. Assume that Alice generates a TMST state with $n$ thermal photons, and sends one mode to Bob over a distance $L$ through open air, with a thermal background characterized by $N_{\text{th}}$ thermal photons. Then, the resulting state is what we call the ``asymmetric'' state
\begin{equation}\label{CM_asym}
\Sigma_{\text{Asym}}=(1+2n)\begin{pmatrix} \cosh 2r\mathbb{1}_{2} &  \sqrt{1-\eta_{\text{eff}}} \sinh 2r \sigma_{z}  \\ 
\sqrt{1-\eta_{\text{eff}}} \sinh 2r \sigma_{z} & \left[\varrho\eta_{\text{eff}} + (1-\eta_{\text{eff}})\cosh 2r\right]\mathbb{1}_2 \end{pmatrix},
\end{equation}
where $\eta_{\text{eff}}=1- e^{-\mu L}(1-\eta_\text{ant})$ represents the combined reflectivities of the antenna $\eta_{\text{ant}}$ and of the environment $\eta_{\text{env}}$. A sketch of the layout that leads to this kind of state can be seen in Fig.~\ref{fig5_2}~(a). With this, using Eq.~\eqref{eq:ptseigen}, we compute the partially transposed symplectic eigenvalue,
\begin{equation}
\tilde{\nu}_{-}^{\text{out}} = \tilde{\nu}_{-}^{\text{in}} + \left( \frac{1}{2} + N_{\text{th}}\right)\eta_{\text{eff}}
\end{equation}
for very low reflectivities, $\eta_{\text{eff}} N_{\text{th}} \ll 1$, with $\tilde{\nu}_{-}^{\text{in}} = (1+2n)e^{-2r}$. Note that, by reducing the reflectivity of the antenna, the impact of thermal noise is reduced, and the partially transposed symplectic eigenvalue approaches that of the input state. In this extreme case, entanglement is fully preserved. 

Let us use the partially transposed symplectic eigenvalue to compute the limit of entanglement. We use the negativity as a measure of Gaussian entanglement, such that this limit occurs for $\tilde{\nu}_{-}=1$. This constitutes a bound on the reflectivity; all smaller values of $\eta_{\text{eff}}$ will result in entanglement preservation. This result is
\begin{equation}\label{ref_bound}
\eta_\text{max} = \frac{1}{1+\frac{N_{\text{th}}}{1+\frac{2n(1+n)}{1-(1+2n)\cosh(2r)}}},
\vspace{0.15cm}
\end{equation}
together with the conditions $n < e^{-r}\sinh(r)$ and $r > 0$. With this bound, the maximum distance entanglement can survive is
\begin{equation}
L_\text{max} = -\frac{1}{\mu}\log(1-\eta_\text{max}).
\end{equation} 
Imagine that TMST states are generated in the cryostat at $50$ mK temperature, with thermal photons $n \sim 10^{-2}$,  and squeezing $r = 1$. In open air, at $300$ K, the number of thermal photons is $N_{\text{th}} \sim 1250$. Assuming a perfect antenna ($\eta_{\text{ant}}=0$), the maximum distance the state can travel before entanglement completely degrades is $L_\text{max} \sim 550$ m.

As a different approach to the entangled resource, we assume that a TMST state is generated at an intermediate spot between both parties, and that each mode is sent through an antenna and travels some distance $L_i$, with $i=\{1,2\}$, before reaching Alice and Bob. Then, each mode will see an effective reflectivity of $\eta^{(i)}_{\text{eff}}= 1 - e^{-\mu L_i}(1-\eta_\text{ant})$, combining the effects of the antenna and the environment. We assume for simplicity that $L_1+L_2 = L$, where $L$ is the linear distance between Alice and Bob. The covariance matrix of such a state, which we refer to as ``symmetric''
%\begin{equation}\label{CM_sym}
%{\small\Sigma_{\text{Sym}} =(1+2n)\begin{pmatrix} \left[\varrho\eta^{(1)}_{\text{eff}}  + \left(1-\eta^{(1)}_{\text{eff}}\right)\cosh 2r\right]\mathbb{1}_2 & \sqrt{\left(1-\eta^{(1)}_{\text{eff}}\right)\left(1-\eta^{(2)}_{\text{eff}}\right)} \sinh 2r \sigma_{z}  \\ 
%\sqrt{\left(1-\eta^{(1)}_{\text{eff}}\right)\left(1-\eta^{(2)}_{\text{eff}}\right)} \sinh 2r \sigma_{z} & \left[ \varrho\eta^{(2)}_{\text{eff}}  + \left(1-\eta^{(2)}_{\text{eff}}\right)\cosh 2r\right]\mathbb{1}_2 \end{pmatrix}},
%\end{equation}
\begin{eqnarray}\label{CM_sym}
\nonumber c'_{i} &=& \varrho\eta^{(i)}_{\text{eff}}  + \left(1-\eta^{(i)}_{\text{eff}}\right)\cosh 2r , \\
s' &=& \sqrt{\left(1-\eta^{(1)}_{\text{eff}}\right)\left(1-\eta^{(2)}_{\text{eff}}\right)} \sinh 2r , \\
\nonumber \Sigma_{\text{Sym}} &=& (1+2n)\begin{pmatrix}  c'_{1}\mathbb{1}_2 & s' \sigma_{z} \\  s' \sigma_{z} & c'_{2}\mathbb{1}_2 \end{pmatrix},
\end{eqnarray}
and corresponds to the layout represented in Fig.~\ref{fig5_2}~(b). With this state, the maximum distance entanglement can survive is $L_\text{max} \sim 480$ m.

Throughout this chapter, we refer to these states as the (asymmetric and/or symmetric) lossy TMST states, the bare states, or the TMST states distributed through open air. 

Furthermore, we could also consider the specific attenuation caused by the presence of water vapor in the environment~\cite{Ho2004}. This would lead to higher attenuation coefficients, thus reducing the distances that entanglement can survive. For an average water vapor density, these distances are 450 and 390 m for asymmetric and symmetric states, respectively. They become 400 m for asymmetric states and 350 m for symmetric states in a maximum water vapor density scenario.

\subsection{Wireless Quantum Teleportation Fidelities}\label{section:teleportation}

In this section, we compute the average teleportation fidelity for different resource states. In all cases, the teleported state is a coherent state $\ket{\alpha_0}\bra{\alpha_0}$. 

\subsubsection{Two-mode squeezed vacuum resource}
The case of a TMSV state is particularly simple, as we can simply plug its covariance matrix into Eq.~\eqref{eq:Gaussian-fidelity},
\begin{equation}
\overline{F}_{{\text{TMSV}}} = \frac{1+\lambda}{2}.
\end{equation}
When symmetric $2k$-photon subtraction is performed, the formula for Gaussian average fidelity can no longer be invoked. The results for $k=1,2$ (2PS and 4PS, respectively) are:
\begin{eqnarray}
\overline{F}_{\text{2PS}} &=& \left(1-\lambda\tau + \frac{\lambda^{2}\tau^{2}}{2}\right)\frac{(1+\lambda\tau)^{3}}{2(1+\lambda^{2}\tau^{2})}, \\
\nonumber \overline{F}_{\text{4PS}} &=& \frac{(1+\lambda\tau )^{5}\left[ 8-\lambda\tau(2-\lambda\tau)(8-3\lambda\tau(2-\lambda\tau)) \right]}{16(1+4\lambda^{2}\tau^2+\lambda^{4}\tau^4)}.
\end{eqnarray}
%%%%%%%%%%%%%%%%%%%%%%%%%%%%%%%%%%%%%%%%%%%%%%%%
\begin{figure}[t]
\centering
\includegraphics[width=0.75 \textwidth]{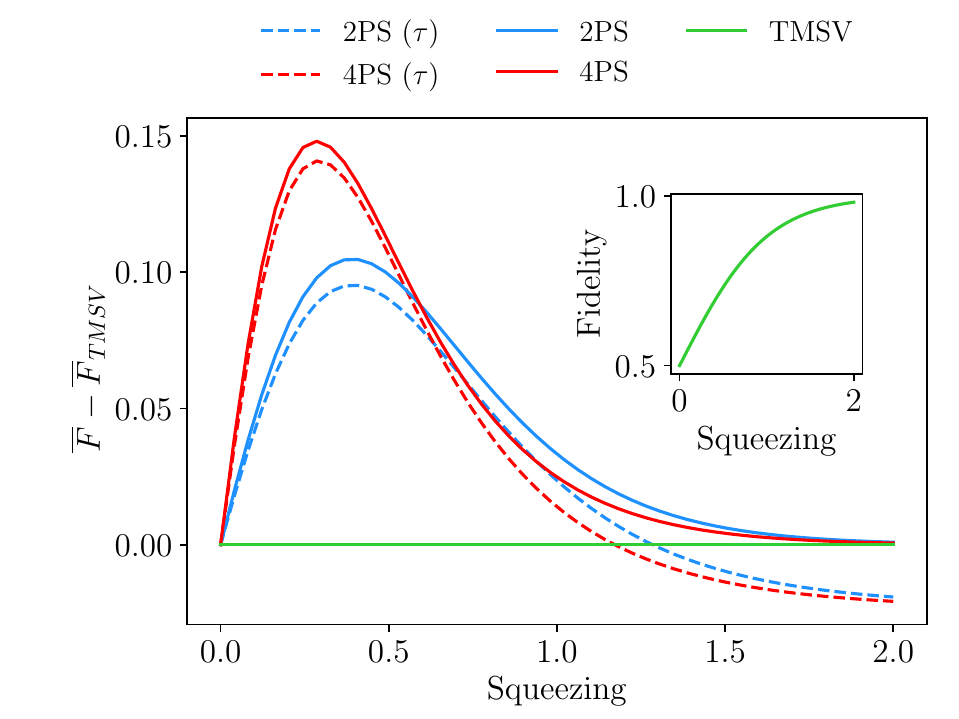}
\caption[Average fidelity of CV quantum teleportation of an unknown coherent state using a TMSV state with PS]{\textbf{Average fidelity of CV quantum teleportation of an unknown coherent state using a TMSV state with photon subtraction}, with respect to the initial squeezing parameter. We subtract the average fidelity associated with a TMSV state resource from the average fidelities of two-photon-subtracted (2PS, blue) and four-photon-subtracted (4PS, red) TMSV states. Curves associated with probabilistic photon subtraction appear dashed, whereas the solid curves are associated with heuristic photon subtraction. The green curve describes the TMSV case, which delimits the no-gain line, above which any point represents an improvement in fidelity due to photon subtraction. In the inset, we plot the average fidelity associated with a TMSV state against the initial squeezing parameter. We have considered the transmissivity of the beam splitters involved in probabilistic photon subtraction to be $\tau=0.95$.}
\label{fig5_3}
\end{figure}
%%%%%%%%%%%%%%%%%%%%%%%%%%%%%%%%%%%%%%%%%%%%%%%%

In Fig.~\ref{fig5_3}, we represent the result of subtracting the fidelity associated with the bare TMSV state to those associated with two-photon-subtracted (2PS, blue) and four-photon-subtracted (4PS, red) TMSV states. Fidelity differences associated with heuristic photon subtraction appear as solid lines, whereas those associated with probabilistic photon subtraction appear dashed. The green solid line represents the no-gain line, above which any PS state presents an advantage in fidelity. Note that photon subtraction works better for low squeezing, and as we increase it, we see that using the TMSV state as a resource for teleportation renders a higher fidelity than probabilistic photon subtraction, while heuristic photon subtraction tends to the TMSV result. 

\subsubsection{Two-mode squeezed thermal resource}
We now study the teleportation fidelity associated with a two-mode squeezed thermal state, sent through a lossy and noisy channel defined by the combination of the antenna and an environment with $N_\text{th}$ photons. By defining $\Sigma_{A}=\alpha\mathbb{1}_{2}$, $\Sigma_{B}=\beta\mathbb{1}_{2}$, and $\varepsilon_{AB}=\gamma \sigma_{z}$, we can write the average fidelity as
\begin{equation}
\overline{F}_{\text{TMST}} = \frac{1}{1+\frac{1}{2}\left(\alpha+\beta-2\gamma\right)}.
\end{equation}
If we consider the composition of $k$ teleportation protocols where each of the parties involved is separated by $L/k$, with $L$ the total distance aimed to cover. The final average fidelity is then given by
\begin{equation}
\overline{F}^{(k)}_{\text{TMST}} = \frac{1}{1+\left(k-\frac{1}{2}\right)\left(\alpha+\beta-2\gamma\right)},
\end{equation}
such that $\overline{F}_{\text{TMST}}>\overline{F}^{(k)}_{\text{TMST}}$ for $k>1$. Since the composition of teleportation protocols does not improve the overall fidelity, we study entanglement distillation and entanglement swapping in search for such gain. However, this fidelity composition may improve the overall fidelity when diffraction effects at the termination of the antenna come into play, which will reduce the reach of entanglement from the hundreds to the tens of meters.  

In Table~\ref{table1} we present the parameters we use to represent the different fidelity curves in this section.
\begin{table}
\centering
\begin{tabular}{|l|c|c|}
\hline
Parameter &Symbol &Value\\
\hline
Losses per unit of length  &$\mu$ & $1.44\times 10^{-6} \text{m}^{-1}$\\
%Signal wavelength in vacuum & $\lambda$  & 6 cm\\
Atmospheric temperature & $T$ &  300 K\\
Mean photon number & $N_\text{th}$ & 1250\\
Squeezing parameter & $r$ &  1\\
Thermal photon number (signal) & $n$ & $10^{-2}$ \\
Transmission coefficient & $\tau$ &  0.95\\
Antenna reflectivity & $\eta_{\text{ant}}$ & 0\\
\hline
\end{tabular}
\caption{Parameters for a terrestrial (1 atm of pressure, temperature of 300 K)  two-mode squeezed thermal state generated at a 50 mK cryostat, for a frequency of 5 GHz. These parameter values correspond to an Earth-based quantum teleportation scenario.}
\label{table1}
\end{table}

\subsubsubsection{Asymmetric case}
Assume that Alice generates a TMST state and sends one of the modes to Bob. Then, the covariance matrix of the state, given in Eq.~\eqref{CM_asym}, is characterized by
\begin{eqnarray}
\nonumber \alpha &=& (1+2n)\cosh 2r, \\
\beta &=& (1+2N_{\text{th}})\eta_{\text{eff}}  + (1+2n)(1-\eta_{\text{eff}})\cosh 2r, \\
\nonumber \gamma &=& (1+2n)\sqrt{1-\eta_{\text{eff}}}\sinh 2r,
\end{eqnarray}
which results in an average fidelity
\begin{eqnarray}
\nonumber \overline{F}_{\text{TMST}} &=& \bigg[ 1 + \left(\frac{1}{2}+N_{\text{th}}\right)\eta_{\text{eff}} + \left(\frac{1}{2}+n\right)(2-\eta_{\text{eff}})\cosh2r\\
&-& (1+2n)\sqrt{1-\eta_{\text{eff}}}\sinh2r \bigg]^{-1},
\end{eqnarray}
with $\eta_{\text{eff}} = 1-e^{-\mu L}(1-\eta_{\text{ant}})$. 

\subsubsubsection{Symmetric case}
In this case, we consider that the resource state is generated at an intermediate point between Alice and Bob, and is sent to both of them, such that now both modes are affected by the lossy and noisy channel described above. The covariance matrix of this state, presented in Eq.~\eqref{CM_sym}, is characterized by
\begin{eqnarray}
\nonumber \alpha &=& (1+2N_{\text{th}})\eta_{\text{eff}}  + (1+2n)(1-\eta_{\text{eff}})\cosh 2r, \\
\beta &=& (1+2N_{\text{th}})\eta_{\text{eff}}  + (1+2n)(1-\eta_{\text{eff}})\cosh 2r, \\
\nonumber \gamma &=& (1+2n)(1-\eta_{\text{eff}})\sinh 2r
\end{eqnarray}
where we have assumed $L_{1}=L_{2}=L/2$, and thus $\eta^{(1)}_{\text{eff}} = \eta^{(2)}_{\text{eff}} = \eta_{\text{eff}} = 1-e^{-\mu\frac{L}{2}}(1-\eta_{\text{ant}})$ . Then, the average fidelity can be written as
\begin{eqnarray}
\nonumber \overline{F}_{\text{TMST}} &=& \bigg[ 1 + (1+2N_{\text{th}})\eta_{\text{eff}}+(1+2n)(1-\eta_{\text{eff}})\cosh2r \\
&-& (1+2n)(1-\eta_{\text{eff}})\sinh2r \bigg]^{-1}.
\end{eqnarray}
Note that, for short distances, the fidelities associated with the asymmetric and symmetric states coincide. That is, at first order in $\mu L\ll 1$, and with $\eta_{\text{ant}}=0$, 
\begin{equation}
\overline{F}_{\text{TMST}} \approx \left[ 1 + (1+2N_{\text{th}})\frac{\mu L}{2}+(1+2n)\left(1-\frac{\mu L}{2}\right)e^{-2r}\right]^{-1}.
\end{equation}
When considering a lossy antenna, we observe higher entanglement degradation in the symmetric state due to the fact that both modes of the state are output by an antenna, whereas only one mode of the asymmetric state goes through it. Although $\sqrt{\eta_{\text{ant}}}$ can theoretically be reduced below $10^{-9}$~\cite{GonzalezRaya2020}, this leads to a slightly lower fidelity in the case of the symmetric state. In the figures appearing in this section, however, we consider $\eta_{\text{ant}}=0$ for simplicity. 

\subsubsection{Fidelity with photon subtraction}

If we consider a symmetric two-photon-subtraction process, in which the desired resource has lost a single photon in each mode, the average fidelity becomes
\begin{equation}
\overline{F}_{\text{2PS}} = \frac{4\left[ 2(5-\alpha\beta)\gamma^{2} - 4\gamma^{3} + \gamma^{4} + (\alpha-1)(\beta-1)(4(1+\gamma) + (\alpha+1)(\beta+1))\right]}{(2+\alpha+\beta-2\gamma)^{3}((\alpha-1)(\beta-1)+\gamma^{2})}.
\end{equation}
This is the heuristic case; in the probabilistic case, the fidelity reads
\begin{eqnarray}
\overline{F}_{\text{2PS}} &=& \frac{1}{4}\left[ 1+\tau\frac{-\alpha\beta + (1+\gamma)^{2} +((1-\alpha)(1-\beta)-\gamma^{2})\tau}{(1+\alpha)(1+\beta)-\gamma^{2}-(\alpha\beta-(1-\gamma)^{2})\tau}\right]^{3} \times \\
\nonumber && \left[1 + \frac{(1-\alpha\beta+\gamma^{2})^{2}-(\alpha-\beta)^{2}+4\gamma^{2}+4\gamma((1-\alpha)(1-\beta)-\gamma^{2})\tau}{(1-\alpha\beta+\gamma^{2}+((1-\alpha)(1-\beta)-\gamma^{2})\tau)^{2}-(\alpha-\beta)^{2}+4\gamma^{2}}\right],
\end{eqnarray}
and the success probability
\begin{equation}\label{eq:success-prob}
%P = 4(1-\tau)^{2}\frac{\left[1-\alpha\beta+\gamma^{2}+((1-\alpha)(1-\beta)-\gamma^{2})\tau\right]^{2}-(\alpha-\beta)^{2}+4\gamma^{2}}{\left[(1+\alpha)(1+\beta)-\gamma^{2}+2(1-\alpha\beta+\gamma^{2})\tau+((1-\alpha)(1-\beta)-\gamma^{2})\tau^{2}\right]^{3}}. 
P = 4(1-\tau)^{2}\frac{\left[1-\alpha\beta+\gamma^{2}+((1-\alpha)(1-\beta)-\gamma^{2})\tau\right]^{2}-(\alpha-\beta)^{2}+4\gamma^{2}}{\left[(1+\tau)^{2} + (\alpha+\beta)(1-\tau^{2}) + (\alpha\beta-\gamma^{2})(1-\tau)^{2}\right]^{3}}. 
\end{equation}

%%%%%%%%%%%%%%%%%%%%%%%%%%%%%%%%%%%%%%%%%%%%%%%%
\begin{figure}[t]
\centering
\includegraphics[width=0.75 \textwidth]{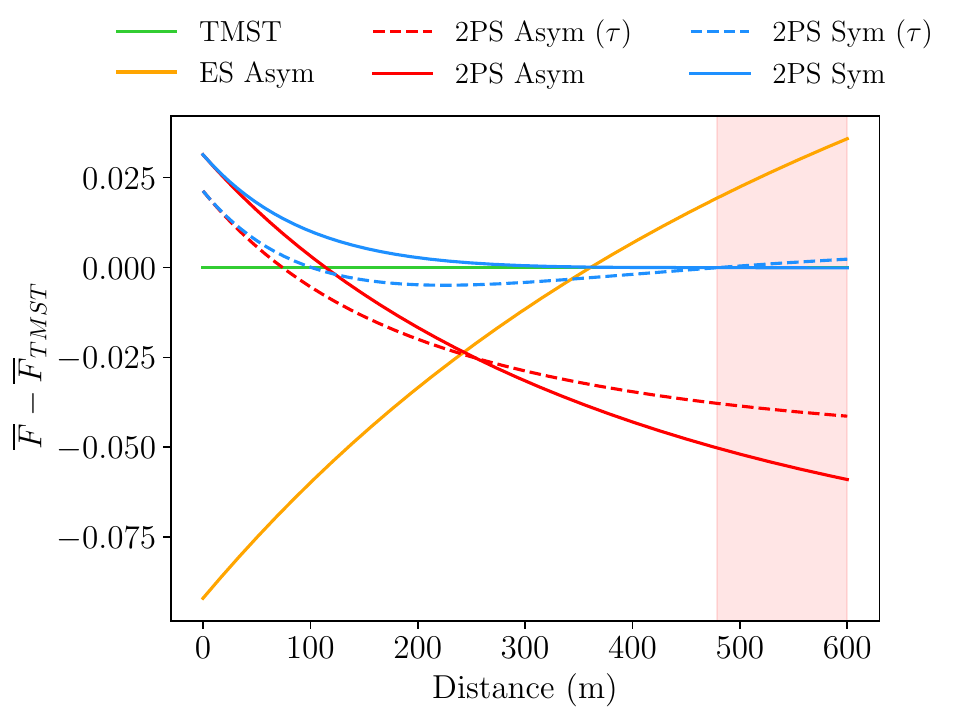}
\caption[Average fidelity of CV quantum teleportation of an unknown coherent state using an entangled resource distributed through open air, with photon subtraction and entanglement swapping]{\textbf{Average fidelity of CV quantum teleportation of an unknown coherent state using an entangled resource distributed through open air, with photon subtraction and entanglement swapping}, represented versus the traveled distance. We subtract the average fidelity associated with the TMST state distributed through open air (green), from the average fidelities of the two-photon-subtracted asymmetric (2PS asym) and symmetric (2PS sym) states, represented in red and blue, respectively, as well as from the average fidelity of the entanglement-swapped asymmetric (ES asym) state, in orange. We represent the states resulting from  probabilistic photon subtraction (dashed), as well as heuristic photon subtraction (solid). The pale red background represents the region where the fidelity is below the maximum classical fidelity of $1/2$, and the quantum advantage is lost. The green line then shows no gain, and any point above it corresponds to an improvement in fidelity. Parameters are $n=10^{-2}$, $N_{\text{th}}=1250$, $r=1$, $\mu=1.44\times^{-6}$ m$^{-1}$, $\eta_{\text{ant}}=0$, $\tau=0.95$.}
\label{fig5_4}
\end{figure}
%%%%%%%%%%%%%%%%%%%%%%%%%%%%%%%%%%%%%%%%%%%%%%%%

\begin{figure}[hbtp]
\centering
\includegraphics[width=\textwidth]{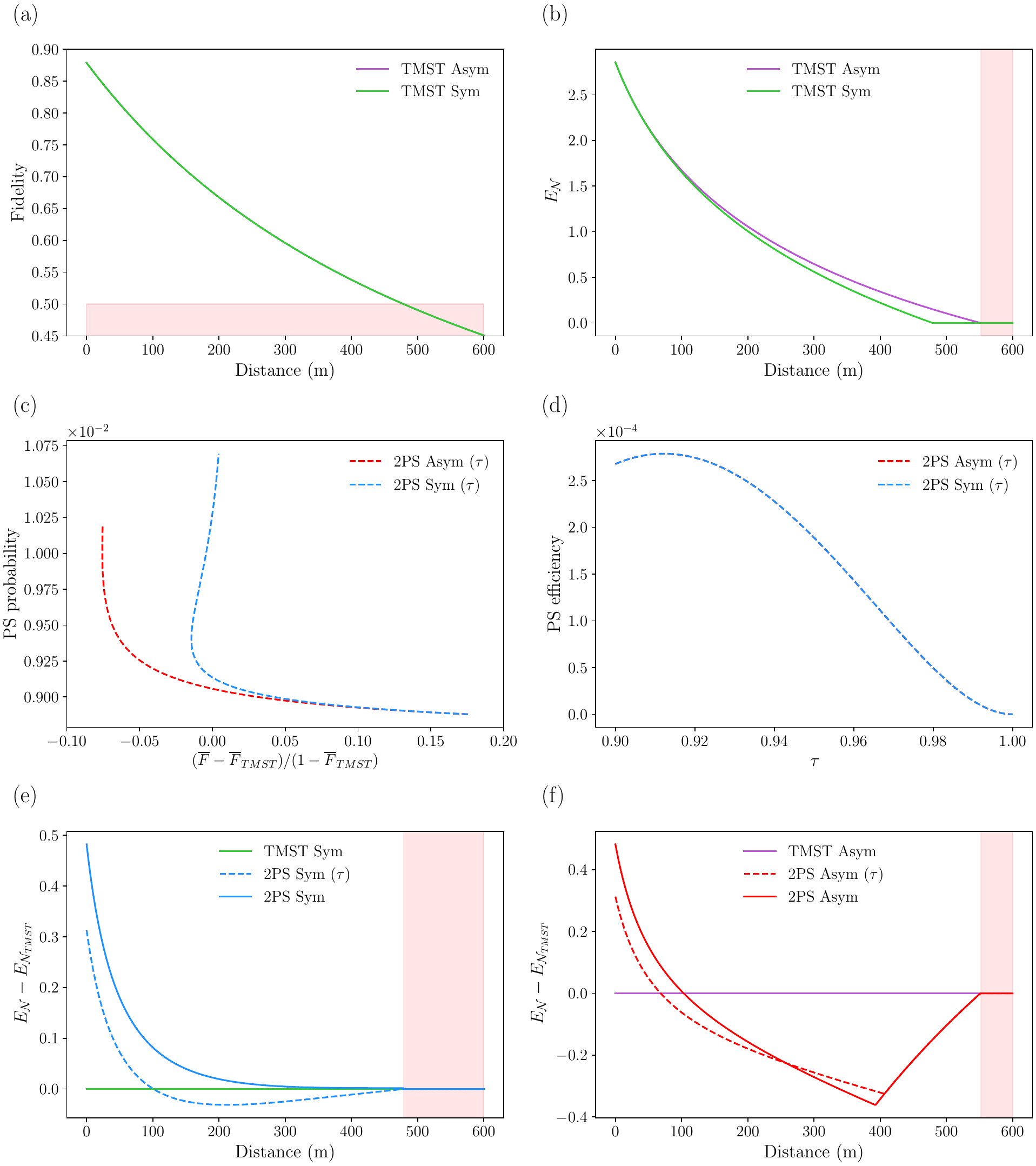}
\caption[Attributes of the symmetric and asymmetric TMST states distributed through open air]{\textbf{Attributes of the symmetric and asymmetric TMST states distributed through open air}. (a), (b) Average fidelity of the CV quantum teleportation protocol of an unknown coherent state and logarithmic negativiy $E_{\mathcal{N}}=\log_{2}(2\mathcal{N}+1)$, respectively, for the lossy TMST symmetric (green) and asymmetric (purple) states, represented against the traveled distance. (c) Success probability of two-photon subtraction on lossy TMST symmetric (blue) and asymmetric (red) states, against the fidelity gain compared to the lossy TMST state, which is larger for higher fidelities. (d) Efficiency of two-photon subtraction on TMST symmetric (blue) and asymmetric (red) states at $x=0$, against the transmissivity $\tau\in[0.9,1]$. (e), (f) Logarithmic negativity of the probabilistic (dashed) and heuristic (solid) re-Gaussified two-photon-subtracted symmetric (2PS sym, blue) and asymmetric (2PS asym, red) states, respectively, minus the logarithmic negativity of the corresponding lossy TMST symmetric (TMST sym, green) and asymmetric (TMST asym, purple) states. Parameters are $n=10^{-2}$, $N_{\text{th}}=1250$, $r=1$, $\mu=1.44\times 10^{-6}$ m$^{-1}$, $\eta_{\text{ant}}=0$, $\tau=0.95$.}
\label{fig5_5}
\end{figure}

In Fig.~\ref{fig5_4} we represent the difference in fidelities associated with a CV open-air quantum teleportation protocols for an unknown coherent state, using two-mode squeezed thermal states distributed through open air as a resource, against the traveled distance. We subtract the fidelity associated with the bare resource (TMST) to those related to 2PS symmetric (blue) and asymmetric (red) states, as well as ES (orange) states. We consider both heuristic (solid lines) and probabilistic (dashed lines, labeled $\tau$) photon subtraction. In Fig.~\ref{fig5_5}~(a), we can see the fidelity associated with the bare resource, knowing that it coincides for the symmetric and asymmetric states in the region $\mu L\ll1$. The solid green line represents the no-gain line, above which any point represents an improvement in fidelity over the bare state. The former gives an enhancement for short distances, whereas the latter helps extend the point where the classical limit is reached. One of the reasons the gain related to photon subtraction is lost might be the increase of thermal photons in the state, which occurs for increasing $L$. This happens because, as photon losses are more relevant, the cost of doing photon subtraction is higher: if we subtract thermal photons, the entanglement hardly increases, whereas if we subtract photons from the signal, entanglement decreases. Using a pale red background, we represent the region in which the fidelity associated with the bare resource reaches the maximum classical value of $1/2$.

In Fig.~\ref{fig5_5}, we represent various features of the two-mode squeezed thermal states distributed through open air: (a) average fidelity, which coincides for the symmetric and asymmetric states for $\mu L\ll 1$; (b) logarithmic negativity $E_{\mathcal{N}}=\log_{2}(2\mathcal{N}+1)$ of the symmetric (green) and asymmetric (purple) states; (c) success probability of photon subtraction (see Eq.~\eqref{eq:success-prob}) for symmetric (blue, dashed) and asymmetric (red, dashed) states, against $(\overline{F}_{\text{2PS}}-\overline{F}_{\text{TMST}})/(1-\overline{F}_{\text{TMST}})$, which represents the gain in fidelity of the photon-subtraction schemes, weighted to show larger values when the gain occurs at larger fidelities; (d) efficiency of photon subtraction at $x=0$, computed as $P(\overline{F}_{\text{2PS}}-\overline{F}_{\text{TMST}})$, against different values of the transmissivity, with $\tau\in[0.9,1]$. Note that greater fidelity gains come at lower success probabilities for photon subtraction, which can be reflected in the efficiency (of the order of $10^{-4}$). The latter achieves maximum values for a transmissivity of $\tau\approx 0.92$, and goes to zero with the probability, as $\tau$ goes to 1.
%%%%%%%%%%%%%%%%%%%%%%%%%%%%%%%%%%%%%%%%%%%%%%%%
%%%%%%%%%%%%%%%%%%%%%%%%%%%%%%%%%%%%%%%%%%%%%%%%
In an attempt to explain the crossing that occurs between the PS and bare fidelities, which delimits the region in which photon subtraction results in an enhanced teleportation fidelity, we consider the following approach: we attempt to find the Gaussian state that is related to our non-Gaussian PS state by the same teleportation fidelity. Essentially, we are looking to identify the PS states with Gaussian resources in order to compute the negativities from their covariance matrices, and investigate what happens to entanglement at the points where fidelity with PS states loses its advantage. First, know that the fidelity with probabilistic two-photon subtraction can be written as
\begin{equation}\label{eq:fidelity_2PS}
\overline{F}_{\text{2PS}} = \frac{1+ g}{\sqrt{\det\left[\mathbb{1}_{2}+\frac{1}{2}\tilde{\Gamma}\right]}},
\end{equation}
where $\tilde{\Gamma} \equiv \sigma_{z} \tilde{\Sigma}_{A}\sigma_{z} + \tilde{\Sigma}_{B} -\sigma_{z} \tilde{\varepsilon}_{AB}-\tilde{\varepsilon}_{AB}^\intercal \sigma_{z}$, and $\tilde{\Sigma}_{A}$, $\tilde{\Sigma}_{B}$, and $\tilde{\varepsilon}_{AB}$ are
\begin{eqnarray}\label{2PS_submatrices}
%\nonumber \tilde{\Sigma}_{A} &=& \left[1-2\tau\frac{(1-\alpha)(1+\beta)+\gamma^{2}+((1-\alpha)(1-\beta)-\gamma^{2})\tau}{(1+\alpha)(1+\beta)-\gamma^{2}+2(1-\alpha\beta+\gamma^{2})\tau+((1-\alpha)(1-\beta)-\gamma^{2})\tau^{2}}\right]\mathbb{1}_{2}, \\
%\nonumber \tilde{\Sigma}_{B} &=& \left[1-2\tau\frac{(1+\alpha)(1-\beta)+\gamma^{2}+((1-\alpha)(1-\beta)-\gamma^{2})\tau}{(1+\alpha)(1+\beta)-\gamma^{2}+2(1-\alpha\beta+\gamma^{2})\tau+((1-\alpha)(1-\beta)-\gamma^{2})\tau^{2}}\right]\mathbb{1}_{2}, \\
%\tilde{\varepsilon}_{AB} &=& \frac{4\tau\gamma}{(1+\alpha)(1+\beta)-\gamma^{2}+2(1-\alpha\beta+\gamma^{2})\tau+((1-\alpha)(1-\beta)-\gamma^{2})\tau^{2}}\sigma_{z}.
\nonumber \tilde{\Sigma}_{A} &=& \left[1-2\tau\frac{(1-\alpha)(1+\beta)+\gamma^{2}+((1-\alpha)(1-\beta)-\gamma^{2})\tau}{(1+\tau)^{2} + (\alpha+\beta)(1-\tau^{2}) + (\alpha\beta-\gamma^{2})(1-\tau)^{2}}\right]\mathbb{1}_{2}, \\
\nonumber \tilde{\Sigma}_{B} &=& \left[1-2\tau\frac{(1+\alpha)(1-\beta)+\gamma^{2}+((1-\alpha)(1-\beta)-\gamma^{2})\tau}{(1+\tau)^{2} + (\alpha+\beta)(1-\tau^{2}) + (\alpha\beta-\gamma^{2})(1-\tau)^{2}}\right]\mathbb{1}_{2}, \\
\tilde{\varepsilon}_{AB} &=& \frac{4\tau\gamma}{(1+\tau)^{2} + (\alpha+\beta)(1-\tau^{2}) + (\alpha\beta-\gamma^{2})(1-\tau)^{2}}\sigma_{z}.
\end{eqnarray}
Here, $g$ is the result of integrating all the non-Gaussian corrections to the characteristic function, which enforces the non-Gaussianity of the state resulting from photon subtraction (see section~\ref{sec_check} for the general expression). We split the terms in the previous equation and write
\begin{eqnarray}
\nonumber && \frac{1}{\sqrt{\det\left[\mathbb{1}_{2}+\frac{1}{2}\tilde{\Gamma}\right]}} = \frac{1}{2}\left[ 1+\tau\frac{-\alpha\beta + (1+\gamma)^{2} +((1-\alpha)(1-\beta)-\gamma^{2})\tau}{(1+\alpha)(1+\beta)-\gamma^{2}-(\alpha\beta-(1-\gamma)^{2})\tau}\right], \\
&& 1+g = \frac{1}{2}\left[ 1+\tau\frac{-\alpha\beta + (1+\gamma)^{2} +((1-\alpha)(1-\beta)-\gamma^{2})\tau}{(1+\alpha)(1+\beta)-\gamma^{2}-(\alpha\beta-(1-\gamma)^{2})\tau}\right]^{2} \times \\
\nonumber && \left[1 + \frac{(1-\alpha\beta+\gamma^{2})^{2}-(\alpha-\beta)^{2}+4\gamma^{2}+4\gamma((1-\alpha)(1-\beta)-\gamma^{2})\tau}{(1-\alpha\beta+\gamma^{2}+((1-\alpha)(1-\beta)-\gamma^{2})\tau)^{2}-(\alpha-\beta)^{2}+4\gamma^{2}}\right].
\end{eqnarray}
If we define a matrix $G = (1+g)\mathbb{1}_{2}$ with $G^{-1} = \frac{1}{1+g}\mathbb{1}_{2}$, then we can write $1+g = \sqrt{\det G}$, which leads to 
\begin{equation}
\frac{1+ g}{\sqrt{\det\left[\mathbb{1}_{2}+\frac{1}{2}\tilde{\Gamma}\right]}} = \frac{1}{\sqrt{\det\left[\left(\mathbb{1}_{2}+\frac{1}{2}\tilde{\Gamma}\right)G^{-1}\right]}}.
\end{equation}
By rearranging the terms resulting from the matrix product, we can obtain
\begin{equation}
\left(\mathbb{1}_{2}+\frac{1}{2}\tilde{\Gamma}\right)G^{-1} = \mathbb{1}_{2} + \frac{1}{2}\left(\frac{\tilde{\Gamma} - 2g\mathbb{1}_{2}}{1+g}\right) \equiv \mathbb{1}_{2} + \frac{1}{2}\tilde{\tilde{\Gamma}},
\end{equation}
where we have defined $\tilde{\tilde{\Gamma}} = \frac{\tilde{\Gamma} - 2g\mathbb{1}_{2}}{1+g}$. Now, we want to incorporate the non-Gaussian corrections into the covariance matrix of the effective Gaussian state by using the formula
\begin{equation}
\tilde{\tilde{\Gamma}} =  \sigma_{z}\tilde{\tilde{\Sigma}}_{A} \sigma_{z} + \tilde{\tilde{\Sigma}}_{B} -  \sigma_{z}\tilde{\tilde{\varepsilon}}_{AB} - \tilde{\tilde{\varepsilon}}_{AB}^{\intercal} \sigma_{z}.
\end{equation}
We refer to the resulting state as the ``re-Gaussified'' state. Then, we define
\begin{eqnarray}\label{2PS_sym_eff_submatrices}
\nonumber \tilde{\tilde{\Sigma}}_{A} &=& \frac{1}{1+g}\left(\tilde{\Sigma}_{A} - g\mathbb{1}_{2}\right), \\
\tilde{\tilde{\Sigma}}_{B} &=& \frac{1}{1+g}\left(\tilde{\Sigma}_{B} - g\mathbb{1}_{2}\right), \\
\nonumber \tilde{\tilde{\varepsilon}}_{AB} &=& \frac{1}{1+g}\tilde{\varepsilon}_{AB}.
\end{eqnarray}
These represent the submatrices of a covariance matrix $\tilde{\tilde{\Sigma}}$ if
\begin{equation}
\left|\sqrt{\det\tilde{\Sigma}} - g(2+\tilde{\alpha}+\tilde{\beta}) -1\right| \geq (1+g)\left| \tilde{\alpha} - \tilde{\beta}\right|
\end{equation}
is satisfied. This condition both ensures the positivity of the covariance matrix and that the uncertainty relation is satisfied. For this, we have assumed that $\tilde{\Sigma}_{A}=\tilde{\alpha}\mathbb{1}_{2}$, $\tilde{\Sigma}_{B}=\tilde{\beta}\mathbb{1}_{2}$, and $\tilde{\varepsilon}_{AB} = \tilde{\gamma} \sigma_{z}$. The problem is that this convention only works for the symmetric state, and not for the asymmetric one. For the latter, we write
\begin{eqnarray}
\nonumber \tilde{\tilde{\Sigma}}_{A} &=& \frac{1}{1+g}\left(\tilde{\Sigma}_{A} - kg\mathbb{1}_{2}\right), \\
\tilde{\tilde{\Sigma}}_{B} &=& \frac{1}{1+g}\left(\tilde{\Sigma}_{B} - (2-k)g\mathbb{1}_{2}\right), \\
\nonumber \tilde{\tilde{\varepsilon}}_{AB} &=& \frac{1}{1+g}\tilde{\varepsilon}_{AB}.
\end{eqnarray}
Since we have seen that a symmetric re-Gaussified state is viable, we impose the same balanced partition on the re-Gaussification of the asymmetric state. From $\tilde{\tilde{\Sigma}}_{A}=\tilde{\tilde{\Sigma}}_{B}$, we obtain $k = 1+(\tilde{\alpha}-\tilde{\beta})/2g$, which leads to the submatrices
\begin{eqnarray}\label{2PS_asym_eff_submatrices}
\nonumber \tilde{\tilde{\Sigma}}_{A} &=& \frac{1}{1+g}\left(\frac{\tilde{\Sigma}_{A}+\tilde{\Sigma}_{B}}{2} - g\mathbb{1}_{2}\right), \\
\tilde{\tilde{\Sigma}}_{B} &=& \frac{1}{1+g}\left(\frac{\tilde{\Sigma}_{A}+\tilde{\Sigma}_{B}}{2} - g\mathbb{1}_{2}\right), \\
\nonumber \tilde{\tilde{\varepsilon}}_{AB} &=& \frac{1}{1+g}\tilde{\varepsilon}_{AB}.
\end{eqnarray}
The condition these terms need to satisfy is
\begin{equation}
\left|\sqrt{\det\tilde{\Sigma}} + \frac{1}{4}(\tilde{\alpha}-\tilde{\beta})^{2} -g(\tilde{\alpha}+\tilde{\beta}) + g^{2} - 1 \right| \geq 0,
\end{equation}
which is naturally met. In a similar fashion, we can write the fidelity with heuristic two-photon subtraction as
\begin{equation}
\overline{F} = \frac{1+ h}{\sqrt{\det\left[\mathbb{1}_{2}+\frac{1}{2}\Gamma\right]}},
\end{equation}
and identify $h$ as the non-Gaussian corrections to the fidelity; we can then mask them as corrections to the covariance matrix of a Gaussian state with the same fidelity. We do this by defining 
\begin{equation}
\tilde{\Gamma} = \frac{\Gamma-2h\mathbb{1}_{2}}{1+h} \equiv \sigma_{z} \tilde{\Sigma}_{A} \sigma_{z} + \tilde{\Sigma}_{B} - \sigma_{z} \tilde{\varepsilon}_{AB}-\tilde{\varepsilon}_{AB}^{\intercal} \sigma_{z}.
\end{equation}
For a symmetric Gaussian resource we define
\begin{eqnarray}\label{H2PS_sym_eff_submatrices}
\nonumber \tilde{\Sigma}_{A} &=& \frac{1}{1+h}\left(\Sigma_{A} - h\mathbb{1}_{2}\right), \\
\tilde{\Sigma}_{B} &=& \frac{1}{1+h}\left(\Sigma_{B} - h\mathbb{1}_{2}\right), \\
\nonumber \tilde{\varepsilon}_{AB} &=& \frac{1}{1+h}\varepsilon_{AB},
\end{eqnarray}
whereas, if the resource is asymmetric, we require $\tilde{\Sigma}_{A}=\tilde{\Sigma}_{B}$, such that
\begin{eqnarray}\label{H2PS_asym_eff_submatrices}
\nonumber \tilde{\Sigma}_{A} &=& \frac{1}{1+h}\left(\frac{\Sigma_{A}+\Sigma_{B}}{2} - h\mathbb{1}_{2}\right), \\
\tilde{\Sigma}_{B} &=& \frac{1}{1+h}\left(\frac{\Sigma_{A}+\Sigma_{B}}{2} - h\mathbb{1}_{2}\right), \\
\nonumber \tilde{\varepsilon}_{AB} &=& \frac{1}{1+h}\varepsilon_{AB}.
\end{eqnarray}
These ``re-Gaussified'' covariance matrices need to satisfy positivity and the uncertainty principle, meaning that $|\sqrt{\det\tilde{\Sigma}}-1|\geq |\tilde{\alpha}-\tilde{\beta}|$, assuming that we can write $\tilde{\Sigma}_{A}=\tilde{\alpha}\mathbb{1}_{2}$, $\tilde{\Sigma}_{B}=\tilde{\beta}\mathbb{1}_{2}$, and $\tilde{\varepsilon}_{AB} = \tilde{\gamma} \sigma_{z}$. Furthermore, if $\Sigma_{A}=\alpha\mathbb{1}_{2}$, $\Sigma_{B}=\beta\mathbb{1}_{2}$, and $\varepsilon_{AB} = \gamma \sigma_{z}$, this condition can be expressed as
\begin{equation}
\left|\sqrt{\det\Sigma}-h(2+\alpha+\beta)-1 \right| \geq (1+h)|\alpha-\beta|
\end{equation}
for a symmetric state, and as
\begin{equation}
\left|\sqrt{\det\Sigma}+\frac{1}{4}(\alpha-\beta)^{2} - h(\alpha+\beta) + h^{2} - 1 \right| \geq 0
\end{equation}
for an asymmetric one. In section~\ref{sec_check}, a graphical proof that these conditions are met is provided.

As a result of these redefinitions, we effectively mask the non-Gaussian corrections in the expression of the fidelity as further corrections to the submatrices of the covariance matrix of an entangled resource, which is now Gaussian, while maintaining the same fidelity we obtained with the PS states. This treatment has shown that we are using a resource that, in the regions in which photon subtraction is beneficial, shows higher entanglement than the bare resource. This is expected given that, among all possible states with the same covariance matrix, entanglement is minimized by Gaussian states~\cite{Wolf2006}.

In Fig.~\ref{fig5_5}~(e) and Fig.~\ref{fig5_5}~(f), we subtract the logarithmic negativity $E_{\mathcal{N}}=\log_{2}(2\mathcal{N}+1)$ of the bare resource (TMST) from those of the heuristic (solid) and the probabilistic (dashed) 2PS states. In Fig.~\ref{fig5_5}~(e), we display the symmetric states, and in Fig.~\ref{fig5_5}~(f), the asymmetric ones. Note that the gain in negativity is lost around the same points as the gain in fidelity. As discussed before, the fidelities corresponding to the symmetric and asymmetric states are equal at first order in $\mu L\ll 1$, and the same behavior can be observed initially in the negativities of both states (see Fig.~\ref{fig5_5}~(b)). However, while the points at which the fidelities of the symmetric and asymmetric states reach the classical limit differs by centimeters, the points at which entanglement is lost for these states differ by tens of meters. This region where negativity is lost is highlighted with a pale red background. Any point above the green and purple line represents an improvement in negativity for the re-Gaussified PS symmetric and asymmetric states, respectively. Although the entanglement in the asymmetric state reaches further, the symmetric photon-subtraction protocol we envision works better when applied on the symmetric state. The logarithmic negativity of heuristic PS states presents a 46\% increase with respect to the value for the bare state at $x=0$, while probabilistic PS states only present an initial gain of 28\%.

\subsubsection{Fidelity with entanglement swapping}
We consider the case in which both Alice and Bob produce two-mode squeezed states, and each sends one mode to Charlie, who is equidistantly located from the two parties. Then, he performs entanglement swapping using the two modes he has received, which have been degraded by thermal noise and photon losses. If Alice and Bob use the remaining entangled resource they share for teleporting an unknown coherent state, the fidelity of the protocol will be given by
\begin{equation}
\overline{F}_{\text{es}} = \frac{1}{1+\alpha-\frac{\gamma^{2}}{\beta}}, 
\end{equation}
where now we have
\begin{eqnarray}
\nonumber \alpha &=& (1+2n)\cosh 2r, \\
\beta &=& (1+2N_{\text{th}})\eta_{\text{eff}}  + (1+2n)(1-\eta_{\text{eff}})\cosh 2r, \\
\nonumber \gamma &=& (1+2n)\sqrt{1-\eta_{\text{eff}}}\sinh 2r,
\end{eqnarray}
and $\eta_{\text{eff}} = 1-e^{-\mu L/2}(1-\eta_{\text{ant}})$, since the total distance has been reduced by half due to the presence of a third, equidistant party. 

This fidelity is represented as the orange curve in Fig.~\ref{fig5_4}, where it shows a gain in fidelity for large distances, right before the classical limit of $\overline{F}=0.5$ is reached. The extended distance represents 14\% of the maximum distance for the bare TMST state. This will be advantageous when the distance at which the classical limit occurs can be extended, for example in the case of quantum communication between satellites.  

\subsection{Positivity and Uncertainty Principle for Covariance Matrices}\label{sec_check}
%%%%%%%%%%%%%%%%%%%%%%%%%%%%%%%%%%%%%%%%%%%%%%%%
\begin{figure}[h!]
\centering
\includegraphics[width=0.75 \textwidth]{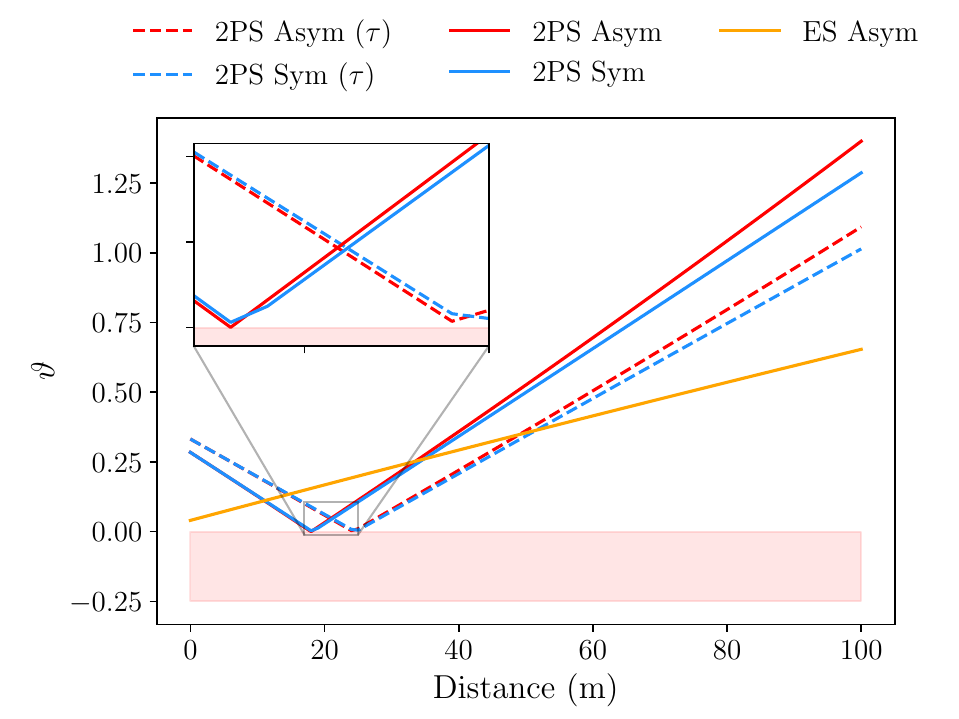}
\caption[Graphical check of positivity and uncertainty principle for covariance matrices]{\textbf{Graphical check of positivity and uncertainty principle for covariance matrices} obtained in this chapter. The quantity in Eq.~\eqref{check} is represented against the travelled distance, computed from each covariance matrix derived in this chapter. If this is positive, it proves that the submatrices used to compute it characterize a covariance matrix satisfying positivity and the uncertainty principle. In orange, we represent the curve associated with the submatrices in Eqs.~\eqref{ES_submatrices} that result from entanglement swapping. The blue and red solid curves correspond to the heuristic 2PS ``re-Gaussified'' symmetric and asymmetric states, respectively, described in Eqs.~\eqref{H2PS_sym_eff_submatrices} and~\eqref{H2PS_asym_eff_submatrices}. The blue and red dashed curves correspond to the probabilistic 2PS ``re-Gaussified'' symmetric and asymmetric states, respectively, described in Eqs.~\eqref{2PS_sym_eff_submatrices} and~\eqref{2PS_asym_eff_submatrices}. Inset: enlarged view of the region of short distances, in which we observe that the condition $\vartheta > 0$ is still met.}
\label{fig5_6}
\end{figure}
%%%%%%%%%%%%%%%%%%%%%%%%%%%%%%%%%%%%%%%%%%%%%%%%
In this section, we discuss the conditions that a covariance matrix must satisfy in order for it to describe a quantum state. Then, we apply this criterion to the covariance matrices presented in this chapter, obtained after photon subtraction and entanglement swapping. The first condition is the positivity of the covariance matrix
\begin{equation}
\Sigma = \begin{pmatrix} \Sigma_{A} & \varepsilon_{AB} \\ \varepsilon_{AB}^{\intercal} & \Sigma_{B} \end{pmatrix} > 0,
\end{equation}
and the second one is preservation of the uncertainty principle,
\begin{equation}
\begin{pmatrix} \Sigma_{A} & \varepsilon_{AB} \\ \varepsilon_{AB}^{\intercal} & \Sigma_{B} \end{pmatrix} + i\begin{pmatrix} \Omega & 0 \\ 0 & \Omega \end{pmatrix} \geq 0.
\end{equation}
If we consider $\Sigma_{A}=\alpha\mathbb{1}_{2}$, $\Sigma_{B}=\beta\mathbb{1}_{2}$, and $\varepsilon_{AB}=\gamma\sigma_{z}$, the positivity condition reduces to
\begin{eqnarray}
\nonumber \alpha &>& 0, \\
\det\Sigma &>& 0,
\end{eqnarray}
whereas the uncertainty principle can be written as
\begin{eqnarray}
\nonumber \alpha &\geq& 1, \\
\det\Sigma &\geq& \alpha^{2} + \beta^{2} - 2\gamma^{2} - 1.
\end{eqnarray}
Note that the latter imposes a more restrictive condition. Given that any covariance matrix requires $\alpha\geq 1$ and $\beta\geq 1$, we can summarize all conditions as
\begin{equation}\label{check}
\vartheta \equiv \left|\sqrt{\det\Sigma} - 1\right| - \left|\alpha-\beta\right| \geq 0.
\end{equation}
In Fig.~\ref{fig5_6}, we investigate whether this condition is satisfied for different modified covariance matrices by representing $\vartheta$ against the traveled distance: submatrices in Eqs.~\eqref{ES_submatrices} due to entanglement swapping (orange); in Eqs.~\eqref{H2PS_sym_eff_submatrices} and~\eqref{H2PS_asym_eff_submatrices} due to re-Gaussified heuristic photon subtraction (symmetric shown with a blue line, asymmetric shown with a red line); in Eqs.~\eqref{2PS_sym_eff_submatrices} and~\eqref{2PS_asym_eff_submatrices}, due to re-Gaussified probabilistic photon subtraction (symmetric shown with a blue dashed line, asymmetric shown with a red dashed line). Note that all five cases satisfy both positivity and uncertainty principle conditions, confirming that they are indeed covariance matrices. In the inset, we present an enlarged view of the short distance behavior, where the curves approach the region in which $\vartheta < 0$ (highlighted in a pale red background). As we can see, even in that area $\vartheta > 0$ is satisfied. 

\subsection{Experimental Limitations to Photocounting and Homodyning with Microwaves}\label{section:homodyning_and_photocounting}
In this section, we review current advances on photocounting and homodyne detection techniques with microwave quantum technologies. These techniques are vital for photon subtraction, as well as for entanglement swapping and quantum teleportation, which are the processes described in this chapter. We also investigate different sources of error that affect them; by using parameters taken from recent experimental benchmarks in microwave quantum technologies, we are able to estimate how inefficiencies and imperfections surrounding microwave photocounting and homodyne detection affect photon subtraction, quantum teleportation and entanglement swapping. We believe that this provides a closer relation to state-of-the-art experiments with quantum microwaves. 

\subsubsection{Photodetection}
Traditionally, the problem of detecting microwaves has been the low energy of the signals when compared to the optical regime. Any of the entanglement distillation protocols we have discussed will require some kind of photodetection scheme. In particular, for photon subtraction, a photocounter for microwave photons is required. In the current landscape of microwave quantum technologies, there have been recent proposals for nondemolition detection of itinerant single microwave photons~\cite{Kono2018,Besse2018,Lescanne2020} in circuit-QED setups, with detection efficiencies ranging from $58\%$ to $84\%$. Based on similar setups, a photocounter has been proposed~\cite{Dassonneville2020} that can detect up to three microwave photons. 

This device is able to catch an incoming wavepacket in a buffer resonator, which is then transferred into the memory by means of pumping a Josephson ring modulator. Then, the information about the number of photons in the memory is transferred to a transmon qubit, which is coupled to the memory modes, and from there it is read bit by bit. Consequently, this photocounter requires previous knowledge on the waveform and the arrival time of the incoming mode to be detected. Furthermore, this device is not characterized by a single quantum efficiency; rather, the detection efficiency varies depending on the number of photons. That is, $99\%$ for zero photons, $76\%\pm3\%$ for a single photon, $71\%\pm3\%$ for two photons and $54\%\pm2\%$ for three, assuming a dark count probability of $3\%\pm0.2\%$ and a dead time of $4.5\,\mu$s.

First, let us introduce a parameter for the efficiency of the microwave photodetectors. In Ref.~\cite{Dassonneville2020}, a circuit-QED-based microwave photon counter was presented, which could detect between zero and three photons, with fidelities ranging from 99 \% to 54 \%. Such a device is particularly useful for the photon-subtraction scheme investigated in the chapter, in which we only consider single-photon subtraction in each mode of a bipartite entangled state. Then, we need to look at the success probability of detecting a single photon, which in this experiment is 76 \%. An imperfect detector can be modeled as a pure-loss channel, which is represented by a beam splitter that mixes the signal traveling towards the detector with a vacuum state, and whose transmissivity determines the efficiency. In this case, we have $\tau_{\text{detector}} = 0.76$. This characterizes the detection probability in the heuristic photon-subtraction case, but in the probabilistic description, this will be given by $\tau_{\text{total}} = \tau_{\text{detector}}\tau = 0.72$, since we have considered $\tau=0.95$.

Taking into account the detector efficiency, we observe that the maximum negativity associated with the re-Gaussified PS states is, at least, 47 \% of the maximum negativity of the PS states with ideal detector efficiency. This means that, taking into account this source of error, the negativity of the states obtained through this entanglement distillation technique is cut by more than half. Furthermore, these values are below the negativity of the bare state in the ideal case, which means that performing photon subtraction leads to entanglement degradation. In order for it to be advantageous, with the parameters we have considered throughout this chapter, we would need to have detection efficiencies above 85 \% for the heuristic protocol, and above 90 \% for the probabilistic one. 

%Qubit T_{1} = $7.1\,\mu$s and T_{2} = $13.6\,\mu$s.

\subsubsection{Homodyne detection}
Homodyne detection allows one to extract information about a single quadrature. It can be used to perform CV-Bell measurements, i.e., a projective measurement in a maximally-quadrature-entangled basis for CV states. One way to perform Bell measurements with propagating CV states is to use the analog feedforward technique, as demonstrated in Ref.~\cite{Fedorov2021}. This approach requires operating two additional phase-sensitive amplifiers in combination with two hybrid rings and a directional coupler, which effectively implements a projection operation for conjugate quadratures of propagating electromagnetic fields. An alternative, more conventional approach can be implemented by adapting microwave single-photon detectors to the well-known optics homodyning techniques.

As we have seen, entanglement swapping provides an advantage if this measurement scheme is used without averaging over the results (single-shot homodyning), whereas the Braunstein-Kimble quantum teleportation protocol assumes that this average is performed, given an unknown coherent state. In theory, single-shot homodyning can be implemented by using quantum-limited superconducting amplifiers and standard demodulation techniques~\cite{Eichler2012}. However, some fundamental aspects of the ``projectiveness'' of this operation and its importance for the Bell detection measurements or for photon subtraction are still unclear and must be verified.

%Now, let us discuss homodyne detection, a measurement technique that is widely used in many quantum communication protocols, and in this article is required for quantum teleportation and for entanglement swapping. Homodyne detection aims at extracting information about the quadratures of a signal by mixing it with a coherent light source with a large number of photons in a balanced beam splitter, and then measuring the photocurrents at both ends. From the difference between the currents, we can get information about either quadrature of the signal. This measurement works ideally if the source has an infinite number of photons (infinite gain), which is not generally the case. Therefore, we can consider the case of finite gain in homodyne detection. 

We are interested in the case of finite gain homodyne detection. The theoretical description of these measurements corresponds to a projection onto a state that is infinitely-squeezed in $x$ (or in $p$) in phase space. That is, an eigenstate of the position operator (or the momentum operator) whose eigenvalue corresponds to the signal's $x$ (or $p$) quadrature value. In the symplectic formalism, this measurement operator has a covariance matrix 
\begin{equation}
\Upsilon = \begin{pmatrix} e^{-2\xi} & 0 \\ 0 & e^{2\xi} \end{pmatrix} \equiv \begin{pmatrix} \frac{1}{\sqrt{G}} & 0 \\ 0 & \sqrt{G} \end{pmatrix}.
\end{equation}
In the limit $G\rightarrow\infty$, we will recover the usual homodyne detection scheme. We have obtained that the fidelity of teleporting an unknown coherent state with $|\theta|^{2}$ photons using a bipartite entangled state with covariance matrix
\begin{equation}
\Sigma = \begin{pmatrix} \alpha\mathbb{1}_{2} & \gamma \sigma_{z} \\ \gamma \sigma_{z} & \beta\mathbb{1}_{2} \end{pmatrix}
\end{equation}
is given by
%\begin{eqnarray}
%&& \overline{F} = \frac{2\left[2+\frac{1}{\sqrt{G}}(1+\alpha)\right]}{4\left(1+\frac{\alpha+\beta-2\gamma}{2}\right)+\frac{1}{\sqrt{G}}\left[\alpha(5+\beta) + \beta -(\gamma-1)(\gamma+5)\right] + \frac{2}{G}(1+\alpha)} \times \\
%\nonumber && \text{exp}\left[-\frac{\frac{2}{G}(1-\alpha+\gamma)^{2}|\theta|^{2}}{\left[2+\frac{1}{\sqrt{G}}(1+\alpha)\right]\left[ 4\left(1+\frac{\alpha+\beta-2\gamma}{2}\right)+\frac{1}{\sqrt{G}}\left[\alpha(5+\beta) + \beta -(\gamma-1)(\gamma+5)\right] + \frac{2}{G}(1+\alpha)\right]}\right].
%\end{eqnarray}
\begin{eqnarray}
\nonumber && \Phi = \frac{2\left[2+\frac{1}{\sqrt{G}}(1+\alpha)\right]}{4\left(1+\frac{\alpha+\beta-2\gamma}{2}\right)+\frac{1}{\sqrt{G}}\left[\alpha(5+\beta) + \beta -(\gamma-1)(\gamma+5)\right] + \frac{2}{G}(1+\alpha)}, \\
&& \overline{F} =\Phi \, \text{exp}\left[-\Phi\left( \frac{1-\alpha+\gamma}{1+2\sqrt{G}+\alpha}\right)^{2}|\theta|^{2}\right].
\end{eqnarray}

In the limit $G\rightarrow\infty$ we recover $\overline{F} = \left(1+\frac{\alpha+\beta-2\gamma}{2}\right)^{-1}$, which is the usual result. 
Notice that, while the average teleportation fidelity for an unknown coherent state with ideal homodyne detection does not depend on the value of the displacement for said state, we find that the first order corrections do include this dependence in the value of $\theta$. The average fidelity associated with a resource with increasing entanglement asymptotically tends to 1 when considering ideal homodyne detection. In this case, it tends to the value $\frac{1}{1 + \frac{1}{\sqrt{G}}}$, which becomes closer to 1 as $G$ increases. 

In a recent paper, CV quantum teleportation in the microwave regime was performed~\cite{Fedorov2021}, where the optimal gain considered was 21 dB, which implies that $1/G\approx0.008$. Using this value, and considering we want to teleport a vacuum state ($\theta=0$), we observe that the fidelity reaches the maximum classical fidelity at 434 m for the asymmetric state, and at 429 m for the symmetric one, while this distance is 480 m with ideal homodyne detection for both kinds of states. 

Here, we also consider the effect of finite-gain homodyne detection on the states that result from entanglement swapping. As a generalization of Eq.~\ref{ES_submatrices}, these states can be characterized by a covariance matrix with submatrices
\begin{eqnarray}
\nonumber \tilde{\Sigma}_{A} &=& \left( \alpha - \frac{\gamma^{2}\left( 1 + \frac{1}{\sqrt{G}}2\beta + \frac{1}{G}\right)}{2\left[\beta + \frac{1}{\sqrt{G}}\left(1+\beta^{2}\right) + \frac{\beta}{G}\right]}\right)\mathbb{1}_{2}, \\
\tilde{\Sigma}_{D} &=& \left( \alpha - \frac{\gamma^{2}\left( 1 + \frac{1}{\sqrt{G}}2\beta + \frac{1}{G}\right)}{2\left[\beta + \frac{1}{\sqrt{G}}\left(1+\beta^{2}\right) + \frac{\beta}{G}\right]}\right)\mathbb{1}_{2}, \\
\nonumber \tilde{\varepsilon}_{AD} &=& \frac{\gamma^{2}\left( 1 - \frac{1}{G}\right)}{2\left[\beta + \frac{1}{\sqrt{G}}\left(1+\beta^{2}\right) + \frac{\beta}{G}\right]}\sigma_{z}.
\end{eqnarray}
With entanglement swapping, the maximum classical fidelity is reached at 416 m, which is smaller than the reach of the bare states taking into account finite-gain homodyne detection. This is natural, since the effects of the finite gain come both from entanglement swapping and from quantum teleportation. Nevertheless, we have seen that finite-gain effects are not significantly detrimental to the measures we have computed in this chapter, and this means that, if larger optimal gains can be engineered, errors can then be reduced. At the end of the day, we have observed that entanglement distillation and entanglement swapping suffer from errors associated with photon counting and homodyne detection. However, we believe that these errors can be easily overcome by technological improvements. Furthermore, by the time all the pieces necessary for experiments in open-air microwave quantum communication arrive, we expect these errors to be further reduced. Meanwhile, entanglement distribution and quantum teleportation with microwaves, in the realistic open-air scenario, are still viable despite the errors we considered in this section.

However, further developments in the field of microwave quantum technologies are needed. Efficient information retrieval from an open-air distribution of microwave quantum states is a key component of open-air quantum communications, which requires the design of a receiver antenna. The device achieving this target may resemble that in chapter~\ref{sec3}, but it calls for a different type of termination into open air in order to, for instance, reduce diffraction losses. Since the lack of an amplification protocol considerably limits the entanglement transmission distance through open air, it seems necessary to develop a theory of quantum repeaters for microwave signals, following the ideas shown in Ref.~\cite{DiCandia2015}. To this end, entanglement distillation and entanglement swapping techniques discussed in this chapter are useful. 

Since superconducting circuits naturally work in the microwave regime, it is desirable to explore realizations of photon subtraction that use devices specific to this technology. In particular, a possible deterministic photon-subtraction scheme can be studied, making use of circuit QED for nondemolition detection of itinerant microwave photons~\cite{Lescanne2020}. In this paper, the detection of a previously unknown microwave photon is triggered by a transmon qubit jumping to its excited state, indicating a successful photon-subtraction event.

%%%%%%%%%%%%%%%%%%%%%%%%
% REDEFINE TITLE FORMAT
%%%%%%%%%%%%%%%%%%%%%%%%

% \titleformat{\section}[display]
% {\vspace*{150pt}
% \bf\Huge}
% {\hspace{-72pt}{\textcolor{grey}{\thesection}} \hspace{38pt} {\textcolor{white}{#1}}}
% {0pt}
% {}
% []
% \titlespacing*{\section}{108pt}{10pt}{60pt}[20pt]
%
% %Intro
% \titleformat{\section}[display]
% {\vspace*{190pt}
% \bfseries\sffamily \huge}
% {\begin{picture}(0,0)\put(-50,-25){\textcolor{grey}{\thesection}}\end{picture}}
% {0pt}
% {\textcolor{white}{#1}}
% []
% \titlespacing*{\section}{80pt}{10pt}{50pt}[20pt]
%
% %ch 2
% \titleformat{\section}[display]
% {\vspace*{190pt}
% \bfseries\sffamily \huge}
% {\begin{picture}(0,0)\put(-50,-28){\textcolor{grey}{\thesection}}\end{picture}}
% {0pt}
% {\textcolor{white}{#1}}
% []
% \titlespacing*{\section}{80pt}{10pt}{50pt}[20pt]
%
% %ch 3, 4 y 5
\titleformat{\section}[display]
{\vspace*{190pt}
\bfseries\sffamily \LARGE}
{\begin{picture}(0,0)\put(-50,-35){\textcolor{grey}{\thesection}}\end{picture}}
{0pt}
{\textcolor{white}{#1}}
[]
\titlespacing*{\section}{80pt}{10pt}{40pt}[20pt]

% %Conclusions
% %Julen conclusions
% \titleformat{\section}[display]
% {\vspace*{190pt}
% \bfseries\sffamily \huge}
% {\begin{picture}(0,0)\put(-50,-25){\textcolor{grey}{\thesection}}\end{picture}}
% {0pt}
% {\textcolor{white}{#1}}
% []
% \titlespacing*{\section}{80pt}{10pt}{50pt}[20pt]

%%%%%%%%%%%%%%%%%%%%%%
% CHAPTER 6
%%%%%%%%%%%%%%%%%%%%%%

\section[Microwave \& Optical Quantum Communication with Satellites]{Microwave \& Optical Quantum \\ Communication with Satellites}
\label{sec6}

% \thiswatermark{\put(1,-282){\color{l-grey}\rule{84pt}{88pt}}
% \put(84,-282){\color{grey}\rule{410pt}{88pt}}}
%
%
% %Julen ch 2
% \thiswatermark{\put(1,-308){\color{l-grey}\rule{70pt}{64pt}}
% \put(70,-308){\color{grey}\rule{297pt}{64pt}}}
% \fancyhead[RO]{\leftmark}

%Julen ch 3
\thiswatermark{\put(1,-302){\color{l-grey}\rule{70pt}{60pt}}
\put(70,-302){\color{grey}\rule{297pt}{60pt}}}

% %Julen ch 4
% \thiswatermark{\put(1,-302){\color{l-grey}\rule{70pt}{60pt}}
% \put(70,-302){\color{grey}\rule{297pt}{60pt}}}
% \fancyhead[RO]{\small \leftmark}
%
%
% %Julen ch 5
% \thiswatermark{\put(1,-302){\color{l-grey}\rule{70pt}{60pt}}
% \put(70,-302){\color{grey}\rule{297pt}{60pt}}}
% \fancyhead[RO]{\leftmark}

% SATELLITE QUANTUM COMMUNICATION
\lettrine[lines=2, findent=3pt,nindent=0pt]{A}{n} important application of the technology and the protocols described in the previous two chapters is quantum communication between ground stations and satellites. The global communication network relies on satellite repeaters to distribute information across the earth. While this is optimized for classical signals, quantum signals will not fare the same. Due to classicalization of signals by introducing thermal pollution, amplification cannot be used in quantum communication. Therefore, a key step in the development of global quantum communication networks is understanding the main loss mechanisms for signals propagating through free space. These include diffraction, atmospheric attenuation, and turbulence. In the case of microwave signals, the main sources of loss are diffraction and thermal radiation; therefore, an advantageous situation for microwave is quantum communication between satellites in the same orbit, where the effect of thermal noise is highly reduced. Wireless microwaves have also been studied for CV QKD with mobile devices in short-range scenarios (see Ref.~\cite{Pirandola2021_3}, sec. III D).

Turbulence effects, caused by small variations of temperature and pressure in the atmosphere, affect optical signals, but not microwaves. In the weak turbulence regime, these suffer two distinct effects: beam broadening and beam-centroid wandering. The effects on turbulence on classical signals have been well studied~\cite{Fante1975,Fante1980}, but also on the quantum regime~\cite{Vasylyev2016,Liorni2019,Vasylyev2019}, after it was demonstrated that the non-classicality of signals can be preserved~\cite{Vasylyev2012}. The QKD capabilities of quantum states propagating through turbulent media has also been addressed, establishing links between ground stations~\cite{Pirandola2021_2} and between ground stations and satellites~\cite{Pirandola2021}. All these works have provided insight into the limitations for the involvement of satellites in quantum communications~\cite{Pirandola2020,Sidhu2021}. We aim to contribute by considering the effects of atmospheric attenuation with turbulence on quantum states, how entanglement is degraded, and their efficiency for performing quantum teleportation, between ground stations and satellites. 

\subsection{Inter-Satellite Microwave Quantum Communication}
Given the security inherent to quantum-based communication protocols, many of the motivations for the use of submillimiter microwaves, i.e., frequencies in the range 30-300 GHz, which is a trend in classical communication between satellites orbiting low earth orbits (LEOs), fade away, and it seems reasonable to aim at maximizing the distances between linked satellites~\cite{Sanz2018}. 

We consider a greatly simplified model for free-space microwave communication, assuming unpolarized signals and hence ignoring the effects of scintillation and polarization rotation, among others. This means that whenever we discuss entanglement, it will be understood that we are talking about particle number entanglement. Polarization entanglement, even if perhaps more natural when considering the physics of antennae, is lost whenever the signal enters a coplanar waveguide, hence making it not a good candidate for quantum communication between one-dimensional superconducting chips. Moreover, we assume that the communication is done within the same altitude, i.e., that the two satellites are in similar orbits, which is typically the case when building satellite constellations. This means that the atmospheric absorption, if any, will remain constant during the time of flight of the signals. Additionally, we ignore Doppler effects caused by relative speeds between the orbits.

There are four main families of satellite orbits: GEO, HEO, MEO, and LEO, corresponding to geosynchronous, high, medium, and low earth orbits, respectively. It is customary to define LEOs as orbits with altitudes in the range 700-2000 km; MEOs would then range between 2000-35786 km; and HEOs in 35 786-$d_M/2$, where $d_M$ is the distance from the Earth to the Moon. The seemingly arbitrary altitude separating MEOs and HEOs is actually the average altitude for which the period equals one sidereal day (23 h 56 m 4 s), and this is precisely where GEOs sit. This altitude is more than 3 times the point at which the exosphere, the last layer of the atmosphere, is observed to fade. GEOs and HEOs are hence ``true'' free-space orbits, in the sense that there is hardly any gas, and temperature is dominated by the cosmic microwave background --which peaks at 2.7 K. The MEO region is the least populated one, since it is home to the Van Allen belts, which contain charged particles moving at relativistic speeds due to the magnetic field of the Earth, and that can destroy unshielded objects. LEOs, on the other hand, are `cheap' orbits, where most of the satellites orbiting our planet live. Their low altitudes simplify the problems arising from delays between earth-based stations and the satellites.

In this section we will be concerned only with two satellites orbiting either the same GEO or the same LEO, as a simple case study of expected losses and entanglement degradation. There are essentially two kinds of loss one must take into account: atmospheric loss and free-space path loss (FSPL). Total loss will then be simply given by
\begin{equation}
    L = L_\text{A} L_{\text{FSPL}}.
\end{equation}

Atmospheric absorption loss is caused by light-matter interactions. These strongly depend on the altitude of the orbits considered, among other parameters such as polarization, frequency, or weather conditions. Atmospheric loss can range from almost negligible (up in the exosphere and beyond) to very significant in the lower layers of the atmosphere, especially when water droplets and dust are present. Atmospheric loss has to be taken into account when considering the case of up- and downlinks, i.e., when linking a satellite with an earth-based station. However, for relatively high altitudes--that is, any altitude where there are satellites--absorption loss is so low in microwaves that it can be taken to vanish as a first approximation, so we set $L_\text{A} =1$.

FSPL is due to the inevitable spreading of a signal in three dimensions; they are often referred to as geometric losses. FSPL is maximal when there is no beam-constraining mechanism, such as a wave guide, or a set of focalizing lenses, i.e., when the signal spreads isotropically: $L_\text{FSPL}=\left({\lambda}/{4\pi z } \right)^{-2}$.

%%%%%%%%%%%%%%%%%%%%%%%%%%%%%%%%%%%%%%%%%%%%%%%%
\begin{figure}[t]
\centering
\includegraphics[width=0.95 \textwidth]{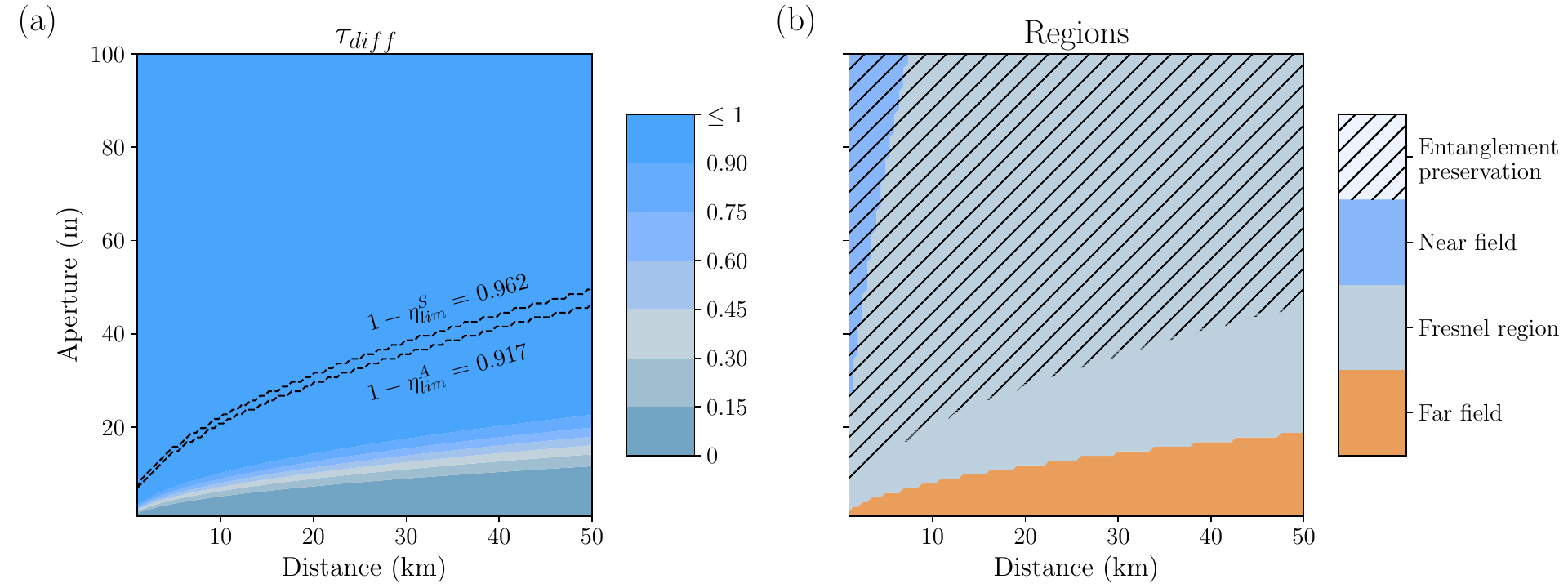}
\caption[Entanglement preservation in different regions of free space and its relation with diffraction-induced signal transmissivity:]{\textbf{Entanglement preservation in different regions of free space and its relation with diffraction-induced signal transmissivity:} (a) Contour plot of the transmissivity associated with diffraction, $\tau_\text{diff}$, against the aperture radius of the antenna and the traveled distance. We can observe that losses are greatly reduced with the aperture of the antenna. (b) Contour plot of the regions of free space as delimited by the relation between the aperture radius of the antenna and the distance at which the signal is observed: near field (blue) $2\varpi_{0} > (z\sqrt{\lambda}/0.62)^{2/3}$, Fresnel (gray) $\sqrt{z \lambda/2} < 2\varpi_{0} < (z\sqrt{\lambda}/0.62)^{2/3}$, and far field (orange) $2\varpi_{0} < \sqrt{z\lambda/2}$. With dashed lines, we represent the region where entanglement can be preserved. Parameters are $\lambda = 6 \text{ cm}$, $a_{R}=2\varpi_{0}$.}
\label{fig6_1}
\end{figure}
%%%%%%%%%%%%%%%%%%%%%%%%%%%%%%%%%%%%%%%%%%%%%%%%

Suppose that two comoving satellites are separated by a linear distance $z$, and that the emitter sends a quasimonochromatic signal with power $P_e$ centered at frequency $\nu=\omega/2\pi = c/\lambda$. The receiver gets a power $P_r$ such that their ratio defines a transmission coefficient that is the product of the loss and gains (or directivities) of the antennae. The resulting equation for long, `far-field' distances is sometimes referred to as Friis' equation \cite{Friis1946, Hogg1993, Shaw2013}, which is the compromise between gain (or directivity) and loss:
\begin{equation}
\frac{P_e}{P_r} =\frac{D_e D_d}{L_\text{A} L_{\text{FSPL}}}= D_e D_r  \left(\frac{\lambda}{4\pi z } \right)^2 \equiv\tau_\text{path}.
\end{equation}
Here $D_e$ and $D_r$ are the directivities of the emitter and receiver antennas, and we set $L_\text{A}=1$ as discussed before. The directivity of an antenna is the maximized gain in power in some preferred direction with respect to a hypothetical isotropic antenna, at a fixed distance from the source, and assuming that the total radiation power is the same for both antennas: $D = \max_{\theta, \phi} D(\theta, \phi)$. It is a quantity that strongly depends on the geometry design, but that can be enhanced in a discrete fashion by means of antenna arrays. Indeed, given $N$ identical antennas with directivity gain $D(\theta, \phi)$, a phased array consists of an array of such antennas, each preceded by a controlled phase shifter. This diffraction problem essentially gives $D_\text{array}(\theta, \phi) = A^2_N(\bm{\varepsilon}) D(\theta, \phi)$, where $A_N$ is the so-called $N$-array factor that symbolically depends on the phases via some vector $\bm{\varepsilon}$~\cite{Balanis2005}. In three dimensions, phase arrays are two-dimensional grids of antennas, so that the main lobe of the resulting signal becomes as sharp as possible. We assume that we have an array of small coplanar antennas as discussed in chapter~\ref{sec4}, adding up to a radiation pattern mimicking that of a parabolic antenna. We also assume that both emitter and receiver have the same design, $D_e = D_r \equiv D$ with
\begin{equation}
D=\left(\frac{\pi a}{\lambda}\right)^2 e_a
\end{equation}
where $0 \leq e_a \leq 1$ is the aperture efficiency, defined as the ratio between the effective aperture $A_e$, and the area of the antenna's actual aperture, $A_\text{phys}$, and $a$ is the diameter of the parabola, such that $A_{\text{phys}}=\pi a^{2}/4$. With this, the parabolic path transmissivity becomes
\begin{equation}
\tau_\text{path} = \left( \frac{\pi a^2 e_a }{4 z \lambda}\right)^{2}.
\end{equation}
The effect of path losses can alternatively be described by a diffraction mechanism, affecting the spot size of the signal beam,
\begin{equation}
\varpi = \varpi_{0}\sqrt{\left(1-\frac{z}{R_{0}}\right)^{2} + \left(\frac{z}{z_{R}} \right)^{2}},
\end{equation}
given an initial spot size $\varpi_{0}$, curvature of the beam $R_{0}$, and Rayleigh range $z_{R}=\pi\varpi_{0}^{2}/\lambda$. Given the aperture radius $a_{R}$ of the receiver antenna, the diffraction-induced transmissivity can be computed as~\cite{Pirandola2021,Pirandola2021_2}
\begin{equation}
\tau_{\text{diff}} = 1 - e^{-2a_{R}^{2}/\varpi^{2}}.
\end{equation}
Note that, in the far-field approximation, we can recover the result for $\tau_\text{path}$, by substituting the beam spot size $\varpi_{0}$ by the intensity spot size $\varpi_{0}/\sqrt{2}$,
\begin{equation}
\tau_{\text{diff}} \approx \left( \frac{\pi\varpi_{0}a_{R}}{\lambda z}\right)^{2},
\end{equation}
and by setting $a_{R}=\varpi_{0}=a/2$, $R_{0}=z$, and assuming that $e_{a}=1$.

Setting $\lambda = 6 \text{ cm}$ and $a_{R}=2\varpi_{0}$, we plot the transmissivity associated with diffraction versus the distance $z$ for different values of the aperture $\varpi_{0}$ in Fig.~\ref{fig6_1}~(a), observing that losses are reduced as a result of an increase in the aperture.

We address entanglement preservation in TMST states distributed through open air by considering that the dominant source of error will be diffraction, as opposed to attenuation, which we describe by means of a beam splitter with a thermal input. We introduce $N_{\text{th}}\sim 11$ as the number of thermal photons in the environment at $2.7$ K. Considering this loss mechanism, entanglement preservation is achieved for reflectivities that satisfy $\eta<(1+N_{\text{th}})^{-1}\sim0.083$ for lossy TMST asymmetric states, and $\eta<[1+N_{\text{th}}(1+\coth r)]^{-1}\sim0.038$ for lossy TMST symmetric states, assuming that $n\approx 0$ and $\tau = 1-\eta$. Given this diffraction channel, entanglement is preserved in the regime $a_{R}\varpi_{0}/z > (\lambda/\pi)\sqrt{-\frac{1}{2}\log\eta_{\text{lim}}} \sim 0.024$, for $\lambda=6$ cm and $\eta_{\text{lim}}=0.038$. This implies that, for two satellites that are separated by $z=1$ km, the product of the apertures of emitter and receiver antennae must be $a_{R}\varpi_{0} > 25 \text{ m}^{2}$ in order to have entanglement preservation. In Fig.~\ref{fig6_1}~(b), we represent the regions of free space as delimited by the relation between the distance at which the signal is detected and the aperture of the emitting antenna, taking $a_{R}=2\varpi_{0}$, and depicting the region in which entanglement is preserved with a dashed line. This shows that the radius of the antennae of emitter and receiver satellites will be large, as is usually the case for microwave communications. In order to correct the effects of diffraction with microwaves, it would also be useful to study focalizing techniques and the incorporation of beam collimators.

\subsection{Free-space Optical Quantum Communication}

In this section, we investigate the effect of free-space turbulence on the propagation of quantum states generated in the optical regime, and how entanglement is degraded in this process. We assume two parties attempt to share an entangled state, distributed through open air, to perform quantum teleportation. We then look at the Braunstein-Kimble teleportation protocol~\cite{Braunstein1998} for continuous-variable (CV) Gaussian states. Particularly, we consider that we initially have TMSV states, and use them to teleport a coherent state. We also look at the negativity of these states after free space attenuation. We investigate different instances of quantum communication: ground station to satellite (uplink), satellite to ground station (downlink), and we also consider the placement of an intermediate station (intermediate), either to generate states, or to refocus the beam, at an intermediate location. We follow by studying the limits for entanglement distribution and quantum teleportation with microwave signals, and compare them with optical signals, in a bad weather situation. We observe that the distances are highly reduced due to diffraction and thermal noise, as we would expect fro microwaves. We conclude by investigating entanglement distribution and quantum teleportation in horizontal paths, between two ground stations, and between two satellites. 

%%%%%%%%%%%%%%%%%%%%%%%%%%
\begin{figure}[t]
\centering
{\includegraphics[width=0.85 \textwidth]{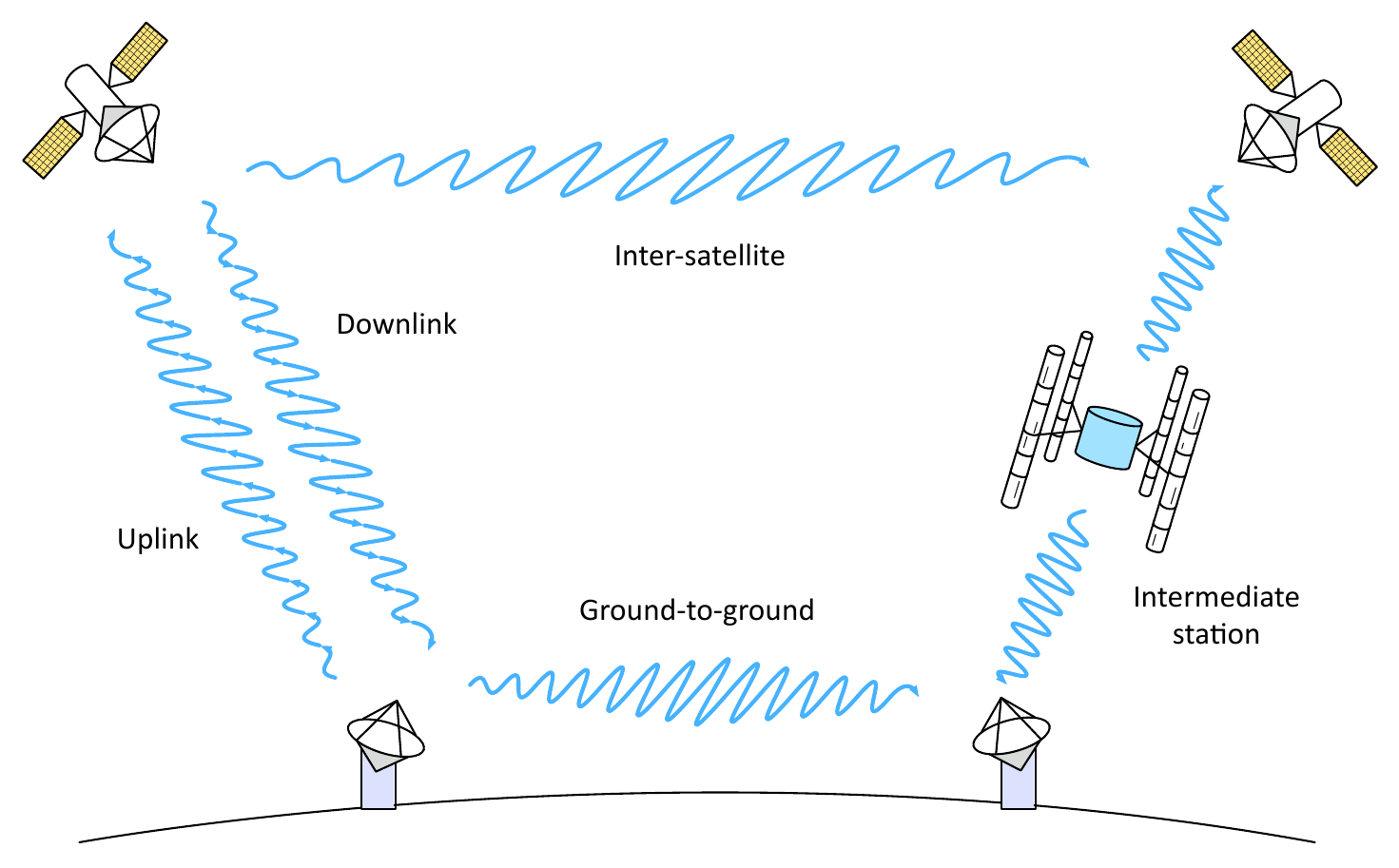}}
\caption[Graphical representation of optical quantum communication scenarios]{\textbf{Optical quantum communication scenarios} that we have studied in this section. We have investigated downlink and uplink channels, between a ground station and a satellite, both directly and with an intermediate station. We have also studied horizontal paths, between two ground stations and between two satellites.}
\label{fig6_2}
\end{figure}
%%%%%%%%%%%%%%%%%%%%%%%%%%

The different quantum communication scenarios studied here are depicted in Fig.~\ref{fig6_2}.

\subsubsection{Loss mechanism}

%%%%%%%%%%%%%%%%%%%%%%%%%%
\begin{figure}[t]
\centering
{\includegraphics[width=0.75 \textwidth]{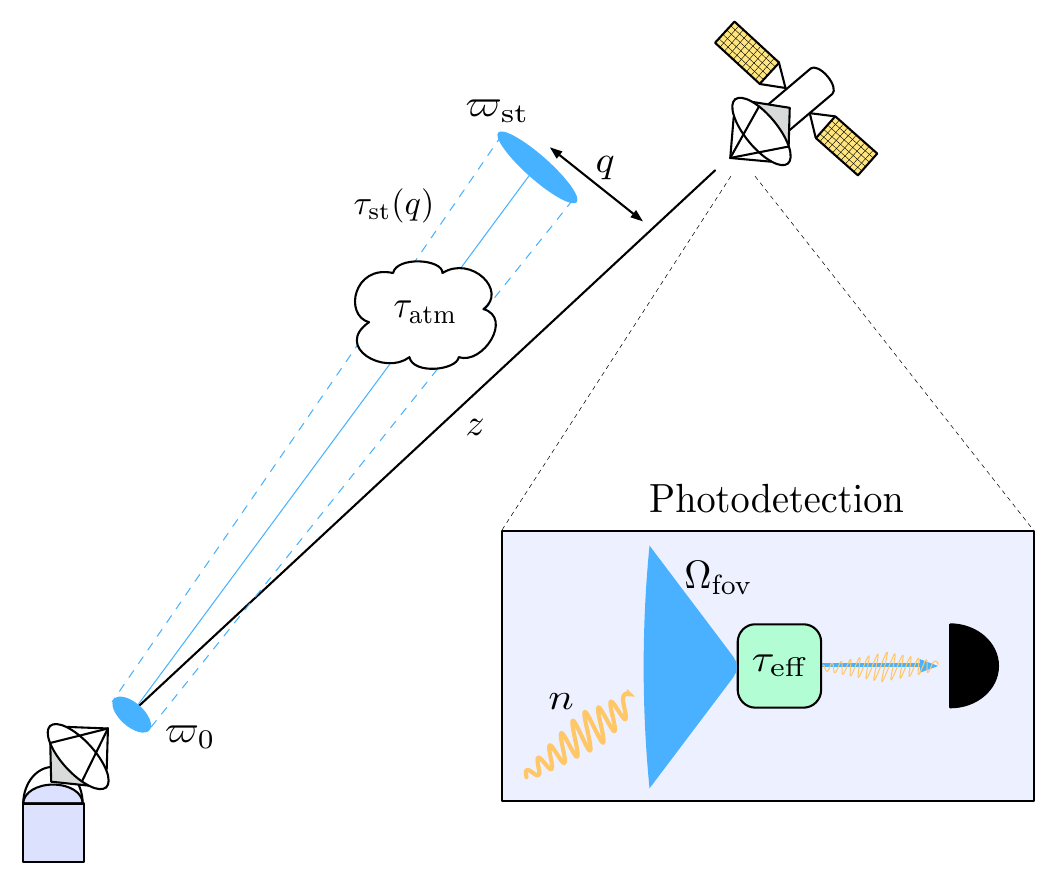}}
\caption[Sketch describing a quantum communication channel between a ground station and a satellite]{\textbf{Quantum communication channel between a ground station and a satellite}. A Gaussian beam is generated at the ground with an initial waist $\varpi_{0}$, and propagates a distance $z$ through free space, where it suffers from diffraction and turbulence effects, as well as atmospheric absorption. These mechanisms induce transmissivities $\tau_{\text{st}}$ and $\tau_{\text{atm}}$, respectively. Apart from the broadening of the beam waist, $\varpi_{\text{st}}$, caused by turbulence, we also have wandering of the beam centroid, quantified by the distance $q$. The efficiency of the photodetectors is represented by the transmissivity $\tau_{\text{eff}}$. Given the field of view, $\Omega_{\text{fov}}$, there is a mean number of thermal photons $n$ detected by the receiver.}
\label{fig6_3}
\end{figure}
%%%%%%%%%%%%%%%%%%%%%%%%%%

We consider that we have a ground station at altitude $h_{0}$, and a satellite of orbit $R_{0}$ and distance from the surface of the Earth $h\equiv R_{0}-R_{\text{E}}$, where $R_{\text{E}}$ is the radius of the Earth. Then the distance between the ground station, that sees the satellite at an angle $\theta$, and the satellite is
\begin{equation}\label{sec6_distance}
z = \sqrt{\left( \Delta h \right)^{2}+2\Delta h R + R^{2}\cos^{2}\theta} - R\cos\theta,
\end{equation}
where we have defined
\begin{eqnarray}\label{sec6_altitude}
\nonumber \Delta h &=& \sqrt{R^{2} + z^{2} + 2z R\cos\theta} - R, \\
R &=& R_{\text{E}} + h_{0}.
\end{eqnarray}
In this section, we have considered zenith communication, i.e. $\theta=0$.

In order to understand the limitations of entanglement distribution and quantum teleportation in free space, we need a comprehensive study of loss mechanisms. We will describe them through attenuation channels with transmissivity $\tau_{i}$, which act on the modes of a given quantum state as
\begin{equation}
\hat{a} \longrightarrow \sqrt{\tau_{i}}\hat{a} + \sqrt{1-\tau_{i}}\hat{a}_{\text{th}}.
\end{equation}
Here, we will consider that these attenuation channels incorporate a thermal mode from the environment, represented here by $\hat{a}_{\text{th}}$. If we assume that the quantum state is propagating through a homogeneous thermal environment, then the composite effect of $N$ attenuation channels is represented by the action of a single one whose effective transmissivity is $\tau = \prod_{i=1}^{N}\tau_{i}$. 

We sketch the general quantum communication scenario in Fig.~\ref{fig6_3}. We will consider the combined effects of different loss mechanisms that apply to signals in the optical regime propagating through free space.These mechanisms have been identified in previous works studying quantum communication links involving ground stations and satellites~\cite{Vasylyev2019,Pirandola2021_2,Pirandola2021}.

\subsubsubsection{Diffraction}
We consider the effects of diffraction in signals propagating through free space. We assume a quasi-monochromatic bosonic mode represented by a Gaussian beam with wavelength $\lambda$, curvature radius of the wave-front $R_{0}$, and initial waist $\varpi_{0}$. For a focused beam, $R_{0}$ is equal to the distance between transmitter and receiver, whereas for a collimated beam, it is set at infinity. The receiver aperture is $a_{R}$, and $z_{R}=\pi\varpi_{0}^{2}/\lambda$ is the Rayleigh range, such that the far-field regime is defined by $z\gg z_{R}$, for a transmission distance $z$.

The transmissivity induced by diffraction is given by
\begin{equation}\label{sec6_tau_diffraction}
\tau_{\text{diff}} = 1 - e^{-2\left(a_{R}/\varpi_{z}\right)^{2}},
\end{equation}
where $\varpi_{z}$ is the waist of the beam at a distance $z$~\cite{Svelto2010}, 
\begin{equation}
\varpi_{z}^{2} = \varpi_{0}^{2}\left[ \left( 1-\frac{z}{R_{0}}\right)^{2} + \left( \frac{z}{z_{R}}\right)^{2}\right].
\end{equation}
Here we will work with collimated beams, for which we have
\begin{equation}
\varpi_{z}^{2} = \varpi_{0}^{2}\left[ 1 + \left( \frac{z}{z_{R}}\right)^{2}\right].
\end{equation}
Notice that losses associated to diffraction will be larger when $a_{R}\ll\varpi_{z}$.

\subsubsubsection{Atmospheric attenuation}
The transmissivity affected by atmospheric attenuation of signals at a fixed altitude is given by
\begin{equation}
\tau_{\text{atm}} = \text{exp}\left[-\mu_{0}z e^{-h/\tilde{h}}\right],
\end{equation}
where $\mu_{0} = N_{0}\sigma$ is the extinction factor, $N_{0}$ is the density of particles, $\sigma=\sigma_{\text{abs}}+\sigma_{\text{sca}}$ is the cross section associated with absorption and scattering, and $\tilde{h}=6600$ m is a scale factor~\cite{Vasylyev2019}. At sea level, and for $\lambda=800\text{ nm}$, we have $\mu_{0}=5\times 10^{-6}\text{ m}^{-1}$.

Naturally, this needs to be adapted to variable altitudes. For that, we will use Eqs.~\eqref{sec6_distance} and~\eqref{sec6_altitude}, considering $h_{0}$ to be negligible. Then, we can write
\begin{equation}
\tau_{\text{atm}} = e^{-\mu_{0}g(h,\theta)},
\end{equation}
where we have defined
\begin{equation}
g(h,\theta) = \int_{0}^{z(h,\theta)} \diff y e^{-h(y,\theta)/\tilde{h}}.
\end{equation}

\subsubsubsection{Detector efficiency and thermal background}
As another source of loss, we can consider that we may have inefficient detectors. We will take, as the lowest value, $\tau_{\text{eff}}=0.4$~\cite{Liorni2019}, whereas the maximum possible one is $\tau_{\text{eff}}=1$. We will refer to the latter as the ideal case. Nevertheless, we consider that the signal traveling through the link will acquire excess noise that will be caught in the detectors, characterized by a thermal state that introduces $\tau_{\text{eff}}n$ thermal photons into our TMSV state. Furthermore, we consider that the effective number of thermal photons that the signal acquires in the path is the one that can be effectively captured by the detectors. We compute this using~\cite{Pirandola2021}
\begin{eqnarray}\label{sec6_thermal_background}
\nonumber G_{R} &=& \Delta\lambda \Delta t\Omega_{\text{fov}}a_{R}^{2}, \\
N_{BB} &=& 2c\lambda^{-4}\left[ e^{hc/(\lambda k_{B}T)}-1\right]^{-1}, \\
\nonumber n &=& G_{R}N_{BB},
\end{eqnarray}
where $G_{R}$ is the photon collection parameter, $\Delta\lambda$ and $\Delta t$ are the spectral filter and the time bandwidth, respectively, $\Omega_{\text{fov}}$ is the field of view of the receiver, and $N_{BB}$ is the number of thermal photons, quantified by the black-body formula, in units of $G_{R}^{-1}$. Furthermore, $c$ is the speed of light, $\lambda$ is the wavelength of the signal, $h$ is Planck's constant, $k_{B}$ is Boltzmann's constant, $T$ is the temperature, and finally $n$ is the average thermal photon number.

We take $\Delta\lambda=1$ nm, $\Delta t=10$ ns, $\Omega_{\text{fov}}=10^{-10}$ sr and $a_{R}=40$ cm. Then, we are left with average thermal-photon number $n_{\text{day}}^{\text{down}}=0.30$ and $n_{\text{night}}^{\text{down}}=3.40\times10^{-6}$ for a daytime and nighttime downlink, respectively, and $n_{\text{day}}^{\text{up}}=0.22$ and $n_{\text{night}}^{\text{up}}=5.43\times10^{-7}$ for a daytime and nighttime uplink, respectively.

\subsubsubsection{Turbulence}
Let us now look at the effects of turbulence. We aim at working in the weak-turbulence regime, in which the effects of scintillation are ignored. This regime can be characterized using the spherical-wave coherence length,
\begin{eqnarray}
\nonumber \rho_{0} &=& [1.46 k^{2}Y_{0}(z)]^{-3/5}, \\
Y_{0}(z) &=& \int_{0}^{z} \diff \xi \left( 1 - \frac{\xi}{z}\right)^{5/3}C^{2}_{n}(h(\xi,\theta))
\end{eqnarray}
through the following formula
\begin{equation}
z \lesssim k\left(\text{min}\{2a_{R},\rho_{0}\}\right)^{2},
\end{equation}
for a beam with wavenumber $k=2\pi/\lambda$, propagation distance $z$, and refraction index structure constant $C_{n}^{2}$. The latter, in the Hufnagel-Valley model of atmospheric turbulence~\cite{Hufnagel1964,Valley1980}, reads
\begin{equation}
C^{2}_{n} = 5.94\times10^{-53}\left( \frac{v}{27}\right)^{2}h^{10}e^{-h/1000} + 2.7\times 10^{-16}e^{-h/1500} + A e^{-h/100},
\end{equation}
and it measures the strength of the fluctuations in the refraction index caused by spatial variations of temperature and pressure. In this chapter, we consider $v=21\text{ m}/\text{s}$ for the wind speed, and $A_{\text{day(night)}}=2.75 (1.7)\times10^{-14} \text{ m}^{-2/3}$ for daytime (nighttime) values. At constant altitude, we see that $\rho_{0}=(0.548 k^{2} C_{n}^{2}z)^{-3/5}$. If we consider an uplink, we can use the above formula, but for a downlink, we need to substitute $\xi\rightarrow z-\xi$ in the structure constant.

In the weak-turbulence regime, we can distinguish between two sources of errors, caused by the interaction of the beam with vortices, or \textit{eddies}, of different sizes: beam broadening and beam wandering~\cite{Fante1975,Fante1980}. Beam broadening is caused by eddies smaller than the beam waist, and acts on a fast time scale. This will replace $\varpi_{z}$ by some short-term waist $\varpi_{\text{st}}$, leading to the modified diffraction-induced transmissivity
\begin{equation}
\tau_{\text{st}} = 1 - e^{-2\left(a_{R}/\varpi_{\text{st}}\right)^{2}}.
\end{equation}
Here, we can write
\begin{equation}
\varpi_{\text{st}}^{2} \simeq \varpi_{z}^{2} + 2\left( \frac{\lambda z}{\pi\rho_{0}}\right)^{2}(1-\phi)^{2},
\end{equation}
where $\phi = 0.33(\rho_{0}/\varpi_{0})^{1/3}$. In the weak-turbulence regime, we find that $\phi\ll1$, and thus we can approximate $(1-\phi)^{2} \approx 1-0.66(\rho_{0}/\varpi_{0})^{1/3}$~\cite{Yura1973}. 

Beam wandering is caused by eddies larger than the beam waist, and act on a slow time scale. This causes the beam to deflect by randomly displacing its center, leading to a wandering of the waist. This random displacement will be assumed to follow a Gaussian probability distribution with variance $\sigma^{2}$, which will be composed of the large-scale turbulence $\sigma^{2}_{\text{TB}}$ and the pointing error $\sigma^{2}_{\text{P}}$ variances. The long-term waist of the beam can be approximated by
\begin{equation}
\varpi_{\text{lt}}^{2} \simeq \varpi_{z}^{2} + 2\left( \frac{\lambda z}{\pi\rho_{0}}\right)^{2},
\end{equation}
and it is related to its short-term counterpart through 
\begin{equation}
\sigma_{\text{TB}}^{2} = \varpi_{\text{lt}}^{2} - \varpi_{\text{st}}^{2} \simeq \frac{0.1337\lambda^{2}z^{2}}{\varpi_{0}^{1/3}\rho_{0}^{5/3}}.
\end{equation}
The beam centroid wanders with total variance $\sigma^{2}=\sigma_{\text{TB}}^{2}+\sigma_{\text{P}}^{2}$, and we will take $\sigma_{\text{P}}=10^{-6}z$. We define $q$ as the distance between the beam centroid and the original center (horizontally-aligned with the transmitter and the receiver), also known as deflection. Following Ref.~\cite{Vasylyev2012}, we assume that this value takes a Gaussian random walk following the Weibull distribution
\begin{equation}
P_{\text{WB}}(q) = \frac{q}{\sigma^{2}}e^{-\frac{q^{2}}{2\sigma^{2}}}.
\end{equation}
Then, the maximum value of the transmissivity occurs for $q=0$,
\begin{equation}
\tau_{\text{max}} = \tau(q=0) = \tau_{\text{st}}(q=0)\tau_{\text{atm}}\tau_{\text{eff}}.
\end{equation}
However, for each instantaneous value of $q$, there will be an instantaneous $\tau(q)\leq\tau_{\text{max}}$ happening with a probability $P(\tau)$. The transmissivity associated to diffraction modified by this behavior is then~\cite{Vasylyev2012, Pirandola2021}
\begin{equation}
\tau_{\text{st}}(q) = e^{-4(q/\varpi_{\text{st}})^{2}}Q_{0}\left(\frac{2s^{2}}{\varpi_{\text{st}}^{2}}, \frac{4 q a_{R}}{\varpi_{\text{st}}^{2}}\right),
\end{equation}
where $Q_{0}(x,y)=\frac{e^{x}}{2x}\int_{0}^{y}\diff t \, te^{-t^{2}/4x}I_{0}(t)$ is an incomplete Weber integral and $I_{0}$ is the modified Bessel function of the first kind, of order $0$. We can express
\begin{equation}
\tau(q) = \tau_{\text{max}}e^{-\left(q/q_{0}\right)^{\kappa}},
\end{equation}
where we have defined
\begin{eqnarray}
\nonumber \tau_{\text{st}}^{\text{far}} &=& \frac{2a_{R}^{2}}{\varpi_{\text{st}}^{2}}, \\
\Lambda_{n}(x) &=& e^{-2x}I_{n}(2x), \\
\nonumber \kappa &=& \frac{4\tau_{\text{st}}^{\text{far}}\Lambda_{1}\left(\tau_{\text{st}}^{\text{far}}\right)}{1-\Lambda_{0}\left(\tau_{\text{st}}^{\text{far}}\right)}\left[ \log\left( \frac{2\tau_{\text{st}}}{1-\Lambda_{0}\left(\tau_{\text{st}}^{\text{far}}\right)}\right)\right]^{-1}, \\
\nonumber q_{0} &=& a_{R}\left[ \log\left( \frac{2\tau_{\text{st}}}{1-\Lambda_{0}\left(\tau_{\text{st}}^{\text{far}}\right)}\right)\right]^{-1/\kappa}.
\end{eqnarray}
The probability distribution over $q$ induces another probability distribution over $\tau$, 
\begin{equation}
P(\tau) = \frac{q_{0}^{2}}{\kappa\sigma^{2}\tau}\left( \log\frac{\tau_{\text{max}}}{\tau}\right)^{\frac{2}{\kappa}-1}\exp\left[ -\frac{q_{0}^{2}}{2\sigma^{2}}\left( \log\frac{\tau_{\text{max}}}{\tau}\right)^{\frac{2}{\kappa}}\right].
\end{equation}
This function can be obtained from the Weibull distribution $P_{\text{WB}}(q)$ by using 
\begin{equation}
P(\tau) = \left. p(q|\sigma)\right|_{q=q(\tau)} \left| \frac{\diff q}{\diff \tau}\right|
\end{equation}
together with
\begin{equation}
q = q_{0}\left( \log\frac{\tau_{\text{max}}}{\tau}\right)^{\frac{1}{\kappa}}.
\end{equation}

%~\cite{Dong2010,Usenko2012}
\subsubsection{Entanglement distribution and quantum teleportation}
The quantum channel, once characterized by transmissivity $\tau$, is now described by the ensemble $\mathcal{E}=\{\mathcal{E}_{\tau},P(\tau)\}$, where the quantum channel $\mathcal{E}_{\tau}$ is selected at random with probability density $P(\tau)$. This is called a fading channel. We will use this to describe the degradation of entanglement on states propagating through free space, which we will quantify through the negativity of the covariance matrix of Gaussian states, and through the average fidelity of teleporting an unknown coherent state using the entangled resources. We will consider two-mode squeezed states as a typical case of bipartite CV entangled states. Since two-mode squeezed states are Gaussian, and the fading channel we consider is Gaussian-preserving, we can use the covariance matrix formalism to describe the evolution of the state. This will provide the obvious advantages of using finite-dimensional matrices to work with infinite-dimensional operators, but it will also lead to a convenient description of fading channels. Consider a two-mode Gaussian state with vanishing first moments and covariance matrix in normal form given by
\begin{equation}\label{sec6_CM}
\Sigma = \begin{pmatrix} \alpha\mathbb{1}_{2} & \gamma \sigma_{z} \\ \gamma \sigma_{z} & \beta\mathbb{1}_{2} \end{pmatrix},
\end{equation}
and consider a single-mode environment described by a Gaussian state with covariance matrix $E=m\mathbb{1}_{2}$. For example, for a daytime downlink, this state is characterized by $m=1+2\tau_{\text{eff}}n_{\text{day}}^{\text{down}}$. We assume that the second mode is the one being transmitted through open air, and therefore it is affected by the fading channel. Keeping only the transmitted contribution, we obtain
\begin{equation}\label{sec6_CM_asym}
\Sigma' = \begin{pmatrix} \alpha\mathbb{1}_{2} & \left\langle\sqrt{\tau}\right\rangle\gamma \sigma_{z} \\ \left\langle\sqrt{\tau}\right\rangle\gamma \sigma_{z} & \left[ \langle\tau\rangle\beta + \left(1-\langle\tau\rangle\right)m \right] \mathbb{1}_{2} \end{pmatrix}.
\end{equation}
This description assumes that turbulence is a fast process, compared with the detection speed. Let us look at how this result can be derived. First, see that the Wigner function of the state that results from applying the fading channel is
\begin{equation}
W'(x,p) = \int_{0}^{\tau_{\text{max}}}\diff \tau P(\tau)W_{\tau}(x,p).
\end{equation}
Here, the Wigner function $W_{\tau}(x,p)$ results from the modification of the quadrature operators $\hat{x}$ and $\hat{p}$ by the quantum channel instance $\varepsilon_{\tau}$,
\begin{eqnarray}
\nonumber \hat{x} &\rightarrow& \hat{x}_{\tau} = \sqrt{\tau}\hat{x} + \sqrt{1-\tau}\hat{x}_{\text{E}}, \\
\hat{p} &\rightarrow& \hat{p}_{\tau} = \sqrt{\tau}\hat{p} + \sqrt{1-\tau}\hat{p}_{\text{E}},
\end{eqnarray}
which get mixed with the quadrature operators of the state of an environment. In this formalism, the expectation value of the operator $\hat{A}\hat{B}$ is computed as
\begin{eqnarray}
\nonumber && \langle\hat{A}\hat{B}\rangle = \int \diff x\diff p A B W'(x,p) \\
\nonumber &&=  \int_{0}^{\tau_{\text{max}}}\diff\tau P(\tau) \int \diff x\diff p A B W_{\tau}(x,p) \\
&&= \int_{0}^{\tau_{\text{max}}}\diff \tau P(\tau)\langle\hat{A}\hat{B}\rangle_{\tau}.
\end{eqnarray}
This result implies that we can replace the elements of the covariance matrix of the state resulting from the fading channel by the weighted integral of the expectation values resulting from each channel instance~\cite{Dong2010,Usenko2012}. The later looks as follows, for the second moments of quadrature operators:
\begin{eqnarray}
\nonumber \hat{x}^{2} &\rightarrow& \hat{x}_{\tau}^{2} = \tau\hat{x}^{2} + (1-\tau)\hat{x}_{\text{E}}^{2} + \sqrt{\tau(1-\tau)}\{\hat{x},\hat{x}_{\text{E}}\}, \\
\hat{p}^{2} &\rightarrow& \hat{p}_{\tau}^{2} = \tau\hat{p}^{2} + (1-\tau)\hat{p}_{\text{E}}^{2} + \sqrt{\tau(1-\tau)}\{\hat{p},\hat{p}_{\text{E}}\}, \\
\nonumber \{\hat{x},\hat{p}\} &\rightarrow& \{\hat{x}_{\tau},\hat{p}_{\tau}\} = \tau\{\hat{x},\hat{p}\} + (1-\tau)\{\hat{x}_{\text{E}},\hat{p}_{\text{E}}\} + \sqrt{\tau(1-\tau)}\left(\{\hat{x},\hat{p}_{\text{E}}\} + \{\hat{p},\hat{x}_{\text{E}}\}\right).
\end{eqnarray}
For the complete fading channel, we will have to make the replacement
\begin{eqnarray}
\nonumber \tau &\longrightarrow& \langle\tau\rangle = \int_{0}^{\tau_{\text{max}}}\diff\tau P(\tau)\tau, \\
\sqrt{\tau} &\longrightarrow& \left\langle\sqrt{\tau}\right\rangle = \int_{0}^{\tau_{\text{max}}}\diff\tau P(\tau)\sqrt{\tau}.
\end{eqnarray}
Therefore we observe only an average characterization of the channel through $\langle\tau\rangle$ and $\langle\sqrt{\tau}\rangle$. If we considered that the detectors were much faster than the turbulence, then we would obtain $\tau$ instead of $\langle\tau\rangle$, and we would have to average the obtained quantity afterwards. In this scenario, the quantum teleportation fidelity would be
\begin{equation}
\overline{F} = \int_{0}^{\tau_{\text{max}}}\diff\tau P(\tau)\overline{F}(\tau).
\end{equation}
We refer to this as the slow-turbulence regime. In contrast, the teleportation fidelity in the fast-turbulence regime is $\overline{F}\left(\langle\tau\rangle\right)$.

For an entangled Gaussian resource that has the covariance matrix in Eq.~\eqref{sec6_CM}, the average fidelity of teleporting an unknown coherent state is $\overline{F}=\left[ 1+\frac{1}{2}(\alpha+\beta-2\gamma) \right]^{-1}$. Now, if we introduce the effect of the fast fading channel, we see that
\begin{equation}
\overline{F} = \left\{ 1 + \frac{1}{2}\left[ \alpha + \langle\tau\rangle\beta + \left( 1 - \langle\tau\rangle\right)m -2\langle\sqrt{\tau}\rangle\gamma \right] \right\}^{-1},
\end{equation}
while for the slow fading channel, the average is computed numerically. The other quantity we are interested in is the negativity of the covariance matrix, a measure of entanglement for bipartite Gaussian states~\cite{Serafini2004}. We will use the smallest symplectic eigenvalue of the partially-transposed covariance matrix, as the condition $\tilde{\nu}_{-}<1$ defines the region of entanglement. For the one in Eq.~\eqref{sec6_CM}, we can write it as
\begin{equation}
\tilde{\nu}_{-} = \frac{\alpha+\beta-\sqrt{(\alpha-\beta)^{2}+4\gamma^{2}}}{2},
\end{equation}
such that $\tilde{\nu}_{-}<1$ can be expressed as $(\alpha-1)(\beta-1)<\gamma^{2}$. 

Both the teleportation fidelity and the negativity are reduced because the entanglement of the state degrades as it propagates through free space. The degradation is more severe with increasing distance, as the transmissivity of the fading channel decreases. Here, we investigate the teleportation fidelity and the negativity associated with a TMSV state with covariance matrix
\begin{equation}\label{sec6_TMSV}
\Sigma = \begin{pmatrix} \cosh2r\mathbb{1}_{2} & \sinh2r \sigma_{z} \\ \sinh2r \sigma_{z} & \cosh2r\mathbb{1}_{2} \end{pmatrix},
\end{equation}
where $r$ is the squeezing parameter, and it is directly related with the (initial) negativity through $\tilde{\nu}_{-}=e^{-2r}$, meaning no entanglement for $r=0$, and infinite entanglement for $r\rightarrow\infty$. The teleportation fidelity associated with using a TMSV state is $\overline{F} = \left( 1+e^{-2r} \right)^{-1}$~\cite{Pirandola2006}, and it reaches the maximum classical fidelity of $1/2$ for no entanglement ($r=0$), while approaching 1 for infinite entanglement ($r\rightarrow\infty$). 

%%%%%%%%%%%%%%%%%%%%%%%%%%
\begin{figure}[h!]
\centering
{\includegraphics[width=\textwidth]{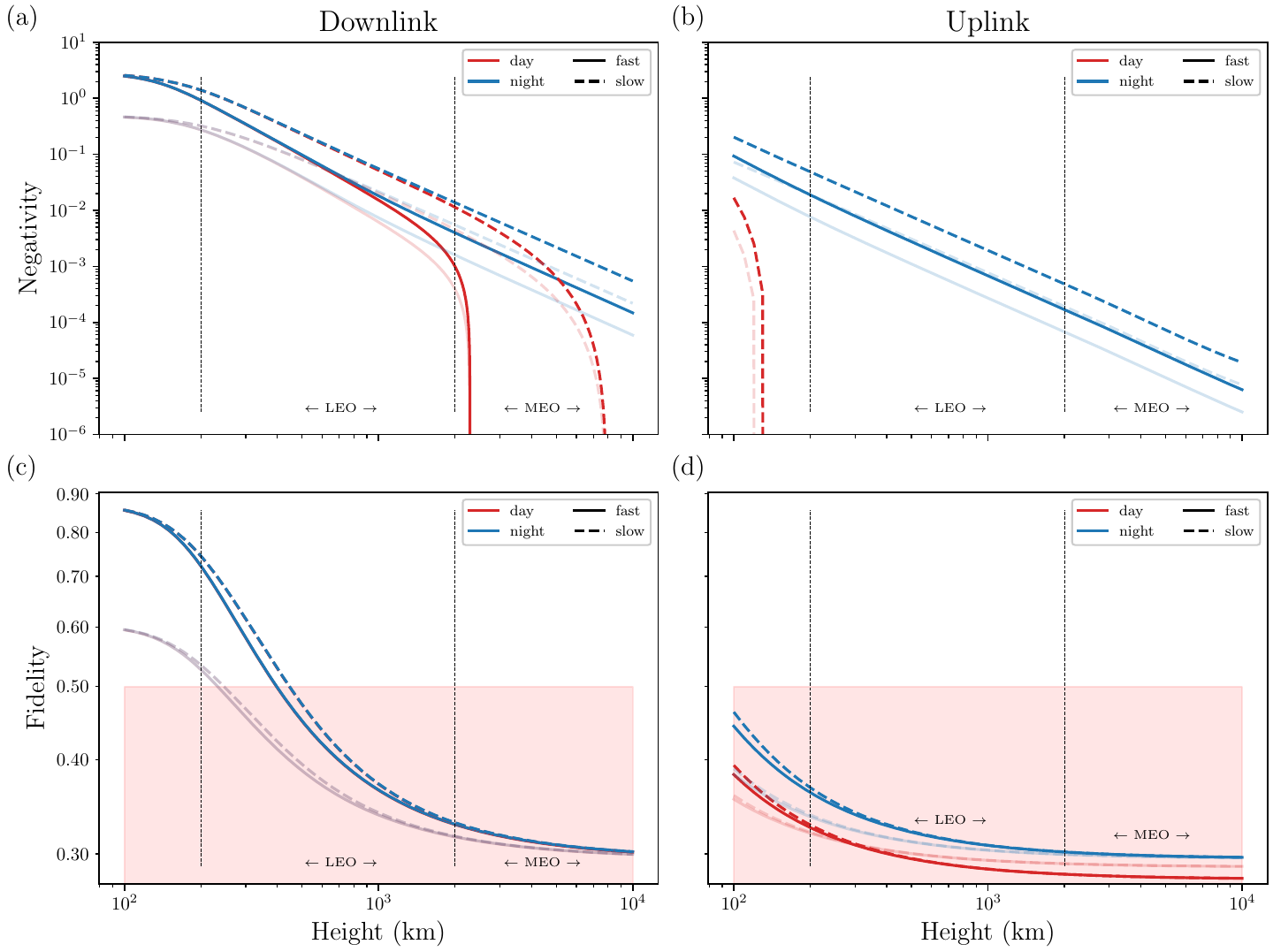}}
\caption[Uplink and downlink quantum communication between a ground station and a satellite]{\textbf{Uplink and downlink quantum communication between a ground station and a satellite} using a TMSV state distributed through free space, which has undergone a loss mechanism comprising diffraction, atmospheric extinction, detector inefficiency and free-space turbulence, for a signal with wavelength $\lambda=800$ nm, squeezing parameter $r=1$, and waist $\varpi_{0}=20$ cm, assuming the receiver has an antenna with aperture $a_{R}=40$ cm. We represent the negativity for (a) a downlink and (b) an uplink. We also represent the fidelity of quantum teleportation for coherent states using this entangled resource, for (c) a downlink and (d) an uplink. Dashed lines represent the regime of slow turbulence, and solid lines represent the regime of fast turbulence, when comparing them to the velocity of the detectors. Nighttime and daytime thermal noise is taken into account in the blue and red curves, respectively. In the case of the downlink fidelity, note that the results which incorporate daytime (red) and nighttime (blue) thermal noise coincide. In full color, we present the results for perfect detector efficiency, $\tau_{\text{eff}}=1$, whereas the high-transparency curves correspond to $\tau_{\text{eff}}=0.4$.}
\label{fig6_4}
\end{figure}
%%%%%%%%%%%%%%%%%%%%%%%%%%

In Figs.~\ref{fig6_4}~(a),~(b), we represent the negativity of a TMSV state with initial squeezing $r=1$ against the height of the link. Fig.~\ref{fig6_4}~(a) shows the results for a downlink, and Fig.~\ref{fig6_4}~(b) illustrates an uplink. In solid lines, we can see the results of a fast-turbulence scenario, whereas the dashed lines represent a slow-turbulence one. Furthermore, blue and red lines incorporate nighttime and daytime thermal noise, respectively. In full color, we can see the values associated with perfect detector efficiency, $\tau_{\text{eff}}=1$, whereas the lines with high transparency correspond to faulty detectors with $\tau_{\text{eff}}=0.4$. We can observe that the negativity is reduced exponentially with the distance, and we see better results for a downlink than for an uplink. In vertical lines, we mark zones associated to different orbital altitudes. These are the low-Earth orbit (LEO), from 200 km to 2000 km and the medium-Earth orbit (MEO), from 2000 km to 42164 km. Orbits from 42164 km on are known as geostationary orbits. 

Figs.~\ref{fig6_4}~(c),~(d) show the fidelity of a quantum teleportation protocol for coherent states, that uses TMSV states distributed through (c) a downlink or (d) an uplink through free space. The degradation of the entanglement of this state is due to the various loss mechanisms that comprise the fading channel: diffraction, atmospheric attenuation, detector inefficiency and turbulence. This degradation is responsible for the deterioration of the teleportation fidelity, which depends only on the entangled resource that is consumed. The slow-turbulence regime is represented by dashed lines, while the fast-turbulence regime is represented by solid lines. The red ones incorporate daytime thermal noise, whereas the blue ones consider nighttime thermal noise. Perfect detector efficiency ($\tau_{\text{eff}}=1$) is represented by full-color lines, while an imperfect detector ($\tau_{\text{eff}}=0.4$) was considered in the high-transparency lines. Here, we observe that only quantum teleportation protocols through a downlink in the LEO region can produce fidelities above the maximum classical result~\cite{Braunstein2001}; all instances worse than this are enclosed in a pale red background. Notice that, in Fig.~\ref{fig6_4}~(c), results for daytime and nighttime thermal noise coincide, both in the perfect and imperfect detector scenarios. This also happens for short distances in Fig.~\ref{fig6_4}~(a).

%We want to clarify that we are considering the satellite involved in quantum teleportation to be fixed for the duration of the protocol. {\color{red} Give some estimation of the duration of the protocol.}

\subsubsubsection{Intermediate station for state generation}
We have observed that the effects of turbulence are more severe in the atmosphere, and have stronger effects on signals that have not suffered diffraction. Therefore, the scenario in which we have an uplink path presents bleaker hopes for free-space entanglement distribution. Nevertheless, we investigate a scenario in which there is an intermediate station connecting the ground station and the satellite, and we consider that TMSV states can be generated at this intermediate station. Our goal is to observe whether there is an increase in the entanglement available when the distance that the signals travel through free space is reduced. This already presents an advantage, because now the uplink does not start at the Earth, but at a given orbit, and the turbulence effects are highly reduced. 
%%%%%%%%%%%%%%%%%%%%%%%%%%
\begin{figure}[h!]
\centering
{\includegraphics[width=\textwidth]{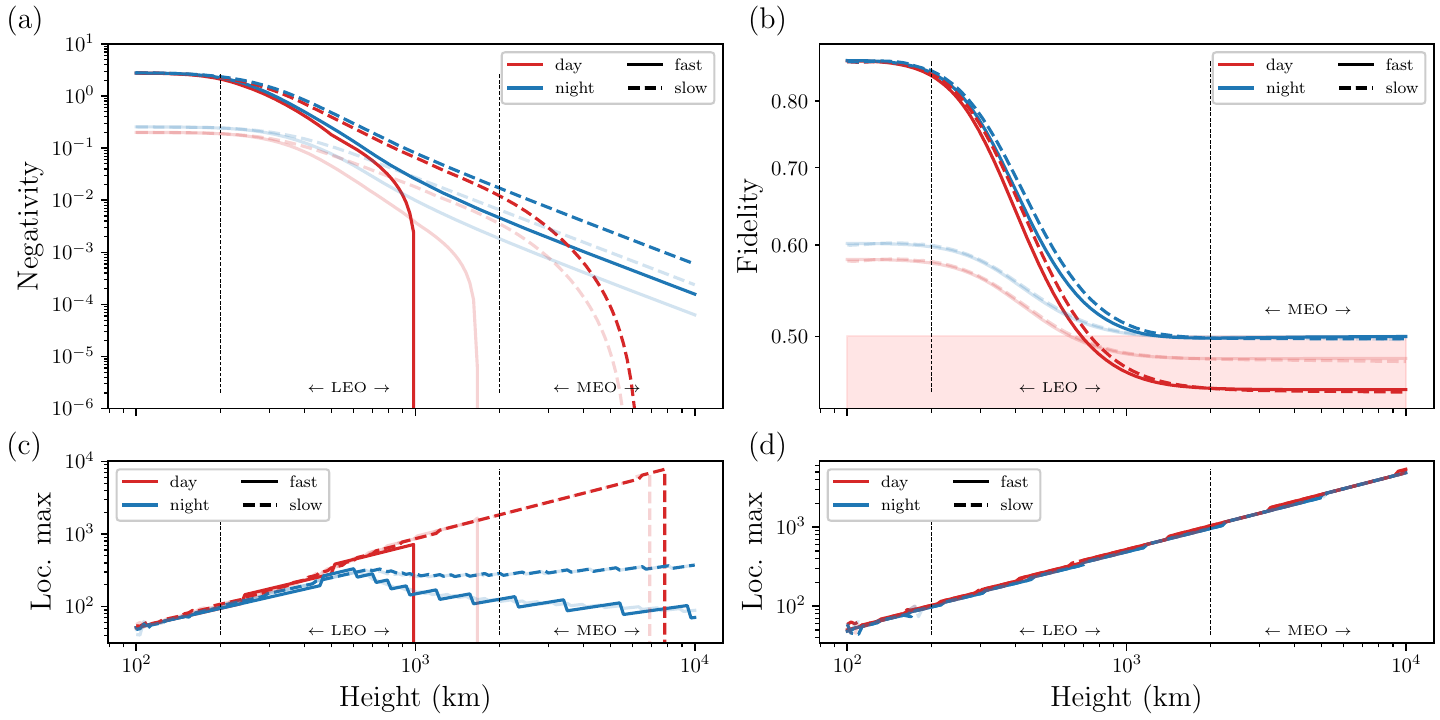}}
\caption[Uplink and downlink quantum communication between a ground station and a satellite, using an intermediate station for state generation]{\textbf{Uplink and downlink quantum communication between a ground station and a satellite, using an intermediate station for state generation}. We consider that one mode of the state is sent to the ground station, and the other to the satellite. (a) Negativity of transmitted TMSV states in free-space communication against the height of the satellite, with respect to the ground station. We study a signal with wavelength $\lambda=800$ nm, squeezing parameter $r=1$, and initial waist $\varpi_{0}=20$ cm, sent to a receiver that has an antenna of radius $a_{R}=40$ cm. This signal is subject to a loss mechanism composed of diffraction, atmospheric extinction, detector inefficiency and free-space turbulence, described by a fading channel. In this case, the results that incorporate daytime (red) and nighttime (blue) thermal noise coincide. We distinguish between the results obtained in the slow-turbulence and fast-detection regime, in dashed lines, and the fast-turbulence and slow-detection regime, in solid lines. The results for perfect detector efficiency, $\tau_{\text{eff}}=1$, appear in full color, whereas the transparent curves correspond to $\tau_{\text{eff}}=0.4$. (b) Fidelity of teleporting an unknown coherent state using these entangled resources. We represent the optimal position of the intermediate station, for the different turbulence conditions, that achieve the maximum possible negativities in (c), and those that lead to maximum fidelities, in (d).}
\label{fig6_5}
\end{figure}
%%%%%%%%%%%%%%%%%%%%%%%%%%
In this case, the covariance matrix of the two-mode Gaussian state, after a single application of the fading channel, is
\begin{equation}\label{sec6_CM_sym}
\Sigma' = \begin{pmatrix} \left[\tau_{\text{d}}\alpha+(1-\tau_{\text{d}})m_{\text{d}}\right]\mathbb{1}_{2} & \sqrt{\tau_{\text{d}}\tau_{\text{u}}}\gamma \sigma_{z} \\ \sqrt{\tau_{\text{d}}\tau_{\text{u}}}\gamma \sigma_{z} & \left[ \tau_{\text{u}}\beta + \left(1-\tau_{\text{u}}\right)m_{\text{u}} \right] \mathbb{1}_{2} \end{pmatrix},
\end{equation} 
where we define by $\tau_{\text{d(u)}}$ the transmissivity of the fading channel describing signal propagation through the downlink (uplink). After multiple applications of the fading channel, in the case of fast turbulence and slow detection, we will have that the negativity and the teleportation fidelity can be averaged as $\mathcal{N}=\mathcal{N}\left(\langle\tau_{\text{d}}\rangle,\langle\tau_{\text{u}}\rangle\right)$ and $\overline{F} = \overline{F}\left(\langle\tau_{\text{d}}\rangle,\langle\tau_{\text{u}}\rangle\right)$, respectively. On the opposite regime, slow turbulence and fast detection, these averages are computed as
\begin{eqnarray}
\nonumber \mathcal{N} &=& \int_{0}^{\tau_{\text{d}^{\text{max}}}} \diff\tau_{\text{d}}P(\tau_{\text{d}}) \int_{0}^{\tau_{\text{u}^{\text{max}}}} \diff\tau_{\text{u}} P(\tau_{\text{u}}) \mathcal{N}\left(\tau_{\text{d}},\tau_{\text{u}}\right), \\
\overline{F} &=& \int_{0}^{\tau_{\text{d}^{\text{max}}}} \diff\tau_{\text{d}}P(\tau_{\text{d}}) \int_{0}^{\tau_{\text{u}^{\text{max}}}} \diff\tau_{\text{u}} P(\tau_{\text{u}}) \overline{F}\left(\tau_{\text{d}},\tau_{\text{u}}\right).
\end{eqnarray}

In Fig.~\ref{fig6_5}~(a), we represent the negativity of the final state, considering an optimal placement of the intermediate station, for each value of the total height. These optimal points are shown in Fig.~\ref{fig6_5}~(c). We can observe that the results are improved, with respect to both the downlink and the uplink. We can also observe this improvement, especially with respect to the uplink, and remarkably for high altitudes, in Fig.~\ref{fig6_5}~(b). Here, we represent the fidelity of teleporting an unknown coherent state using TMSV states, generated at the intermediate station, and having both modes distributed through the noisy and turbulent links. Only the fidelities with nighttime thermal noise remain above the maximum classical fidelity of $1/2$, while the accumulated thermal noise in daytime links leads to fidelities that fall below this limit at altitudes in the LEO region. We can see that the limit is extended with respect to the downlink, and the fidelity for an uplink never achieved values above it. Therefore, the generation of entangled states in an intermediate station between the ground station and the satellite greatly improves the teleportation fidelity. 

In the case of the negativity with an intermediate station, we observe an improvement especially in the case of ideal detectors; for imperfect ones, represented by $\tau_{\text{eff}}=0.4$, the results do not differ significantly from those of the downlink. This is because, for an intermediate station, we are considering now two detection events, instead of one, which enhances the error in the case of imperfect detectors. 

\begin{figure}[h!]
\centering
{\includegraphics[width=\textwidth]{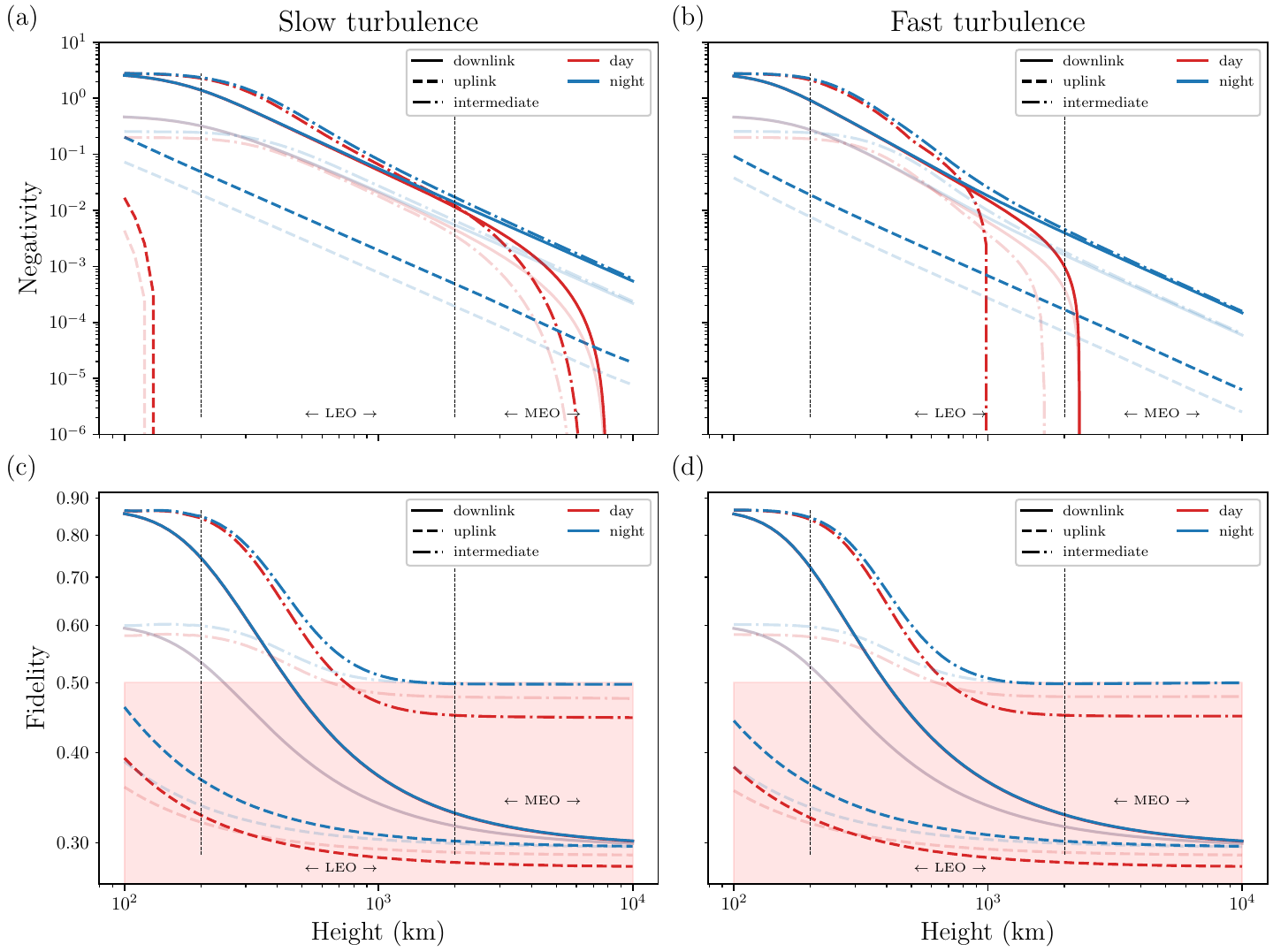}}
\caption[Ground-to-satellite uplink and downlink quantum communication comparison with intermediate station]{\textbf{Ground-to-satellite uplink and downlink quantum communication comparison with intermediate station}, where these states are generated. We study a signal of wavelength $\lambda=800$ nm and initial waist $\varpi_{0}=20$ cm, in a TMSV state with squeezing parameter $r=1$, sent to a receiver that has an antenna of radius $a_{R}=40$ cm, and that is subject to a loss mechanism composed of diffraction, atmospheric extinction, detector inefficiency and free-space turbulence, and described by a fading channel. In a solid line, we plot the quantities associated to a downlink; those corresponding to an uplink appear dashed, and the dashed-dotted lines describe the case of an intermediate station. In full color we represent the results for perfect detector efficiency, $\tau_{\text{eff}}=1$, whereas imperfect detection, $\tau_{\text{eff}}=0.4$, is marked by the transparent curves. Negativity of the state after the fading channel in the slow-turbulence and fast-detection regime (a), and in the fast-turbulence and slow-detection regime (b), against the height of the complete link. The average fidelity of teleporting an unknown coherent state, using the entangled states that result from the fading process, in the slow-turbulence and fast-detection regime (c), and in the fast-turbulence and slow-detection regime (d), against the height of the complete link.}
\label{fig6_6}
\end{figure}

These comparisons are illustrated in Figs.~\ref{fig6_6}~(a) and (b). On the contrary, the results for the teleportation fidelity are highly improved with an intermediate station, and extend also to the case of imperfect detectors, as can be seen in Figs.~\ref{fig6_6}~(c) and (d). Although, for imperfect detectors, fidelities with daytime thermal noise can go below the maximum classical fidelity. 

In Fig.~\ref{fig6_6}, we present the different negativities and fidelities, for fast and slow turbulence regimes. We compare the case of a downlink, an uplink, and the combination required by an intermediate station, against the height of the link. We can observe in Fig.~\ref{fig6_6}~(a), (b) that the negativity, in the case of an intermediate station is larger that a single dowlink/uplink, but only in the ideal case; when we consider inefficient detectors ($\tau_{\text{eff}}=0.4$), this gain is not so clear. As we increase the height of the link, this gain is not significant with respect to the downlink, although it remains relevant against the uplink. In Fig.~\ref{fig6_6}~(c), (d) the fidelity of teleporting an unknown coherent state is much better with an intermediate station, with respect to either a downlink or an uplink. Specially, we can highlight its partial saturation at the maximum classical fidelity value. 

\begin{figure}[h!]
\centering
{\includegraphics[width=0.75 \textwidth]{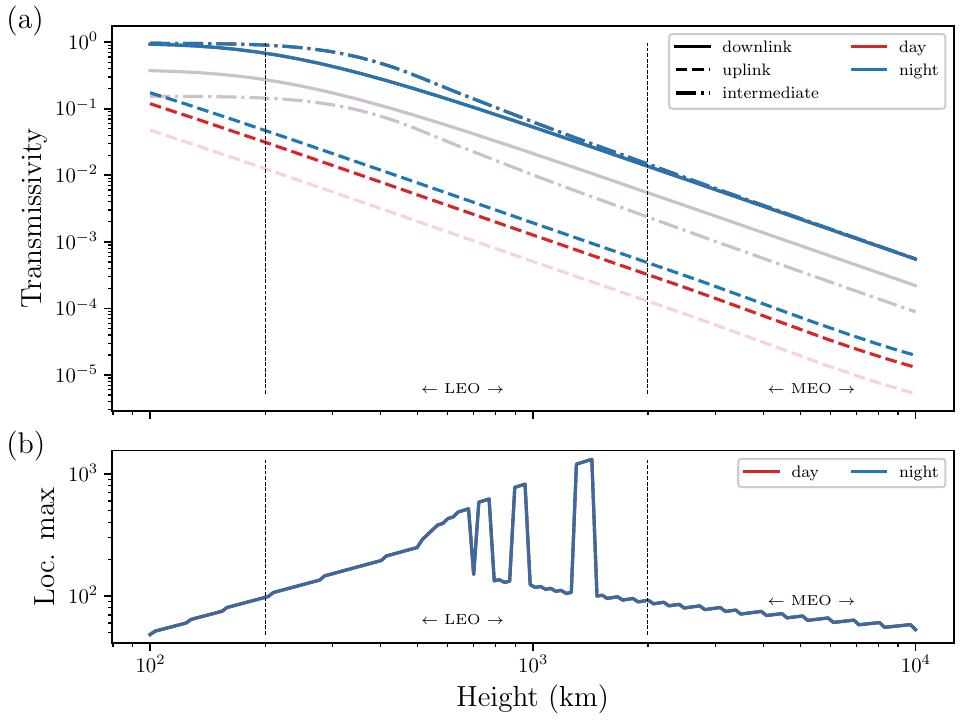}}
\caption[Transmissivity comparison between ground-to-satellite uplink and downlink and intermediate station quantum communication]{\textbf{Transmissivity comparison between ground-to-satellite uplink and downlink and intermediate station quantum communication} (a) Average transmissivity of a fading channel modelling a loss mechanism present in a link connecting a ground station and a satellite, and composed of diffraction, atmospheric extinction, detector inefficiency and free-space turbulence. In a solid line, we plot the quantities associated to a downlink; those corresponding to an uplink appear dashed, and the dashed-dotted lines describe the case of an intermediate station. In full color we represent the results for perfect detector efficiency, $\tau_{\text{eff}}=1$, whereas imperfect detection, $\tau_{\text{eff}}=0.4$, is marked by the high-transparency curves. In blue, we represent the result associated with nighttime thermal noise, whereas those associated to daytime thermal noise appear in red. (b) Optimal positions of the intermediate station, in order to maximize the transmissivity, against the height of the link.}
\label{fig6_7}
\end{figure}

We observe that the transmissivity in the case of the intermediate station is only higher than that of the downlink in the ideal case; when we have imperfect detectors, since there are now two detection events, the transmissivity is always worse. This can be observed in Fig.~\ref{fig6_7}, where we represent the transmissivity induced by downlink, uplink, and intermediate-station scenarios. We consider nighttime and daytime noise, again seeing that transmissivities associated to downlink and intermediate-station communication coincide. The same thing happens in Fig.~\ref{fig6_6}. Although we can see this behavior in the negativity plots, the fidelity behaves different. Of course, it would be natural to assume that we obtain good results for the fidelity because we are optimizing the placement of the intermediate station and keeping the highest fidelity at each altitude. And rightly so, but the improvement difference in the negativity and the fidelity is due to the fact that the states generated at the intermediate station and distributed through a downlink to Earth and through an uplink to a satellite are more symmetric. On the other hand, in the case of a single downlink or uplink, one of the modes was kept and the other was sent through free space, resulting in a covariance matrix that was highly asymmetric (see Eq.~\eqref{sec6_CM_asym}). Given two Gaussian quantum states with the same negativity, the one whose covariance matrix is more symmetric shows higher teleportation fidelity. 

%This is because the state is more symmetric in the case of the intermediate station, and this improves the teleportation fidelity, such that for a perfectly symmetric state, the teleportation fidelity if $\overline{F}=1/(1+\tilde{\nu}_{-})$, given a partially-transposed symplectic eigenvalue $\tilde{\nu}_{-}$ that completely characterizes the negativity. 

%%%%%%%%%%%%%%%%%%%%%%%%%%
%%%%%%%%%%%%%%%%%%%%%%%%%%

Take the covariance matrix in Eq.~\eqref{sec6_CM}, and assume it represents an asymmetric state. Here, we are referring to symmetry in the second moments of modes A and B, and not in the sense that the covariance matrix is symmetric. The partially-transposed symplectic eigenvalue of this asymmetric covariance matrix is 
\begin{equation}
\tilde{\nu}_{-}^{\text{A}} = \frac{\alpha+\beta-\sqrt{(\alpha-\beta)^{2}+4\gamma^{2}}}{2},
\end{equation}
and the associated teleportation fidelity is
\begin{equation}
\overline{F}_{\text{A}} = \frac{1}{1+(\alpha+\beta-2\gamma)/2}.
\end{equation}
For a symmetric Gaussian state with covariance matrix 
\begin{equation}
\Sigma_{\text{S}} = \begin{pmatrix} \delta\mathbb{1}_{2} & \varepsilon \sigma_{z} \\ \varepsilon \sigma_{z} & \delta\mathbb{1}_{2} \end{pmatrix},
\end{equation}
we have
\begin{eqnarray}
\nonumber \tilde{\nu}_{-}^{\text{S}} &=& \delta-\varepsilon, \\
\overline{F}_{\text{S}} &=& \frac{1}{1+\delta-\varepsilon}.
\end{eqnarray}
If these two states have the same negativity, then
\begin{equation}
\frac{\alpha+\beta-\sqrt{(\alpha-\beta)^{2}+4\gamma^{2}}}{2} = \delta - \varepsilon,
\end{equation}
and we can write 
\begin{equation}
\overline{F}_{\text{S}} = \frac{1}{1+\frac{\alpha+\beta-\sqrt{(\alpha-\beta)^{2}+4\gamma^{2}}}{2}}.
\end{equation}
Claiming that the fidelity with the symmetric state is higher than that with the asymmetric state amounts to checking that
\begin{equation}
\alpha+\beta-\sqrt{(\alpha-\beta)^{2}+4\gamma^{2}} < \alpha+\beta-2\gamma,
\end{equation}
and this is always true for $\alpha\neq\beta$. This statement works for a perfectly symmetric state, but we can study an extension for more general covariance matrices. We take
\begin{equation}
\Sigma_{1} = \begin{pmatrix} \alpha_{1}\mathbb{1}_{2} & \gamma_{1} \sigma_{z} \\ \gamma_{1} \sigma_{z} & \beta_{1}\mathbb{1}_{2} \end{pmatrix}, \qquad \Sigma_{2} = \begin{pmatrix} \alpha_{2}\mathbb{1}_{2} & \gamma_{2} \sigma_{z} \\ \gamma_{2} \sigma_{z} & \beta_{2}\mathbb{1}_{2} \end{pmatrix},
\end{equation}
assuming $\alpha_{i}>\beta_{i}$ ($i\in\{1,2\}$) for convenience, and expand up to first order in $\alpha_{i}-\beta_{i}\ll 1$. We can say that, if the states represented by these two covariance matrices have the same negativity, then state 1 shows higher teleportation fidelity for an unknown coherent state if
\begin{equation}
\alpha_{1}-\beta_{1} < \sqrt{\frac{\gamma_{1}}{\gamma_{2}}}(\alpha_{2}-\beta_{2}).
\end{equation}
This also works the other way around; for two states with the same teleportation fidelity, state 1 shows lower entanglement if its covariance matrix elements satisfy the above condition. Furthermore, we could fix $\gamma_{1}=\gamma_{2}=\gamma$, and see that for higher orders of the expansion $\alpha_{i}-\beta_{i}\ll 1$, we obtain that $\alpha_{1}-\beta_{1} < \alpha_{2}-\beta_{2}$ if 
\begin{equation}
\sqrt{(\alpha_{1}-\beta_{1})^{2} +  (\alpha_{2}-\beta_{2})^{2}} < 4\gamma 
\end{equation}
is satisfied. Therefore, we have shown that for two states with the same negativity, the one that is more symmetric will result in higher teleportation fidelity. 
%%%%%%%%%%%%%%%%%%%%%%%%%%
%%%%%%%%%%%%%%%%%%%%%%%%%%

\begin{figure}[t]
\centering
{\includegraphics[width=\textwidth]{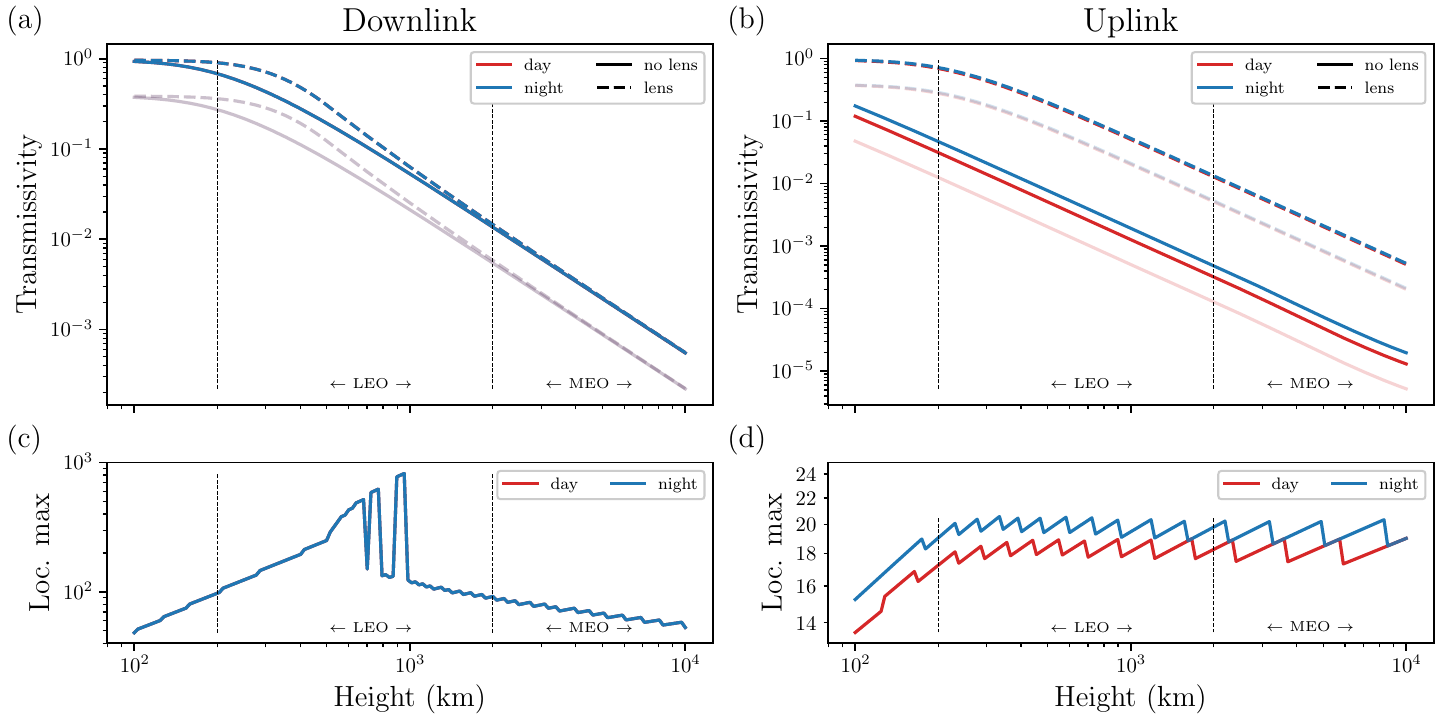}}
\caption[Average transmissivity of a turbulent fading channel connecting a ground station and a satellite]{\textbf{Average transmissivity of a turbulent fading channel connecting a ground station and a satellite}, composed of diffraction, atmospheric extinction, detector inefficiency and free-space turbulence. In a solid line, we plot the quantities associated an unaltered channel, and in a dashed line, we represent the cases in which a lens has been placed in a mid point of the link to reduce beam broadening. In full color we represent the results for perfect detector efficiency, $\tau_{\text{eff}}=1$, whereas imperfect detection, $\tau_{\text{eff}}=0.4$, is marked by the transparent curves. In blue, we represent the result associated with nighttime thermal noise, whereas those associated to daytime thermal noise appear in red. We represent the results associated to a downlink in (a), with the optimal location of the lens, in order to maximize the transmissivity, given in (c). The results for nighttime thermal noise fall on top of those for daytime thermal noise. The results associated to an uplink are represented in (b), with optimal positions of the lens shown in (d).}
\label{fig6_8}
\end{figure}

\subsubsubsection{Intermediate station for beam focusing}
Here, we consider using the intermediate station as a point where the signal is refocused, in an attempt to reduce the effects of diffraction and turbulence. This could improve the transmissivity of the downlink, but especially that of the uplink, where the turbulence effects are more damaging.  This is what we observe in Fig.~\ref{fig6_8}, where we represent the transmissivity of the fading channel describing the propagation through the link, against the total height. In Fig.~\ref{fig6_8}~(a) the transmissivity for a downlink is improved in the ideal case, similarly to how it was improved by generating the states in the intermediate station; in this case, however, we consider that the sender generates both modes, and thus only have one detector at the receiver. Furthermore, notice that in Fig.~\ref{fig6_8}~(c) the optimal location of the intermediate lens is very similar to the optimal position of the intermediate station in Fig.~\ref{fig6_7}~(b). This emphasizes the statement that an intermediate station and an intermediate lens contribute about equally to improving the transmissivity of the channel, considering a downlink. However, when we see the case of an uplink in Fig.~\ref{fig6_8}~(b), we notice that it is improved greatly, achieving values above the transmissivity of the downlink. This is because the optimal locations of the focusing lens, represented in Fig.~\ref{fig6_8}~(d), all fall in the tens of kilometres, very close to the ground station, in order to reduce the effects of turbulence inside the atmosphere. As a last remark, see that the results the results for daytime and nighttime thermal noise coincide for certain ranges, in Fig.~\ref{fig6_8}, both in downlink and uplink scenarios. 
%%%%%%%%%%%%%%%%%%%%%%%%%%
%%%%%%%%%%%%%%%%%%%%%%%%%%

\subsubsection{Microwave slant links}
We aim at expanding the results shown in this chapter by considering the attenuation of microwave quantum signals in free-space propagation. The major difference with the model for signals in the optical regime will be the omission of turbulence effects. Given the wavelengths for microwaves, on the order of centimetres, we can see that they will not be affected by the fluctuations that lead to turbulence for optical signals. Nevertheless, also because of the long wavelengths, microwaves will be highly affected by diffraction. By proposing a loss mechanism composed of diffraction, atmospheric attenuation and detector inefficiency, we aim at investigating the limits for entanglement distribution and quantum teleportation with microwaves in free space. With the diffraction-induced transmissivity given in Eq.~\eqref{sec6_tau_diffraction}, and assuming ideal detector efficiency $\tau_{\text{eff}}=1$, we describe the absorption-induced transmissivity along a slant path of zenit angle $\theta$, starting at altitude $h_{0}$ and ending at $h$ by
\begin{equation}
\tau_{\text{atm}} = \exp\left[ -\text{sec}\theta\int_{h_{0}}^{h}\diff h'\mu(h')\right].
\end{equation}
the atmospheric absorption coefficient $\mu(h')=\mu_{o}+\mu_{w}(h')$ represents the combined attenuation due to oxygen and water vapor. The former can be considered constant inside the atmosphere, but the latter will depend on the variation of the water concentration with the altitude. The specific coefficients are~\cite{Ho2004}
\begin{eqnarray}
\nonumber \mu_{o} &=& 1.44\times10^{-3} \text{ km}^{-1}, \\
\mu_{w}(h') &=& 4.44\times10^{-5} p_{0}e^{-\frac{h'}{2}} \text{ km}^{-1}, 
\end{eqnarray}
where $p_{0}$ is the water-vapor density, whose average ground value is $7.5 \text{ g/m}^{3}$, at 5 GHz. These frequencies present one of the lowest attenuation profiles among microwaves~\cite{ITU-R}, and therefore make them suitable for telecommunications independent of the weather conditions. However, the main sources of loss for microwave signals are diffraction and the thermal background. 

Due to the bright thermal background that microwave present at room temperatures, these states are generated at cryogenic temperatures; nevertheless, we consider that the squeezing operations are applied to a thermal state, and not to an ideal vacuum state, which leads to the more realistic TMST state. Our choice of entangled resource describes a TMST state, characterized by $n = 10^{-2}$ average number of thermal photons per mode, and squeezing parameter $r=1$. 

In order for these states to remain entangled when distributed through free space, we need the transmissivity of the channel to satisfy
\begin{equation}
\tau > \frac{(m-1)(c-1)}{(m-c)(c-1)+s^{2}},
\end{equation}
assuming that $m>c$, for a state represented by the covariance matrix in Eq.~\eqref{sec6_CM_asym} with $\alpha=\beta=c$ and $\gamma=s$. If the state is symmetric, and its covariance matrix resembles that in Eq.~\eqref{sec6_CM_sym}, with $\tau_{\text{d}}\approx\tau_{\text{u}}\equiv\tau$, this condition turns to
\begin{equation}
\tau > \frac{m-1}{m-c+s}.
\end{equation}
Considering identical initial resources (see Eq.~\eqref{sec6_TMSV}), this condition is always more restrictive for symmetric ($\tau>0.9997$) than for asymmetric states ($\tau>0.9992$, given the states studied here).

We can reduce the effects of thermal noise if we assume that we know the time of arrival of the signal, and therefore by using Eq.~\eqref{sec6_thermal_background}. In Ref.~\cite{Pirandola2021_3}, the limits for short-range microwave QKD were studied, using as parameters $\Delta\, t \, \Delta \, \nu\simeq 1$. This lead to $G_{R}\simeq\lambda^{2}\Omega_{\text{fov}}a_{R}^{2}/c$ and, by taking $\lambda=6$ cm, $\Omega_{\text{fov}}=10^{-4}$ sr and $a_{R}=2$ m, the number effective number of thermal photons becomes $n\simeq 266$ at 288 K. 

%%%%%%%%%%%%%%%%%%%%%%%%%%
\begin{figure}[h!]
\centering
{\includegraphics[width=0.75 \textwidth]{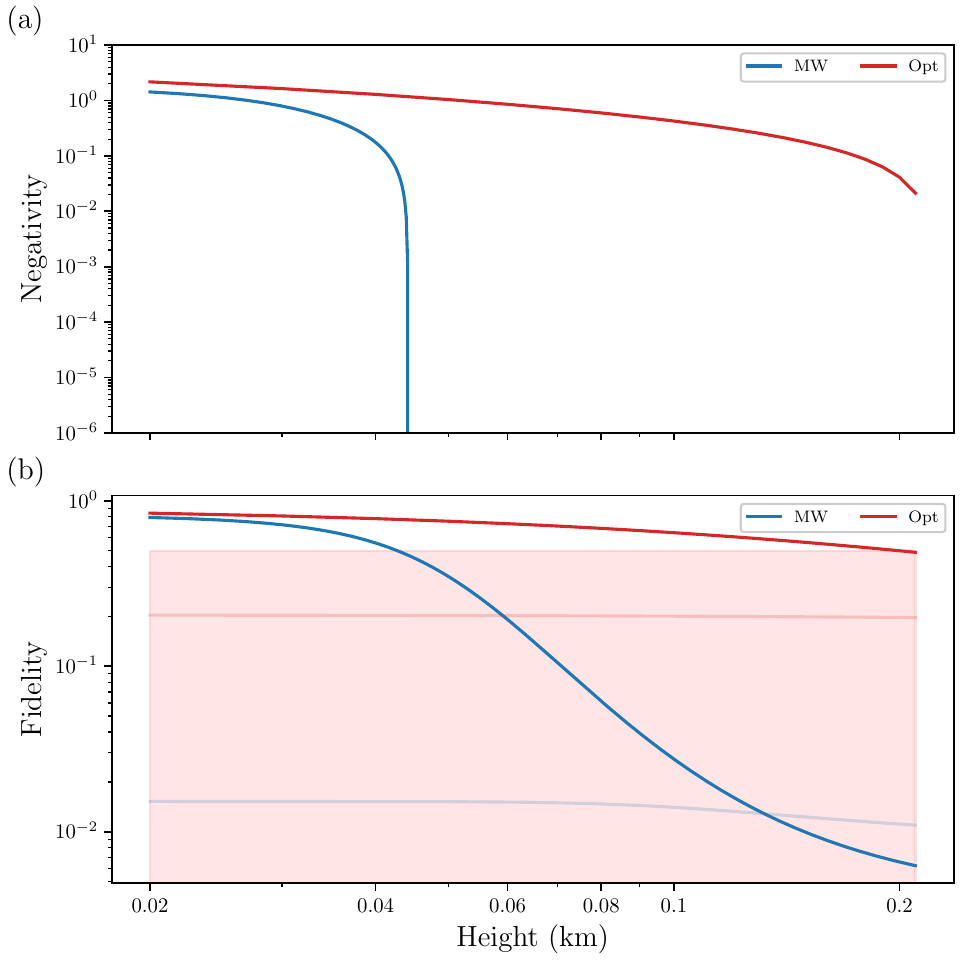}}
\caption[Performance comparison between microwave and optical signals in free space quantum communications under severe weather conditions]{\textbf{Performance comparison between microwave and optical signals in free space quantum communications under severe weather conditions}. We represent the negativity (a) and the quantum teleportation fidelity (b) for TMSV states generated at a ground station at an altitude of 10 m, and where one of the modes is sent through an uplink. We consider the signal has squeezing parameter $r=1$, and initial waist $\varpi_{0}=1$ m, assuming the receiver has an antenna of radius $a_{R}=2$ m. In red, we represent the results associated to a signal in the optical regime, with wavelength $\lambda=800$ nm and zero thermal photons, whereas in blue we represent the results for microwave signals with wavelength $\lambda=6$ cm and $n=10^{-2}$ thermal photons. For optical signals, the thermal noise coming from the environment is characterized by $13.57$ photons, whereas for microwave signals, we have $266$ photons. In full color, we present the results for perfect detector efficiency, $\tau_{\text{eff}}=1$, whereas the transparent curves correspond to $\tau_{\text{eff}}=0.4$. The pale red background represents the region in which the teleportation fidelity falls below the maximum classical value.}
\label{fig6_9}
\end{figure}
%%%%%%%%%%%%%%%%%%%%%%%%%%

With this, the condition for entanglement preservation on asymmetric states becomes $\tau>0.996$. Then, we see that the entanglement-distribution limit is 44 m, while the fidelity reaches the classical limit at 43 m. In this case, the asymmetry between both modes of the state distributed through free space does not lead to a significant difference between entanglement preservation and quantum teleportation distances. For symmetric states, the condition for symmetric states is $\tau>0.998$. Considering an intermediate station for state generation, the entanglement-distribution and quantum teleportation limit extends to 49 m. On the other hand, an intermediate station for beam refocusing leads to a limit for entanglement preservation at 52 m, whereas the teleportation fidelity reaches the classical limit at 49 m. 

As we can observe, microwave quantum communication is highly limited by diffraction and thermal noise. However, inside the atmosphere, the attenuation suffered by microwaves in severe weather conditions is inferior to that suffered by optical signals. Let us look at an example, and compare the performance of signals in both regimes. To account for the effects of rain on atmospheric attenuation and visibility, we set $\mu_{0} = 3.4\times 10^{-4} \text{ m}^{-1}$~\cite{Kaushal2017} and, in the Hufnagel-Valley turbulence model, we now write $A_{\text{day(night)}}=3.15 (2.15)\times10^{-14} \text{ m}^{-2/3}$~\cite{Liorni2019}. This exemplifies adverse meteorological conditions for optical signals, which is a convenient scenario for a comparison between microwave and optical. In the microwave regime, we need to set the water-vapor density to $p_{0} = 12 \text{ g/m}^{3}$~\cite{Ho2004}.

We observe that, when the link starts on the ground, microwaves can only do as well as optical for a short distance, and then they worsen. This can be observed in Fig.~\ref{fig6_9}, where we represent the negativity (a) and the teleportation fidelity (b) associated to a TMSV state distributed through free space. The effects of diffraction remain severe on microwave signals. These results show that microwave quantum communication can be appropriate for inter-satellite quantum communications. There, the conditions for entanglement preservation become $\tau>0.706$ for asymmetric states, and $\tau>0.847$ for symmetric ones, with an effective number of thermal photons $n = 2.39$. 

\subsubsection{Horizontal paths}
For the sake of completeness, we investigate the effects that free-space propagation through turbulent media inside the atmopshere has on the negativity of TMSV states, and how it affects the fidelity of a quantum teleportation protocol that uses these states as resources, in order to teleport an unknown coherent state. We consider a scenario in which TMSV states are distributed between two ground stations, at an altitude of $h=30$ m, each station having an receiving antenna with $a_{R}=5$ cm of aperture radius, and able to generate quasi-monochromatic beams with wavelength $\lambda= 800$ nm and $\varpi_{0}=5$ cm of initial waist. 
%%%%%%%%%%%%%%%%%%%%%%%%%%
\begin{figure}[h!]
\centering
{\includegraphics[width=\textwidth]{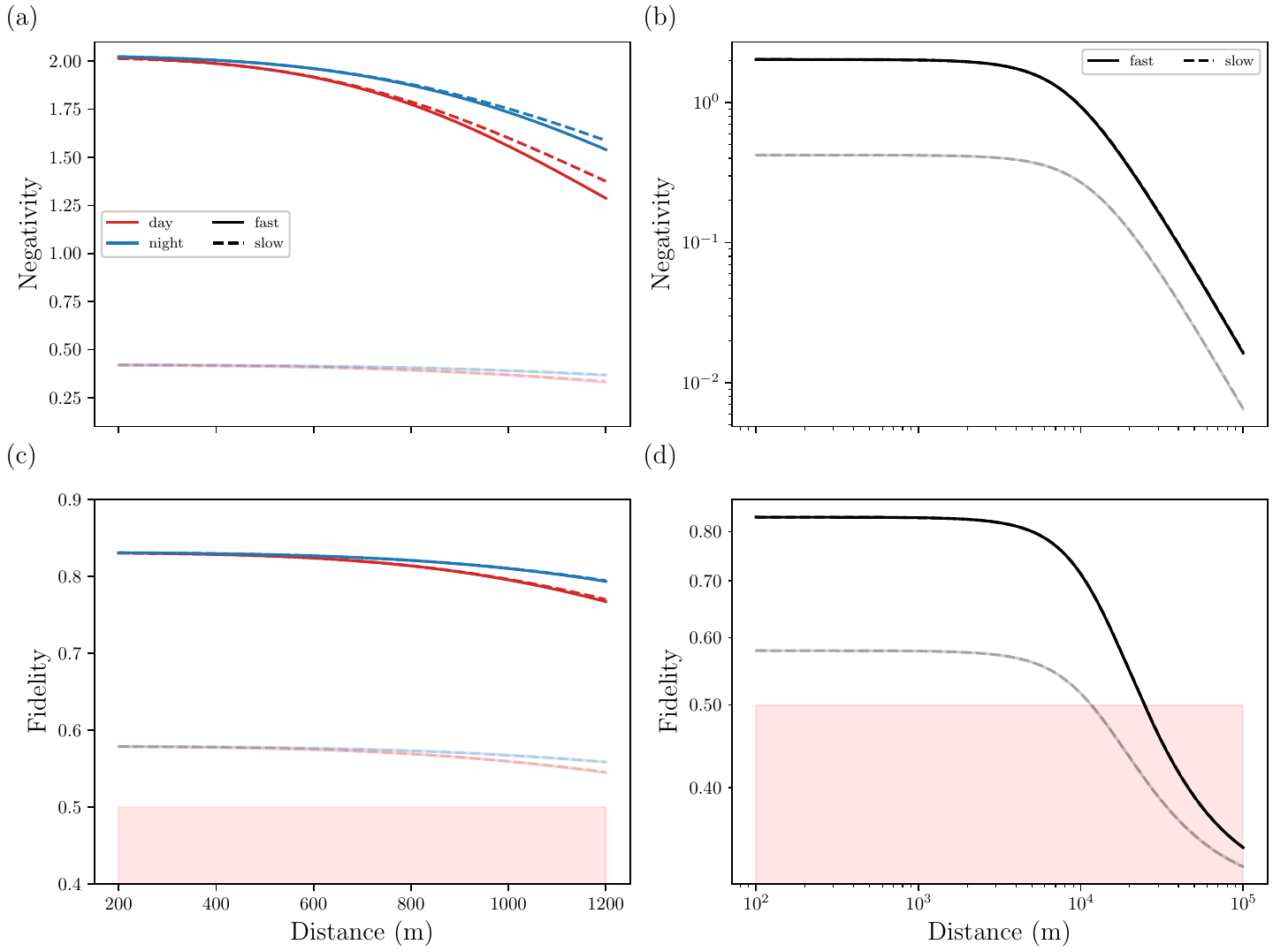}}
\caption[Quantum communication through horizontal paths]{\textbf{Quantum communication through horizontal paths} with TMSV states distributed through free space. Ground-to-ground station quantum communication is studied through the negativity~(a) and the quantum teleportation fidelity~(c), for TMSV states subject to a loss mechanism composed of diffraction, atmospheric extinction, detector inefficiency and free-space turbulence. Inter-satellite quantum communication is studied through the negativity~(b) and the quantum teleportation fidelity~(d), where now the only loss mechanisms relevant are diffraction, pointing errors, and detector inefficiency. We consider the signal has wavelength $\lambda=800$ nm, squeezing parameter $r=1$, and initial waist $\varpi_{0}=5$ cm, assuming the receiver has an antenna of radius $a_{R}=5$ cm. In red, we represent the results that incorporate daytime thermal noise, whereas the blue lines consider nighttime thermal noise. The dashed lines correspond to the instance of slow turbulence and fast detection, and the solid lines correspond to fast turbulence and slow detection. In full color, we present the results for perfect detector efficiency, $\tau_{\text{eff}}=1$, whereas the transparent curves correspond to $\tau_{\text{eff}}=0.4$.}
\label{fig6_10}
\end{figure}
%%%%%%%%%%%%%%%%%%%%%%%%%%
In this situation, since the altitude is fixed, and for a wind speed of $v=21 \text{ m}/\text{s}$, the refraction index structure constant is $C_{n}^{2}=2.06(1.29)\times10^{-14} \text{ m}^{-2/3}$ for daytime (nighttime) values. We characterize the excess noise in the detectors by $n_{\text{day}}=4.75\times10^{-3}$ thermal photons for daytime events, and $n_{\text{night}}=4.75\times10^{-8}$ thermal photons for nighttime events.

We represent the results of entanglement distribution and quantum teleportation with TMSV states between two ground stations in Fig.~\ref{fig6_10}. Daytime (nighttime) results are shown in red (blue), and the solid (dashed) curves correspond to fast (slow) turbulence. The high-transparency curves show the results for inefficient detectors, with $\tau_{\text{eff}}=0.4$, whereas the curves in full color correspond to ideal detection, with $\tau_{\text{eff}}=1$. In Fig.~\ref{fig6_10}~(a), we show the negativity of the TMSV state, with squeezing parameter $r=1$, against the traveled distance. We show the average fidelity of quantum teleportation using these states, distributed through free space, in Fig.~\ref{fig6_10}~(b). We observe that, even in the low detector-efficiency case, entanglement is preserved, and therefore quantum teleportation fidelity is still higher than the maximum classical fidelity achievable, marked in a pale red background in Fig.~\ref{fig6_10}~(b).

The range of distances chosen to represent these quantities corresponds to the ``sweet spot'' $200\leq z\leq1066$, were the weak-turbulence expansion used here is approximately correct~\cite{Pirandola2021_2}. 

In Fig.~\ref{fig6_10}~(b) and~(d), we represent the negativity~(b) and the quantum teleportation fidelity~(d) for the same TMSV states, between two satellites in the same orbit. In this scenario, the only relevant sources of noise are diffraction, pointing errors, and detector inefficiency. Also, we are considering that the excess noise in the detectors is characterized by $n=8.48\times10^{-9}$ thermal photons. Here, the solid lines are associated to fast turbulence and slow detection, whereas the dashed lines describe slow-turbulence and fast-detection results. Notice that these appear overlapped. 

This approach is quite different from the one we took in the previous section, where we looked at entanglement preservation distances depending on the size of the antenna.

%%%%% Conclusions

In the microwave regime, we have studied the requirements for entanglement preservation in the size of the antennae involved, for quantum communication between satellites in the same orbit. As expected, due to the larger wavelengths of signals in this frequency regime, antennae must reach a few meters in diameter. With state-of-the-art experimental parameters, entanglement cannot be preserved in the far-field regime. 

%We have then moved on to the optical regime, where we have studied the effects of diffraction, atmospheric attenuation, detector inefficiency and turbulence on quantum signals propagating through free space, between a ground station and a satellite. More precisely, we have investigated the effects of these loss mechanisms combined, and described as a fading quantum channel, acting on two-mode squeezed vacuum states, which are a paradigmatic example of entangled Gaussian quantum states. We have observed the degradation of entanglement through the negativity of the state, and looked at the fidelity of performing quantum teleportation with the remaining entangled resource, both after downlink and uplink communications, and for satellites in different orbits. We concluded that the best case occurs when we use a downlink, i.e. when the bipartite states are generated in the satellite and one of the modes is sent down to the ground station. The uplink represents the worst case because turbulence effects, which are more drastic inside the atmosphere, distort the waist of the beam and displace the focusing point; when considering the whole path, these errors have a higher impact on a beam that is starting its path. 

We have then moved on to the optical regime, where we have studied the effects of diffraction, atmospheric attenuation, detector inefficiency and turbulence on quantum signals propagating through free space, between a ground station and a satellite. We have observed the degradation of entanglement in TMSV states that propagate through free space, and looked at the fidelity of performing quantum teleportation with the remaining entangled resource, both after downlink and uplink communications, and for satellites in different orbits. We concluded that the best case occurs when we use a downlink, i.e. when the bipartite states are generated in the satellite and one of the modes is sent down to the ground station. The uplink represents the worst case because turbulence effects, which are more drastic inside the atmosphere, distort the waist of the beam and displace the focusing point; when considering the whole path, these errors have a higher impact on a beam that is starting its path. 

We have also considered the introduction of an intermediate station; we first investigated a scenario in which the states were generated there, and one mode was then sent to the ground station through a downlink, while the other was sent to the satellite through an uplink. Considering that now the uplink does not start inside the atmosphere, the results for the negativity were slightly better than those for the downlink in the simple case, provided an optimal placement of the intermediate station. Furthermore, the results for the fidelity were highly improved because the generation of states in an intermediate station leads to states that are almost symmetric. As we discussed, for two Gaussian states with the same entanglement, the one that presents a more symmetric covariance matrix will have a higher teleportation fidelity, in the well-known Braunstein-Kimble quantum teleportation protocol. The second intermediate-station scenario we considered was one were the beam could be refocused, but in a simple downlink or uplink. The uplink showed a higher improvement than the downlink, because a refocusing station can help mitigate the combined effects of diffraction and turbulence, which as we discussed earlier, are more severe on more ideal beams.  

We have followed by studying a similar free-space loss mechanism for microwave signals, which are largely affected by diffraction and thermal noise. Although atmospheric absorption and turbulence effects can be neglected, the distances for entanglement distribution and effective quantum teleportation are highly reduced with respect to the optical case. In a bad weather scenario, we observed that microwave and optical signals yielded a similar performance for short distances, microwaves then leading to worse results as we separated from the source, mainly due to diffraction.

We have concluded by showing the limits of entanglement distribution and quantum teleportation through horizontal paths, in ground-to-ground scenarios, where turbulence effects are present, and inter-satellite quantum communication, where we have mostly diffraction and pointing errors. Between satellites, the loss mechanism is reduced to diffraction and beam wandering, and therefore entanglement and quantum teleportation fidelity can be preserved for longer distances than in horizontal paths between ground stations, where atmospheric absorption and turbulence come into play.

%We believe these results are relevant in the development of quantum communications in free space, establishing limits based on experimental parameters to the realizability of quantum information transfer between Earth and satellites. Provided that classical secure channels are well-established, the installation of adequate quantum channels represents the next step towards developing quantum communication networks, that can lead to advances such as the quantum internet.

%%%%%%%%%%%%%%%%%%%%%%%%
% REDEFINE TITLE FORMAT
%%%%%%%%%%%%%%%%%%%%%%%%

% \titleformat{\section}[display]
% {\vspace*{150pt}
% \bf\Huge}
% {\hspace{-72pt}{\textcolor{grey}{\thesection}} \hspace{38pt} {\textcolor{white}{#1}}}
% {0pt}
% {}
% []
% \titlespacing*{\section}{108pt}{10pt}{60pt}[20pt]
%
% %Intro
% \titleformat{\section}[display]
% {\vspace*{190pt}
% \bfseries\sffamily \huge}
% {\begin{picture}(0,0)\put(-50,-25){\textcolor{grey}{\thesection}}\end{picture}}
% {0pt}
% {\textcolor{white}{#1}}
% []
% \titlespacing*{\section}{80pt}{10pt}{50pt}[20pt]
%
% %ch 2
% \titleformat{\section}[display]
% {\vspace*{190pt}
% \bfseries\sffamily \huge}
% {\begin{picture}(0,0)\put(-50,-28){\textcolor{grey}{\thesection}}\end{picture}}
% {0pt}
% {\textcolor{white}{#1}}
% []
% \titlespacing*{\section}{80pt}{10pt}{50pt}[20pt]
%
% %ch 3, 4 y 5
\titleformat{\section}[display]
{\vspace*{190pt}
\bfseries\sffamily \LARGE}
{\begin{picture}(0,0)\put(-50,-35){\textcolor{grey}{\thesection}}\end{picture}}
{0pt}
{\textcolor{white}{#1}}
[]
\titlespacing*{\section}{80pt}{10pt}{40pt}[20pt]

% %Conclusions
% %Julen conclusions
% \titleformat{\section}[display]
% {\vspace*{190pt}
% \bfseries\sffamily \huge}
% {\begin{picture}(0,0)\put(-50,-25){\textcolor{grey}{\thesection}}\end{picture}}
% {0pt}
% {\textcolor{white}{#1}}
% []
% \titlespacing*{\section}{80pt}{10pt}{50pt}[20pt]

%\input{snp/fancySection2.tex}

%%%%%%%%%%%%%%%%%%%%%%
% CHAPTER 7
%%%%%%%%%%%%%%%%%%%%%%

\section[Microwave Quantum Local Area Networks]{Microwave Quantum Local \\ Area Networks }
\label{sec7}

% \thiswatermark{\put(1,-282){\color{l-grey}\rule{84pt}{88pt}}
% \put(84,-282){\color{grey}\rule{410pt}{88pt}}}
%
%
% %Julen ch 2
% \thiswatermark{\put(1,-308){\color{l-grey}\rule{70pt}{64pt}}
% \put(70,-308){\color{grey}\rule{297pt}{64pt}}}
% \fancyhead[RO]{\leftmark}

%Julen ch 3
\thiswatermark{\put(1,-302){\color{l-grey}\rule{70pt}{60pt}}
\put(70,-302){\color{grey}\rule{297pt}{60pt}}}

% %Julen ch 4
% \thiswatermark{\put(1,-302){\color{l-grey}\rule{70pt}{60pt}}
% \put(70,-302){\color{grey}\rule{297pt}{60pt}}}
% \fancyhead[RO]{\small \leftmark}
%
%
% %Julen ch 5
% \thiswatermark{\put(1,-302){\color{l-grey}\rule{70pt}{60pt}}
% \put(70,-302){\color{grey}\rule{297pt}{60pt}}}
% \fancyhead[RO]{\leftmark}

% MICROWAVE QLAN FOR DISTRIBUTED QUANTUM COMPUTING

\lettrine[lines=2, findent=3pt,nindent=0pt]{T}{he} paradigm of distributed computing attempts to distribute a processing task among multiple processing units. These units, which perform different parts of the computation, can be close together, forming a local area network, or physically distant, and connected through a wide area network. Therefore, scalability is not an issue, and hence redundancy can be considered less parasitic. In fact, the latter is beneficial to prevent the system from failing completely when one of the units does. Applications of distributed computing include telecommunication networks, the World Wide Web, or cloud computing for scientific purposes, among others. The latter refers to the coordinated strategy of dividing a problem in different tasks, which are solved in different computers, communicated with each other. 

This logic can be applied to the design of current quantum computers, which suffer from scalability problems, including connectivity issues, fabrication errors, or lack of controllability, among others. The term NISQ (Noisy intermediate-scale quantum)~\cite{Preskill2018} is an adjective describing the current quantum computing landscape, far away from quantum error correction: small quantum processors with noisy qubits deprived of fault tolerance. However, there exist quantum algorithms specifically design for NISQ devices~\cite{Bharti2022}, such as the variational quantum eigensolver and the quantum approximate optimization algorithm. 

Therefore, quantum computing in the NISQ era can benefit from a distributed configuration. By disseminating the workload between different medium-size quantum processors~\cite{Beals2013,DiAdamo2021}, distributed quantum computing~\cite{Cuomo2020,Gyongyosi2021} can attempt to reduce the scalability overhead. In order to achieve it, it is necessary an efficient transmission of quantum information between the different processing units. If we have a qubit, a straightforward technique to communicate the information it holds to another unit is to use discrete-variable (DV) quantum teleportation of its quantum state. Experiments in microwave quantum teleportation with DVs have not surpassed the 90 \% fidelity~\cite{Baur2012,Steffen2013}, nor have remote entanglement generation ones~\cite{Kurpiers2018,Magnard2020}. The problem is that DV states are very sensitive to losses. Therefore, we are interested in exploring the use of CV states as the entangled resources, due to their higher resilience to photon losses when compared to DV entangled states. This has been considered in many different works, which used either TMSV states~\cite{Ide2001,Takeda2013,Takeda2014,Lie2019,He2022,He2022_2} or Schr\"{o}dinger cat states~\cite{Ulanov2017,Podoshvedov2019} as the entangled resources. The latter have also been used in quantum-repeater protocols~\cite{Brask2010,Lim2016}, as well as GKP states~\cite{Fukui2021}.

In this chapter, we study the fidelity of teleporting unknown qubit states employing different resources. In the Braunstein-Kimble quantum teleportation protocol, the fidelity does not depend on the displacement of the coherent state, only on its second moments. In this case, the fidelity will depend on the amplitudes of the qubit, and therefore we will particularly focus on the fidelity for teleporting an average qubit. Therefore, we want to find the best teleportation protocol for an average qubit, using a CV Gaussian state as the entangled resource. We compute the fidelity of the Braunstein-Kimble quantum teleportation protocol with a two-mode Gaussian quantum state, which involves homodyne detection, i.e. a projection into the maximally-entangled basis for CV states, between the initial state and a mode of the entangled resource. We attempt to improve the results by considering a 2PS resource. Then, we consider the DV quantum teleportation protocol, in which we also project onto a maximally-entangled basis, this time in the subspace of two qubits, what is normally known as a Bell measurement. The displacement applied on the remaining mode, characteristic of CV teleportation, is replaced by a single-qubit projection. 

We are interested in the extension of the single-qubit case to a multi-qubit setting, in which we use a two-mode entangled resource to teleport each qubit. This presents many difficulties; in the case that the state of the full system is separable, the fidelity will just be the product of the fidelities of teleporting each individual mode; otherwise if the state is entangled, the entanglement will be teleported with some loss, as our Gaussian resource will not present infinite entanglement. 

%Another relevant subject is the loss suffered by the entangled modes when they are distributed between different processing units, as well as the measurement imperfections. The latter comprise the effects of finite-gain homodyne detection and the inefficiency of photon-number-resolving measurements. 

\subsection{Single-Qubit Quantum Teleportation}
In this section, we study quantum teleportation of a single qubit state. We look at the teleportation fidelity, and take an average for qubit states uniformly distributed on the Bloch sphere, since we will restrict this analysis to pure states. We will investigate CV and hybrid approaches, and compare them with the DV case. We also investigate the distribution of entangled states between different processing units and compute the teleportation fidelities associated with the resulting resources. 

\subsubsection{CV quantum teleportation}
We consider a CV quantum teleportation protocol in which we aim at teleporting a single qubit state from one processor to another, using an entangled resource shared between both. We take an initial qubit state $|\psi\rangle = a|0\rangle+b|1\rangle$, with characteristic function 
\begin{equation}
\chi(\theta) = \left( 1 + \vec{\theta}^{\intercal}\vec{y} - \frac{|b|^{2}}{2}\vec{\theta}^{\intercal}\Omega^{\intercal}\mathbb{1}_{2}\Omega\vec{\theta}\right) \exp\left[ -\frac{1}{4}\vec{\theta}^{\intercal}\Omega^{\intercal}\mathbb{1}_{2}\Omega\vec{\theta}\right],
\end{equation}
where we have defined $\vec{y}^{\intercal} = \frac{1}{\sqrt{2}}\begin{pmatrix} a\bar{b} - \bar{a}b & i(a\bar{b} + \bar{a}b) \end{pmatrix}$. The entangled resource is a two-mode Gaussian state with null first moments and a covariance matrix given by
\begin{equation}
\Sigma = \begin{pmatrix} \Sigma_{A} & \varepsilon_{AB} \\ \varepsilon_{AB}^{\intercal} & \Sigma_{B} \end{pmatrix},
\end{equation}
this its characteristic function is
\begin{equation}
%\chi_{AB}(\alpha,\beta) = \exp\left[ -\frac{1}{4} \left( \vec{\alpha}^{\intercal}\Omega^{\intercal} \Sigma_{A}\Omega \vec{\alpha} +\vec{\alpha}^{\intercal}\Omega^{\intercal} \varepsilon_{AB}\Omega \vec{\beta} + \vec{\beta}^{\intercal} \Omega^{\intercal} \varepsilon_{AB}^{\intercal} \Omega \vec{\alpha} + \vec{\beta}^{\intercal} \Omega^{\intercal} \Sigma_{B}\Omega \vec{\beta} \right)\right].
\chi_{AB}(\alpha,\beta) = \exp\left[ -\frac{\vec{\alpha}^{\intercal}\Omega^{\intercal} \Sigma_{A}\Omega \vec{\alpha} +\vec{\alpha}^{\intercal}\Omega^{\intercal} \varepsilon_{AB}\Omega \vec{\beta} + \vec{\beta}^{\intercal} \Omega^{\intercal} \varepsilon_{AB}^{\intercal} \Omega \vec{\alpha} + \vec{\beta}^{\intercal} \Omega^{\intercal} \Sigma_{B}\Omega \vec{\beta} }{4} \right].
\end{equation}
The fidelity of teleporting this qubit state is
\begin{equation}
\overline{F}(a,b) = \frac{1}{\sqrt{\det X}}\left\{ 1 - |b|^{2}\tr X^{-1} - \vec{y}^{\intercal}X^{-1}\vec{y} + \frac{|b|^{4}}{4}3\left( \tr\ X^{-1}\right)^{2} - \frac{|b|^{4}}{\det X} \right\},
\end{equation}
with $X = \Omega^{\intercal}\left[ \mathbb{1}_{2}+\frac{1}{2}\left( \sigma_{z}\Sigma_{A}\sigma_{z} + \Sigma_{B} - \sigma_{z}\varepsilon_{AB} - \varepsilon_{AB}^{\intercal}\sigma_{z}\right)\right]\Omega$. 

%As usual, we will consider a two-mode squeezed vacuum state as the entangled resource. In Fig.~\ref{fig1} we represent this fidelity against the amplitude of the qubit state and the squeezing of the resource. 

Notice that this quantity depends on the amplitudes of the qubit; however, for $b=0$, we recover the well-known fidelity for teleporting a vacuum or a coherent state. Nevertheless, we can obtain a more general formula for the fidelity; assuming that we do not have information about the state we want to teleport, we average over all possible qubit amplitudes. For that, we have to draw values from a uniform distribution on the Bloch sphere, also known as a Haar distribution. A transformation from Cartesian to spherical coordinates leads to the identifications
\begin{eqnarray}
\nonumber x &=& r\sin\theta\cos\varphi, \\
y &=& r\sin\theta\sin\varphi, \\
\nonumber z &=& r\cos\theta,
\end{eqnarray}
with $r>0$, $\theta\in[0,\pi]$ and $\varphi\in[0,2\pi]$. Using this convention, the state of a qubit in the Bloch sphere can be expressed as
\begin{equation}
|\psi\rangle = \cos\frac{\theta}{2}|0\rangle + \sin\frac{\theta}{2}e^{i\varphi}|1\rangle,
\end{equation}
taking $r=1$, since we want to deal with pure states. Let us define $u$ as a random value drawn from an uniform distribution that produces values between 0 and 1; then, in order to obtain an uniform distribution of states in the Bloch sphere, we need to sample according to~\cite{Simon2015}
\begin{eqnarray}
\nonumber \theta &\longrightarrow& \arccos(1-2u), \\
\varphi &\longrightarrow& 2\pi u.
\end{eqnarray}
Since we have identified $a = \cos\frac{\theta}{2}$ and $b = \sin\frac{\theta}{2}e^{i\varphi}$, this is equivalent to replacing
\begin{eqnarray}
\nonumber a &\longrightarrow& \sqrt{1-u}, \\
b &\longrightarrow& \sqrt{u}e^{i\varphi}.
\end{eqnarray}
Therefore, we can replace the sampling by integrals over $u$ and $\varphi$, such that
\begin{eqnarray}
\nonumber &&\frac{1}{2\pi}\int_{0}^{2\pi} \diff\varphi \int_{0}^{1} \diff u \, |b|^{2} = \frac{1}{2}, \\
&& \frac{1}{2\pi}\int_{0}^{2\pi} \diff\varphi \int_{0}^{1} \diff u \, |b|^{4} = \frac{1}{3}, \\
\nonumber &&\frac{1}{2\pi}\int_{0}^{2\pi} \diff\varphi \int_{0}^{1} \diff u \, \vec{y}^{\intercal}X^{-1}\vec{y} = -\frac{1}{6}\tr X^{-1}.
\end{eqnarray}
Consequently, we obtain
\begin{equation}
\overline{F} = \frac{1}{\sqrt{\det X}}\left\{ 1 - \frac{1}{3}\left( \tr X^{-1} + \frac{1}{\det X}\right)+ \frac{1}{4}\left( \tr X^{-1}\right)^{2} \right\}.
\end{equation}

We now consider single-photon (heuristic) subtraction in each mode of the entangled resource, and compute the teleportation fidelity for the same qubit. The characteristic function of the PS Gaussian resource can be seen in Eq.~\eqref{2PS_heur_CF}. Using this resource to teleport a qubit, we find that
\begin{eqnarray}
\nonumber \overline{F}(a,b) &=& \frac{1}{E_{0}\sqrt{\det X}}\bigg\{ E_{0} + \tr\left( X^{-1}E_{1}\right) + 3\tr\left( X^{-1}E_{2}^{A}\right)\tr\left( X^{-1}E_{2}^{B}\right) \\
\nonumber &-& E_{0} |b|^{2} \tr X^{-1} - E_{0} \vec{y}^{\intercal} X^{-1}\vec{y} - |b|^{2} \left[ 3\tr X^{-1} \tr\left( X^{-1}E_{1}\right) - \frac{2}{\det X}\tr E_{1} \right] \\
\nonumber &+& E_{0} \frac{|b|^{4}}{4} \left[ 3\left( \tr X^{-1}\right)^{2} - \frac{4}{\det X}\right] - \tr\left( X^{-1}E_{1}\right) \vec{y}^{\intercal} X^{-1}\vec{y} - 2\vec{y}^{\intercal} W_{X,E_{1}}\vec{y} \\
\nonumber &+& \frac{|b|^{4}}{4} \left[ 15\left( \tr X^{-1}\right)^{2}\tr\left( X^{-1}E_{1}\right) -  \frac{12}{\det X} \left( \tr\left( X^{-1}E_{1}\right) + \tr X^{-1} \tr E_{1} \right)\right] \\
\nonumber &-& |b|^{2} \Big[ 15\tr X^{-1} \tr\left( X^{-1}E_{2}^{A}\right)\tr\left( X^{-1}E_{2}^{B}\right) - \frac{2}{\det X}\tr\left( \Omega^{\intercal}E_{2}^{A}\Omega E_{2}^{B}\right) \\
\nonumber &-& \frac{6}{\det X} \left( \tr\left( X^{-1}E_{2}^{B}\right)\tr E_{2}^{A} + \tr\left( X^{-1}E_{2}^{A}\right)\tr E_{2}^{B} + \tr X^{-1} \tr\left( \Omega^{\intercal}E_{2}^{A}\Omega E_{2}^{B}\right) \right) \Big] \\
\nonumber &-& \vec{y}^{\intercal} \Big[ 15 X^{-1} \tr\left( X^{-1}E_{2}^{A}\right)\tr\left( X^{-1}E_{2}^{B}\right) - X^{-1}\frac{6}{\det X} \tr\left( \Omega^{\intercal}E_{2}^{A}\Omega E_{2}^{B}\right) \\
\nonumber &-& \frac{6}{\det X} \tr\left( X^{-1}E_{2}^{A}\right) \Omega^{\intercal}E_{2}^{B}\Omega - \frac{6}{\det X} \tr\left( X^{-1}E_{2}^{B}\right) \Omega^{\intercal}E_{2}^{A}\Omega \Big] \vec{y} \\
\nonumber &+& \frac{|b|^{4}}{4} \Big[ 105 \left( \tr X^{-1} \right)^{2}\tr\left( X^{-1}E_{2}^{A}\right)\tr\left( X^{-1}E_{2}^{B}\right) \\
\nonumber &-& \frac{30}{\det X} \Big( 2\tr\left( X^{-1}E_{2}^{A}\right)\tr\left( X^{-1}E_{2}^{B}\right) + 2\tr X^{-1} \tr E_{2}^{A} \tr\left( X^{-1}E_{2}^{B}\right) \\
\nonumber &+& 2\tr X^{-1} \tr E_{2}^{B}\tr\left( X^{-1}E_{2}^{A}\right) + \left( \tr X^{-1}\right)^{2}\tr\left( \Omega^{\intercal}E_{2}^{A}\Omega E_{2}^{B}\right) \Big) \\
&+& \frac{24}{\left(\det X\right)^{2}} \left( \tr\left( \Omega^{\intercal}E_{2}^{A}\Omega E_{2}^{B}\right) + \tr E_{2}^{A}\tr E_{2}^{B} \right) \Big] \bigg\},
\end{eqnarray}
where we have used the definitions in Eq.~\eqref{PS_QT_fid_def}. 

%This formula is also valid for the quantum teleportation fidelity for a resource with photon addition; we just need to change the definitions of the matrices that describe the photon-added state. 

\subsubsection{Hybrid quantum teleportation}
We consider here the DV quantum teleportation protocol of a single qubit state, while using a CV Gaussian state, as done in Ref.~\cite{He2022}. This means that, instead of homodyne detection, we will project the input state and the first mode of the entangled resource into a Bell state. Then, instead of applying a displacement on the second mode of the entangled state, we will project it into a different state depending on the measurement, as shown below
\begin{eqnarray}
\nonumber |\Phi_{\pm}\rangle &=& \frac{|0,0\rangle \pm |1,1\rangle}{\sqrt{2}} \quad \longrightarrow \quad \sigma_{\Phi_{\pm}} = |0\rangle\langle 0| \pm |1\rangle\langle 1|, \\
|\Psi_{\pm}\rangle &=& \frac{|0,1\rangle \pm |1,0\rangle}{\sqrt{2}} \quad \longrightarrow \quad \sigma_{\Psi_{\pm}} = |0\rangle\langle 1| \pm |1\rangle\langle 0|.
\end{eqnarray}
Therefore, if we measure in the $|\Phi_{\pm}\rangle$ basis, the remaining state will be
\begin{equation}
\rho_{\text{out}}^{B} = \frac{1}{P_{\Phi_{\pm}}}\sigma_{\Phi_{\pm}} \langle\Phi_{\pm}|\rho_{\text{in}}\otimes\rho_{AB}|\Phi_{\pm}\rangle\sigma_{\Phi_{\pm}}^{\dagger},
\end{equation}
and we obtain the teleportation fidelity
\begin{eqnarray}
\nonumber \overline{F}_{\Phi}(a,b) &=& \frac{1}{P_{\Phi_{\pm}}\sqrt{\det\left( \mathbb{1}_{2}+\Sigma_{A}\right)\det X}}\bigg\{ \left( 1-|b|^{2}\tr\left(\mathbb{1}_{2}+\Sigma_{A}\right)^{-1}\right)\left( 1-\frac{|b|^{2}}{2}\tr X^{-1}\right) \\
\nonumber &+& \vec{y}^{\intercal}X^{-1}\Omega^{\intercal}\varepsilon_{AB}^{\intercal}(\mathbb{1}_{2}+\Sigma_{A})^{-1}\Omega \sigma_{z} \vec{y} - \frac{|b|^{4}}{2\det X}\tr\left( \varepsilon_{AB}^{\intercal} W_{\mathbb{1}_{2}+\Sigma_{A},\mathbb{1}_{2}}\varepsilon_{AB}\right) \\
&-& \frac{|b|^{2}}{2}\tr\left( X^{-1}\Omega^{\intercal}\varepsilon_{AB}^{\intercal} W_{\mathbb{1}_{2}+\Sigma_{A},\mathbb{1}_{2}}\varepsilon_{AB}\Omega\right)\left( 1-\frac{3|b|^{2}}{2}\tr X^{-1}\right) \bigg\},
\end{eqnarray}
where we have defined 
\begin{eqnarray}
\nonumber P_{\Phi_{\pm}}&=& \frac{1}{\sqrt{\det\left( \mathbb{1}_{2}+\Sigma_{A}\right)\det X}}\bigg\{ \frac{1}{2}(1-|b|^{2}\tr\left(\mathbb{1}_{2}+\Sigma_{A}\right)^{-1})\left(4 - \tr X^{-1}\right) \\
\nonumber &-& |b|^{2}\tr\left( X^{-1}\Omega^{\intercal}\varepsilon_{AB}^{\intercal} W_{\mathbb{1}_{2}+\Sigma_{A},\mathbb{1}_{2}}\varepsilon_{AB}\Omega\right)\left( 1-\frac{3}{4}\tr X^{-1}\right) \\
&-& \frac{|b|^{2}}{2\det X}\tr\left( \varepsilon_{AB}^{\intercal} W_{\mathbb{1}_{2}+\Sigma_{A},\mathbb{1}_{2}}\varepsilon_{AB}\right) \bigg\}, \\
\nonumber X &=& \frac{1}{2}\Omega^{\intercal}\left[ \mathbb{1}_{2} + \Sigma_{B} - \varepsilon_{AB}^{\intercal}\left(\mathbb{1}_{2}+\Sigma_{A}\right)^{-1}\varepsilon_{AB}\right]\Omega.
\end{eqnarray}
On the contrary, if we measure in the basis of $|\Psi_{\pm}\rangle$, we get
\begin{eqnarray}
\nonumber \overline{F}_{\Psi}(a,b) &=& \frac{1}{P_{\Psi_{\pm}}\sqrt{\det\left( \mathbb{1}_{2}+\Sigma_{A}\right)\det X}}\bigg\{ \left( 1-|a|^{2}\tr\left(\mathbb{1}_{2}+\Sigma_{A}\right)^{-1}\right)\left( 1-\frac{|a|^{2}}{2}\tr X^{-1}\right) \\
\nonumber &+& \vec{y}^{\intercal}X^{-1}\Omega^{\intercal}\varepsilon_{AB}^{\intercal}(\mathbb{1}_{2}+\Sigma_{A})^{-1}\Omega \sigma_{z} \vec{y} - \frac{|a|^{4}}{2\det X}\tr\left( \varepsilon_{AB}^{\intercal} W_{\mathbb{1}_{2}+\Sigma_{A},\mathbb{1}_{2}}\varepsilon_{AB}\right) \\
&-& \frac{|a|^{2}}{2}\tr\left( X^{-1}\Omega^{\intercal}\varepsilon_{AB}^{\intercal} W_{\mathbb{1}_{2}+\Sigma_{A},\mathbb{1}_{2}}\varepsilon_{AB}\Omega\right)\left( 1-\frac{3|a|^{2}}{2}\tr X^{-1}\right) \bigg\},
\end{eqnarray}
with
\begin{eqnarray}
\nonumber P_{\Psi_{\pm}}&=& \frac{1}{\sqrt{\det\left( \mathbb{1}_{2}+\Sigma_{A}\right)\det X}}\bigg\{ \frac{1}{2}(1-|a|^{2}\tr\left(\mathbb{1}_{2}+\Sigma_{A}\right)^{-1})\left(4 - \tr X^{-1}\right) \\
\nonumber &-& |a|^{2}\tr\left( X^{-1}\Omega^{\intercal}\varepsilon_{AB}^{\intercal} W_{\mathbb{1}_{2}+\Sigma_{A},\mathbb{1}_{2}}\varepsilon_{AB}\Omega\right)\left( 1-\frac{3}{4}\tr X^{-1}\right) \\
&-& \frac{|a|^{2}}{2\det X}\tr\left( \varepsilon_{AB}^{\intercal} W_{\mathbb{1}_{2}+\Sigma_{A},\mathbb{1}_{2}}\varepsilon_{AB}\right) \bigg\}.
\end{eqnarray}
%%%%%%%%%%%%%%%%%%%%%%%%%%%%%%%
\begin{figure}[h]
\centering
\includegraphics[width=0.75\textwidth]{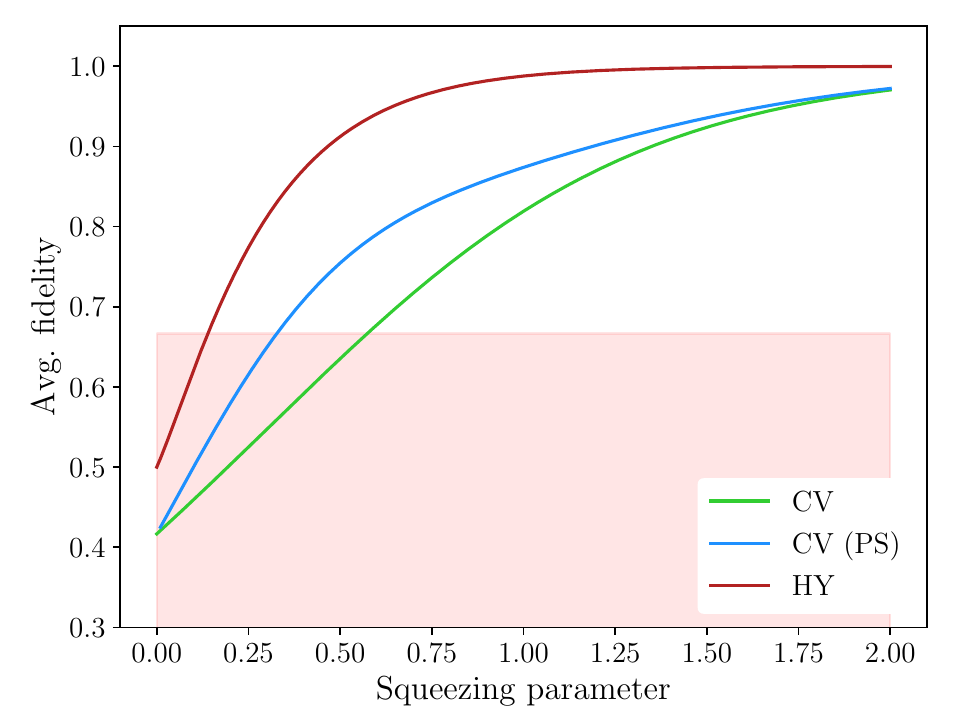}
\caption[Average CV and hybrid (HY) quantum teleportation fidelity of a qubit using a TMSV state]{\textbf{Average CV and hybrid (HY) quantum teleportation fidelity of a qubit using a TMSV state}. Taking the qubit state $|\psi\rangle = a|0\rangle + \sqrt{1-a^{2}}e^{i\varphi}|1\rangle$, we average over all possible amplitudes and phases, uniformly distributed in the Bloch sphere, and represent the fidelity against the squeezing of the entangled resource. In blue and in green, we represent the CV quantum teleportation fidelities with and without photon subtraction, respectively. In red, we show the fidelities corresponding to the hybrid approach, with Bell projections onto states $|\Phi_{\pm}\rangle$ and $|\Psi_{\pm}\rangle$. The red pale background indicates the fidelities that can be achieved with a classical strategy, with a maximum value of 2/3.}
\label{fig7_1}
\end{figure}
%%%%%%%%%%%%%%%%%%%%%%%%%%%%%%%
In Fig.~\ref{fig7_1}, we compare the CV and DV quantum teleportation approaches, using a TMSV state as the entangled resource. We represent the quantum teleportation fidelity for a qubit, taking $a=\sqrt{1-u}$ and $b=\sqrt{u}e^{i\varphi}$, and averaging over $u\in[0,1]$ and $\varphi\in[0,2\pi]$ uniformly. In blue and in green, we represent the fidelities associated with CV strategies, with and without photon subtraction, respectively. In these cases, the strategy involves homodyne detection and displacement of the remaining state. The hybrid strategies, whose fidelity is shown in red, use Bell measurements $|\Psi_{\pm}\rangle$ and $|\Phi_{\pm}\rangle$, and the corresponding projections, $\sigma_{\Psi_{\pm}}$ and $\sigma_{\Phi_{\pm}}$, on the remaining state. Both these DV strategies lead to the same fidelity once we average over all possible qubit configurations. In a red pale background, we show the region of fidelities that can be obtained with a classical strategy. The maximum fidelity that can be obtained for teleporting a qubit state with classical means is $2/3$~\cite{Popescu1994}. Recall that this value was $1/2$ for the teleportation of a coherent state. We can observe that the hybrid strategies show better results for the average fidelity, reaching over 90 \% for squeezing parameters around 0.5, in contrast with the CV strategies, which require squeezing over 1.35 to reach these fidelities. Nevertheless, all of them tend to 1 for larger squeezing. Among the CV strategies, we see that photon subtraction brings an advantage, as it increases the entanglement of the resource. In the hybrid case, both Bell projections, $|\Psi_{\pm}\rangle$ and $|\Phi_{\pm}\rangle$, show the same fidelity; that is why we just represent one curve. 

Similar to what we discussed in chapter~\ref{sec5} concerning photon subtraction, the advantage of the hybrid strategy lies in being non-deterministic. This process is quite inefficient, because the probability of projecting a TMSV state onto the Bell basis is low. In Ref.~\cite{He2022}, this measurement is proposed by using the quantum scissors~\cite{Pegg1998}, which were discussed in chapter~\ref{sec2} of this Thesis in the context of entanglement distillation. Here, this technique is used for projection synthesis; by using single-photon generation and detection, they are able to truncate coherent states to obtain a qubit in a superposition state. If we use this process to truncate a TMSV state onto the Bell state $(|00\rangle+|11\rangle)/\sqrt{2}$, the success probability will be $\lambda^{2}/(2(1+\lambda)^{2})$, where $\lambda=\tanh r$. For a typical squeezing parameter $r=1$, this probability is $0.09$. If we take into account the 0.76 single-photon detection probability taken from the microwave photocounter in Ref.~\cite{Dassonneville2020}, discussed in chapter~\ref{sec5}, we find are left with a 0.07 probability that the Bell state projection is successful. 

\subsubsection{Quantum teleportation with losses}
\begin{figure}[h]
\centering
\includegraphics[width=0.75\textwidth]{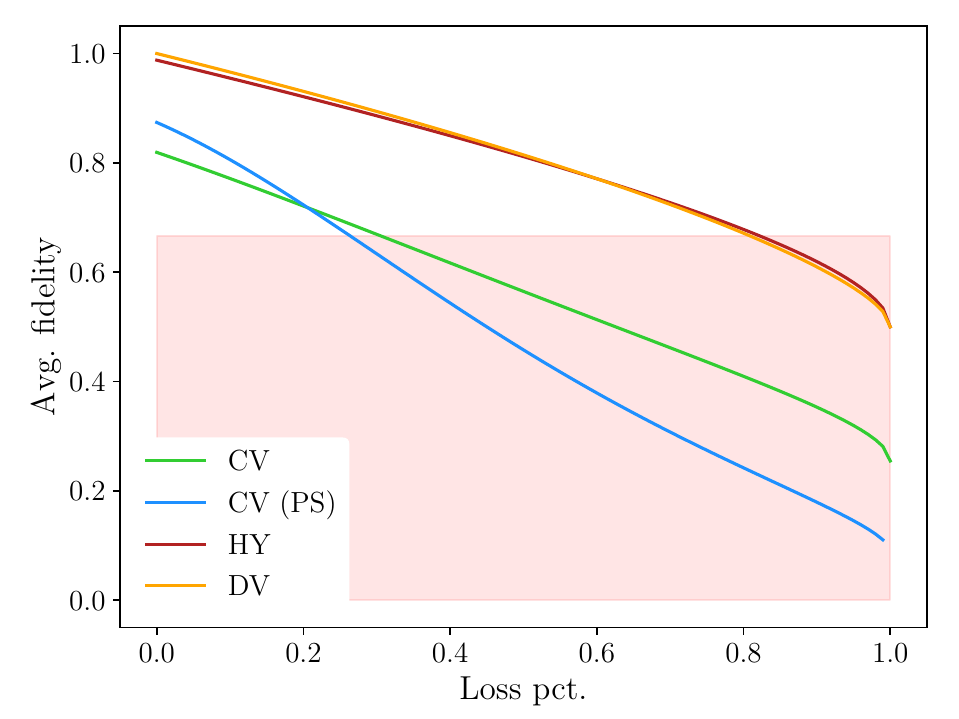}
\caption[Average lossy CV and hybrid (HY) quantum teleportation fidelity of a qubit using a TMSV state, and lossy DV quantum teleportation fidelity using a Bell state]{\textbf{Average lossy CV and hybrid (HY) quantum teleportation fidelity of a qubit using a TMSV state, and lossy DV quantum teleportation fidelity using a Bell state}. We consider that one of the modes of the entangled resource is subject to a pure-loss channel, and represent the fidelity against the percentage of loss, after takin the average over all possible qubit configurations. We fix the squeezing of the TMSV at $r=1$. In blue and in green, we represent the CV quantum teleportation fidelities with and without photon subtraction, respectively. In red, we show the fidelities corresponding to the hybrid approach, and in orange, we represent the results of a DV approach with a Bell state as the resource. The red pale background indicates the fidelities that can be achieved with a classical strategy, with a maximum value of 2/3.}
\label{fig7_2}
\end{figure}

We take into account the losses suffered by the modes distributed among different processors by considering a pure loss channel applied to one of the modes of the TMSV state. Then, we compute the quantum teleportation fidelities for a single qubit, and average over all possible amplitudes, showing them in Fig.~\ref{fig7_2}. The results of a CV quantum teleportation strategy are shown in green (without photon subtraction) and in blue (with photon subtraction). In red, we represent the results of a hybrid quantum teleportation strategy. Additionally, we consider a DV strategy using a Bell state as the entangled resource which has suffered a pure-loss channel in one of the modes. The resulting fidelity is represented in orange. The pale red background indicates the region of fidelities that can be obtained with a classical strategy. Again, we observe that the hybrid strategy lead to better fidelities than a CV strategy. The latter reaches the maximum classical fidelity of 2/3 for a 26 \% losses for a resource with photon subtraction, and at 30 \% losses for the bare resource, whereas the former does it for 82 \% in the worst case. Notably, this limit is 81 \% for the DV case, in which the fidelity behaves quite closely to the hybrid case.
%%%%%%%%%%%%%%%%%%%%%%%%%%%%%%%
%%%%%%%%%%%%%%%%%%%%%%%%%%%%%%%
We want to explore a more realistic case; given that the attenuation factor in superconducting coaxial cables is $\mu \sim 10^{-3} \text{ m}^{-1}$, we can see how the degradation of entanglement affects the quantum teleportation fidelity. Let us assume that we generated TMSV states with $r=1$; the attenuation suffered by the mode that travels between cryostats is modelled by a beam splitter, with reflectivity $\eta = 1-e^{-\mu L}$, $L$ being the travelled distance, that couples the signal mode and a thermal mode with average thermal photons $N_\text{th}=10^{-2}$. In the symplectic formalism, we see that this transformation is quite simple; if $\Sigma_{\text{th}} =(1+2N_{\text{th}})\mathbb{1}_{2}$, we obtain
\begin{eqnarray}
\nonumber \Sigma_{A} &\longrightarrow& \Sigma_{A} = \cosh2r \mathbb{1}_{2}, \\
\Sigma_{B} &\longrightarrow& (1-\eta)\Sigma_{B} + \eta\Sigma_{\text{th}}= \left[(1-\eta)\cosh2r + \eta(1+2N_{\text{th}})\right]\mathbb{1}_{2}, \\
\nonumber \varepsilon_{AB} &\longrightarrow& \sqrt{1-\eta}\varepsilon_{AB} = \sqrt{1-\eta}\sinh2r \sigma_{z}. 
\end{eqnarray}
%%%%%%%%%%%%%%%%%%%%%%%%%%%%%%%
\begin{figure}[h]
\centering
\includegraphics[width=\textwidth]{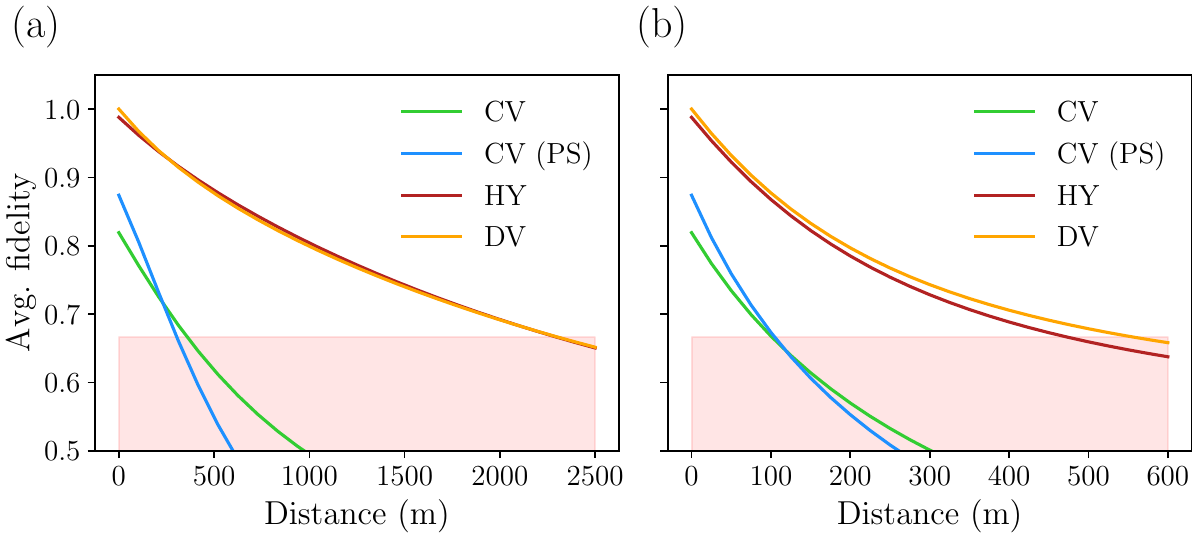}
\caption[Average quantum teleportation fidelity for a qubit using entanglement distributed through a cryolink and through open air, using CV and hybrid (HY) strategies with a TMSV state, and a DV strategy with a Bell state]{\textbf{Average quantum teleportation fidelity for a qubit using entanglement distributed through a cryolink and through open air, using CV and hybrid (HY) strategies with a TMSV state, and a DV strategy with a Bell state}. Both the CV and the hybrid strategies use a TMSV state with $r=1$, while the DV case uses a Bell state. We consider that one of the modes of the entangled resource suffers losses into a thermal environment with average thermal photons $N_{\text{th}}=10^{-2}$ inside the cryolink, and $N_{\text{th}}=1250$ in open air, and represent the average fidelity against the distance between the two units. (a) Entanglement is distributed through a cryolink, where the attenuation factor is $\mu=10^{-3} \text{ m}^{-1}$, and the environment is characterized by $N_{\text{th}} = 10^{-2}$ average thermal photons. (b) Entanglement is distributed through open air, where the attenuation factor is $\mu=1.44\times10^{-6} \text{ m}^{-1}$, and the environment is characterized by $N_{\text{th}} = 1250$ average thermal photons. In blue and in green, we represent the CV quantum teleportation fidelities with and without photon subtraction, respectively. In red, we show the fidelities corresponding to the hybrid approach, and in orange, we represent the results of a DV approach, with a Bell state as the resource. The red pale background indicates the fidelities that can be achieved with a classical strategy, with a maximum value of 2/3.}
\label{fig7_3}
\end{figure}
%%%%%%%%%%%%%%%%%%%%%%%%%%%%%%%
The case of a Bell state in which one of the modes undergoes such a transformation is a bit more complicated. Starting from the state $(|00\rangle+|11\rangle)/\sqrt{2}$, we end up with
\begin{eqnarray}
\rho(a,b) &=& \frac{|a|^{2} + \eta N_{\text{th}}}{(1+\eta N_{\text{th}})^{2}}|0\rangle\langle0| + \frac{\sqrt{1-\eta}}{(1+\eta N_{\text{th}})^{2}} \left(a\bar{b} |0\rangle\langle1| + \bar{a}b |1\rangle\langle0|\right) \\
\nonumber &+& \frac{1}{(1+\eta N_{\text{th}})^{2}}\left[ |a|^{2} \eta(1+N_{\text{th}}) + |b|^{2}\frac{1+\eta(n-1+\eta N_{\text{th}}^{2}(2-\eta))}{1+\eta N_{\text{th}}} \right] |1\rangle\langle1|.
\end{eqnarray}
We can see these results in Fig.~\ref{fig7_3}~(a). In Fig.~\ref{fig7_3}~(b), we represent the case of open-air entanglement distribution, with $\mu = 1.44\times10^{-6} \text{ m}^{-1}$ and $N_\text{th}=1250$. We can observe that the hybrid and the DV strategies lead to better results than the CV ones, behaving the hybrid and DV ones quite similarly. It is rather surprising to observe how similarly are TMSV and Bell states affected by the same attenuation channel; we would have expected entanglement in Bell states to degrade much faster than in TMSV states.

With CV strategies, the fidelity reaches its maximum classical value at 300 m through a cryolink, and at 80 m through open air, whereas for the hybrid strategy, it is 2.3 km through a cryolink and 475 m through open air. Finally, for the DV strategy, these distances are 2.3 km through a cryolink, and 550 m through open air.  

\subsection{Two-Qubit Quantum Teleportation}
In this section, we investigate the extension of the aforementioned protocols to the case in which we have a two-qubit state, such as
\begin{equation}
|\psi\rangle = c_{00}|00\rangle + c_{01}|01\rangle + c_{10}|10\rangle + c_{11}|11\rangle. 
\end{equation}
Similar to what we did for a single qubit, we need to characterize the amplitudes of this qubit in terms of the angles of the equivalent of a Bloch sphere for two qubits. For pure states, this object can be described by three spheres~\cite{Wie2014}, characterized by the coordinates
\begin{equation}
\begin{matrix} x_{1} = \sin\theta_{1}\cos\varphi_{1} \\ y_{1} = \sin\theta_{1}\sin\varphi_{1} \\ z_{1} = \cos\theta_{1} \end{matrix} \qquad \begin{matrix} x_{e} = y_{1}\sin\chi\cos\xi_{1} \\ y_{e} = y_{1}\sin\chi\sin\xi_{1} \\ z_{e} = y_{1}\cos\chi \end{matrix} \qquad \begin{matrix} x_{2} = \sin\theta_{2}\cos\varphi_{2} \\ y_{2} = \sin\theta_{2}\sin\varphi_{2} \\ z_{2} = \cos\theta_{2} \end{matrix}
\end{equation}
The first set corresponds to the base sphere, the second one to the entanglement sphere, and the third one to the fibre sphere; additionally, there is a phase parameter $\xi_{2}$. While $\varphi_{i}\in[0,2\pi]$, $\xi_{i}\in[0,2\pi]$, and $\theta_{i}\in[0,\pi]$, for $i\in\{1,2\}$, we have $\chi\in[0,\pi/2]$. In order to obtain a uniform distribution, we need to sample from a uniform distribution in the same manner we discussed above, such that
\begin{eqnarray}
\nonumber \theta_{i}, \, \chi &\longrightarrow& \arccos(1-2u), \\
\varphi_{i}, \, \xi_{i} &\longrightarrow& 2\pi u.
\end{eqnarray}
With this description, we can characterize the amplitudes of the two qubits described above, as presented in Ref.~\cite{Wie2014},
\begin{eqnarray}
\nonumber c_{00} &=& \cos\frac{\theta_{1}}{2}\cos\frac{\theta_{2}}{2}e^{i\xi_{2}}, \\
c_{01} &=& \cos\frac{\theta_{1}}{2}\sin\frac{\theta_{2}}{2}e^{i(\varphi_{2}-\xi_{2})}, \\
\nonumber c_{10} &=& \sin\frac{\theta_{1}}{2}\left[ (\cos\varphi_{1} + i\sin\varphi_{1}\cos\chi)\cos\frac{\theta_{2}}{2} + i\sin\varphi_{1}\sin\chi\sin\frac{\theta_{2}}{2} e^{i(\xi_{1}-\varphi_{2})}\right]e^{i\xi_{2}}, \\
\nonumber c_{11} &=& \sin\frac{\theta_{1}}{2}\left[ (\cos\varphi_{1} + i\sin\varphi_{1}\cos\chi)\sin\frac{\theta_{2}}{2} - i\sin\varphi_{1}\sin\chi\cos\frac{\theta_{2}}{2} e^{i(\xi_{1}-\varphi_{2})}\right]e^{i(\varphi_{2}-\xi_{2})}.
\end{eqnarray}
Notice that the condition for separability in this state, which is $c_{00}c_{11}=c_{01}c_{10}$, implies $\sin\theta_{1}\sin\varphi_{1}\sin\chi$. This quantity is zero if either the radius of the entanglement sphere or the angle $\chi$ are zero.

We proceed to teleport the two-qubit state $|\psi\rangle$ using two TMSV states with equal squeezing, in the CV and hybrid cases. If the state was separable, we would expect the fidelity to be the square of the one obtained for the single-qubit scenario, since we are just performing two independent teleportation protocols. However, due to the possibility of the state being entangled, we expected the average fidelity to be different from this; it is not the case. Since Alice and Bob are sharing two entangled modes, but both Alice's or Bob's modes are not entangled among themselves, it is as if we were just teleporting two qubits independently. This result cannot change based on information about the state to be teleported which we do not know. 

Another remark we want to make is that the fidelity is not an appropriate measure for the success of the protocol as we increase the number of qubits. The fidelity between two $N$-qubit states tends to zero as $N$ increases, so we should find a measure that remained constant when we increased $N$.

In this chapter, we have investigated the quantum teleportation of qubit states between different quantum processors, in a distributed quantum computing environment. We focused on TMSV states as the CV resources, and explored both the CV and DV quantum teleportation protocols. We referred to the latter as the hybrid protocol, in which we used a TMSV state to teleport a qubit through the DV protocol, consisting of Bell state projections and single-qubit rotations. This process, non-deterministic and quite inefficient, led to better results than the CV one. We introduced losses in the entangled resources and computed the fidelities, under pure-loss and thermal attenuation channels, now comparing with a DV Bell state as well. The results of the purely DV and the hybrid approaches fare similarly, whereas we had expected Bell states to degrade much faster by the quantum channels considered. All the fidelities represented here result from averaging from a Haar distribution all possible qubit states in the Bloch sphere. We do the same for a two-qubit scenario, and find that the average fidelity is just the square of fidelity for teleporting a single qubit. This means that entanglement does not play a role, and it is as if we were teleporting the two qubits independently.

Obtaining efficient transfer of quantum information between different processing units will entail quantum error correction on the entangled resources. When increasing the number of qubits that are teleported, we need the protocol to show robustness. Furthermore, for a proper description it makes sense to use a different measure other than the fidelity, which goes to zero with increasing number of qubits.

%%%%%%%%%%%%%%%%%%%%%%%%
% REDEFINE TITLE FORMAT
%%%%%%%%%%%%%%%%%%%%%%%%

%\input{snp/fancySection1.tex}

%%%%%%%%%%%%%%%%%%%%%%
% CHAPTER 8
%%%%%%%%%%%%%%%%%%%%%%

%\section[Quantum Communication Roadmap]{Quantum Communication Roadmap}
%\label{sec8}

%\input{snp/doubleLineWaterMark.tex}

%\input{chap/chapter8.tex}

%%%%%%%%%%%%%%%%%%%%%%%%
% REDEFINE TITLE FORMAT
%%%%%%%%%%%%%%%%%%%%%%%%

% \titleformat{\section}[display]
% {\vspace*{150pt}
% \bf\Huge}
% {\hspace{-72pt}{\textcolor{grey}{\thesection}} \hspace{38pt} {\textcolor{white}{#1}}}
% {0pt}
% {}
% []
% \titlespacing*{\section}{108pt}{10pt}{60pt}[20pt]
%
% %Intro
% \titleformat{\section}[display]
% {\vspace*{190pt}
% \bfseries\sffamily \huge}
% {\begin{picture}(0,0)\put(-50,-25){\textcolor{grey}{\thesection}}\end{picture}}
% {0pt}
% {\textcolor{white}{#1}}
% []
% \titlespacing*{\section}{80pt}{10pt}{50pt}[20pt]
%
% %ch 2
% \titleformat{\section}[display]
% {\vspace*{190pt}
% \bfseries\sffamily \huge}
% {\begin{picture}(0,0)\put(-50,-28){\textcolor{grey}{\thesection}}\end{picture}}
% {0pt}
% {\textcolor{white}{#1}}
% []
% \titlespacing*{\section}{80pt}{10pt}{50pt}[20pt]
%
% %ch 3, 4 y 5
% \titleformat{\section}[display]
% {\vspace*{190pt}
% \bfseries\sffamily \LARGE}
% {\begin{picture}(0,0)\put(-50,-35){\textcolor{grey}{\thesection}}\end{picture}}
% {0pt}
% {\textcolor{white}{#1}}
% []
% \titlespacing*{\section}{80pt}{10pt}{40pt}[20pt]

%Conclusions
%Julen conclusions
\titleformat{\section}[display]
{\vspace*{190pt}
\bfseries\sffamily \huge}
{\begin{picture}(0,0)\put(-50,-25){\textcolor{grey}{\thesection}}\end{picture}}
{0pt}
{\textcolor{white}{#1}}
[]
\titlespacing*{\section}{80pt}{10pt}{50pt}[20pt]

%%%%%%%%%%%%%%%%%%%%%%
% CONCLUSIONS
%%%%%%%%%%%%%%%%%%%%%%

\section{Conclusions}
\fancyhead[LE]{\rightmark}

% \thiswatermark{\put(1,-241){\color{l-grey}\rule{84pt}{48pt}}
% \put(84,-241){\color{grey}\rule{410pt}{48pt}}}

%JUlEN INTRO
\thiswatermark{\put(1,-280){\color{l-grey}\rule{70pt}{42pt}}
\put(70,-280){\color{grey}\rule{297pt}{42pt}}}

% 
% %Jilen conclusions
% \fancyhead[LE]{\rightmark}
%
% \thiswatermark{\put(1,-280){\color{l-grey}\rule{70pt}{42pt}}
% \put(70,-280){\color{grey}\rule{297pt}{42pt}}}

\lettrine[lines=2, findent=3pt,nindent=0pt]{I}{n} this Thesis, we explore the feasibility of propagating quantum microwaves, in the form of Gaussian states, as resources for quantum communication and quantum metrology protocols in open air with the current advances in superconducting circuit technology. 
%%%%%%%%%%%%%%%

%%%%% Chapter 2 %%%%%
In chapter \ref{sec2}, we have introduced the formalism of continuous variables through Gaussian states. We have shown how the symplectic formalism can be conveniently used to represent infinite-dimensional states and operators using finite vectors and matrices. By means of the displacement vector and the covariance matrix, Gaussian states can be completely characterized, and features like purity, negativity and separability can be computed. An $N$-mode Gaussian state can be described by a positive and symmetric $2N\times2N$ covariance matrix, which, by virtue of Williamson's theorem, can be brought into a block-diagonal normal form, and also admits a diagonalization into symplectic eigenvalues. We have provided transformations to two such instances in the two-mode case. We have shown that the symplectic formalism is also convenient to describe all-Gaussian evolutions and measurements. 

Gaussian states are not only convenient in their description; as we have shown, they can provide an advantage over classical strategies in quantum teleportation. The advantage in this technique relies on quantum entanglement, and for that we have also explored entanglement distillation protocols with Gaussian states. Given the impossibility to distill entanglement with Gaussian operations, we have used the characteristic function formalism to describe the states after photon subtraction, one of the most resource-wise efficient entanglement distillation techniques. Following this, we have discussed entanglement swapping, a protocol that attempts to transform two entangled states, shared pair-wisely by three parties, into a single entangled state shared by the two previously-unconnected parties. Both entanglement distillation and entanglement swapping can be crucial for quantum repeater protocols; in contrast with distillation, swapping is Gaussian preserving. 

%%%%% Chapter 3 %%%%%
 In chapter~\ref{sec3}, we have investigated the limits of purifying Gaussian states using Gaussian operations. We have shown that a single mode of a two-mode Gaussian state can be completely purified, however causing the resulting state to be separable. Therefore, we have studied different partial purification protocols where, with a single copy or two copies of the initial state, and with different combinations of Gaussian-preserving operations and measurements, we traded entanglement for purity. 

%We tested the resulting states as resources in quantum illumination, a quantum sensing protocol in which a two-mode entangled state is used to determine the presence of a low-reflectivity object by sending one of the modes, and then measuring the reflected photons with the mode kept in the laboratory. The quantum advantage in this protocol is measured with the Fisher information, 

We tested the resulting states as resources in quantum illumination, using the inverse of the Cram\'{e}r-Rao bound, which indicates the minimum measurement error in estimating the variance of a target observable, as an efficiency measure. The partially-purified states not only have fewer photons, but they also present larger quantum Fisher information than the initial TMST states; therefore, equating the amount of resources lead to reducing the error by up to 1.5. We used this protocol as a case study, where entanglement is not the only resource required for a quantum advantage, but there could also be other quantum metrology protocols where purity is relevant.

%%%%% Chapter 4 %%%%%

In chapter \ref{sec4}, we have addressed the state of the art of superconducting quantum devices working in the microwave regime, which are involved in different stages of quantum communication, such as state generation or amplification. We described how parametric amplification, aided by JPAs, can be used to generate entangled resources, and how these have to be shielded from thermal microwave radiation by working at cryogenic temperatures. Thus, the necessity for an antenna, matching the cryostat and the open air. Knowing that previous studies using similar architectures had failed to detect entanglement in open air, we investigated a simple quantum antenna, a finite inhomogeneous transmission line with an impedance that changes with the position. The main difference from classical antennae is the lack of an amplification feature, which can degrade quantum correlations. 

We have studied entanglement preservation in the transmission of two-mode squeezed thermal states from the cryostat into open air, and found that maximizing entanglement transmission implies minimizing the reflectivity of the antenna. We have seen that a cavity with a linear impendance, despite being an analytically-solvable case, does not provide a low-enough reflectivity. Nevertheless, we were able to use this result to introduce an optimization problem; the antenna was split into infinitesimally-small slices of linear impedance, and the frontier points were optimized to reduce the reflectivity. We have found numerical values down to $10^{-9}$, for a global impedance function resembling an exponential. 

%We have also proposed an analytical ansatz for this function.

To conclude, we have observed a high-sensitivity of the optimal impedance to potential fabrication errors. We have introduced errors proportional to the value of the impedance in each point, and found that the negativity of the output state drops to zero when these errors are larger than 3 \%.

In chapter \ref{sec5}, we have studied the feasibility of microwave entanglement distribution in open air with two-mode squeezed states. We have taken these as resources for the Braunstein-Kimble quantum-teleportation protocol, adapted to microwave technology, reviewing the steps involved in this process and the possible experimental realization. In chapter~\ref{sec4}, we already discussed two key two key steps in this process, which are the generation of two-mode squeezed states using JPAs, that was experimentally demonstrated in Ref.~\cite{Fedorov2018}, and the formulation of an antenna model for optimal transmission of these states into open air, described in Ref.~\cite{GonzalezRaya2020}. As the next step, we have addressed the degradation of entanglement in open air for two-mode squeezed thermal states; by identifying absorption and thermalization of the signal as the main loss mechanism, using experimental parameters for the photon losses per unit length, we were able to estimate the maximum distance that entanglement can be preserved in different weather conditions. These distances fluctuated from 550 m in ideal weather conditions, to 400 in high-humidity environments, with asymmetric states. With symmetric resources, the distances range from 480 to 350 m.

We have tested the states distributed through open air, including those states after entanglement distillation and entanglement swapping discussed in chapter~\ref{sec2}, as resources for quantum teleportation of an unknown coherent state, following the Braunstein-Kimble protocol. The PS states perform better than the bare resources for short distances, for longer distances the ES ones can extend the reach of teleportation. We have concluded that the fidelities reach of the maximum classical value when their corresponding resources lose entanglement. To compute the negativity of the PS states, which are non-Gaussian, we have proposed a re-Gaussification trick. We masked the non-Gaussian corrections to the fidelity as corrections to the submatrices of the covariance matrix. We have also checked that the resulting covariance matrices are positive and satisfy the uncertainty principle. 

The operations discussed in this chapter require either homodyne detection or photon subtraction; therefore, we have discussed the state of the art of these two crucial techniques, given the current microwave quantum technologies available. We have also attempted to quantify the error introduced in each of these operations by imperfect photocounting, and finite-power homodyne detection. We have found that, for finite-power homodyne detection, the fidelity of quantum teleportation fidelity state depends not only on the gain, but also on the number of photons of the coherent state we want to teleport. 

%%%%% Chapter 6 %%%%%

In chapter \ref{sec6}, we have studied the applicability and efficiency of the techniques discussed in this Thesis for quantum communication between satellites, a field where, the reach of entanglement can be greatly increased, given the low absorption rates. We have studied this limit for signals in the microwave regime, showing that the sizes of the emitting and receiving antennae must be larger in order for the entanglement preservation to reach outside of the near-field. This behaviour is caused by diffraction, which we consider as the main source of loss in this environment. With it, entanglement can reach up to a km, with a receiving antenna with radius of 5 m. 

As the main advances in satellite quantum communication, and a few ground-breaking experiments, have used signals in the optical regime, we have investigated the limits for entanglement preservation and quantum teleportation in this frequency range. We focused on the effects of diffraction, atmospheric attenuation, turbulence, and detector inefficiency on various communication scenarios: ground station to ground station (ground-to-ground), ground station to satellite (uplink), satellite to ground station (downlink), and satellite to satellite (intersatellite). Being the action of weak turbulence inside the atmosphere the main source of loss, we observe that the downlink presents the most favorable results; in this regime, the quantum advantage of teleportation over a classical strategy can be obtained for satellites in the LEO region (from 200 up to 2000 km from the Earth's surface). In adverse weather conditions, we found out that microwaves perform equally or worse than optics, even though the disadvantages the latter present. Therefore, for the further development of microwave quantum communication in free space, proper directivity control has to be developed. With similar intentions, we also investigated here the placement of intermediate stations between ground and satellite, both for state generation, and for beam refocusing, which have shown promising results; as we expected, they improved substantially versus uplink communications, and performed similar to downlink ones. 

%%%%% Chapter 7 %%%%%
In chapter \ref{sec7}, we looked at the difficulties surrounding quantum computing architectures at the moment, and how a distributed configuration could be serviceable. Therefore, we looked at quantum teleportation of qubit states as a way of connecting different processing units. Making use of CV entangled states, we looked at the teleportation fidelities using the Braunstein-Kimble and the DV protocols; the latter, a hybrid approach, led to higher fidelities, as homodyne detection was replaced by a Bell-state projective measurement. We average the final results over all possible qubit configurations, taken from a uniform distribution on the Bloch sphere, also known as Haar distribution.

Studying losses on one mode of the entangled resources, we find that both a hybrid and a DV approach yield similar results, reaching the maximum classical fidelity of 2/3 at around 80 \% pure loss. The former combines a TMSV state and the DV protocol, while the latter uses a two-qubit Bell state. Investigating a channel with thermal loss, we find that these the hybrid and DV results grow apart, the latter being more resilient to thermalization, both in a cryolink and in open air. Finally, we investigate the teleportation of two qubit states using a pair of TMSV states. Despite accounting for the possibility of the two qubits to be entangled, this does not make a difference, since we are two independent teleportation channels;  therefore, we obtain the squared average fidelity for teleporting a single qubit.

%%%%%%%%%%%%%%%

As a whole, this Thesis analyzes the advantages and limitations of performing quantum communication and quantum sensing with propagating quantum microwaves in the form of Gaussian states. It proposes an improvement on quantum illumination using partially-purified Gaussian states, and it brings insight onto the process of entanglement distribution for quantum teleportation, describing the process of state generation in the current landscape of superconducting technology, as well as the design of an antenna for open-air transmission and the inefficiencies of the measurement techniques involved. It explores applications to quantum networks, from satellite links to local area networks, and with current experimental parameters, it establishes a recipe for understanding the technological and the physical overheads. Our efforts are meant to spur the development of quantum technologies working in the microwave regime for the development of wireless quantum communication networks.

\titleformat{\section}[display]
{\vspace*{150pt}
\bfseries\sffamily \Huge}
{\begin{picture}(0,0)\put(-64,-31){\textcolor{grey}{\thesection}}\end{picture}}
{0pt}
{\textcolor{black}{#1}}
[]
\titlespacing*{\section}{80pt}{10pt}{50pt}[0pt]

\section*{Appendices}
\phantomsection
\addcontentsline{toc}{section}{Appendices}

\fancyhead[RO]{APPENDICES}

\cleardoublepage

\appendix

\fancyhead[RO]{\leftmark}

\section{Gaussian integrals}\label{app_A}
In this appendix, we provide the formulas for various species of Gaussian integrals, which we have derived in the calculation of the different teleportation fidelities using Gaussian states with photon subtraction. Some of these have also been used to obtain the fidelity of teleporting a qubit using a Gaussian quantum state. 

Please note that, to obtain these formulas, we have assumed that we are dealing with $2\times 2$ invertible, symmetric matrices. Therefore, the identities we present below work for these types of matrices, and an extension to higher dimensions is not trivial. 
First of all, to obtain these integrals we have repeatedly used
\begin{eqnarray}
\nonumber \partial_{\mu} \left. \frac{1}{\sqrt{\det(X-\mu Y)}} \right |_{\mu=0} &=& \frac{\tr(X^{-1}Y)}{2\sqrt{\det X}}, \\
\partial_{\mu} \left. (X-\mu Y)^{-1} \right |_{\mu=0} &=& W_{X,Y},
\end{eqnarray}
where we have defined the function 
\begin{equation}
W_{X,M} = X^{-1}\tr(X^{-1}M) - \frac{\Omega^{\intercal} M \Omega}{\det X},
\end{equation}
which has the following properties
\begin{eqnarray}
\nonumber \tr\left( X^{-1}W_{X^{-1},M}\right) &=& \tr\left( XM\right), \\
\nonumber \tr\left( X^{-1}M\right) &=& \tr\left( XW_{X,M}\right), \\
\nonumber \det\left( X^{-1}W_{X^{-1},M}\right) &=& \det\left( M X\right), \\
 W_{X^{-1},W_{X,M}} &=& M, \\
\nonumber M &=& X W_{X,M} X, \\
\nonumber M &=& X^{-1} W_{X^{-1},M} X^{-1}, \\
\nonumber W_{X,aA+bB} &=& aW_{X,A} + bW_{X,B}.
\end{eqnarray}
In deriving these, we have used the identities
\begin{eqnarray}
\nonumber \det(A+B) &=& \det A\left[ 1 +\tr(A^{-1}B)\right] +\det B, \\
(A+B)^{-1} &=& \frac{\Omega(A+B)\Omega^{T}}{\det(A+B)},
\end{eqnarray}
which only work for $2\times 2$ invertible symmetric matrices, as mentioned above.

The collection of Gaussian integrals we have used is
%\begin{landscape}
\begin{eqnarray}
&\bullet& \int \diff^{2}\alpha e^{-\frac{1}{2}\vec{\alpha}^{\intercal} X \vec{\alpha}+\vec{\alpha}^{\intercal} \vec{J}} = \frac{\pi}{\sqrt{\det X}}e^{\frac{1}{2}\vec{J}^{\intercal}X^{-1}\vec{J}},\\
\nonumber &\bullet& \int \diff^{2}\alpha \vec{\alpha}^{\intercal} M \vec{\alpha} e^{-\frac{1}{2}\vec{\alpha}^{\intercal} X \vec{\alpha} + \vec{\alpha}^{\intercal} \vec{J}} = \frac{\pi}{\sqrt{\det X}} \left[ \tr\left( X^{-1}M \right) + \vec{J}^{\intercal}W_{X,M}\vec{J} \right] e^{\frac{1}{2}\vec{J}^{\intercal} X^{-1} \vec{J}} ,\\
\nonumber &\bullet& \int \diff^{2}\alpha \vec{\alpha}^{\intercal}\vec{G} e^{-\frac{1}{2}\vec{\alpha}^{\intercal} X \vec{\alpha}+\vec{\alpha}^{\intercal} \vec{J}} = \frac{\pi}{2\sqrt{\det X}}\left( \vec{G}^{\intercal}X^{-1} \vec{J} + \vec{J}^{\intercal} X^{-1} \vec{G} \right)e^{\frac{1}{2}\vec{J}^{\intercal} X^{-1}\vec{J}} ,\\
\nonumber &\bullet& \int \diff^{2}\alpha \left(\vec{\alpha}^{\intercal} M \vec{\alpha}\right) \left(\vec{\alpha}^{\intercal}\vec{G}\right) e^{-\frac{1}{2}\vec{\alpha}^{\intercal} X \vec{\alpha}+\vec{\alpha}^{\intercal} \vec{J}} = \frac{\pi}{\sqrt{\det X}}\Big[ \vec{G}^{\intercal}W_{X,M} \vec{J} + \vec{J}^{\intercal}W_{X,M}\vec{G} \\
\nonumber && + \frac{1}{2}\left( \tr(X^{-1}M) + \vec{J}^{\intercal}W_{X,M}\vec{J}\right)\left(\vec{G}^{\intercal}X^{-1} \vec{J} + \vec{J}^{\intercal}X^{-1} \vec{G}\right) \Big]e^{\frac{1}{2}\vec{J}^{\intercal}X^{-1}\vec{J}},\\
\nonumber &\bullet& \int \diff^{2}\alpha \left(\vec{\alpha}^{\intercal} M \vec{\alpha}\right) \left(\vec{\alpha}^{\intercal} P \vec{\alpha}\right) e^{-\frac{1}{2}\vec{\alpha}^{\intercal} X \vec{\alpha}+\vec{\alpha}^{\intercal} \vec{J}} = \frac{\pi}{\sqrt{\det X}}\Big[ 3\tr(X^{-1}M)\tr(X^{-1}P) \\
\nonumber && - \frac{2}{\det X}\tr(\Omega^{\intercal} P\Omega M) + \vec{J}^{\intercal}W_{X,M}\vec{J}\vec{J}^{\intercal}W_{X,P}\vec{J} + \vec{J}^{\intercal}\Big( 3W_{X,M}\tr(X^{-1}P)  \\
\nonumber && + 3W_{X,P}\tr(X^{-1}M) - \frac{2X^{-1}}{\det X}\tr(\Omega^{\intercal} P\Omega M)\Big)\vec{J} \Big]e^{\frac{1}{2}\vec{J}^{\intercal}X^{-1}\vec{J}}, \\
\nonumber &\bullet& \int \diff^{2}\alpha \left(\vec{\alpha}^{\intercal}\vec{G}\right) \left(\vec{\alpha}^{\intercal}\vec{K}\right) e^{-\frac{1}{2}\vec{\alpha}^{\intercal} X \vec{\alpha}+\vec{\alpha}^{\intercal} \vec{J}} = \frac{\pi}{2\sqrt{\det X}}\Big[ \vec{G}^{\intercal}X^{-1}\vec{K} + \vec{K}^{\intercal}X^{-1}\vec{G} \\
\nonumber && + \frac{1}{2}\left(\vec{G}^{\intercal}X^{-1} \vec{J} + \vec{J}^{\intercal}X^{-1} \vec{G}\right)\left(\vec{K}^{\intercal}X^{-1} \vec{J} + \vec{J}^{\intercal}X^{-1} \vec{K}\right) \Big]e^{\frac{1}{2}\vec{J}^{\intercal}X^{-1}\vec{J}}, 
\end{eqnarray}

\begin{eqnarray}
\nonumber &\bullet& \int \diff^{2}\alpha \left(\vec{\alpha}^{\intercal}\vec{G}\right) \left(\vec{\alpha}^{\intercal}\vec{K}\right) \left(\vec{\alpha}^{\intercal}M\vec{\alpha}\right) e^{-\frac{1}{2}\vec{\alpha}^{\intercal} X \vec{\alpha}+\vec{\alpha}^{\intercal} \vec{J}} = \\
\nonumber && \frac{\pi}{\sqrt{\det X}}\bigg\{\frac{1}{2}\left[ \tr\left( X^{-1}M \right) + \vec{J}^{\intercal}W_{X,M}\vec{J} \right] \Big[ \vec{G}^{\intercal}X^{-1}\vec{K} + \vec{K}^{\intercal}X^{-1}\vec{G} \\
\nonumber && + \frac{1}{2}\left( \vec{J}^{\intercal}X^{-1}\vec{K} + \vec{K}^{\intercal}X^{-1}\vec{J} \right)\left( \vec{J}^{\intercal}X^{-1}\vec{G} + \vec{G}^{\intercal}X^{-1}\vec{J} \right) \Big] + \vec{G}^{\intercal}W_{X,M} \vec{K} \\
\nonumber && + \vec{K}^{\intercal}W_{X,M} \vec{G} + \frac{1}{2}\left(\vec{G}^{\intercal}X^{-1} \vec{J} + \vec{J}^{\intercal}X^{-1} \vec{G}\right)\left(\vec{K}^{\intercal}W_{X,M} \vec{J} + \vec{J}^{\intercal}W_{X,M} \vec{K}\right) \\
\nonumber && + \frac{1}{2}\left(\vec{K}^{\intercal}X^{-1} \vec{J} + \vec{J}^{\intercal}X^{-1} \vec{K}\right)\left(\vec{G}^{\intercal}W_{X,M} \vec{J} + \vec{J}^{\intercal}W_{X,M} \vec{G}\right) \bigg\}e^{\frac{1}{2}\vec{J}^{\intercal}X^{-1}\vec{J}}, \\
&\bullet& \int \diff^{2}\alpha \left(\vec{\alpha}^{\intercal}M\vec{\alpha}\right)  \left(\vec{\alpha}^{\intercal}P\vec{\alpha}\right)  \left(\vec{\alpha}^{\intercal}Q\vec{\alpha}\right) e^{-\frac{1}{2}\vec{\alpha}^{\intercal} X \vec{\alpha}+\vec{\alpha}^{\intercal} \vec{J}} = \frac{\pi}{\sqrt{\det X}}\times\\
\nonumber && \bigg\{ 15\tr(X^{-1}M)\tr(X^{-1}P)\tr(X^{-1}Q) - \frac{6}{\det X}\Big[ \tr(X^{-1}Q)\tr(\Omega^{\intercal}M\Omega P) \\
\nonumber && + \tr(X^{-1}M)\tr(\Omega^{\intercal}Q\Omega P) + \tr(X^{-1}P)\tr(\Omega^{\intercal}Q\Omega M)\Big] \\
\nonumber && + \vec{J}^{\intercal}\Big[ 15\tr(X^{-1}M)\tr(X^{-1}P)W_{X,Q} + 15\tr(X^{-1}M)\tr(X^{-1}Q)W_{X,P} \\
\nonumber && + 15\tr(X^{-1}P)\tr(X^{-1}Q)W_{X,M} -\frac{6}{\det X}\Big( \tr(\Omega^{\intercal}P\Omega M)W_{X,Q}  \\
\nonumber && + \tr(\Omega^{\intercal}Q\Omega M)W_{X,P} + \tr(\Omega^{\intercal}Q\Omega P)W_{X,M} \Big) -\frac{6 X^{-1}}{\det X}\Big( \tr(\Omega^{\intercal}P\Omega M)\tr(X^{-1}Q) \\
\nonumber && + \tr(\Omega^{\intercal}Q\Omega M)\tr(X^{-1}P) + \tr(\Omega^{\intercal}Q\Omega P)\tr(X^{-1}M) \Big) \Big] \vec{J} \\
\nonumber && + 5\tr(X^{-1}Q) \left( \vec{J}^{\intercal}W_{X,M}\vec{J}\right)\left( \vec{J}^{\intercal}W_{X,P}\vec{J}\right) \\
\nonumber && + 5\tr(X^{-1}P) \left( \vec{J}^{\intercal}W_{X,M}\vec{J}\right)\left( \vec{J}^{\intercal}W_{X,Q}\vec{J}\right) \\
\nonumber && + 5\tr(X^{-1}M) \left( \vec{J}^{\intercal}W_{X,Q}\vec{J}\right)\left( \vec{J}^{\intercal}W_{X,P}\vec{J}\right) \\
\nonumber && -\frac{2}{\det X}\left( \vec{J}^{\intercal}X^{-1}\vec{J}\right) \Big[  \left( \vec{J}^{\intercal}W_{X,Q}\vec{J}\right) \tr(\Omega^{\intercal}M\Omega P) \\
\nonumber && + \left( \vec{J}^{\intercal}W_{X,M}\vec{J}\right) \tr(\Omega^{\intercal}Q\Omega P) + \left( \vec{J}^{\intercal}W_{X,P}\vec{J}\right) \tr(\Omega^{\intercal}Q\Omega M)\Big] \\
\nonumber && + \left( \vec{J}^{\intercal}W_{X,M}\vec{J}\right) \left( \vec{J}^{\intercal}W_{X,P}\vec{J}\right) \left( \vec{J}^{\intercal}W_{X,Q}\vec{J}\right)\bigg\}e^{\frac{1}{2}\vec{J}^{\intercal}X^{-1}\vec{J}}.
\end{eqnarray}
%\end{landscape}
Note that the following integrals do not present a source term:
\begin{eqnarray}
\nonumber &\bullet& \int \diff^{2}\alpha \left(\vec{\alpha}^{\intercal}\vec{G}\right) \left(\vec{\alpha}^{\intercal}\vec{K}\right) \left(\vec{\alpha}^{\intercal}M\vec{\alpha}\right)\left(\vec{\alpha}^{\intercal}P\vec{\alpha}\right) e^{-\frac{1}{2}\vec{\alpha}^{\intercal} X \vec{\alpha}} = \frac{\pi}{\sqrt{\det X}}\times \\
\nonumber && \bigg\{ \vec{G}^{\intercal}\Big[ \frac{15}{2} X^{-1}\tr\left( X^{-1}M \right)\tr\left( X^{-1}P\right) -\frac{3}{\det X}X^{-1} \tr\left( \Omega^{\intercal}P\Omega M \right) \\
\nonumber && -\frac{3}{\det X}\Omega^{\intercal}P\Omega\tr\left( X^{-1}M \right) -\frac{3}{\det X}\Omega^{\intercal}M\Omega\tr\left( X^{-1}P \right) \Big] \vec{K} \\
\nonumber && \vec{K}^{\intercal}\Big[ \frac{15}{2} X^{-1}\tr\left( X^{-1}M \right)\tr\left( X^{-1}P\right) -\frac{3}{\det X}X^{-1} \tr\left( \Omega^{\intercal}P\Omega M \right) \\
\nonumber && -\frac{3}{\det X}\Omega^{\intercal}P\Omega\tr\left( X^{-1}M \right) -\frac{3}{\det X}\Omega^{\intercal}M\Omega\tr\left( X^{-1}P \right) \Big] \vec{G} \bigg\}, \\
 &\bullet& \int \diff^{2}\alpha \left(\vec{\alpha}^{\intercal}M\vec{\alpha}\right)  \left(\vec{\alpha}^{\intercal}P\vec{\alpha}\right)  \left(\vec{\alpha}^{\intercal}Q\vec{\alpha}\right) \left(\vec{\alpha}^{\intercal}R\vec{\alpha}\right)e^{-\frac{1}{2}\vec{\alpha}^{\intercal} X \vec{\alpha}} = \frac{\pi}{\sqrt{\det X}}\times\\
\nonumber && \bigg\{ 105\tr(X^{-1}M)\tr(X^{-1}P)\tr(X^{-1}Q)\tr(X^{-1}R) \\
\nonumber && - \frac{30}{\det X}\Big[ \tr(X^{-1}Q)\tr(X^{-1}R)\tr(\Omega^{\intercal}M\Omega P) + \tr(X^{-1}M)\tr(X^{-1}R)\tr(\Omega^{\intercal}Q\Omega P) \\
\nonumber && + \tr(X^{-1}P)\tr(X^{-1}R)\tr(\Omega^{\intercal}Q\Omega M) + \tr(X^{-1}P)\tr(X^{-1}Q)\tr(\Omega^{\intercal}R\Omega M)\\
\nonumber && + \tr(X^{-1}M)\tr(X^{-1}Q)\tr(\Omega^{\intercal}R\Omega P) + \tr(X^{-1}M)\tr(X^{-1}P)\tr(\Omega^{\intercal}R\Omega Q)\Big] \\
\nonumber && + \frac{12}{(\det X)^{2}} \Big[ \tr(\Omega^{\intercal}R\Omega Q)\tr(\Omega^{\intercal}M\Omega P) + \tr(\Omega^{\intercal}R\Omega M)\tr(\Omega^{\intercal}Q\Omega P) \\
\nonumber && + \tr(\Omega^{\intercal}R\Omega P)\tr(\Omega^{\intercal}Q\Omega M) \Big] \bigg\}.
\end{eqnarray}

%\begin{figure}[hbt]
% \centering
%\includegraphics[width=\textwidth]{img/app/experiment/alltimetraces.pdf}
%\caption[Pulse sequences in the experiment]{Pulse sequences in the experiment. Pulse sequences for a single two-mode $(a)$, three-mode $(b)$, and four-mode $(c)$ Trotter step. Used gates are entangling gates as well as single-qubit microwave, idle and detuning gates.
%The legend is in the bottom right.}
%\label{fig:experiment_pulseseq}
%\end{figure}

\section{Step-by-step quantum teleportation}\label{app_B}
In this section, we derive the famous formula for the fidelity of teleporting an unknown coherent state using a two-mode Gaussian quantum state. We also derive the formulas for a general two-mode Gaussian state with heuristic and with probabilistic photon subtraction. 

All teleportation protocols require the involved parties to share an entangled state. Moreover, they require the sender to make homodyne detection measurements, communicating the results to the receiver through a classical channel, who makes a displacement in his state depending on the outcome of said measurements. Given a shared entangled state $\rho_{AB}$ and an initial state $\rho_{\text{in}}$ to be teleported, the state that the receiver has after the homodyne measurements is
\begin{equation}
\sigma_{B}(x,p) = \frac{1}{P_{B}(x,p)} \tr_{TA}\left[ \rho^{\text{in}}_{T}\otimes \rho_{AB}\Pi(x,p)_{TA}\right],
\end{equation}
with $P_{B}(x,p) = \tr_{TAB}\left[ \rho^{\text{in}}_{T}\otimes \rho_{AB}\Pi(x,p)_{TA}\right]$. Now, this expectation value over the teleported ($T$) and the senders ($A$) modes is computed as
\begin{equation}
\tr_{TA}\left[ \rho^{\text{in}}_{T}\otimes \rho_{AB}\Pi(x,p)_{TA}\right] = \frac{1}{2\pi}\int_{-\infty}^{\infty}\int_{-\infty}^{\infty}\diff y \diff y' e^{ip(y-y')}\langle x+y'|\rho_{\text{in}}|x+y\rangle_{T} \langle y'|\rho_{AB}| y\rangle_{A}.
\end{equation}
Once we have computed $\sigma_{B}$, we need to compute the outcoming state after the receiver applies the displacements,
\begin{equation}
\rho_{\text{out}} = \int_{-\infty}^{\infty} \diff x \int_{-\infty}^{\infty} \diff p P_{B}(x,p)\hat{D}_{B}(\xi)\sigma_{B}(x,p)\hat{D}_{B}(-\xi).
\end{equation}
The average fidelity is computed as
\begin{equation}
\bar{F} = \tr[\rho_{\text{in}}\rho_{\text{out}}].
\end{equation}

\subsection{Coherent + Gaussian bipartite state}
Assume two parties share an entangled two-mode gaussian state $\rho_{AB}$ with covariance matrix
\begin{equation}
\Sigma = \begin{pmatrix} \Sigma_{A} & \varepsilon_{AB} \\ \varepsilon_{AB}^{\intercal} & \Sigma_{B} \end{pmatrix},
\end{equation}
and they want to use this resource to teleport a given quantum state $\rho^{\text{in}}_{T}$. In this case, we will choose a coherent state, with covariance matrix $\Sigma_{\text{coh}}=\mathbb{1}_{2}$ and displacement vector $\vec{\alpha}_{0}$. 

The first step is to compute Bobs reduced state after the homodyne measurement, for which we write the density matrices in terms of this corresponding characteristic function,
\begin{equation}
\rho = \frac{1}{\pi}\int \diff^{2}\alpha \chi(\alpha) \hat{D}(-\alpha),
\end{equation}
where $\hat{D}(\alpha)=e^{\alpha \hat{a}^{\dagger}-\bar{\alpha}\hat{a}}$ is the displacement operator. The characteristic function is then obtained as $\chi(\alpha)=\tr[\rho \hat{D}(\alpha)]$, and that of a gaussian state can be constructed from the covariance matrix $\Sigma$ and the displacement vector $\vec{d}$, such that
\begin{equation}
\chi(\alpha) = \exp\left[ -\frac{1}{4}\vec{\alpha}^{\intercal}\Omega^{\intercal}\Sigma\Omega\vec{\alpha} -i\vec{\alpha}^{\intercal}\Omega\vec{d} \right]
\end{equation}
with $\Omega = \begin{pmatrix} 0 & 1 \\ -1 & 0 \end{pmatrix}$ being the symplectic matrix. The reduced state that Bob obtains can then be written as
\begin{eqnarray}
\nonumber && \tr_{TA}\left[ \rho^{\text{in}}_{T}\otimes \rho_{AB}\Pi(x,p)_{TA}\right] = \frac{1}{2\pi^{4}} \int \diff^{2}\alpha_{1}\int \diff^{2}\alpha_{2}\int \diff^{2}\alpha_{3} \chi_{T}^{\text{in}}(\alpha_{1})\chi_{AB}(\alpha_{2},\alpha_{3})\times \\
\nonumber && \hat{D}_{B}(-\alpha_{3})  \int_{-\infty}^{\infty}\int_{-\infty}^{\infty} \diff y \diff y' e^{ip(y-y')}\langle x+y'|\hat{D}_{T}(-\alpha_{1})|x+y\rangle_{T} \langle y'|\hat{D}_{A}(-\alpha_{2})| y\rangle_{A}.
\end{eqnarray}
Notice that we can write
\begin{equation}
\hat{D}(-\alpha) = e^{\bar{\alpha}\hat{a}-\alpha \hat{a}^{\dagger}} = e^{-\sqrt{2}i\mathbb{Im}\alpha \hat{x}+\sqrt{2}i\mathbb{Re}\alpha \hat{p}} = e^{i\mathbb{Re}\alpha \mathbb{Im}\alpha}e^{\sqrt{2}i\mathbb{Re}\alpha \hat{p}}e^{-\sqrt{2}i\mathbb{Im}\alpha \hat{x}},
\end{equation}
which leads to
\begin{eqnarray}
\nonumber \langle y' |\hat{D}(-\alpha)| y\rangle &=& e^{i\mathbb{Re}\alpha \mathbb{Im}\alpha} \langle y'| e^{\sqrt{2}i\mathbb{Re}\alpha \hat{p}}e^{-\sqrt{2}i\mathbb{Im}\alpha \hat{x}} |y\rangle = e^{i\mathbb{Im}\alpha(\mathbb{Re}\alpha-\sqrt{2}y)} \langle y'| e^{\sqrt{2}i\mathbb{Re}\alpha \hat{p}} |y\rangle \\
\nonumber &=& e^{i\mathbb{Im}\alpha(\mathbb{Re}\alpha-\sqrt{2}y)} \langle y'|y - \sqrt{2}\mathbb{Re}\alpha \rangle = e^{i\mathbb{Im}\alpha(\mathbb{Re}\alpha-\sqrt{2}y)} \delta(y'-y+\sqrt{2}\mathbb{Re}\alpha),
\end{eqnarray}
where we have used the fact that $e^{-ix\hat{p}}|y\rangle = |x+y\rangle$. Then, if we use this result on the integrals over $y$, $y'$, we obtain
\begin{eqnarray}
&& \int_{-\infty}^{\infty}\int_{-\infty}^{\infty} \diff y \diff y' e^{ip(y-y')} e^{i\mathbb{Im}\alpha_{1}(\mathbb{Re}\alpha_{1}-\sqrt{2}(x+y))} e^{i\mathbb{Im}\alpha_{2}(\mathbb{Re}\alpha_{2}-\sqrt{2}y)} \times \\
\nonumber && \delta(x+y'-x-y+\sqrt{2}\mathbb{Re}\alpha_{1})  \delta(y'-y+\sqrt{2}\mathbb{Re}\alpha_{2}) = \int_{-\infty}^{\infty} \diff y e^{-i\sqrt{2}y(\mathbb{Im}\alpha_{1}+\mathbb{Im}\alpha_{2})}  \\
\nonumber && e^{i\sqrt{2}(p\mathbb{Re}\alpha_{2}-x\mathbb{Im}\alpha_{1})} e^{i(\mathbb{Im}\alpha_{1}\mathbb{Re}\alpha_{1}+\mathbb{Im}\alpha_{2}\mathbb{Re}\alpha_{2})} \delta(\sqrt{2}\mathbb{Re}\alpha_{1}-\sqrt{2}\mathbb{Re}\alpha_{2}) = 2\pi \times \\
\nonumber && e^{i\sqrt{2}(p\mathbb{Re}\alpha_{2}-x\mathbb{Im}\alpha_{1})} e^{i(\mathbb{Im}\alpha_{1}\mathbb{Re}\alpha_{1}+\mathbb{Im}\alpha_{2}\mathbb{Re}\alpha_{2})} \delta(\sqrt{2}\mathbb{Re}\alpha_{1}-\sqrt{2}\mathbb{Re}\alpha_{2}) \delta(\sqrt{2}\mathbb{Im}\alpha_{1}+\sqrt{2}\mathbb{Im}\alpha_{2}).
\end{eqnarray}
Notice that we get an extra factor of $\frac{1}{2}$ coming from Dirac deltas, since $\delta(kx)=\delta(x)/|k|$. Basically, the Dirac delta functions we obtain imply that $\alpha_{2}=\bar{\alpha}_{1}$. Then, going back to the reduced state,
\begin{eqnarray}
\nonumber && \tr_{TA}\left[ \rho^{\text{in}}_{T}\otimes \rho_{AB}\Pi(x,p)_{TA}\right] = \frac{1}{\pi^{3}} \int \diff^{2}\alpha_{1}\int \diff^{2}\alpha_{2}\int \diff^{2}\alpha_{3} \chi_{T}^{\text{in}}(\alpha_{1})\chi_{AB}(\alpha_{2},\alpha_{3})\times \\
\nonumber && \hat{D}_{B}(-\alpha_{3}) e^{i\sqrt{2}(p\mathbb{Re}\alpha_{2}-x\mathbb{Im}\alpha_{1})} e^{i(\mathbb{Im}\alpha_{1}\mathbb{Re}\alpha_{1}+\mathbb{Im}\alpha_{2}\mathbb{Re}\alpha_{2})} \delta(\sqrt{2}\mathbb{Re}\alpha_{1}-\sqrt{2}\mathbb{Re}\alpha_{2}) \times \\
\nonumber && \delta(\sqrt{2}\mathbb{Im}\alpha_{1}+\sqrt{2}\mathbb{Im}\alpha_{2})  =  \frac{1}{2\pi^{3}} \int \diff^{2}\alpha_{1}\int \diff^{2}\alpha_{3} \chi_{T}^{\text{in}}(\alpha_{1})\chi_{AB}(\bar{\alpha}_{1},\alpha_{3})\hat{D}_{B}(-\alpha_{3}) \times \\
&& e^{i\sqrt{2}(p\mathbb{Re}\alpha_{1}-x\mathbb{Im}\alpha_{1})}.
\end{eqnarray}
Let us explicitly compute the product of the characteristic functions,
\begin{eqnarray}
\nonumber \chi_{T}^{\text{in}}(\alpha_{1})\chi_{AB}(\bar{\alpha}_{1},\alpha_{3}) &=& \exp\bigg\{ -\frac{1}{4} \big[ \vec{\alpha}_{1}^{\intercal}\Omega^{\intercal} \left(\mathbb{1}_{2}+\sigma_{z} \Sigma_{A} \sigma_{z}\right)\Omega \vec{\alpha}_{1} - \vec{\alpha}_{1}^{\intercal}\Omega^{\intercal} \sigma_{z} \varepsilon_{AB}\Omega \vec{\alpha}_{3} \\
&-& \vec{\alpha}_{3}^{\intercal} \Omega^{\intercal} \varepsilon_{AB}^{\intercal} \sigma_{z} \Omega \vec{\alpha}_{1} + \vec{\alpha}_{3}^{\intercal} \Omega^{\intercal} \Sigma_{B}\Omega \vec{\alpha}_{3} \big] - i\vec{\alpha}^{\intercal}_{1}\Omega\vec{\alpha}_{0} \bigg]\},
\end{eqnarray}
knowing that $\vec{\alpha}_{2}=\sigma_{z}\vec{\alpha}_{1}$, $\Omega \sigma_{z} = -\sigma_{z}\Omega$ and $\Omega^{\intercal}\Omega=\mathbb{1}_{2}$. Here, we have identified $\vec{\alpha}_{i}^{\intercal} = \sqrt{2}\begin{pmatrix}\mathbb{Re}\alpha_{i} & \mathbb{Im}\alpha_{i}\end{pmatrix} = \begin{pmatrix} x_{i} & p_{i} \end{pmatrix}$. Notice that this will mean that $\diff^{2}\alpha_{i} = \frac{1}{2}\diff x_{i} \diff p_{i}$ Furthermore, we write
\begin{equation}
e^{i\sqrt{2}(p\mathbb{Re}\alpha_{1}-x\mathbb{Im}\alpha_{1})} = e^{i \vec{\alpha}_{1}^{\intercal}\Omega\vec{\xi}}
\end{equation}
with $\vec{\xi}^{\intercal} = \begin{pmatrix} x & p \end{pmatrix}$. Joining everything together, we obtain 
\begin{eqnarray}
\nonumber && \chi_{T}^{\text{in}}(\alpha_{1})\chi_{AB}(\bar{\alpha}_{1},\alpha_{3})e^{i\sqrt{2}(p\mathbb{Re}\alpha_{1}-x\mathbb{Im}\alpha_{1})} = \exp\bigg\{ -\frac{1}{4} \Big[ \vec{\alpha}_{1}^{\intercal}\Omega^{\intercal}(\mathbb{1}_{2} + \sigma_{z} \Sigma_{A} \sigma_{z})\Omega \vec{\alpha}_{1}\\
&& -  \vec{\alpha}_{1}^{\intercal}\Omega^{\intercal} \sigma_{z} \varepsilon_{AB}\Omega \vec{\alpha}_{3} - \vec{\alpha}_{3}^{\intercal} \Omega^{\intercal} \varepsilon_{AB}^{\intercal} \sigma_{z} \Omega \vec{\alpha}_{1} + \vec{\alpha}_{3}^{\intercal} \Omega^{\intercal} \Sigma_{B}\Omega \vec{\alpha}_{3} \Big] -i\vec{\alpha}_{1}^{\intercal}\Omega (\vec{\alpha}_{0}-\vec{\xi}) \bigg\}.
\end{eqnarray}
Since $\vec{\alpha}_{1}^{\intercal}\Omega^{\intercal} \sigma_{z} \varepsilon_{AB}\Omega \vec{\alpha}_{3} = \vec{\alpha}_{3}^{\intercal} \Omega^{\intercal} \varepsilon_{AB}^{\intercal} \sigma_{z} \Omega \vec{\alpha}_{1}$, we can simplify things by
\begin{eqnarray}
\nonumber && \chi_{T}^{\text{in}}(\alpha_{1})\chi_{AB}(\bar{\alpha}_{1},\alpha_{3})e^{i\sqrt{2}(p\mathbb{Re}\alpha_{1}-x\mathbb{Im}\alpha_{1})} = \exp\bigg\{ -\frac{1}{4} \big[ \vec{\alpha}_{1}^{\intercal}\Omega^{\intercal}(\mathbb{1}_{2} + \sigma_{z} \Sigma_{A} \sigma_{z})\Omega \vec{\alpha}_{1} \\
&& + \vec{\alpha}_{3}^{\intercal} \Omega^{\intercal} \Sigma_{B}\Omega \vec{\alpha}_{3} \big] + \vec{\alpha}_{1}^{\intercal}\Omega^{\intercal} \left[ \frac{1}{2}\sigma_{z} \varepsilon_{AB}\Omega \vec{\alpha}_{3} +i(\vec{\alpha}_{0}-\vec{\xi})\right] \bigg\}.
\end{eqnarray}
We then define $M = \frac{1}{2}\Omega^{\intercal}(\mathbb{1}_{2} + \sigma_{z} \Sigma_{A} \sigma_{z})\Omega$ and $\vec{J} = \frac{1}{2}\Omega^{\intercal} \sigma_{z} \varepsilon_{AB}\Omega \vec{\alpha}_{3} +i\Omega^{\intercal}(\vec{\alpha}_{0}-\vec{\xi})$, and proceed to solve the integral over $\alpha_{1}$,
\begin{eqnarray}
\nonumber && \int \diff^{2}\alpha_{1} \chi_{T}^{\text{in}}(\alpha_{1})\chi_{AB}(\bar{\alpha}_{1},\alpha_{3})e^{i\sqrt{2}(p\mathbb{Re}\alpha_{1}-x\mathbb{Im}\alpha_{1})} \\
\nonumber &=&  \int \diff^{2}\alpha_{1} e^{-\frac{1}{2}\vec{\alpha}_{1}^{\intercal}M\vec{\alpha}_{1}+\vec{\alpha}_{1}^{\intercal}\vec{J}}e^{-\frac{1}{4}\vec{\alpha}_{3}^{\intercal} \Omega^{\intercal} \Sigma_{B}\Omega \vec{\alpha}_{3}} = \frac{\pi}{\sqrt{\det M}} e^{\frac{1}{2}\vec{J}^{\intercal}M^{-1}\vec{J}}e^{-\frac{1}{4}\vec{\alpha}_{3}^{\intercal} \Omega^{\intercal} \Sigma_{B}\Omega \vec{\alpha}_{3}}.
\end{eqnarray}
Then, we write the reduced state for Bob as
\begin{equation}
\sigma_{B}(x,p) = \frac{1}{2\pi^{2}P(x,p)\sqrt{\det M}}\int \diff^{2}\alpha_{3}e^{\frac{1}{2}\vec{J}^{\intercal}M^{-1}\vec{J}}e^{-\frac{1}{4}\vec{\alpha}_{3}^{\intercal} \Omega^{\intercal} \Sigma_{B}\Omega \vec{\alpha}_{3}}\hat{D}_{B}(-\alpha_{3}).
\end{equation}
We compute the normalization by using $\tr [\hat{D}(\alpha)] = \pi\delta(\alpha)$, such that
\begin{equation}
P(x,p) = \frac{1}{2\pi\sqrt{\det M}}e^{-\frac{1}{2}(\vec{\alpha}_{0}-\vec{\xi})^{\intercal}M^{-1}(\vec{\alpha}_{0}-\vec{\xi})}.
\end{equation}
After Bob performs a displacement on his reduced state depending on the outcome of the homodyne measurement Alice performed, given by $\xi$, the outcoming state is 
\begin{equation}
\rho_{B}^{\text{out}} = \int_{-\infty}^{\infty} \diff x \int_{-\infty}^{\infty} \diff p P(x,p) \hat{D}_{B}(\xi)\sigma_{B}(x,p)\hat{D}_{B}(-\xi).
\end{equation}
We use the fact that
\begin{equation}
\hat{D}_{B}(\xi)\hat{D}_{B}(-\alpha_{3})\hat{D}_{B}(-\xi) = e^{\alpha_{3}\bar{\xi}-\bar{\alpha}_{3}\xi}\hat{D}_{B}(-\alpha_{3}),
\end{equation}
and we write
\begin{equation}
\rho_{B}^{\text{out}} = \frac{1}{2\pi^{2}\sqrt{\det M}} \int_{-\infty}^{\infty} \diff x \int_{-\infty}^{\infty} \diff p \int \diff^{2}\alpha_{3}e^{\frac{1}{2}\vec{J}^{\intercal}M^{-1}\vec{J}}e^{-\frac{1}{4}\vec{\alpha}_{3}^{\intercal} \Omega^{\intercal} \Sigma_{B}\Omega \vec{\alpha}_{3}}e^{-i\vec{\alpha}_{3}^{\intercal}\Omega\vec{\xi}}\hat{D}_{B}(-\alpha_{3}).
\end{equation}
Here, we can identify the characteristic function associated to $\rho_{B}^{\text{out}}$ as
\begin{equation}
\chi_{B}^{\text{out}}(\alpha_{3}) = \frac{1}{2\pi\sqrt{\det M}} \int_{-\infty}^{\infty} \diff x \int_{-\infty}^{\infty} \diff p e^{\frac{1}{2}\vec{J}^{\intercal}M^{-1}\vec{J}}e^{-\frac{1}{4}\vec{\alpha}_{3}^{\intercal} \Omega^{\intercal} \Sigma_{B}\Omega \vec{\alpha}_{3}}e^{-i\vec{\alpha}_{3}^{\intercal}\Omega\vec{\xi}},
\end{equation}
and hence we can attempt to solve this integral. First, we will combine the exponentials
\begin{eqnarray}
&& \frac{1}{2}\vec{J}^{\intercal}M^{-1}\vec{J}-\frac{1}{4}\vec{\alpha}_{3}^{\intercal} \Omega^{\intercal} \Sigma_{B}\Omega \vec{\alpha}_{3}-i\vec{\alpha}_{3}^{\intercal}\Omega\vec{\xi} =\\
\nonumber &&  \frac{1}{2}\vec{G}^{\intercal} M^{-1} \vec{G} -\frac{1}{4} \vec{\alpha}_{3}^{\intercal} \Omega^{\intercal} \Sigma_{B}\Omega \vec{\alpha}_{3} -\frac{1}{2}\vec{\xi}^{\intercal}\Omega M^{-1}\Omega^{\intercal}\vec{\xi} + i(M^{-1}\vec{G}-\vec{\alpha}_{3})^{\intercal}\Omega\vec{\xi},
\end{eqnarray}
where we have defined $\vec{G} = \frac{1}{2}\Omega^{\intercal} \sigma_{z} \varepsilon_{AB}\Omega \vec{\alpha}_{3}+i\Omega^{\intercal}\vec{\alpha}_{0}$, and integrate over $\xi$
\begin{eqnarray}
&& \int_{-\infty}^{\infty} \diff x \int_{-\infty}^{\infty} \diff p e^{-\frac{1}{2}\vec{\xi}^{\intercal}\Omega M^{-1}\Omega^{\intercal}\vec{\xi}+ i(M^{-1}\vec{G}-\vec{\alpha}_{3})^{\intercal}\Omega\vec{\xi}} =\\
\nonumber &&  \frac{2\pi}{\sqrt{\det M^{-1}}}e^{-\frac{1}{2}(M^{-1}\vec{G}-\vec{\alpha}_{3})^{\intercal}\Omega^{\intercal} (\Omega M^{-1}\Omega^{\intercal})^{-1}\Omega(M^{-1}\vec{G}-\vec{\alpha}_{3})}.
\end{eqnarray}
where we have used the fact that $\det(\Omega M\Omega^{\intercal})=\det M$. Knowing that $(\Omega M\Omega^{\intercal})^{-1} = \Omega M^{-1}\Omega^{\intercal}$, we write
\begin{equation}
\chi_{B}^{\text{out}}(\alpha_{3}) = e^{\frac{1}{2}\vec{G}^{\intercal} M^{-1} \vec{G}} e^{-\frac{1}{4}\vec{\alpha}_{3}^{\intercal} \Omega^{\intercal} \Sigma_{B}\Omega \vec{\alpha}_{3}} e^{-\frac{1}{2}(M^{-1}\vec{G}-\vec{\alpha}_{3})^{\intercal}M(M^{-1}\vec{G}-\vec{\alpha}_{3})},
\end{equation}
which simplifies to
\begin{equation}
\chi_{B}^{\text{out}}(\alpha_{3}) = e^{-\frac{1}{4}\vec{\alpha}_{3}^{\intercal} \Omega^{\intercal} (\mathbb{1}_{2} + \sigma_{z}\Sigma_{A}\sigma_{z} + \Sigma_{B} -\sigma_{z}\varepsilon_{AB}-\varepsilon_{AB}^{\intercal}\sigma_{z})\Omega \vec{\alpha}_{3}-i\vec{\alpha}_{3}^{\intercal}\Omega\vec{\alpha}_{0}}.
\end{equation}
Finally, the average fidelity of the teleportation protocol is computed as
\begin{equation}
\bar{F} = \tr[\rho^{\text{in}}\rho^{\text{out}}] = \frac{1}{\pi^{2}}\int \diff^{2}\alpha_{3}\diff^{2}\beta \chi^{\text{in}}(\beta)\chi^{\text{out}}(\alpha_{3})\tr[\hat{D}(-\beta)\hat{D}(-\alpha_{3})],
\end{equation}
and by using $\tr[\hat{D}(-\beta)\hat{D}(-\alpha_{3})]=\pi\delta(\alpha_{3}+\beta)$, we arrive at
\begin{equation}
\bar{F} = \tr[\rho^{\text{in}}\rho^{\text{out}}] = \frac{1}{\pi}\int \diff^{2}\beta \chi^{\text{in}}(-\beta)\chi^{\text{out}}(\beta).
\end{equation}
Introducing the characteristic functions, we obtain
\begin{equation}
\bar{F} = \frac{1}{\pi}\int \diff^{2}\beta e^{-\frac{1}{4}\vec{\beta}^{\intercal}\mathbb{1}_{2}\vec{\beta} + i\vec{\beta}^{\intercal}\Omega\vec{\alpha}_{0}}e^{-\frac{1}{4}\vec{\beta}^{\intercal} \Omega^{\intercal} (\mathbb{1}_{2} + \sigma_{z} \Sigma_{A} \sigma_{z} + \Sigma_{B} -\sigma_{z} \varepsilon_{AB}-\varepsilon_{AB}^{\intercal}\sigma_{z})\Omega \vec{\beta}-i\vec{\beta}^{\intercal}\Omega\vec{\alpha}_{0}}.
\end{equation}
Eventually, the average fidelity is given by
\begin{eqnarray}
\bar{F} &=& \frac{1}{\pi}\int \diff^{2}\beta e^{-\frac{1}{2}\vec{\beta}^{\intercal} \Omega^{\intercal} [\mathbb{1}_{2} + \frac{1}{2}(\sigma_{z}\Sigma_{A}\sigma_{z} + \Sigma_{B} -\sigma_{z}\varepsilon_{AB}-\varepsilon_{AB}^{\intercal}\sigma_{z})]\Omega \vec{\beta}} \\
\nonumber &=& \frac{1}{\sqrt{\det[\mathbb{1}_{2} + \frac{1}{2}(\sigma_{z}\Sigma_{A}\sigma_{z} + \Sigma_{B} -\sigma_{z}\varepsilon_{AB}-\varepsilon_{AB}^{\intercal}\sigma_{z})]}}.
\end{eqnarray}

%%%%%%%%%%%%%%%%%%%%%%%%%%%%%%%%%%%%%%%%%%%%%%%%%%%%%%%%%%%%%%%%%%%%%
%%%%%%%%%%%%%%%%%%%%%%%%%%%%%%%%%%%%%%%%%%%%%%%%%%%%%%%%%%%%%%%%%%%%%

\subsubsection{Heuristic photon subtraction}
In this case, we can start from the reduced state that Bob has after Alice has performed the homodyne measurements,
\begin{eqnarray}
\sigma_{B}(x,p) &=& \frac{1}{P(x,p)} \tr_{TA}\left[ \rho^{\text{in}}_{T}\otimes \rho_{AB}\Pi(x,p)_{TA}\right] \\
\nonumber &=&  \frac{1}{2\pi^{3}} \int \diff^{2}\alpha_{1}\int \diff^{2}\alpha_{3} \chi_{T}^{\text{in}}(\alpha_{1})\chi_{AB}(\bar{\alpha}_{1},\alpha_{3})\hat{D}_{B}(-\alpha_{3}) e^{i \vec{\alpha}_{1}^{\intercal}\Omega\vec{\xi}},
\end{eqnarray}
where the characteristic functions are again
%\begin{eqnarray}
%\chi_{T}^{\text{in}}(\alpha_{1}) &=& \exp\left[ -\frac{1}{4}\vec{\alpha}_{1}^{\intercal}\Omega^{\intercal}\mathbb{1}_{2}\Omega\vec{\alpha}_{1} - i\vec{\alpha}^{\intercal}_{1}\Omega\vec{\alpha}_{0} \right], \\
%\nonumber \chi_{AB}(\bar{\alpha}_{1},\alpha_{3}) &=& \exp\left[ -\frac{1}{4} \left( \vec{\alpha}_{1}^{\intercal}\Omega^{\intercal} \sigma_{z} A \sigma_{z}\Omega \vec{\alpha}_{1} - \vec{\alpha}_{1}^{\intercal}\Omega^{\intercal} \sigma_{z} C\Omega \vec{\alpha}_{3} - \vec{\alpha}_{3}^{\intercal} \Omega^{\intercal} C^{\intercal} \sigma_{z} \Omega \vec{\alpha}_{1} + \vec{\alpha}_{3}^{\intercal} \Omega^{\intercal} B\Omega \vec{\alpha}_{3} \right)\right],
%\end{eqnarray}
\begin{eqnarray}
\nonumber \chi_{T}^{\text{in}}(\alpha_{1})\chi_{AB}(\bar{\alpha}_{1},\alpha_{3}) &=& \exp\bigg\{ -\frac{1}{4} \big[ \vec{\alpha}_{1}^{\intercal}\Omega^{\intercal} \left(\mathbb{1}_{2}+\sigma_{z} \Sigma_{A} \sigma_{z}\right)\Omega \vec{\alpha}_{1} - \vec{\alpha}_{1}^{\intercal}\Omega^{\intercal} \sigma_{z} \varepsilon_{AB}\Omega \vec{\alpha}_{3} \\
&-& \vec{\alpha}_{3}^{\intercal} \Omega^{\intercal} \varepsilon_{AB}^{\intercal} \sigma_{z} \Omega \vec{\alpha}_{1} + \vec{\alpha}_{3}^{\intercal} \Omega^{\intercal} \Sigma_{B}\Omega \vec{\alpha}_{3} \big] - i\vec{\alpha}^{\intercal}_{1}\Omega\vec{\alpha}_{0} \bigg]\},
\end{eqnarray}
The one for the coherent state we attempt to teleport will not change, while the photon subtraction process modifies the one for the bipartite Gaussian state as
\begin{eqnarray}
\nonumber && \chi_{AB}(\bar{\alpha}_{1},\alpha_{3}) \rightarrow N \left[ \partial^{2}_{x_{1}} + \partial^{2}_{p_{1}} +\frac{x_{1}^{2}}{4} + \frac{p_{1}^{2}}{4} + x_{1}\partial_{x_{1}} + p_{1}\partial_{p_{1}} +1 \right]\times \\
&& \left[ \partial^{2}_{x_{3}} + \partial^{2}_{p_{3}} +\frac{x_{3}^{2}}{4} + \frac{p_{3}^{2}}{4} + x_{3}\partial_{x_{3}} + p_{3}\partial_{p_{3}} +1 \right] \chi_{AB}(\bar{\alpha_{1}},\alpha_{3}).
\end{eqnarray}
Notice that we are applying photon subtraction to the characteristic function $\chi_{AB}(\bar{\alpha}_{1},\alpha_{3})$ in which we have already applied ``half'' Homodyne detection, simply because when integrating over $y$ and $y'$ of the position basis in which the POVM is expressed, we get a Dirac delta which sets $\alpha_{2}=\bar{\alpha}_{1}$, and thus $\vec{\alpha}_{2} = \sigma_{z}\vec{\alpha}_{1}$. Then, we can apply photon subtraction on $\chi_{AB}(\alpha_{2},\alpha_{3})$, where the complex variable $\alpha_{1}$ is reserved for the coherent state that Alice wants to teleport, and then compute the integral over $\alpha_{2}$ with the delta functions. Alternatively, we can directly apply photon subtraction on $\chi_{AB}(\bar{\alpha}_{1},\alpha_{3})$, which is what we do here. Both procedures should in the end be equivalent.

Assume that, after applying the derivatives we obtain something that can be written as
\begin{eqnarray}
\nonumber && N \Big[ \big( m_{B} + \vec{\alpha}_{3}^{\intercal} M_{B} \vec{\alpha}_{3} + \vec{\alpha}_{1}^{\intercal}M_{BC}\vec{\alpha}_{3} + \vec{\alpha}_{1}^{\intercal} \sigma_{z} M_{C}\sigma_{z} \vec{\alpha}_{1} \big) \big( m_{A} + \vec{\alpha}_{1}^{\intercal} M_{A} \vec{\alpha}_{1} \\
\nonumber && + \vec{\alpha}_{1}^{\intercal}M_{AC}\vec{\alpha}_{3} + \vec{\alpha}_{3}^{\intercal} M_{C} \vec{\alpha}_{3} \big) + m_{C} + \vec{\alpha}_{1}^{\intercal}M_{AC}\Omega^{\intercal} \varepsilon_{AB} \sigma_{z}\Omega\vec{\alpha}_{1} + 2\vec{\alpha}_{3}^{\intercal}M_{C}\left(\mathbb{1}_{2} - \Omega^{\intercal} \Sigma_{B}\Omega\right)\vec{\alpha}_{3} \\
&&+ \vec{\alpha}_{1}^{\intercal}\left[M_{AC}\left(\mathbb{1}_{2}-\Omega^{\intercal} \Sigma_{B}\Omega\right) + \Omega^{\intercal} \sigma_{z}\varepsilon_{AB}\Omega M_{C}\right]\vec{\alpha}_{3} \Big] \chi_{AB}(\bar{\alpha_{1}},\alpha_{3}),
\end{eqnarray}
where we have defined 
\begin{eqnarray}
\nonumber m_{A} &=& 1 - \frac{1}{2}\tr \Sigma_{A}, \\
\nonumber m_{B} &=& 1 - \frac{1}{2}\tr \Sigma_{B}, \\
m_{C} &=& \frac{1}{2}\tr \left(\varepsilon_{AB}^{\intercal}\varepsilon_{AB}\right), \\
\nonumber  M_{A} &=& \frac{1}{4}\left( \mathbb{1}_{2} -2\Omega^{\intercal} \sigma_{z}\Sigma_{A}\sigma_{z}\Omega + \Omega^{\intercal} \sigma_{z}\Sigma_{A}^{2}\sigma_{z}\Omega\right),
\end{eqnarray}

\begin{eqnarray}
\nonumber M_{B} &=& \frac{1}{4}\left( \mathbb{1}_{2} -2\Omega^{\intercal} \Sigma_{B}\Omega + \Omega^{\intercal} \Sigma_{B}^{2}\Omega\right), \\
M_{C} &=& \frac{1}{4}\Omega^{\intercal} \varepsilon_{AB}^{\intercal} \varepsilon_{AB}\Omega, \\
\nonumber M_{AC} &=& \frac{1}{2}\left( \Omega^{\intercal} \sigma_{z}\varepsilon_{AB}\Omega - \Omega^{\intercal} \sigma_{z}AC\Omega\right), \\
\nonumber M_{BC} &=& \frac{1}{2}\left( \Omega^{\intercal} \sigma_{z}\varepsilon_{AB}\Omega - \Omega^{\intercal} \sigma_{z}CB\Omega\right).
\end{eqnarray}
Notice that all matrices are symmetric, except for $M_{AC}$ and $M_{BC}$. The normalization constant is given by $N^{-1}=m_{A}m_{B}+m_{C}$. 

Let's start integrating over $\alpha_{1}$. Recall that
\begin{equation}
\chi_{T}^{\text{in}}(\alpha_{1})\chi_{AB}(\bar{\alpha}_{1},\alpha_{3})e^{i\sqrt{2}(p\mathbb{Re}\alpha_{1}-x\mathbb{Im}\alpha_{1})} = e^{-\frac{1}{2}\vec{\alpha}_{1}^{\intercal}X\vec{\alpha}_{1} + \vec{\alpha}_{1}^{\intercal}\vec{J}},
\end{equation}
where we defined $X = \frac{1}{2}\Omega^{\intercal}(\mathbb{1}_{2} + \sigma_{z} \Sigma_{A} \sigma_{z})\Omega$ and $\vec{J} = \frac{1}{2}\Omega^{\intercal} \sigma_{z} \varepsilon_{AB}\Omega \vec{\alpha}_{3} +i\Omega^{\intercal}(\vec{\alpha}_{0}-\vec{\xi})$. First, we integrate the free terms
\begin{eqnarray}
&& \int \diff^{2}\alpha_{1} \Big[ m_{A}m_{B} + m_{C}+\vec{\alpha}_{3}^{\intercal}\left( m_{A}M_{B}+(2+m_{B})M_{C}\right)\vec{\alpha}_{3} \\
\nonumber && + \left(\vec{\alpha}_{3}^{\intercal} M_{B}\vec{\alpha}_{3}\right)\left(\vec{\alpha}_{3}^{\intercal} M_{C}\vec{\alpha}_{3}\right) \Big] e^{-\frac{1}{2}\vec{\alpha}_{1}^{\intercal}X\vec{\alpha}_{1} + \vec{\alpha}_{1}^{\intercal}\vec{J}} = \frac{\pi}{\sqrt{\det X}} \Big[ m_{A}m_{B} + m_{C} \\ 
\nonumber && +\vec{\alpha}_{3}^{\intercal}\left( m_{A}M_{B}+(2+m_{B}M_{C})\right)\vec{\alpha}_{3} + \left(\vec{\alpha}_{3}^{\intercal} M_{B}\vec{\alpha}_{3}\right)\left(\vec{\alpha}_{3}^{\intercal} M_{C}\vec{\alpha}_{3}\right) \Big] e^{\frac{1}{2}\vec{J}^{\intercal}X^{-1}\vec{J}}.
\end{eqnarray}
We continue with the terms 
\begin{eqnarray}
\nonumber && \int \diff^{2}\alpha_{1} \Big[ \vec{\alpha}_{1}^{\intercal}\left( m_{A}M_{BC}+(1+m_{B})M_{AC}\right)\vec{\alpha}_{3} + \left(\vec{\alpha}_{3}^{\intercal} M_{B}\vec{\alpha}_{3}\right)\left(\vec{\alpha}_{1}^{\intercal} M_{AC}\vec{\alpha}_{3}\right) \\
\nonumber && + \left(\vec{\alpha}_{1}^{\intercal} M_{BC}\vec{\alpha}_{3}\right)\left(\vec{\alpha}_{3}^{\intercal} M_{C}\vec{\alpha}_{3}\right) \Big] e^{-\frac{1}{2}\vec{\alpha}_{1}^{\intercal}X\vec{\alpha}_{1} + \vec{\alpha}_{1}^{\intercal}\vec{J}} \\ 
\nonumber && = \frac{\pi}{2\sqrt{\det X}} \Big[ \vec{J}^{\intercal}X^{-1}(m_{A}M_{BC}+(1+m_{B}M_{AC}))\vec{\alpha}_{3} \\
\nonumber && + \vec{\alpha}_{3}^{\intercal}(m_{A}M_{BC}+(1+m_{B}M_{AC}))^{\intercal}X^{-1}\vec{J} \\
\nonumber &&+ \left(\vec{\alpha}_{3}^{\intercal} M_{B}\vec{\alpha}_{3}\right) (\vec{J}^{\intercal}X^{-1}M_{AC}\vec{\alpha}_{3} + \vec{\alpha}_{3}^{\intercal}M_{AC}^{\intercal}X^{-1}\vec{J} ) \\
&& + (\vec{J}^{\intercal}X^{-1}M_{BC}\vec{\alpha}_{3} + \vec{\alpha}_{3}^{\intercal}M_{BC}^{\intercal}X^{-1}\vec{J} )\left(\vec{\alpha}_{3}^{\intercal} M_{C}\vec{\alpha}_{3}\right) \Big] e^{\frac{1}{2}\vec{J}^{\intercal}X^{-1}\vec{J}}.
\end{eqnarray}
Also, we integrate the first order terms
\begin{eqnarray}
\nonumber && \int \diff^{2}\alpha_{1} \Big[ \vec{\alpha}_{1}^{\intercal}\left( m_{A}\sigma_{z}M_{C}\sigma_{z}+m_{B}M_{A}\right)\vec{\alpha}_{1} + \left(\vec{\alpha}_{3}^{\intercal} M_{B}\vec{\alpha}_{3}\right)\left(\vec{\alpha}_{1}^{\intercal} M_{A}\vec{\alpha}_{1}\right) \\
\nonumber && + \left(\vec{\alpha}_{1}^{\intercal}\sigma_{z}M_{C}\sigma_{z}\vec{\alpha}_{1}\right)\left(\vec{\alpha}_{3}^{\intercal} M_{C}\vec{\alpha}_{3}\right) \Big] e^{-\frac{1}{2}\vec{\alpha}_{1}^{\intercal}X\vec{\alpha}_{1} + \vec{\alpha}_{1}^{\intercal}\vec{J}} \\ 
\nonumber && = \frac{\pi}{\sqrt{\det X}} \Big[ \tr\left[ X^{-1}\left( m_{A}\sigma_{z}M_{C}\sigma_{z}+m_{B}M_{A} \right)\right] + \vec{J}^{\intercal}W_{X,m_{A}\sigma_{z}M_{C}\sigma_{z}+m_{B}M_{A}}\vec{J} \\
\nonumber && + \left(\vec{\alpha}_{3}^{\intercal} M_{B}\vec{\alpha}_{3}\right) \left[\tr(X^{-1}M_{A}) \vec{J}^{\intercal}W_{X,M_{A}}\vec{J} \right]\\
&& +  \left[\tr(X^{-1}\sigma_{z}M_{C}\sigma_{z}) + \vec{J}^{\intercal}W_{X,\sigma_{z}M_{C}\sigma_{z}}\vec{J} \right] \left(\vec{\alpha}_{3}^{\intercal} M_{C}\vec{\alpha}_{3}\right) \Big] e^{\frac{1}{2}\vec{J}^{\intercal}X^{-1}\vec{J}}.
\end{eqnarray}
Now, it is time to integrate the second order terms,
\begin{eqnarray}
&& \int \diff^{2}\alpha_{1} \left(\vec{\alpha}_{1}^{\intercal}M_{BC}\vec{\alpha}_{3}\right)\left(\vec{\alpha}_{1}^{\intercal}M_{AC}\vec{\alpha}_{3}\right) e^{-\frac{1}{2}\vec{\alpha}_{1}^{\intercal}X\vec{\alpha}_{1} + \vec{\alpha}_{1}^{\intercal}\vec{J}} =\\
\nonumber && \frac{\pi}{2\sqrt{\det X}} \Big[ \vec{\alpha}_{3}^{\intercal}M_{BC}^{\intercal}X^{-1}M_{AC}\vec{\alpha}_{3} + \vec{\alpha}_{3}^{\intercal}M_{AC}^{\intercal}X^{-1}M_{BC}\vec{\alpha}_{3}^{\intercal} \\
\nonumber && + \frac{1}{2}\left( \vec{\alpha}_{3}^{\intercal}M_{BC}^{\intercal}X^{-1}\vec{J}^{\intercal} + \vec{J}^{\intercal}X^{-1}M_{BC}\vec{\alpha}_{3}\right) \times \\
&& \left( \vec{\alpha}_{3}^{\intercal}M_{AC}^{\intercal}X^{-1}\vec{J} + \vec{J}^{\intercal}X^{-1}M_{AC}\vec{\alpha}_{3}\right) \Big] e^{\frac{1}{2}\vec{J}^{\intercal}X^{-1}\vec{J}},
\end{eqnarray}
and also
\begin{eqnarray}
\nonumber && \int \diff^{2}\alpha_{1} \left(\vec{\alpha}_{1}^{\intercal}\sigma_{z}M_{C}\sigma_{z}\vec{\alpha}_{1}\right)\left(\vec{\alpha}_{1}^{\intercal}M_{A}\vec{\alpha}_{1}\right) e^{-\frac{1}{2}\vec{\alpha}_{1}^{\intercal}X\vec{\alpha}_{1} + \vec{\alpha}_{1}^{\intercal}\vec{J}} =\\
\nonumber && \frac{\pi}{\sqrt{\det X}}\bigg[ 3\tr(X^{-1}M_{A})\tr(X^{-1}\sigma_{z}M_{C}\sigma_{z}) -\frac{2}{\det X}\tr(\Omega^{\intercal} \sigma_{z}M_{C}\sigma_{z}\Omega M_{A}) \\
\nonumber &+& \vec{J}^{\intercal}W_{X,M_{A}}\vec{J}\vec{J}^{\intercal}W_{X,\sigma_{z}M_{C}\sigma_{z}}\vec{J}+  \vec{J}^{\intercal}\bigg( 3W_{X,M_{A}}\tr(X^{-1}\sigma_{z}M_{C}\sigma_{z}) \\
\nonumber && + 3W_{X,\sigma_{z}M_{C}\sigma_{z}}\tr(X^{-1}M_{A}) \\
&& - \frac{2X^{-1}}{\det X}\tr(\Omega^{\intercal} \sigma_{z}M_{C}\sigma_{z}\Omega M_{A})\bigg)\vec{J}  \bigg]e^{\frac{1}{2}\vec{J}^{\intercal}X^{-1}\vec{J}}.
\end{eqnarray}
Finally, we integrate the second order cross terms,
\begin{eqnarray}
\nonumber && \int \diff^{2}\alpha_{1} \left[ \left(\vec{\alpha}_{1}^{\intercal}\sigma_{z}M_{C}\sigma_{z}\vec{\alpha}_{1}\right)\left(\vec{\alpha}_{1}^{\intercal}M_{AC}\vec{\alpha}_{3}\right) + \left(\vec{\alpha}_{1}^{\intercal}M_{BC}\vec{\alpha}_{3}\right)\left(\vec{\alpha}_{1}^{\intercal}M_{A}\vec{\alpha}_{1}\right) \right]e^{-\frac{1}{2}\vec{\alpha}_{1}^{\intercal}X\vec{\alpha}_{1} + \vec{\alpha}_{1}^{\intercal}\vec{J}} \\
\nonumber && = \frac{\pi}{\sqrt{\det X}}\Big[ \vec{\alpha}_{3}^{\intercal}M_{AC}^{\intercal}W_{X,\sigma_{z}M_{C}\sigma_{z}} \vec{J} + \vec{J}^{\intercal}W_{X,\sigma_{z}M_{C}\sigma_{z}}M_{AC}\vec{\alpha}_{3}  \\
&& + \vec{\alpha}_{3}^{\intercal}M_{BC}^{\intercal}W_{X,M_{A}} \vec{J} + \vec{J}^{\intercal}W_{X,M_{A}}M_{BC}\vec{\alpha}_{3} \\
\nonumber && + \frac{1}{2}\left( \tr(X^{-1}\sigma_{z}M_{C}\sigma_{z}) + \vec{J}^{\intercal}W_{X,\sigma_{z}M_{C}\sigma_{z}}\vec{J}\right)\left(\vec{\alpha}_{3}^{\intercal}M_{AC}^{\intercal}X^{-1}\vec{J} + \vec{J}^{\intercal}X^{-1} M_{AC}\vec{\alpha}_{3}\right) \\
\nonumber && + \frac{1}{2}\left( \tr(X^{-1}M_{A}) + \vec{J}^{\intercal}W_{X,M_{A}}\vec{J}\right)\left(\vec{\alpha}_{3}^{\intercal}M_{BC}^{\intercal}X^{-1}\vec{J} + \vec{J}^{\intercal}X^{-1} M_{BC}\vec{\alpha}_{3}\right) \Big]e^{\frac{1}{2}\vec{J}^{\intercal}X^{-1}\vec{J}}.
\end{eqnarray}
In order to simplify things, from now on we will consider that the matrices $M_{AC}$ and $M_{BC}$ are symmetric, which implies that we assume $[\Sigma_{A},\varepsilon_{AB}]=0$ and $[\Sigma_{B},\varepsilon_{AB}]=0$.

Summing together all the terms resulting from the integral, we obtain
\begin{eqnarray}
\nonumber && \frac{\pi e^{\frac{1}{2}\vec{J}^{\intercal}X^{-1}\vec{J}}}{\sqrt{\det X}} \bigg\{ m_{A}m_{B}+m_{C}+\vec{\alpha}_{3}^{\intercal}\left[ m_{A}M_{B}+(2+m_{B}M_{C})\right]\vec{\alpha}_{3} \\
\nonumber && \vec{J}^{\intercal}X^{-1}[m_{A}M_{BC}+(1+m_{B}M_{AC})]\vec{\alpha}_{3} + \left(\vec{\alpha}_{3}^{\intercal} M_{B}\vec{\alpha}_{3}\right)\vec{J}^{\intercal}X^{-1}M_{AC}\vec{\alpha}_{3} \\
\nonumber && + \vec{J}^{\intercal}X^{-1}M_{BC}\vec{\alpha}_{3}\left(\vec{\alpha}_{3}^{\intercal} M_{C}\vec{\alpha}_{3}\right) + \tr\left[ X^{-1}\left( m_{A}\sigma_{z}M_{C}\sigma_{z}+m_{B}M_{A} \right)\right]  \\
\nonumber && + \vec{J}^{\intercal}W_{X,m_{A}\sigma_{z}M_{C}\sigma_{z}+m_{B}M_{A}}\vec{J} + \left(\vec{\alpha}_{3}^{\intercal} M_{B}\vec{\alpha}_{3}\right) \left[\tr(X^{-1}M_{A}) \vec{J}^{\intercal}W_{X,M_{A}}\vec{J} \right] \\
\nonumber && +  \left[\tr(X^{-1}\sigma_{z}M_{C}\sigma_{z}) \vec{J}^{\intercal}W_{X,\sigma_{z}M_{C}\sigma_{z}}\vec{J} \right] \left(\vec{\alpha}_{3}^{\intercal} M_{C}\vec{\alpha}_{3}\right)  + \vec{\alpha}_{3}^{\intercal}M_{BC}X^{-1}M_{AC}\vec{\alpha}_{3} \\
\nonumber && + \left(\vec{J}^{\intercal}X^{-1}M_{BC}\vec{\alpha}_{3}\right) \left( \vec{J}^{\intercal}X^{-1}M_{AC}\vec{\alpha}_{3}\right) + 3\tr(X^{-1}M_{A})\tr(X^{-1}\sigma_{z}M_{C}\sigma_{z}) \\
\nonumber && -\frac{2}{\det X}\tr(\Omega^{\intercal} \sigma_{z}M_{C}\sigma_{z}\Omega M_{A}) + \vec{J}^{\intercal}W_{X,M_{A}}\vec{J}\vec{J}^{\intercal}W_{X,\sigma_{z}M_{C}\sigma_{z}}\vec{J} \\
\nonumber &+& \vec{J}^{\intercal}\Big( 3W_{X,M_{A}}\tr(X^{-1}\sigma_{z}M_{C}\sigma_{z}) + 3W_{X,\sigma_{z}M_{C}\sigma_{z}}\tr(X^{-1}M_{A})  \\
\nonumber && - \frac{2X^{-1}}{\det X}\tr(\Omega^{\intercal} \sigma_{z}M_{C}\sigma_{z}\Omega M_{A})\Big)\vec{J} + 2\vec{J}^{\intercal}W_{X,\sigma_{z}M_{C}\sigma_{z}}M_{AC}\vec{\alpha}_{3} + 2\vec{J}^{\intercal}W_{X,M_{A}}M_{BC}\vec{\alpha}_{3} \\
&& + \left[ \tr(X^{-1}\sigma_{z}M_{C}\sigma_{z}) + \vec{J}^{\intercal}W_{X,\sigma_{z}M_{C}\sigma_{z}}\vec{J}\right]\left(\vec{J}^{\intercal}X^{-1} M_{AC}\vec{\alpha}_{3}\right) \\
\nonumber && + \left(\vec{\alpha}_{3}^{\intercal} M_{B}\vec{\alpha}_{3}\right)\left(\vec{\alpha}_{3}^{\intercal} M_{C}\vec{\alpha}_{3}\right) + \left[ \tr(X^{-1}M_{A}) + \vec{J}^{\intercal}W_{X,M_{A}}\vec{J}\right]\left(\vec{J}^{\intercal}X^{-1} M_{BC}\vec{\alpha}_{3}\right) \bigg\}
\end{eqnarray}
We can regroup these terms by their dependence on $\alpha_{3}$ and $J$,
\begin{eqnarray}
\nonumber H_{0} &=& m_{A}m_{B}+m_{C} +  \tr\left[ X^{-1}\left( m_{A}\sigma_{z}M_{C}\sigma_{z}+m_{B}M_{A} \right)\right] \\
\nonumber &+& 3\tr(X^{-1}M_{A})\tr(X^{-1}\sigma_{z}M_{C}\sigma_{z}) -\frac{2}{\det X}\tr(\Omega \sigma_{z}M_{C}\sigma_{z}\Omega^{T}M_{A}), \\
\nonumber &+& \vec{\alpha}_{3}^{\intercal}\left[ m_{A}M_{B}+(2+m_{B}M_{C}) + M_{BC}X^{-1}M_{AC}\right]\vec{\alpha}_{3} \\
&+& \left(\vec{\alpha}_{3}^{\intercal} M_{B}\vec{\alpha}_{3}\right)\left(\vec{\alpha}_{3}^{\intercal} M_{C}\vec{\alpha}_{3}\right), 
\end{eqnarray}

\begin{eqnarray}
\nonumber \vec{J}^{\intercal} \vec{H}_{1} &=& \vec{J}^{\intercal}X^{-1}\Big[m_{A}M_{BC}+(1+m_{B}M_{AC}) + M_{AC}\tr(X^{-1}\sigma_{z}M_{C}\sigma_{z}) \\
\nonumber &+& M_{BC}\tr(X^{-1}M_{A})\Big]\vec{\alpha}_{3} + 2 \vec{J}^{\intercal}(W_{X,\sigma_{z}M_{C}\sigma_{z}}M_{AC} + W_{X,M_{A}}M_{BC})\vec{\alpha}_{3} \\
&+& \left(\vec{\alpha}_{3}^{\intercal} M_{B}\vec{\alpha}_{3}\right)\vec{J}^{\intercal}X^{-1}M_{AC}\vec{\alpha}_{3} + \vec{J}^{\intercal}X^{-1}M_{BC}\vec{\alpha}_{3}\left(\vec{\alpha}_{3}^{\intercal} M_{C}\vec{\alpha}_{3}\right),
\end{eqnarray}

\begin{eqnarray}
\nonumber \vec{J}^{\intercal}H_{2}\vec{J} &=& \vec{J}^{\intercal}\Big[ W_{X,m_{A}\sigma_{z}M_{C}\sigma_{z}+m_{B}M_{A}} + 3W_{X,M_{A}}\tr(X^{-1}\sigma_{z}M_{C}\sigma_{z}) \\
\nonumber &+& 3W_{X,\sigma_{z}M_{C}\sigma_{z}}\tr(X^{-1}M_{A}) - \frac{2X^{-1}}{\det X}\tr(\Omega^{\intercal} \sigma_{z}M_{C}\sigma_{z}\Omega M_{A})\Big]\vec{J} \\
\nonumber &+& \left(\vec{\alpha}_{3}^{\intercal} M_{B}\vec{\alpha}_{3}\right) \left[\tr(X^{-1}M_{A}) \vec{J}^{\intercal}W_{X,M_{A}}\vec{J} \right] \\
&+& \left[\tr(X^{-1}\sigma_{z}M_{C}\sigma_{z}) \vec{J}^{\intercal}W_{X,\sigma_{z}M_{C}\sigma_{z}}\vec{J} \right] \left(\vec{\alpha}_{3}^{\intercal} M_{C}\vec{\alpha}_{3}\right), \\
\nonumber \vec{J}^{\intercal}\vec{H}_{3} \vec{J}^{\intercal}\vec{K}_{3} &=& \left(\vec{J}^{\intercal}X^{-1}M_{BC}\vec{\alpha}_{3}\right) \left( \vec{J}^{\intercal}X^{-1}M_{AC}\vec{\alpha}_{3}\right), \\
\nonumber H_{4} &=& \left(\vec{J}^{\intercal}W_{X,M_{A}}\vec{J}\right)\left(\vec{J}^{\intercal}W_{X,\sigma_{z}M_{C}\sigma_{z}}\vec{J}\right), \\
\nonumber H_{5} &=& \left( \vec{J}^{\intercal}W_{X,\sigma_{z}M_{C}\sigma_{z}}\vec{J}\right)\left(\vec{J}^{\intercal}\vec{K}_{3}\right) + \left( \vec{J}^{\intercal}W_{X,M_{A}}\vec{J}\right)\left(\vec{J}^{\intercal}\vec{H}_{3}\right).
\end{eqnarray}
Now, we must integrate over all possible results from the Homodyne detection process, $x$ and $p$. Instead, we will integrate over $J_{x}$ and $J_{p}$, such that $\diff x \diff p = \diff J_{x} \diff J_{p}$. Let us begin by integrating $H_{0}$,
\begin{equation}
\int \diff J_{x} \diff J_{p} H_{0} e^{\frac{1}{2}\vec{J}^{\intercal}X^{-1}\vec{J}-\vec{J}^{\intercal}\vec{\alpha}_{3}} = \frac{2\pi}{\sqrt{\det X^{-1}}} e^{-\frac{1}{2}\vec{\alpha}_{3}^{\intercal}X\vec{\alpha}_{3}} H_{0}
\end{equation}
and continue with
\begin{equation}
\int \diff J_{x} \diff J_{p} \vec{J}^{\intercal}\vec{H}_{1} e^{\frac{1}{2}\vec{J}^{\intercal}X^{-1}\vec{J}-\vec{J}^{\intercal}\vec{\alpha}_{3}} = \frac{2\pi}{\sqrt{\det X^{-1}}} e^{-\frac{1}{2}\vec{\alpha}_{3}^{\intercal}X\vec{\alpha}_{3}} \vec{H}_{1}^{\intercal}X\vec{\alpha}_{3}.
\end{equation}
Then, we compute the integral
\begin{eqnarray}
\nonumber && \int \diff J_{x} \diff J_{p} \vec{J}^{\intercal}H_{2}\vec{J} e^{\frac{1}{2}\vec{J}^{\intercal}X^{-1}\vec{J}-\vec{J}^{\intercal}\vec{\alpha}_{3}} = \\
&& \frac{2\pi}{\sqrt{\det X^{-1}}} e^{-\frac{1}{2}\vec{\alpha}_{3}^{\intercal}X\vec{\alpha}_{3}} \left[ -\tr\left(X H_{2}\right) + \vec{\alpha}_{3}^{\intercal}W_{X^{-1},H_{2}}\vec{\alpha}_{3} \right].
\end{eqnarray}
together with
\begin{eqnarray}
\nonumber && \int \diff J_{x} \diff J_{p} \vec{J}^{\intercal}\vec{H}_{3}\vec{J}^{\intercal}\vec{K}_{3} e^{\frac{1}{2}\vec{J}^{\intercal}X^{-1}\vec{J}-\vec{J}^{\intercal}\vec{\alpha}_{3}} = \\
&& \frac{2\pi}{\sqrt{\det X^{-1}}} e^{-\frac{1}{2}\vec{\alpha}_{3}^{\intercal}X\vec{\alpha}_{3}} \left[ -\vec{H}_{3}^{\intercal}X\vec{K}_{3} + \left(\vec{\alpha}_{3}^{\intercal}X\vec{H}_{3}\right)\left(\vec{\alpha}_{3}^{\intercal}X\vec{K}_{3}\right) \right],
\end{eqnarray}
followed by
\begin{eqnarray}
\nonumber && \int \diff J_{x} \diff J_{p} \vec{J}^{\intercal}W_{X,M_{A}}\vec{J}\vec{J}^{\intercal}W_{X,\sigma_{z}M_{C}\sigma_{z}}\vec{J} e^{\frac{1}{2}\vec{J}^{\intercal}X^{-1}\vec{J}-\vec{J}^{\intercal}\vec{\alpha}_{3}} = \frac{2\pi}{\sqrt{\det X^{-1}}} e^{-\frac{1}{2}\vec{\alpha}_{3}^{\intercal}X\vec{\alpha}_{3}} \times \\
\nonumber && \Big[ 3\tr\left(X W_{X,M_{A}}\right)\tr\left(X W_{X,\sigma_{z}M_{C}\sigma_{z}}\right) -\frac{2}{\det X^{-1}}\tr\left( \Omega^{\intercal} W_{X,\sigma_{z}M_{C}\sigma_{z}}\Omega W_{X,M_{A}}\right) \\
\nonumber && + \vec{\alpha}_{3}^{\intercal}W_{X^{-1},W_{X,M_{A}}}\vec{\alpha}_{3}\vec{\alpha}_{3}^{\intercal}W_{X^{-1},W_{X,\sigma_{z}M_{C}\sigma_{z}}}\vec{\alpha}_{3} - \vec{\alpha}_{3}^{\intercal}\Big(3W_{X^{-1},W_{X,M_{A}}}\tr\left(X W_{X,\sigma_{z}M_{C}\sigma_{z}}\right) \\
\nonumber && + 3W_{X^{-1},W_{X,\sigma_{z}M_{C}\sigma_{z}}}\tr\left(X W_{X,M_{A}}\right) -\frac{2X}{\det X^{-1}}\tr\left( \Omega^{\intercal} W_{X,\sigma_{z}M_{C}\sigma_{z}}\Omega W_{X,M_{A}}\right) \Big)\vec{\alpha}_{3} \Big].
\end{eqnarray}
Finally, we compute
\begin{eqnarray}
\nonumber && \int \diff J_{x} \diff J_{p} \left[ \left( \vec{J}^{\intercal}W_{X,\sigma_{z}M_{C}\sigma_{z}}\vec{J}\right)\left(\vec{J}^{\intercal}\vec{K}_{3}\right) + \left( \vec{J}^{\intercal}W_{X,M_{A}}\vec{J}\right)\left(\vec{J}^{\intercal}\vec{H}_{3}\right) \right] \times  \\
\nonumber && e^{\frac{1}{2}\vec{J}^{\intercal}X^{-1}\vec{J}-\vec{J}^{\intercal}\vec{\alpha}_{3}} = \frac{2\pi}{\sqrt{\det X^{-1}}} e^{-\frac{1}{2}\vec{\alpha}_{3}^{\intercal}X\vec{\alpha}_{3}} \bigg\{ -2\vec{K}_{3}^{\intercal}W_{X^{-1},W_{X,\sigma_{z}M_{C}\sigma_{z}}}\vec{\alpha}_{3} \\
&& + \left[ -\tr\left( X W_{X,\sigma_{z}M_{C}\sigma_{z}}\right) + \vec{\alpha}_{3}^{\intercal}W_{X^{-1},W_{X,\sigma_{z}M_{C}\sigma_{z}}}\vec{\alpha}_{3} \right]\vec{K}_{3}^{\intercal}X\vec{\alpha}_{3} \\
\nonumber && -2\vec{H}_{3}^{\intercal}W_{X^{-1},W_{X,M_{A}}}\vec{\alpha}_{3} + \left[ -\tr\left( X W_{X,M_{A}}\right) + \vec{\alpha}_{3}^{\intercal}W_{X^{-1},W_{X,M_{A}}}\vec{\alpha}_{3} \right]\vec{H}_{3}^{\intercal}X\vec{\alpha}_{3} \bigg\}.
\end{eqnarray}
Putting everything together, we find that
\begin{eqnarray}
\nonumber && \chi_{B}^{\text{out}}(\alpha_{3}) = \frac{e^{-\frac{1}{4}\vec{\alpha}_{3}^{\intercal} \Omega^{\intercal} (\mathbb{1}_{2} + \sigma_{z}\Sigma_{A}\sigma_{z} + \Sigma_{B} -\sigma_{z}\varepsilon_{AB}-\varepsilon_{AB}^{T}\sigma_{z})\Omega \vec{\alpha}_{3}-i\vec{\alpha}_{3}^{\intercal}\Omega\vec{\alpha}_{0}}}{m_{A}m_{B} + m_{C}} \times \\
\nonumber && \Big\{ \left[ m_{B} + \vec{\alpha}_{3}^{\intercal} \left(M_{B} + M_{BC} + \sigma_{z}M_{C}\sigma_{z} \right) \vec{\alpha}_{3} \right] \left[ m_{A} + \vec{\alpha}_{3}^{\intercal} \left(M_{A} + M_{AC} + M_{C}\right) \vec{\alpha}_{3} \right] \\
 && m_{C} +  \vec{\alpha}_{3}^{\intercal}\left(2M_{C}+M_{AC}\right)\Omega^{\intercal}\left(\mathbb{1}_{2} + \varepsilon_{AB} \sigma_{z} - \Sigma_{B}\right)\Omega\vec{\alpha}_{3} \Big\} ,
\end{eqnarray}
which is exactly the result of applying the differential operators of photon subtraction to the characteristic function, but changing $\alpha_{1}\rightarrow \alpha_{3}$. Then, we can define
\begin{eqnarray}
\nonumber E_{0} &=& m_{A}m_{B} + m_{C}, \\
\nonumber E_{1} &=& m_{A}\left( M_{B} + \sigma_{z}M_{C}\sigma_{z} + M_{BC}\right) + m_{B}\left( M_{A} + M_{C} + M_{AC}\right) \\
&+& \left(2M_{C}+M_{AC}\right)\Omega^{\intercal}\left( \mathbb{1}_{2} + \sigma_{z}\varepsilon_{AB} - \Sigma_{B} \right)\Omega \\
\nonumber E_{2}^{A} &=& M_{C} + M_{AC} + M_{A}, \\
\nonumber E_{2}^{B} &=& M_{B} + M_{BC} + \sigma_{z}M_{C}\sigma_{z},
\end{eqnarray}
and write a shorter formula, in order to integrate
\begin{eqnarray}
\nonumber \bar{F} &=& \frac{1}{\pi}\int \diff^{2}\beta e^{-\frac{1}{4}\vec{\beta}^{\intercal} \Omega^{\intercal}\Gamma\Omega \vec{\beta}}\frac{1}{E_{0}}\Big( E_{0} + \vec{\beta}^{\intercal}E_{1}\vec{\beta} +  \vec{\beta}^{\intercal}E_{2}^{A}\vec{\beta} \vec{\beta}^{\intercal}E_{2}^{B}\vec{\beta}\Big) \\
\nonumber &=& \frac{1}{E_{0}\sqrt{\det\Gamma}}\Big[ E_{0} + \tr\left(\Omega\Gamma^{-1}\Omega^{T}E_{1}\right) + 3\tr\left( \Omega\Gamma^{-1}\Omega^{T}E_{2}^{A}\right)\tr\left( \Omega\Gamma^{-1}\Omega^{T}E_{2}^{B}\right) \\
&-& \frac{2}{\det\Gamma}\tr\left(\Omega E_{2}^{A}\Omega^{T}E_{2}^{B}\right)\Big].
\end{eqnarray}

%%%%%%%%%%%%%%%%%%%%%%%%%%%%%%%%%%%%%%%%%%%%%%%%%%%%%%%%%%%%%%%%%%%%%
%%%%%%%%%%%%%%%%%%%%%%%%%%%%%%%%%%%%%%%%%%%%%%%%%%%%%%%%%%%%%%%%%%%%%

\subsubsection{Probabilistic photon subtraction (beam splitters \& photocounters)}
Assume two parties want to perform quantum teleportation using a shared bipartite gaussian entangled resource with covariance matrix
\begin{equation}
\Sigma_{AB} = \begin{pmatrix} \Sigma_{A} & \varepsilon_{AB} \\ \varepsilon_{AB}^{\intercal} & \Sigma_{B} \end{pmatrix},
\end{equation}
where $\Sigma_{A}$, $\Sigma_{B}$, and $\varepsilon_{AB}$ are $2\times 2$ symmetric matrices. In order to improve the entanglement of this resource, we want to perform entanglement distillation through a photon subtraction procedure. This implies sending both modes of the state through low-reflectivity beam splitters, where they become mixed with two ancillary modes $C$ and $D$, both vacuum states. The global covariance matrix is then given by
\begin{equation}
\Sigma_{ACBD} = \begin{pmatrix} \Sigma_{A} & 0 & \varepsilon_{AB} & 0 \\ 0 & \mathbb{1}_{2} & 0 & 0 \\ \varepsilon_{AB}^{\intercal} & 0 & \Sigma_{B} & 0 \\ 0 & 0 & 0 & \mathbb{1}_{2} \end{pmatrix},
\end{equation}
and after combining modes $A$ with $C$, and $B$ with $D$ by identical beam splitter with reflectivity $1-\tau$ this matrix becomes
%\begin{eqnarray}
%&& \hat{B}\sigma_{ACBD}\hat{B}^{\dagger} = \\
%\nonumber && \begin{pmatrix} \tau A +(1-\tau)\mathbb{1}_{2} & (\mathbb{1}_{2}-A)\sqrt{\tau(1-\tau)} & \tau C & -C\sqrt{\tau(1-\tau)} \\ (\mathbb{1}_{2}-A)\sqrt{\tau(1-\tau)} & (1-\tau) A +\tau\mathbb{1}_{2} & -C\sqrt{\tau(1-\tau)} & (1-\tau)C \\ \tau C & -C\sqrt{\tau(1-\tau)} & \tau B +(1-\tau)\mathbb{1}_{2} & (\mathbb{1}_{2}-B)\sqrt{\tau(1-\tau)} \\ -C\sqrt{\tau(1-\tau)} & (1-\tau)C & (\mathbb{1}_{2}-B)\sqrt{\tau(1-\tau)} & (1-\tau) B +\tau\mathbb{1}_{2} \end{pmatrix}.
%\end{eqnarray}
\begin{equation}
\hat{B}\sigma_{ACBD}\hat{B}^{\dagger} = \begin{pmatrix} \Sigma_{A}' & \varepsilon_{AB}' \\ \varepsilon_{AB}'^{\intercal} & \Sigma_{B}' \end{pmatrix},
\end{equation}
where we have identified
\begin{eqnarray}
\nonumber \Sigma_{A}' &=& \begin{pmatrix} \tau \Sigma_{A} +(1-\tau)\mathbb{1}_{2} & (\mathbb{1}_{2}-\Sigma_{A})\sqrt{\tau(1-\tau)} \\ (\mathbb{1}_{2}-\Sigma_{A})\sqrt{\tau(1-\tau)} & (1-\tau) \Sigma_{A} +\tau\mathbb{1}_{2} \end{pmatrix}, \\
\Sigma_{B}' &=& \begin{pmatrix} \tau \Sigma_{B} +(1-\tau)\mathbb{1}_{2} & (\mathbb{1}_{2}-\Sigma_{B})\sqrt{\tau(1-\tau)} \\ (\mathbb{1}_{2}-\Sigma_{B})\sqrt{\tau(1-\tau)} & (1-\tau) \Sigma_{B} +\tau\mathbb{1}_{2} \end{pmatrix}, \\
\nonumber \varepsilon_{AB}' &=& \begin{pmatrix} \tau \varepsilon_{AB} & -\varepsilon_{AB}\sqrt{\tau(1-\tau)} \\ -\varepsilon_{AB}\sqrt{\tau(1-\tau)} & (1-\tau)\varepsilon_{AB} \end{pmatrix}.
\end{eqnarray}
Knowing the covariance matrix of the state, we can use it to construct the characteristic function, and express
\begin{eqnarray}
\nonumber \rho_{ABCD} &=& \frac{1}{\pi^{4}}\int \diff^{2}\alpha_{1} \int \diff^{2}\alpha_{2} \int \diff^{2}\beta_{1} \int \diff^{2}\beta_{2} \chi_{ABCD}(\alpha_{1},\alpha_{2},\beta_{1},\beta_{2}) \times \\
&& \hat{D}_{A}(-\alpha_{1})\hat{D}_{B}(-\alpha_{2})\hat{D}_{C}(-\beta_{1})\hat{D}_{D}(-\beta_{2}).
\end{eqnarray}
Now, in order to perform photon subtraction, we need to measure a given number of photons in the reflected arm of each beam splitter. We are interested in two-photon subtraction, in the particular case in which a single photon is reflected on each beam splitter. Then, we need to project the state into the subspace that describes this outcome, 
\begin{eqnarray}
\nonumber && \langle 1,1|\rho_{ABCD}|1,1\rangle_{CD} = \frac{1}{\pi^{4}}\int \diff^{2}\alpha_{1}\int\diff^{2}\alpha_{2}\int\diff^{2}\beta_{1}\int\diff^{2}\beta_{2} \chi_{ABCD}(\alpha_{1},\alpha_{2},\beta_{1},\beta_{2}) \\
&& \hat{D}_{A}(-\alpha_{1})\hat{D}_{B}(-\alpha_{2})\langle 1|\hat{D}_{C}(-\beta_{1})|1\rangle_{C}\langle 1|\hat{D}_{D}(-\beta_{2})|1\rangle_{D}.
\end{eqnarray}
In order to compute this, we need to know that
\begin{equation}
\langle m|\hat{D}(\alpha)|n\rangle = e^{-|\alpha|^{2}/2}\sqrt{\frac{m!}{n!}}\sum_{k=0}^{n}\begin{pmatrix}n\\k\end{pmatrix}\theta(m-k)\frac{\alpha^{m-k}(-\bar{\alpha})^{n-k}}{(m-k)!},
\end{equation}
which for the special case $n=m=1$ leaves
\begin{equation}
\langle 1|\hat{D}(-\alpha)|1\rangle = e^{-|\alpha|^{2}/2}(1-|\alpha|^{2}).
\end{equation}
Remember that, since we had defined $\vec{\alpha} = \sqrt{2}\begin{pmatrix} \mathbb{Re}\alpha & \mathbb{Im}\alpha \end{pmatrix} = \begin{pmatrix} x_{\alpha} & p_{\alpha} \end{pmatrix}$, we can write
\begin{equation}
1 - |\alpha|^{2} = 1 - \left(\mathbb{Re}^{2}\alpha + \mathbb{Im}^{2}\alpha\right) = 1 - \frac{1}{2}\begin{pmatrix} x_{\alpha} & p_{\alpha} \end{pmatrix}\begin{pmatrix} x_{\alpha} \\ p_{\alpha} \end{pmatrix} = 1 - \frac{1}{2}\vec{\alpha}^{\intercal}\mathbb{1}_{2}\vec{\alpha}
\end{equation}
We consider writing the characteristic function as
\begin{eqnarray}
&& \chi_{ABCD}(\alpha_{1},\alpha_{2},\beta_{1},\beta_{2})e^{-\frac{1}{2}|\beta_{1}|^{2}}e^{-\frac{1}{2}|\beta_{2}|^{2}} =\\
\nonumber &&  \exp\Big[ -\frac{1}{2}\vec{\beta}_{1}^{\intercal}X_{A}\vec{\beta}_{1} + \vec{\beta}_{1}^{\intercal}\left(\vec{J}_{A}+H\vec{\beta}_{2}\right) -\frac{1}{2}\vec{\beta}_{2}^{\intercal}X_{B}\vec{\beta}_{2} + \vec{\beta}_{2}^{\intercal}\vec{J}_{B} \\
\nonumber && -\frac{1}{4}\vec{\alpha}_{1}^{\intercal}\Omega^{\intercal}\left( \tau \Sigma_{A}+(1-\tau)\mathbb{1}_{2}\right)\Omega\vec{\alpha}_{1} -\frac{1}{4}\vec{\alpha}_{2}^{\intercal}\Omega^{\intercal}\left( \tau \Sigma_{B}+(1-\tau)\mathbb{1}_{2}\right)\Omega\vec{\alpha}_{2} - \frac{\tau}{2}\vec{\alpha}_{1}^{\intercal}\Omega^{\intercal} \varepsilon_{AB}\Omega\vec{\alpha}_{2} \Big],
\end{eqnarray}
where we have defined
\begin{eqnarray}
\nonumber X_{A} &=& \frac{1}{2}\Omega^{\intercal} \left[ (1-\tau)\Sigma_{A} + (1+\tau)\mathbb{1}_{2} \right]\Omega, \\
\nonumber X_{B} &=& \frac{1}{2}\Omega^{\intercal} \left[ (1-\tau)\Sigma_{B} + (1+\tau)\mathbb{1}_{2} \right]\Omega, \\
\vec{J}_{A} &=& \frac{1}{2}\sqrt{\tau(1-\tau)}\Omega^{\intercal}\left[ (\Sigma_{A}-\mathbb{1}_{2})\Omega\vec{\alpha}_{1}^{T} + \varepsilon_{AB}\Omega\vec{\alpha}_{2} \right], \\
\nonumber \vec{J}_{B} &=& \frac{1}{2}\sqrt{\tau(1-\tau)}\Omega^{\intercal}\left[ \varepsilon_{AB}\Omega\vec{\alpha}_{1} + (\Sigma_{B}-\mathbb{1}_{2})\Omega\vec{\alpha}_{2} \right], \\
\nonumber H &=& -\frac{1}{2}(1-\tau)\Omega^{\intercal} \varepsilon_{AB}\Omega.
\end{eqnarray}
Now we can integrate over the ancillary modes. We will start with mode $C$,
\begin{eqnarray}
&& \int \diff^{2}\beta_{1} \left(1-\frac{1}{2}\vec{\beta}_{1}^{\intercal}\mathbb{1}_{2}\vec{\beta}_{1}\right)e^{-\frac{1}{2}\vec{\beta}_{1}^{\intercal}X_{A}\vec{\beta}_{1} + \vec{\beta}_{1}^{\intercal}\left(\vec{J}_{A}+H\vec{\beta}_{2}\right)} =  \frac{\pi}{\sqrt{\det X_{A}}} \times \\
\nonumber && \left[ 1 - \frac{1}{2}\tr\left(X_{A}^{-1}\right) - \frac{1}{2}\left(\vec{J}_{A}^{\intercal} + \vec{\beta}_{2}^{\intercal}H\right)W_{X_{A},\mathbb{1}_{2}}\left(\vec{J}_{A} + H\vec{\beta}_{2}\right)\right]e^{\frac{1}{2}\left(\vec{J}_{A}^{\intercal} + \vec{\beta}_{2}^{\intercal}H\right)X_{A}^{-1}\left(\vec{J}_{A} + H\vec{\beta}_{2}\right)},
\end{eqnarray}
and then integrate over mode $D$. Let's do this step by step; first, we define
\begin{eqnarray}
\nonumber Y &=& X_{B} - HX_{A}^{-1}H, \\
\vec{K} &=& \vec{J}_{B} + H X_{A}^{-1}\vec{J}_{A}.
\end{eqnarray}
and integrate the free terms,
\begin{eqnarray}
\nonumber && \frac{\pi e^{\frac{1}{2}\vec{J}_{A}^{\intercal}X_{A}^{-1}\vec{J}_{A}}}{\sqrt{\det X_{A}}}\left[ 1 - \frac{1}{2}\tr\left(X_{A}^{-1}\right) - \frac{1}{2}\vec{J}_{A}^{\intercal} W_{X_{A},\mathbb{1}_{2}}\vec{J}_{A}\right]\int \diff^{2}\beta_{2} \left(1-\frac{1}{2}\vec{\beta}_{2}^{\intercal}\mathbb{1}_{2}\vec{\beta}_{2}\right) \times \\
\nonumber && e^{-\frac{1}{2}\vec{\beta}_{2}^{\intercal}Y\vec{\beta}_{2} + \vec{\beta}_{2}^{\intercal}\vec{K}} = \frac{\pi^{2}}{\sqrt{\det X_{A} \det Y}}e^{\frac{1}{2}\vec{J}_{A}^{\intercal}X_{A}^{-1}\vec{J}_{A}+\frac{1}{2}\vec{K}^{\intercal}Y^{-1}\vec{K}}  \times \\
&& \left[ 1 - \frac{1}{2}\tr\left(X_{A}^{-1}\right) - \frac{1}{2}\vec{J}_{A}^{\intercal} W_{X_{A},\mathbb{1}_{2}}\vec{J}_{A}\right] \left[ 1 - \frac{1}{2}\tr Y^{-1} - \frac{1}{2}\vec{K}^{\intercal}W_{Y,\mathbb{1}_{2}}\vec{K}\right].
\end{eqnarray}
We continue with the integral
\begin{eqnarray}
\nonumber && \frac{\pi e^{\frac{1}{2}\vec{J}_{A}^{\intercal}X_{A}^{-1}\vec{J}_{A}}}{\sqrt{\det X_{A}}}\int \diff^{2}\beta_{2} \left(1-\frac{1}{2}\vec{\beta}_{2}^{\intercal}\mathbb{1}_{2}\vec{\beta}_{2}\right)\vec{\beta}_{2}^{\intercal}HW_{X_{A},\mathbb{1}_{2}}\vec{J}_{A}e^{-\frac{1}{2}\vec{\beta}_{2}^{\intercal}Y\vec{\beta}_{2} + \vec{\beta}_{2}^{\intercal}\vec{K}} \\
\nonumber && = \frac{\pi^{2}}{\sqrt{\det X_{A} \det Y}}e^{\frac{1}{2}\vec{J}_{A}^{\intercal}X_{A}^{-1}\vec{J}_{A}+\frac{1}{2}\vec{K}^{\intercal}Y^{-1}\vec{K}} \bigg\{ - \vec{J}_{A}^{\intercal}W_{X_{A},\mathbb{1}_{2}}H W_{Y,\mathbb{1}_{2}}\vec{K} \\
&& \vec{J}_{A}^{\intercal}W_{X_{A},\mathbb{1}_{2}}H Y^{-1}\vec{K}\left[ 1 - \frac{1}{2}\tr Y^{-1} - \frac{1}{2}\vec{K}^{\intercal}W_{Y,\mathbb{1}_{2}}\vec{K}\right]\bigg\},
\end{eqnarray}
and finally compute
\begin{eqnarray}
\nonumber && \frac{\pi e^{\frac{1}{2}\vec{J}_{A}^{\intercal}X_{A}^{-1}\vec{J}_{A}}}{\sqrt{\det X_{A}}}\int \diff^{2}\beta_{2} \left(1-\frac{1}{2}\vec{\beta}_{2}^{\intercal}\mathbb{1}_{2}\vec{\beta}_{2}\right)\vec{\beta}_{2}^{\intercal}HW_{X_{A},\mathbb{1}_{2}}H\vec{\beta}_{2}e^{-\frac{1}{2}\vec{\beta}_{2}^{\intercal}Y\vec{\beta}_{2} + \vec{\beta}_{2}^{\intercal}\vec{K}} \\
\nonumber && = \frac{\pi^{2}}{\sqrt{\det X_{A} \det Y}}e^{\frac{1}{2}\vec{J}_{A}^{\intercal}X_{A}^{-1}\vec{J}_{A}+\frac{1}{2}\vec{K}^{\intercal}Y^{-1}\vec{K}} \bigg\{ -\tr\left( W_{Y,\mathbb{1}_{2}}HW_{X_{A}\mathbb{1}_{2}}H\right) \\
\nonumber && \left[\tr\left(Y^{-1}HW_{X_{A}\mathbb{1}_{2}}H\right) + \vec{K}^{\intercal}W_{Y,HW_{X_{A}\mathbb{1}_{2}}H}\vec{K}\right] \left[ 1 - \frac{1}{2}\tr Y^{-1} - \frac{1}{2}\vec{K}^{\intercal}W_{Y,\mathbb{1}_{2}}\vec{K}\right]  \\
\nonumber && - \vec{K}^{\intercal}\Big[ W_{Y,\mathbb{1}_{2}}\tr\left(Y^{-1}HW_{X_{A}\mathbb{1}_{2}}H\right) + Y^{-1}\tr\left( W_{Y,\mathbb{1}_{2}}HW_{X_{A}\mathbb{1}_{2}}H\right) \\
&& -\frac{\Omega^{\intercal} HW_{X_{A}\mathbb{1}_{2}}H\Omega}{\det Y}\tr Y^{-1} \Big]\vec{K}.
\end{eqnarray}
Now, putting everything together, we can write the final result as
\begin{eqnarray}
\nonumber && \frac{\pi^{2}}{\sqrt{\det X_{A} \det Y}}e^{\frac{1}{2}\vec{J}_{A}^{\intercal}X_{A}^{-1}\vec{J}_{A}+\frac{1}{2}\vec{K}^{\intercal}Y^{-1}\vec{K}}\bigg\{ \left( 1 - \frac{1}{2}\tr Y^{-1} - \frac{1}{2}\vec{K}^{\intercal}W_{Y,\mathbb{1}_{2}}\vec{K}\right)\times \\
\nonumber && \Big(1-\frac{1}{2}\tr X_{A}^{-1} - \frac{1}{2}\vec{J}_{A}^{\intercal}W_{X_{A},\mathbb{1}_{2}}\vec{J}_{A} - \vec{J}_{A}^{\intercal}W_{X_{A},\mathbb{1}_{2}}HY^{-1}\vec{K} -\frac{1}{2}\tr\left(Y^{-1}HW_{X_{A},\mathbb{1}_{2}}H\right) \\
\nonumber && -\frac{1}{2} \vec{K}^{\intercal}W_{Y,HW_{X_{A},\mathbb{1}_{2}}H}\vec{K} \Big) + \vec{J}_{A}^{\intercal}W_{X_{A},\mathbb{1}_{2}}HW_{Y,\mathbb{1}_{2}}\vec{K} + \frac{1}{2}\tr\left(W_{Y,\mathbb{1}_{2}}HW_{X_{A},\mathbb{1}_{2}}H\right) \\
\nonumber && + \frac{1}{2}\vec{K}^{\intercal}\Big[ W_{Y,\mathbb{1}_{2}}\tr\left(Y^{-1}HW_{X_{A},\mathbb{1}_{2}}H\right) + Y^{-1}\tr\left(W_{Y,\mathbb{1}_{2}}HW_{X_{A},\mathbb{1}_{2}}H\right) \\
&& - \frac{\Omega^{\intercal} HW_{X_{A},\mathbb{1}_{2}}H\Omega}{\det Y}\tr Y^{-1} \big]\vec{K} \bigg\}.
\end{eqnarray}
Let's expand the terms in $\alpha_{1}$ and $\alpha_{2}$, since we will need to integrate them later on. In order to keep the expressions as short as possible, let us write
\begin{eqnarray}
\nonumber \vec{K}^{\intercal} &=& \vec{\alpha}_{1}^{\intercal}\Omega K_{1} + \vec{\alpha}_{2}^{\intercal}\Omega K_{2}, \\
\vec{J}_{A}^{\intercal} &=& \vec{\alpha}_{1}^{\intercal}\Omega J_{1} + \vec{\alpha}_{2}^{\intercal}\Omega J_{2},
\end{eqnarray}
where we have defined
\begin{eqnarray}
\nonumber K_{1} &=& \frac{1}{2}\sqrt{\tau(1-\tau)}\left[ \varepsilon_{AB}\Omega^{\intercal} + (\Sigma_{A}-\mathbb{1}_{2})\Omega^{\intercal}X_{A}^{-1}H \right], \\
K_{2} &=& \frac{1}{2}\sqrt{\tau(1-\tau)}\left[ (\Sigma_{B}-\mathbb{1}_{2})\Omega^{\intercal} + \varepsilon_{AB}\Omega^{\intercal}X_{A}^{-1}H \right], \\
\nonumber J_{1} &=& \frac{1}{2}\sqrt{\tau(1-\tau)}(\Sigma_{A}-\mathbb{1}_{2})\Omega^{\intercal}, \\
\nonumber J_{2} &=& \frac{1}{2}\sqrt{\tau(1-\tau)}\varepsilon_{AB}\Omega^{\intercal}.
\end{eqnarray}
Furthermore, we also define
\begin{eqnarray}
\nonumber m_{1} &=& 1-\frac{1}{2}\tr Y^{-1}, \\
\nonumber m_{2} &=& 1-\frac{1}{2}\tr X_{A}^{-1} -\frac{1}{2}\tr\left(Y^{-1}HW_{X_{A},\mathbb{1}_{2}}H\right), \\
\nonumber m_{3} &=& \frac{1}{2}\tr\left(W_{Y,\mathbb{1}_{2}}HW_{X_{A},\mathbb{1}_{2}}H\right), \\
\nonumber P_{1} &=& -\frac{1}{2}\Omega K_{1}W_{Y,\mathbb{1}_{2}}K_{1}^{\intercal}\Omega^{\intercal}, \\
\nonumber P_{2} &=& -\frac{1}{2}\Omega K_{2}W_{Y,\mathbb{1}_{2}}K_{2}^{\intercal}\Omega^{\intercal}, \\
\nonumber P_{12} &=& -\Omega K_{1}W_{Y,\mathbb{1}_{2}}K_{2}^{\intercal}\Omega^{\intercal}, \\
\nonumber Q_{1} &=& -\frac{1}{2}\Omega\left( J_{1}W_{X_{A},\mathbb{1}_{2}}J_{1}^{\intercal} +2J_{1}W_{X_{A},\mathbb{1}_{2}}HY^{-1}K_{1}^{\intercal} + K_{1}W_{Y,HW_{X_{A},\mathbb{1}_{2}}H}K_{1}^{\intercal}\right)\Omega^{\intercal}, \\
\nonumber Q_{2} &=& -\frac{1}{2}\Omega \left(J_{2}W_{X_{A},\mathbb{1}_{2}}J_{2}^{\intercal} +2J_{2}W_{X_{A},\mathbb{1}_{2}}HY^{-1}K_{2}^{\intercal} + K_{2}W_{Y,HW_{X_{A},\mathbb{1}_{2}}H}K_{2}^{\intercal}\right)\Omega^{\intercal}, \\
 Q_{12} &=& -\Omega \Big(J_{1}W_{X_{A},\mathbb{1}_{2}}J_{2}^{\intercal} + J_{1}W_{X_{A},\mathbb{1}_{2}}HY^{-1}K_{2}^{\intercal} + K_{1}Y^{-1}HW_{X_{A},\mathbb{1}_{2}}J_{2}^{\intercal}  \\
\nonumber && + K_{1}W_{Y,HW_{X_{A},\mathbb{1}_{2}}H}K_{2}^{\intercal}\Big)\Omega^{\intercal}, \\
\nonumber R_{1} &=& \frac{1}{2}\Omega\bigg[ J_{1} W_{X_{A},\mathbb{1}_{2}}HW_{Y,\mathbb{1}_{2}}K_{1}^{\intercal} \\
\nonumber &+& K_{1}\Big(W_{Y,\mathbb{1}_{2}}\tr\left(Y^{-1}HW_{X_{A},\mathbb{1}_{2}}H\right) + Y^{-1}\tr\left(W_{Y,\mathbb{1}_{2}}HW_{X_{A},\mathbb{1}_{2}}H\right) \\
\nonumber &-& \frac{\Omega H W_{X_{A},\mathbb{1}_{2}}H\Omega^{\intercal}}{\det Y}\tr Y^{-1} \Big) K_{1}^{\intercal} \bigg]\Omega^{\intercal}, \\
\nonumber R_{2} &=& \frac{1}{2}\Omega\bigg[ J_{2} W_{X_{A},\mathbb{1}_{2}}HW_{Y,\mathbb{1}_{2}}K_{2}^{\intercal} \\
\nonumber &+& K_{2}\Big(W_{Y,\mathbb{1}_{2}}\tr\left(Y^{-1}HW_{X_{A},\mathbb{1}_{2}}H\right) + Y^{-1}\tr\left(W_{Y,\mathbb{1}_{2}}HW_{X_{A},\mathbb{1}_{2}}H\right) \\
\nonumber &-& \frac{\Omega HW_{X_{A},\mathbb{1}_{2}}H\Omega^{\intercal}}{\det Y}\tr Y^{-1} \Big) K_{2}^{\intercal} \bigg]\Omega^{\intercal},
\end{eqnarray}

\begin{eqnarray}
\nonumber R_{12} &=& \frac{1}{2}\Omega\bigg[ J_{1} W_{X_{A},\mathbb{1}_{2}}HW_{Y,\mathbb{1}_{2}}K_{2}^{\intercal} + K_{1} W_{Y,\mathbb{1}_{2}}HW_{X_{A},\mathbb{1}_{2}} J_{2}^{\intercal} \\
\nonumber &+& 2 K_{1}\Big(W_{Y,\mathbb{1}_{2}}\tr\left(Y^{-1}HW_{X_{A},\mathbb{1}_{2}}H\right) + Y^{-1}\tr\left(W_{Y,\mathbb{1}_{2}}HW_{X_{A},\mathbb{1}_{2}}H\right) \\
&-& \frac{\Omega HW_{X_{A},\mathbb{1}_{2}}H\Omega^{\intercal}}{\det Y}\tr Y^{-1} \Big) K_{2}^{\intercal} \bigg]\Omega^{\intercal}.
\end{eqnarray}
Then, we can rewrite the previous result as
\begin{eqnarray}
\nonumber && \frac{\pi^{2}}{\sqrt{\det X_{A} \det Y}} e^{\frac{1}{2}\vec{\alpha}_{1}^{\intercal}\Omega^{\intercal}\left(J_{1}X_{A}^{-1}J_{1}^{\intercal} + K_{1}Y^{-1}K_{1}^{\intercal}\right)\Omega\vec{\alpha}_{1}}\times \\
\nonumber && e^{\frac{1}{2}\vec{\alpha}_{2}^{\intercal}\Omega^{\intercal}\left(J_{2}X_{A}^{-1}J_{2}^{\intercal} + K_{2}Y^{-1}K_{2}^{\intercal}\right)\Omega\vec{\alpha}_{2} + \vec{\alpha}_{1}^{\intercal}\Omega^{\intercal}\left(J_{1}X_{A}^{-1}J_{2}^{\intercal} + K_{1}Y^{-1}K_{2}^{\intercal}\right)\Omega\vec{\alpha}_{2}}\times \\
\nonumber && \Big[ \left( m_{1} + \vec{\alpha}_{1}^{\intercal}P_{1}\vec{\alpha}_{1} + \vec{\alpha}_{2}^{\intercal}P_{2}\vec{\alpha}_{2} + \vec{\alpha}_{1}^{\intercal}P_{12}\vec{\alpha}_{2} \right) \left(m_{2} + \vec{\alpha}_{1}^{\intercal}Q_{1}\vec{\alpha}_{1} + \vec{\alpha}_{2}^{\intercal}Q_{2}\vec{\alpha}_{2} + \vec{\alpha}_{1}^{\intercal}Q_{12}\vec{\alpha}_{2} \right) \\
&& + m_{3} + \vec{\alpha}_{1}^{\intercal}R_{1}\vec{\alpha}_{1} + \vec{\alpha}_{2}^{\intercal}R_{2}\vec{\alpha}_{2} + \vec{\alpha}_{1}^{\intercal}R_{12}\vec{\alpha}_{2} \Big].
\end{eqnarray}
If we recover the exponentials remaining in the characteristic function, and introduce the normalization factor $N$, we can write the characteristic function of the remaining resource as
\begin{eqnarray}
\nonumber && \chi_{AB}^{(-1,-1)}(\alpha_{1},\alpha_{2}) = \frac{N e^{-\frac{1}{4}\left[\vec{\alpha}_{1}^{\intercal}\Omega^{\intercal} \tilde{\Sigma}_{A}\Omega\vec{\alpha}_{1} + \vec{\alpha}_{2}^{\intercal}\Omega^{\intercal} \tilde{\Sigma}_{B}\Omega\vec{\alpha}_{2} + 2\vec{\alpha}_{1}^{\intercal}\Omega^{\intercal} \tilde{\varepsilon}_{AB}\Omega\vec{\alpha}_{2} \right]}}{\sqrt{\det X_{A}\det Y}} \times \\
\nonumber && \Big[ \left( m_{1} + \vec{\alpha}_{1}^{\intercal}P_{1}\vec{\alpha}_{1} + \vec{\alpha}_{2}^{\intercal}P_{2}\vec{\alpha}_{2} + \vec{\alpha}_{1}^{\intercal}P_{12}\vec{\alpha}_{2} \right) \left(m_{2} + \vec{\alpha}_{1}^{\intercal}Q_{1}\vec{\alpha}_{1} + \vec{\alpha}_{2}^{\intercal}Q_{2}\vec{\alpha}_{2} + \vec{\alpha}_{1}^{\intercal}Q_{12}\vec{\alpha}_{2} \right) \\
&& + m_{3} + \vec{\alpha}_{1}^{\intercal}R_{1}\vec{\alpha}_{1} + \vec{\alpha}_{2}^{\intercal}R_{2}\vec{\alpha}_{2} + \vec{\alpha}_{1}^{\intercal}R_{12}\vec{\alpha}_{2} \Big].
\end{eqnarray}
where we have defined
\begin{eqnarray}
\nonumber \tilde{\Sigma}_{A} &=& \tau \Sigma_{A} + (1-\tau)\mathbb{1}_{2} - 2\left( J_{1}X_{A}^{-1}J_{1}^{\intercal} + K_{1}Y^{-1}K_{1}^{\intercal}\right), \\
\tilde{\Sigma}_{B} &=& \tau \Sigma_{B} + (1-\tau)\mathbb{1}_{2} - 2\left( J_{2}X_{A}^{-1}J_{2}^{\intercal} + K_{2}Y^{-1}K_{2}^{\intercal}\right), \\
\nonumber \tilde{\varepsilon}_{AB} &=& \tau \varepsilon_{AB} - 2\left( J_{1}X_{A}^{-1}J_{2}^{\intercal} + K_{1}Y^{-1}K_{2}^{\intercal} \right).
\end{eqnarray}
The normalization constant is given by
\begin{equation}
N = \frac{\sqrt{\det X_{A}\det Y}}{m_{1}m_{2} + m_{3}},
\end{equation}
such that the success probability of this protocol can be computed as $P = 1/N$. Remember from the previous case that after applying Homodyne detection we get $\vec{\alpha}_{1}\rightarrow \sigma_{z}\vec{\alpha}_{1}$, and after integrating over $\alpha_{1}$ and over $\xi$, the factors in front of the integral are transformed only by $\alpha_{1}\rightarrow\alpha_{2}$. This means that we eventually obtain
\begin{eqnarray}
\nonumber && \chi_{B}^{(-1,-1)}(\alpha_{2}) = \frac{N e^{-\frac{1}{4}\vec{\alpha}_{2}^{\intercal}\Omega^{\intercal}\left( \mathbb{1}_{2} + \sigma_{z}\tilde{\Sigma}_{A}\sigma_{z} + \tilde{\Sigma}_{B} - \sigma_{z}\tilde{\varepsilon}_{AB} - \tilde{\varepsilon}_{AB}^{\intercal}\sigma_{z} \right)\Omega\vec{\alpha}_{2} - i\vec{\alpha}_{2}^{\intercal}\Omega\vec{\alpha}_{0}}}{\sqrt{\det X_{A}\det Y}} \times \\
\nonumber && \Big\{ \left[ m_{1} + \vec{\alpha}_{2}^{\intercal}\left(ZP_{1}\sigma_{z} + P_{2} + ZP_{12}\right) \vec{\alpha}_{2}\right] \left[m_{2} + \vec{\alpha}_{2}^{\intercal}\left(ZQ_{1}\sigma_{z} + Q_{2} + ZQ_{12}\right)\vec{\alpha}_{2} \right] \\
&& + m_{3} + \vec{\alpha}_{2}^{\intercal}\left( ZR_{1}\sigma_{z} + R_{2} + ZR_{12} \right)\vec{\alpha}_{2} \Big\}.
\end{eqnarray}
and when the product of this with the characteristic function is integrated to obtain the average fidelity, we obtain
\begin{eqnarray}
\nonumber \bar{F} &=& \frac{1}{\sqrt{\det\tilde{\Gamma}}}\bigg\{ 1 + \frac{1}{m_{1}m_{2}+m_{3}}\bigg[ m_{1}\tr\left[\Omega\tilde{\Gamma}^{-1}\Omega^{T}\left(ZQ_{1}\sigma_{z} + Q_{2} + ZQ_{12}\right)\right] \\
\nonumber &+& m_{2}\tr\left[\Omega\tilde{\Gamma}^{-1}\Omega^{T}\left(ZP_{1}\sigma_{z} + P_{2} + ZP_{12}\right)\right] + \tr\left[\Omega\tilde{\Gamma}^{-1}\Omega^{T}\left(ZR_{1}\sigma_{z} + R_{2} + ZR_{12}\right)\right] \\
\nonumber &+& \tr\left[\Omega\tilde{\Gamma}^{-1}\Omega^{T}\left(ZP_{1}\sigma_{z} + P_{2} + ZP_{12}\right)\right]\tr\left[\Omega\tilde{\Gamma}^{-1}\Omega^{T}\left(ZQ_{1}\sigma_{z} + Q_{2} + ZQ_{12}\right)\right] \\
&+& 2\tr\left[ W_{\Omega\tilde{\Gamma}\Omega^{T},ZP_{1}\sigma_{z} + P_{2} + ZP_{12}}\left(ZQ_{1}\sigma_{z} + Q_{2} + ZQ_{12}\right)\right] \bigg] \bigg\}.
\end{eqnarray}
Here, we have defined $\tilde{\Gamma} = \mathbb{1}_{2} + \frac{1}{2}\left( \sigma_{z}\tilde{\Sigma}_{A}\sigma_{z} + \tilde{\Sigma}_{B} - \sigma_{z}\tilde{\varepsilon}_{AB} -\tilde{\varepsilon}_{AB}^{\intercal}\sigma_{z} \right)$.

%%%%%%%%%%%%%%%%%%%%%%%%%%%%%%%%%%%%%%%%%%%%%%%%%%%%%%%%%%%%%%%%%%%%%
%%%%%%%%%%%%%%%%%%%%%%%%%%%%%%%%%%%%%%%%%%%%%%%%%%%%%%%%%%%%%%%%%%%%%
%%%%%%%%%%%%%%%%%%%%%%%%%%%%%%%%%%%%%%%%%%%%%%%%%%%%%%%%%%%%%%%%%%%%%
%%%%%%%%%%%%%%%%%%%%%%%%%%%%%%%%%%%%%%%%%%%%%%%%%%%%%%%%%%%%%%%%%%%%%
%%%%%%%%%%%%%%%%%%%%%%%%%%%%%%%%%%%%%%%%%%%%%%%%%%%%%%%%%%%%%%%%%%%%%
%%%%%%%%%%%%%%%%%%%%%%%%%%%%%%%%%%%%%%%%%%%%%%%%%%%%%%%%%%%%%%%%%%%%%
\section{Step-by-step entanglement swapping}\label{app_C}
Consider the case in which we have two entangled states, shared by three parties pairwise. That is, between Alice and Charlie, and between Charlie and Bob. Consider that these states are gaussian, with covariance matrices
\begin{equation}
\Sigma_{AB} = \begin{pmatrix} \Sigma_{A} & \varepsilon_{AB} \\ \varepsilon_{AB}^{T} & \Sigma_{B} \end{pmatrix}, \quad \Sigma_{CD} = \begin{pmatrix} \Sigma_{C} & \varepsilon_{CD} \\ \varepsilon_{CD}^{T} & \Sigma_{D} \end{pmatrix},
\end{equation}
and null displacement vectors, meaning that we can write the characteristic function of, for example, the first one as
\begin{equation}
\chi_{AB}(\alpha_{1},\alpha_{2}) = \exp\left[ -\frac{1}{4}\vec{\alpha}_{1}^{\intercal}\Omega^{\intercal} \Sigma_{A}\Omega\vec{\alpha}_{1} -\frac{1}{4}\vec{\alpha}_{2}^{\intercal}\Omega^{\intercal} \Sigma_{B}\Omega\vec{\alpha}_{2} -\frac{1}{2}\vec{\alpha}_{1}^{\intercal}\Omega^{\intercal} \varepsilon_{AB}\Omega\vec{\alpha}_{2}\right].
\end{equation}
With this, we can express the density matrix of the state as
\begin{equation}
\rho_{AB} = \frac{1}{\pi^{2}}\int \diff^{2}\alpha_{1} \diff^{2}\alpha_{2} \chi_{AB}(\alpha_{1},\alpha_{2}) \hat{D}_{A}(-\alpha_{1})\hat{D}_{B}(-\alpha_{2}).
\end{equation}
Now, entanglement swapping is a technique which allows to convert two pairwise entangled states into a single one between initially unconnected parties. By making measurements in a maximally-entangled basis, Charlie is able to transform the entangled resources he shares with Alice and Bob into a single entangled state shared by Alice and Bob. In CV, these measurements are described by Homodyne detection, and its effect on the state is computed as we have done before,
\begin{eqnarray}
\nonumber && \tr_{BC}\left[ \rho_{AB}\otimes \rho_{CD}\Pi(x,p)_{BC}\right] = \frac{1}{\pi^{4}} \int \diff^{2}\alpha_{1}\int \diff^{2}\alpha_{2}\int \diff^{2}\beta_{1}\int \diff^{2}\beta_{2} \times \\
\nonumber && \chi_{AB}(\alpha_{1},\alpha_{2})\chi_{CD}(\beta_{1},\beta_{2})\hat{D}_{A}(-\alpha_{1})\hat{D}_{D}(-\beta_{2}) e^{i\sqrt{2}(p\mathbb{Re}\beta_{1}-x\mathbb{Im}\alpha_{2})} \times \\
\nonumber && e^{i(\mathbb{Im}\alpha_{2}\mathbb{Re}\alpha_{2}+\mathbb{Im}\beta_{1}\mathbb{Re}\beta_{1})} \delta(\sqrt{2}\mathbb{Re}\alpha_{2}-\sqrt{2}\mathbb{Re}\beta_{1}) \delta(\sqrt{2}\mathbb{Im}\alpha_{2}+\sqrt{2}\mathbb{Im}\beta_{1}) \\
\nonumber && = \frac{1}{2\pi^{4}} \int \diff^{2}\alpha_{1}\int \diff^{2}\beta_{2}\int \diff^{2}\alpha_{2} \chi_{ABCD}(\alpha_{1},\alpha_{2},\bar{\alpha}_{2},\beta_{2})\hat{D}_{A}(-\alpha_{1})\hat{D}_{D}(-\beta_{2}) \times \\
&& e^{i\sqrt{2}(p\mathbb{Re}\alpha_{2}-x\mathbb{Im}\alpha_{2})}.
\end{eqnarray}
We can rewrite the characteristic function as
\begin{eqnarray}
&& \chi_{AB}(\alpha_{1},\alpha_{2})\chi_{CD}(\bar{\alpha}_{2},\beta_{2}) = \exp\Big[ -\frac{1}{4}\vec{\alpha}_{1}^{\intercal}\Omega^{\intercal} \Sigma_{A}\Omega\vec{\alpha}_{1} -\frac{1}{4}\vec{\alpha}_{2}^{\intercal}\Omega^{\intercal} \Sigma_{B}\Omega\vec{\alpha}_{2}  \\
\nonumber && -\frac{1}{2}\vec{\alpha}_{1}^{\intercal}\Omega^{\intercal} \varepsilon_{AB}\Omega\vec{\alpha}_{2} - \frac{1}{4}\vec{\alpha}_{2}^{\intercal}\Omega^{\intercal} \sigma_{z}\Sigma_{C}\sigma_{z}\Omega\vec{\alpha}_{2} -\frac{1}{4}\vec{\beta}_{2}^{\intercal}\Omega^{\intercal} \Sigma_{D}\Omega\vec{\beta}_{2} + \frac{1}{2}\vec{\alpha}_{2}^{\intercal}\Omega^{\intercal} \sigma_{z}\varepsilon_{CD}\Omega\vec{\beta}_{2} \Big].
\end{eqnarray}
knowing that, since $\beta_{1} = \bar{\alpha}_{2}$, $\vec{\beta}_{1}=\sigma_{z}\vec{\alpha}_{2}$, with $\Omega \sigma_{z} = -\sigma_{z}\Omega$ and $\Omega^{\intercal}\Omega=\mathbb{1}_{2}$. Here, we have identified $\vec{\alpha}_{i}^{\intercal} = \sqrt{2}\begin{pmatrix}\mathbb{Re}\alpha_{i} & \mathbb{Im}\alpha_{i}\end{pmatrix} = \begin{pmatrix} x_{i} & p_{i} \end{pmatrix}$. Notice that this will mean that $\diff^{2}\alpha_{i} = \frac{1}{2}\diff x_{i} \diff p_{i}$ Furthermore, we write
\begin{equation}
e^{i\sqrt{2}(p\mathbb{Re}\alpha_{2}-x\mathbb{Im}\alpha_{2})} = e^{i \vec{\alpha}_{2}^{\intercal}\Omega\vec{\xi}}
\end{equation}
with $\vec{\xi}^{\intercal} = \begin{pmatrix} x & p \end{pmatrix}$. Joining everything together, we can express
\begin{eqnarray}
\nonumber && \chi_{AB}(\alpha_{1},\alpha_{2})\chi_{CD}(\bar{\alpha}_{2},\beta_{2})e^{i\sqrt{2}(p\mathbb{Re}\alpha_{2}-x\mathbb{Im}\alpha_{2})} =\\
\nonumber && \exp\bigg[ -\frac{1}{4}\vec{\alpha}_{1}^{\intercal}\Omega^{\intercal} \Sigma_{A}\Omega\vec{\alpha}_{1} -\frac{1}{4}\vec{\beta}_{2}^{\intercal}\Omega^{\intercal} \Sigma_{D}\Omega\vec{\beta}_{2} + i\vec{\alpha}_{2}^{\intercal}\Omega\vec{\xi} \\
&& - \frac{1}{4}\vec{\alpha}_{2}^{\intercal}\Omega^{\intercal} \left( \Sigma_{B}+\sigma_{z}\Sigma_{C}\sigma_{z}\right)\Omega\vec{\alpha}_{2}  + \frac{1}{2}\vec{\alpha}_{2}^{\intercal}\Omega^{\intercal} (\sigma_{z}\varepsilon_{CD}\Omega\vec{\beta}_{2}-\varepsilon_{AB}\Omega\vec{\alpha}_{1})\bigg].
\end{eqnarray}
and we can define
\begin{eqnarray}
\nonumber X &=& \frac{1}{2}\Omega^{\intercal}\left(\Sigma_{B}+\sigma_{z}\Sigma_{C}\sigma_{z}\right)\Omega, \\
\vec{J} &=& i\Omega\vec{\xi} + \frac{1}{2}\Omega^{\intercal} (\sigma_{z}\varepsilon_{CD}\Omega\vec{\beta}_{2}-\varepsilon_{AB}\Omega\vec{\alpha}_{1}).
\end{eqnarray}
We will integrate first over $\alpha_{2}$,
\begin{equation}
\int \diff^{2}\alpha_{2} e^{-\frac{1}{2}\vec{\alpha}_{2}^{\intercal}X\vec{\alpha}_{2}+\vec{\alpha}_{2}^{\intercal}\vec{J}} = \frac{\pi}{\sqrt{\det X}}e^{\frac{1}{2}\vec{J}^{\intercal}X^{-1}\vec{J}}.
\end{equation}
After Homodyne detection, need to apply a displacement to the remaining modes, proportional to the result of the measurement. This amounts to computing
\begin{equation}
\hat{D}(\xi)\hat{D}(-\alpha)\hat{D}(-\xi) = e^{\alpha\bar{\xi}-\bar{\alpha}\xi}\hat{D}(-\alpha) = e^{-i\vec{\alpha}^{\intercal}\Omega\vec{\xi}}\hat{D}(-\alpha).
\end{equation}
Now, we can identify the resulting state, conditional on the measurement results $x$ and $p$. The covariance matrix of the resulting state is characterized by
\begin{eqnarray}
\nonumber \Sigma_{A}^{\text{cond}} &=& \Sigma_{A} - \varepsilon_{AB}^{\intercal}\left(\Sigma_{B}+\sigma_{z}\Sigma_{C}\sigma_{z}\right)^{-1}\varepsilon_{AB}, \\
\Sigma_{D}^{\text{cond}} &=& \Sigma_{D} - \varepsilon_{CD}^{\intercal}\sigma_{z}\left(\Sigma_{B}+\sigma_{z}\Sigma_{C}\sigma_{z}\right)^{-1}\sigma_{z}\varepsilon_{CD}, \\
\nonumber \varepsilon_{AD}^{\text{cond}} &=& \varepsilon_{AB}^{\intercal}\left(\Sigma_{B}+\sigma_{z}\Sigma_{C}\sigma_{z}\right)^{-1}\sigma_{z}\varepsilon_{CD},
\end{eqnarray}
where the residual exponents can be grouped into $e^{f(\xi)+g(\xi^{2})}$, such that
\begin{eqnarray}
\nonumber f(\xi) &=& \frac{i}{2}\vec{\xi}^{\intercal}\Omega^{\intercal} X^{-1}\Omega^{\intercal}(\sigma_{z}\varepsilon_{CD}\Omega\vec{\beta}_{2} - \varepsilon_{AB}\Omega\vec{\alpha}_{1}), \\
g(\xi^{2}) &=& -\frac{1}{2}\vec{\xi}^{\intercal}\Omega X^{-1}\Omega^{\intercal}\vec{\xi}.
\end{eqnarray}

%%%%%%%%%%%%%%%%%%%%%%%%%%%%%%%%%%%%%%%%%%%%%%%%%%%%%%%%%%%%%%%%
Now we are in position to integrate over all possible results from the Homodyne measurement, which is equivalent to integrating over $\vec{\xi}$. By a simple change of variables, we introduce $-i\Omega\vec{\xi} = \vec{G} - \vec{J}$, and we integrate over $\vec{J}$, such that
\begin{eqnarray}
\nonumber && \frac{\pi}{\sqrt{\det X}}\int \diff J_{x} \diff J_{p} e^{\frac{1}{2}\vec{J}^{\intercal}X^{-1}\vec{J}-\left(\vec{\alpha}_{1}^{\intercal}+\vec{\beta}_{2}^{\intercal}\right)\vec{J}}e^{\left(\vec{\alpha}_{1}^{\intercal}+\vec{\beta}_{2}^{\intercal}\right)\vec{G}} =\\
&& 2\pi^{2}e^{-\frac{1}{2}\left(\vec{\alpha}_{1}^{\intercal}+\vec{\beta}_{2}^{\intercal}\right)X\left(\vec{\alpha}_{1}+\vec{\beta}_{2}\right)}e^{\left(\vec{\alpha}_{1}^{\intercal}+\vec{\beta}_{2}^{\intercal}\right)\vec{G}}.
\end{eqnarray}
We can combine all the exponents remaining, and group them to obtain
\begin{eqnarray}
\nonumber && -\frac{1}{4}\vec{\alpha}_{1}^{\intercal}\Omega^{\intercal} \Sigma_{A}\Omega\vec{\alpha}_{1} -\frac{1}{4}\vec{\beta}_{2}^{\intercal}\Omega^{\intercal} \Sigma_{D}\Omega\vec{\beta}_{2} -\frac{1}{2}\left(\vec{\alpha}_{1}^{\intercal}+\vec{\beta}_{2}^{\intercal}\right)X\left(\vec{\alpha}_{1}+\vec{\beta}_{2}\right) + \left(\vec{\alpha}_{1}^{\intercal}+\vec{\beta}_{2}^{\intercal}\right)\vec{G} \\
\nonumber && = -\frac{1}{4}\vec{\alpha}_{1}^{\intercal}\Omega^{\intercal}\left[\Sigma_{A} + 2\Omega^{\intercal}X\Omega + 2\varepsilon_{AB} \right]\Omega\vec{\alpha}_{1} -\frac{1}{4}\vec{\beta}_{2}^{\intercal}\Omega^{\intercal}\left[\Sigma_{D} + 2\Omega^{\intercal}X\Omega - 2 \sigma_{z}\varepsilon_{CD}\right]\Omega\vec{\beta}_{2} \\
&& -\frac{1}{2}\vec{\alpha}_{1}^{\intercal}\Omega^{\intercal}\left[2\Omega^{\intercal}X\Omega + \varepsilon_{AB} - \sigma_{z}\varepsilon_{CD} \right]\Omega\vec{\beta}_{2}.
\end{eqnarray}
We write the resulting state after measurement and displacements as
\begin{equation}
\rho_{AD}^{\text{ES}} = \frac{1}{\pi^{2}}\int \diff^{2}\alpha \diff^{2}\beta \chi_{AD}^{es}(\alpha,\beta) \hat{D}_{A}(-\alpha)\hat{D}_{D}(-\alpha),
\end{equation}
and identify the previous exponents as the components of the covariance matrix associated to the remaining state,
\begin{equation}
\Sigma_{AD}^{\text{ES}} = \begin{pmatrix} \tilde{\Sigma}_{A} & \tilde{\varepsilon}_{AD} \\ \tilde{\varepsilon}_{AD}^{T} & \tilde{\Sigma}_{D} \end{pmatrix},
\end{equation}
as follows
\begin{eqnarray}
\nonumber \tilde{\Sigma}_{A} &=& \Sigma_{A} + \Sigma_{B} + \sigma_{z}\Sigma_{C}\sigma_{z} + 2\varepsilon_{AB}, \\
\tilde{\Sigma}_{D} &=& \Sigma_{D} + \Sigma_{B} + \sigma_{z}\Sigma_{C}\sigma_{z} - 2 \sigma_{z} \varepsilon_{CD}, \\
\nonumber \tilde{\varepsilon}_{AD} &=& \Sigma_{B}+\sigma_{z}\Sigma_{C}\sigma_{z} + \varepsilon_{AB} - \sigma_{z}\varepsilon_{CD}.
\end{eqnarray} 
%%%%%%%%%%%%%%%%%%%%%%%%%%%%%%%%%%%%%%%%%%%%%%%%%%%%%%%%%%%%%%%%
%Now, let us consider $\Sigma_{A}=\Sigma_{C}$, $\Sigma_{B}=\Sigma_{D}$, and $\varepsilon_{AB}=\varepsilon_{CD}$, as well as $\Sigma_{A}=a\mathbb{1}_{2}$, $\Sigma_{B}=b\mathbb{1}_{2}$, and $\varepsilon_{AB}=cZ$ in order to simplify things. This leads to the form for the conditional covariance matrix
%\begin{eqnarray}
%\Sigma_{A}^{\text{cond}} &=& \left( a - \frac{c^{2}}{a+b}\right)\mathbb{1}_{2}, \\
%\nonumber \Sigma_{D}^{\text{cond}} &=& \left( b - \frac{c^{2}}{a+b}\right)\mathbb{1}_{2}, \\
%\nonumber \varepsilon_{AD}^{\text{cond}} &=& \frac{c^{2}}{a+b}Z.
%\end{eqnarray} 
%If we do the same for the state averaged over all measurement results we obtain
%\begin{eqnarray}
%\tilde{\Sigma}_{A} &=& \left(2a+b\right)\mathbb{1}_{2} + 2cZ, \\
%\nonumber \tilde{\Sigma}_{D} &=& \left(a+2b-2c\right)\mathbb{1}_{2}, \\
%\nonumber \tilde{\varepsilon}_{AD} &=& \left(a+b-c\right)\mathbb{1}_{2} + cZ.
%\end{eqnarray} 

%%%%%%%%%%%%%%%%%%%%%%%%
% BIBLIOGRAPY
%%%%%%%%%%%%%%%%%%%%%%%%

\gdef\thesubsection{}

\section*{Bibliography}
\fancyhead[RO]{BIBLIOGRAPHY}
\fancyhead[LE]{}
\phantomsection
\addcontentsline{toc}{section}{Bibliography}

% \begin{flushright}
% {\it Nothing comes from nothing.}
%
% -Parmenides
% \end{flushright}

\renewcommand{\refname}{}

\let\oldbibliography\bibliography% Store \bibliography in \oldbibliography

\renewcommand{\bibliography}[1]{{%
\let\section\subsection% Copy \subsection over \section
\oldbibliography{#1}}}% Old \bibliography

%\bibliography{bib/biblioPhDthesis}
\vspace*{-6cm}
\providecommand{\href}[2]{#2}

\end{document}